# High rate nanofluidic energy absorption in porous zeolitic frameworks

Yueting Sun[1,3†*], Sven M.J. Rogge[2†*], Aran Lamaire[2], Steven Vandenbrande[2], Jelle Wieme[2], Clive R. Siviour[1], Veronique Van Speybroeck[2*], Jin-Chong Tan[1*]

[1]Department of Engineering Science, University of Oxford, Parks Road, Oxford OX1 3PJ, United Kingdom.

[2]Center for Molecular Modeling (CMM), Ghent University, Technologiepark 46, 9052 Zwijnaarde, Belgium.

[3]School of Engineering, University of Birmingham, Edgbaston, Birmingham B15 2TT, United Kingdom.

[†]These authors contributed equally to this work

*Corresponding authors: Y.Sun.9@bham.ac.uk ; Sven.Rogge@UGent.be ;
Veronique.VanSpeybroeck@UGent.be ; jin-chong.tan@eng.ox.ac.uk

**Abstract**

Optimal mechanical impact absorbers are reusable and exhibit high specific energy absorption. The forced intrusion of liquid water in hydrophobic nanoporous materials, such as zeolitic imidazolate frameworks (ZIFs), presents an attractive pathway to engineer such systems. However, to harness their full potential, it is crucial to understand the underlying water intrusion and extrusion mechanisms under realistic, high-rate deformation conditions. Herein, we report a critical increase of the energy absorption capacity of confined water-ZIF systems at elevated strain rates. Starting from ZIF-8 as proof-of-concept, we demonstrate that this attractive rate dependence is generally applicable to cage-type ZIFs but disappears for channel-containing zeolites. Molecular simulations reveal that this phenomenon originates from the intrinsic nanosecond timescale needed for critical-sized water clusters to nucleate inside the nanocages, expediting water transport through the framework. Harnessing this fundamental understanding, design rules are formulated to construct effective, tailorable, and reusable impact energy absorbers for challenging new applications.

Energy absorption during mechanical impact plays a crucial role in modern society, from injury prevention and safety measures in industrial settings to cushioning systems that increase user comfort.[1] As current state-of-the-art energy absorption materials rely on processes such as extensive plastic deformation, cell buckling, and viscoelastic dissipation,[2,3] a major dilemma is the conflict between the required high energy density and the desire for reusability, to afford protection from multiple impacts. This challenge motivates the development of efficient energy absorbing systems that are intrinsically recoverable, which requires one to identify and leverage fundamentally new energy absorption mechanisms.

In this regard, the pressurized intrusion of liquid water and aqueous solutions in hydrophobic nanoporous materials such as zeolites and metal-organic frameworks (MOFs) has emerged as a promising mechanism to yield high-performance energy absorbing systems.[4-6] In this process, a hydrostatic pressure forces water to intrude into the hydrophobic nanopores, thereby converting mechanical work into interfacial energy. Given the exceedingly large surface area of MOFs (typically 1,000-10,000 $m^2 g^{-1}$) combined with their highly tuneable framework architecture and chemical composition,[7] MOFs are emerging as an attractive platform for nanofluidic energy absorption. Hitherto, among the huge family of MOFs,[8] a few materials have been identified for this application,[9-13] mainly hydrothermally stable zeolitic imidazolate frameworks (ZIFs) consisting of hydrophobic nanocages.[14-16] However, current research has focused only on their performance under quasi-static loading conditions, *i.e.*, through slow intrusion and extrusion processes with typical strain rates of $10^{-5}$-$10^{-3}$ $s^{-1}$.[4,17] Some studies have started to investigate the influence of loading speed,[10,18,19] but remain far from realistic strain rates, which can exceed $10^3$ $s^{-1}$ for impact-attenuating materials.

Herein, we systematically investigate the response of various promising impact-attenuating MOFs, namely ZIF-8, ZIF-7, ZIF-9, ZIF-67, and ZIF-71, under practically relevant strain rates of up to $10^3$ $s^{-1}$. Hereto, dynamic water intrusion-extrusion experiments using the dedicated high-rate experimental platform depicted in Fig. 1a-c were conducted. Most interestingly, the energy absorption densities of the investigated ZIF materials improve substantially upon increasing strain rate. Molecular dynamics (MD) simulations demonstrate that this beneficial effect originates from the intrinsic nanosecond timescale necessary for water molecules to cluster in the ZIFs' hydrophobic nanocages and to facilitate transport across nanocages. This fundamental timescale, which increases the water intrusion pressure and energy absorption density at higher strain rates, depends on the cage-type geometry of the framework materials and is absent in channel-containing zeolites (ZSM-5, zeolite-$\beta$, mordenite). Based on these findings, four rules are formulated to design efficient and reusable energy-absorbing materials for high-rate mechanical impacts *via* the pressurized liquid intrusion mechanism, identifying ZIFs as a unique class of energy-absorbing materials. These generally applicable design rules are significant to further the development of nanofluidics, which has become a flourishing field over the last decade.[20]

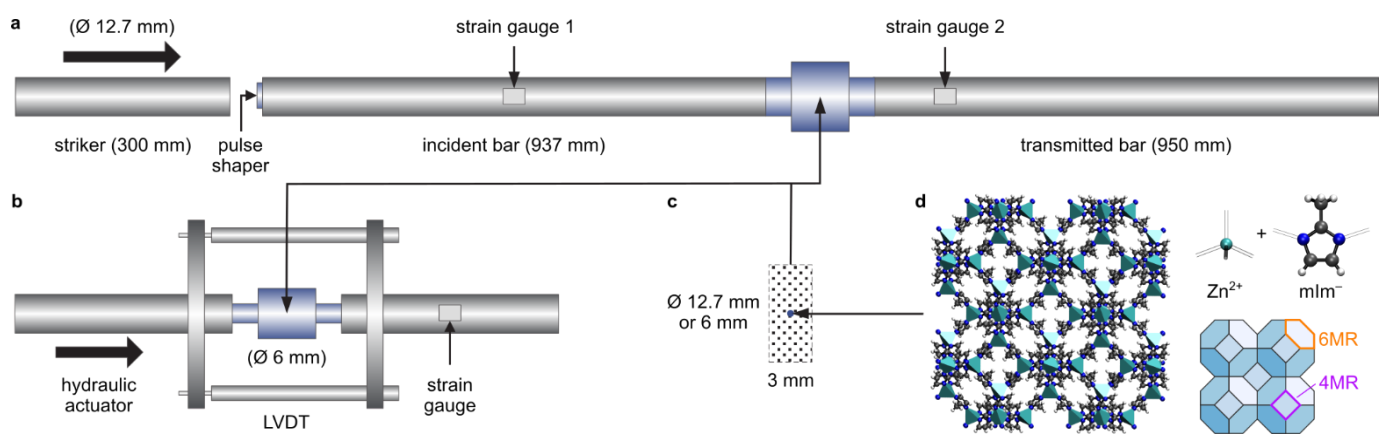

**Fig. 1 | Experimental Setup**. **a**, Split-Hopkinson-Pressure-Bar (SHPB) setup for high-rate experiments ($10^3$ s$^{-1}$). **b**, Hydraulic compression setup for medium-rate experiments (1-$10^2$ s$^{-1}$), with a pair of linear variable differential transformers (LVDTs) for displacement measurement. **c**, Water suspension of ZIF-8, which is sealed in stainless-steel chambers each with a pair of pistons (shown in **a-b**). **d**, Nanoporous framework structure, building blocks, and sodalite topology of ZIF-8 with indication of the 4-membered ring (4MR) and 6-membered ring (6MR) apertures. Low-rate experiments (up to 0.1 s$^{-1}$) are performed on a commercial screw driven testing machine (Instron 5582) using the stainless-steel chamber shown in **b**.

**ZIF-8: Proof-of-concept for the strain rate dependent water intrusion and extrusion**

We started our investigation with ZIF-8,[21] which is arguably the best-known ZIF for liquid intrusion studies.[9,17,18,22,23] ZIF-8 adopts the sodalite (**sod**) topology with relatively narrow apertures comprising 6-membered rings (6MR, aperture size ~3.40 Å) connecting larger internal cages (~11.6 Å, see Fig. 1d and Supplementary Section S1).[15] These two geometrical parameters correspond with the pore-limiting diameter (PLD) and the largest cavity diameter (LCD), respectively.

Fig. 2b shows the water intrusion and extrusion of ZIF-8 at different strain rates, $\dot{\varepsilon}$, encompassing six orders of magnitude from $10^{-3}$ s$^{-1}$ to $10^{3}$ s$^{-1}$. As schematically indicated in Fig. 2a, three stages can be identified during loading. Initially and up to a strain, $\varepsilon$, of about 0.05 (equivalent to a specific volume change $\Delta V$ of ~0.17 cm$^3$ g$^{-1}$), the pressure increases linearly with the reduction in system volume. This is attributed to the elastic compression of the {ZIF-8+water} system, without any pore intrusion owing to ZIF-8's hydrophobicity. Next, the intrusion of water in the ZIF-8 nanocages gives rise to a plateau at the intrusion pressure $P_{in}$, until water molecules occupy the entire accessible pore volume. Afterwards, a linear reduction in system volume with increasing pressure is again observed. Similarly, the unloading curve shows an extrusion plateau, albeit at a lower extrusion pressure, $P_{ex}$, than the intrusion pressure, during which water escapes from the ZIF-8 cages.

As shown in Fig. 2c, the strain rate strongly affects the intrusion pressure, which almost triples from 25 MPa during the quasi-static compression to 70 MPa during the high-rate experiment. In contrast, the extrusion pressure experiences a drop with increasing strain rate. This yet-unidentified behaviour substantially increases the hysteresis and hence absorption capacity at the high loading rate compared to the quasi-static behaviour, eventually absorbing 85% of the mechanical energy stored during the intrusion process compared to only 17% at quasi-static conditions. Consequently, the energy absorption density, $E_{ab}$, is enhanced 17-fold, from ~3 J g$^{-1}$ under quasi-static compression to ~47 J g$^{-1}$ under high-rate loading representative of impact events. In contrast to other size-dependent MOF phenomena,[24,25] this enhanced absorption density can be obtained with different crystal sizes (see Supplementary Section S2.9).

To use this promising {ZIF-8+water} system as a reusable shock absorber, the intruded water molecules should eventually extrude from the framework. Fig. 2d presents five consecutive high-rate experiments, which exhibit a consistent performance subject to multiple impact cycles. Since the applied mechanical pressure is not yet high enough to cause structural amorphization,[26] the molecular structure of ZIF-8 remains intact, as evidenced from the X-ray diffraction patterns (Supplementary Fig. 4). Fig. 2e demonstrates that even when considering 1,000 loading-unloading cycles (at a strain rate of 0.03 $s^{-1}$), the performance reveals only a slight initial drop in the intrusion and extrusion pressure upon recycling. Furthermore, Fig. 2e shows that the system can be fully recovered after a 24 h relaxation (*i.e.*, with mechanical pressure removed), implying that all water molecules extrude from ZIF-8 given sufficient relaxation time. To further confirm the material stability and reusability, Supplementary Section S2.7 demonstrates that ZIF-8 is stable after twenty high-rate intrusion/extrusion cycles or after being immersed in water for over a week.

While the water intrusion rate during these experiments as well as its extrusion rate in the low-rate and medium-rate experiments are externally controlled through the displacement rate, the high-rate SHPB setup of Fig. 1a leads to a free water extrusion process which represents the performance under realistic impact and reveals the intrinsic timescale of water mobility in ZIF-8. Fig. 2c reveals that this intrinsic water extrusion occurs at a much lower rate than its externally driven intrusion process ($10^2$ $s^{-1}$ *vs.* $10^3$ $s^{-1}$, see also Supplementary Fig. 12), corroborating the earlier observation that some water molecules remain in the structure when the system is not allowed to relax sufficiently between different cycles. High-rate experiments with different loading pulses indicate that water extrusion starts at a higher rate, which then gradually decreases, indicating a higher water mobility when more water is present inside the framework (see Supplementary Fig. 13).

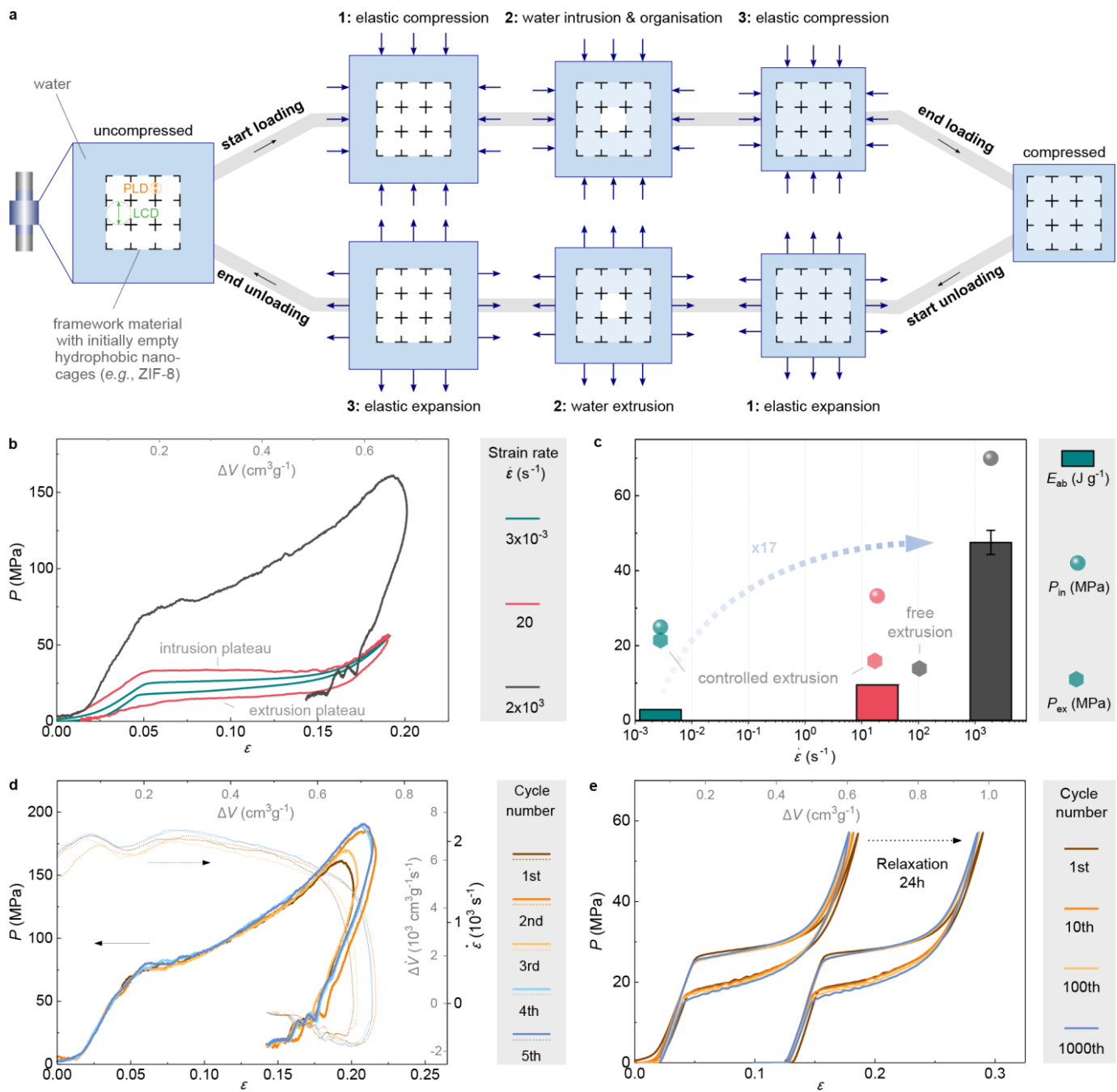

**Fig. 2 | Water intrusion and extrusion of ZIF-8 at low-rate, medium-rate, and high-rate loading conditions. a**, Schematic overview of the different processes during compression (top) and expansion (bottom), including the water intrusion/extrusion processes that give rise to distinct plateaus and indication of the PLD and LCD of the nanoporous material. **b**, Compressive stress-strain curves at three different strain rates corresponding to a specific volume change rate $\Delta\dot{V}$ of $1\times10^{-2}$, 70, and $7\times10^3$ cm$^3$ g$^{-1}$ s$^{-1}$, respectively. The unloading part of the high-rate experiment is uncontrolled (*i.e.,* without external driving force), so only part of the extrusion plateau can be recorded. **c**, Intrusion pressure, extrusion pressure, and energy absorption density under three different loading conditions, plotted as a function of the intrusion and extrusion strain rates measured by the SHPB technique (see also Supplementary Fig. 12). The error bar represents the uncertainty due to the incomplete unloading curve at high strain rate. **d**, Five consecutive

high-rate experiments (~6 min interval between each cycle), showing consistent performance against multiple impacts. The small variation in their responses can be attributed to the strain rate history shown as the dashed lines in the graph. **e**, 1,000 intrusion-extrusion cycles at a strain rate of 0.03 s$^{-1}$ (or 0.1 cm$^3$ g$^{-1}$ s$^{-1}$), confirming the durability of the system. After 24 h relaxation during which the sample was kept under no mechanical pressure, another set of 1,000 cycles was recorded. The intrusion-extrusion cycles after relaxation are horizontally offset by a value of 0.12 for clarity.

**The intrinsic water mobility timescale revealed by molecular dynamics simulations**

These experiments unveil a highly interesting mechanism, in which the ZIF-8 energy absorption capacity critically increases with increasing strain rate. While this is expected to be related to the mobility and reorganisation of water in the nanocages, it is necessary to understand the nanoscale origin of this phenomenon to fully explore the potential of the rate-dependent intrusion-extrusion performance and generalise it towards other materials. To this end, MD simulations were conducted using a fully flexible and *ab initio* derived ZIF-8 force field. This force field is validated in Supplementary Section S3 and complemented by the flexible TIP4P/2005f water model,[27] given its agreement with experimental adsorption isotherms.[16,28,29]

First, grand canonical Monte Carlo simulations were performed, revealing that water saturation in ZIF-8 is obtained at ~80 molecules per unit cell (see Supplementary Section S4.1) or, equivalently, ~40 molecules per cage, in excellent agreement with previous studies.[16,28,29] Subsequently, snapshots at different water loadings were extracted to start separate canonical Monte Carlo simulations. Fig. 3a reveals the distribution of these water molecules in ZIF-8 for the different cross-sections defined in Fig. 3b. At low water loading, distinct crystallographic adsorption sites are detected, which can be compared to the experimental argon adsorption sites by Hobday *et al.*[30] For water, the most favourable adsorption sites are located between the 4MR and 6MR apertures (site Ar-2 in Ref. 30) and near the 4MR apertures (site Ar-4 in Ref. 30). Notably, no water is adsorbed directly inside either the 4MR or 6MR apertures, although the 6MR apertures are the most favourable adsorption sites for argon.[30] At high water loadings, the water molecules agglomerate around the cage centre due to ZIF-8's hydrophobicity, partaking in a hydrogen-bonded cluster that provides more favourable interactions than the ZIF-8 framework.

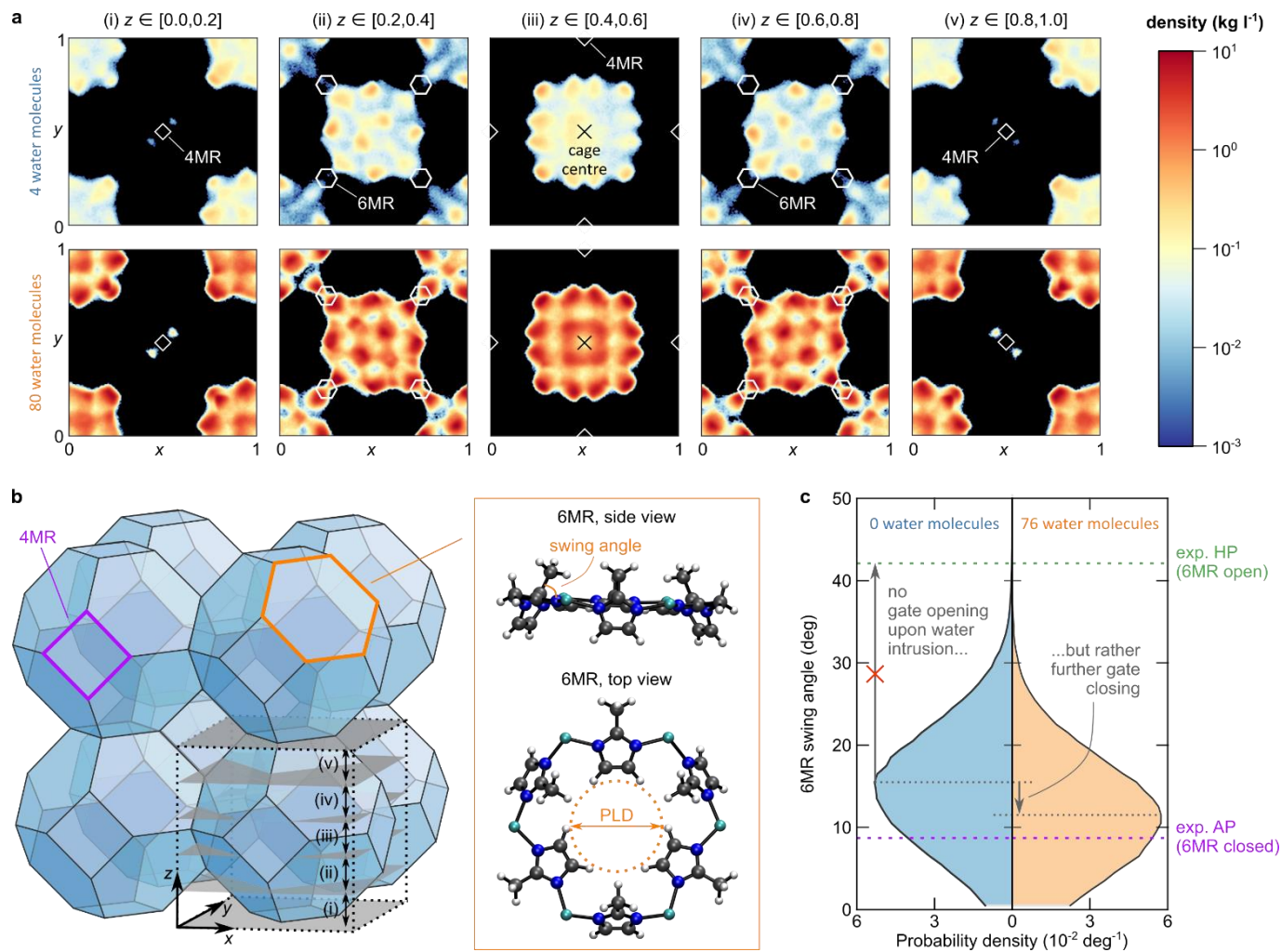

**Fig. 3 | Simulated water distribution in ZIF-8 and its effect on gate opening. a,** Symmetrised water density from a 2×2×2 ZIF-8 simulation cell as obtained from canonical Monte Carlo simulations at 298 K with either 4 or 80 water molecules per unit cell in the ambient pressure (AP) phase. The axes are defined in fractional coordinates and the unit cell is sliced in five equal slabs for clarity (see **b**). The 4MR and 6MR apertures connecting the cages are denoted by (truncated) diamonds and hexagons, respectively. Other water loadings are shown in Supplementary Section S4.2. **b,** The ZIF-8 topology with the cross-sections used for the density plots in **a** as well as the 4MR and 6MR apertures. Atomistic representation of the 6MR aperture, showing in orange the pore limiting diameter (PLD) and 6MR swing angle. **c,** Probability density of the 6MR dihedral swing angle of ZIF-8 during a 5 ns MD simulation at 300 K and 0 MPa with either 0 or 76 water molecules per ZIF-8 unit cell. Experimental AP and HP (high pressure) phases are indicated in purple and green, respectively.[31] Swing angle distributions and X-ray diffraction patterns at other water loadings and increased mechanical pressures are shown in Supplementary Section S5, while the swing angle distributions obtained through *ab initio* MD simulations at various temperatures are shown in Supplementary Section S6.

Subsequently, 300 K MD simulations of fully flexible ZIF-8 structures were performed, confirming that water avoids the 6MR apertures, irrespective of the applied pressure (see Supplementary Section S7). As a result, without a driving force, water molecules seldom hop between ZIF-8 cages. However, an increasing water loading facilitates hopping and hence enables water transport (see Supplementary Section S7.2). This agrees well with the higher water mobility at the onset of the extrusion experiments, when more water is present in the cages (see Supplementary Section S2.2). To understand whether this water mobility can be attributed to structural effects such as gate opening, the swing angle defining the 6MR aperture and the associated PLD (see Fig. 3b) are monitored.[32] As shown in Fig. 3c and Ref. 23, higher water loadings do not lead to the anticipated gate opening but rather close the 6MR aperture even further.

To mimic water mobility more closely, an inhomogeneous water distribution inside a 1×1×2 supercell of ZIF-8 was created, which only contains 42 water molecules in cage 1, as shown in Fig. 4a. During the MD simulation, this water gradient steers the molecules from cage 1 towards the neighbouring cages 2 and 3 through the 6MR apertures, despite their hydrophobicity (see Fig. 4b). After 0.45 ns, six and two water molecules have diffused to cages 2 and 3, respectively (see Fig. 4c(ii)). While the six water molecules in cage 2 form a stable hydrogen-bonded cluster, the two in cage 3 are insufficiently stabilised and diffuse back into cage 1 – against the water gradient. The water cluster inside cage 2 continues to grow as the existing cluster facilitates further hopping from cage 1 to 2. An exponential fit to the number of water molecules in cage 2 reveals that this nucleation process occurs on a nanosecond timescale (see Fig. 4b), independent of the cages in which the cluster nucleates and the ZIF-8 model size (Supplementary Section S8).

To further quantify the free energy barrier of the nucleation process, the umbrella sampling (US) free energy profiles associated with a water molecule transitioning from cage 1 to 2 are shown in Fig. 4d and Supplementary Section S9.3. Herein, besides the water molecule that undergoes the transition, cage 1 contained a critical-sized cluster of five water molecules while cage 2 contained in between zero and five water molecules. Fig. 4d and Supplementary Table 3 indicate that, while the empty cage 2 (CV > 0 Å) is substantially less favourable than cage 1 (CV < 0 Å), with free energy differences $\Delta F$ up to 15 kJ mol$^{-1}$, the transition of additional water molecules is facilitated once a critical-sized cluster of about four water molecules is present in cage 2. Fig. 4d therefore confirms that the slow nucleation of such critical-sized water clusters is crucial to facilitate water diffusion through ZIF-8's hydrophobic cages, in agreement with Fig. 4b.

This observation also explains the substantial increase in intrusion pressure and energy absorption density at higher strain rates in Fig. 2c. Since the intrinsic intrusion process is hypothesised to occur through the nucleation of critical-sized water clusters, the timescale for nucleation can be associated with an intrinsic strain rate (see Supplementary Section S8.3). If the externally applied strain rate is lower than this intrinsic rate, critical-sized water clusters will spontaneously nucleate in cages neighbouring already filled cages, facilitating the further intrusion process. However, if the strain rate exceeds this intrinsic rate, critical-sized water clusters cannot organise in time and additional work needs to be exerted – through an increased input pressure – to help overcome the free energy barrier of the 6MR aperture. This also implies that water molecules remaining in the ZIF-8 cages during the extrusion process due to insufficient relaxation will facilitate the subsequent intrusion process and lower the intrusion pressure, as confirmed by Fig. 2e.

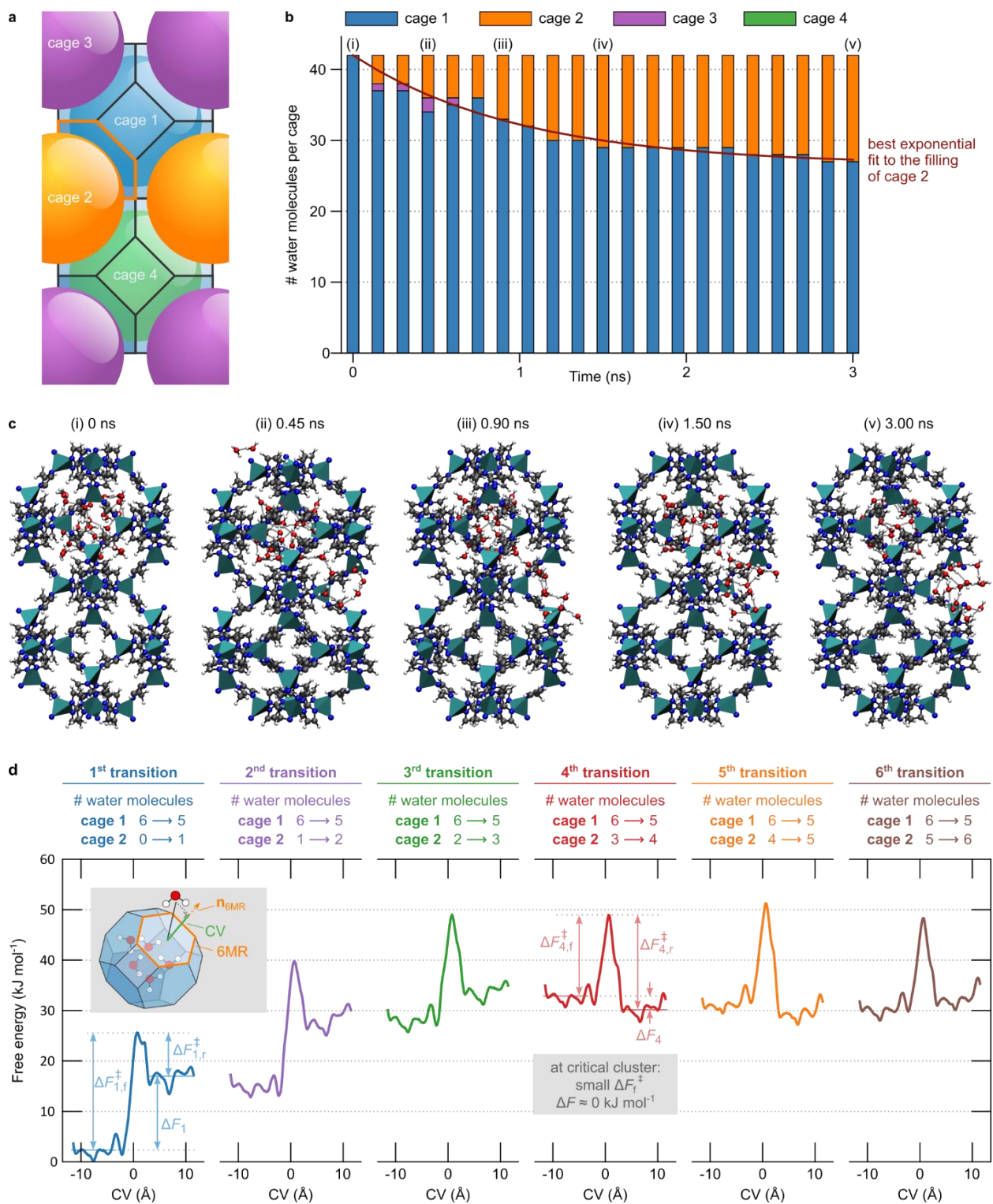

**Fig. 4 | Determining the intrinsic timescale for water mobility in the ZIF-8 nanocages by non-equilibrium MD simulations. a,** A 1×1×2 ZIF-8 supercell with four inequivalent cages connected through 6MR apertures. **b,** Evolution of the number of water molecules per cage at 300 K and 0 MPa when starting from 42 water molecules in cage 1 while all other cages are initially empty. The best exponential fit of $a(1 - e^{-t/\tau})$ to the filling of cage 2 (red line) yields a time constant $\tau \sim 1$ ns (see statistical analysis in Supplementary Section S8). **c,** Visualization of water-filled ZIF-8 structures at five representative points during the simulation; a full water cluster analysis is provided in

Supplementary Section S9.4. **d,** Free energy profiles associated with a water molecule transitioning from cage 1, containing a critical-sized water cluster of five additional molecules, to cage 2, initially containing in between zero (first transition) and five (sixth transition) water molecules. Results obtained through six independent sets of umbrella sampling (US) simulations using a similar collective variable (CV, see figure inset) as in Ref. 33. Additional US simulations are reported in Supplementary Section S9.3.

**Generalisation of the rate effect and design rules**

The here established water intrusion mechanism for ZIF-8 suggests that other materials that are constructed from nanocages connected through hydrophobic narrow apertures could also exhibit the rate-dependent water intrusion behaviour. To derive generally applicable design rules, we repeated the water intrusion experiments on a group of hydrophobic ZIFs, namely ZIF-67, ZIF-7, ZIF-9, and ZIF-71 (see Supplementary Fig. 1), leading to the energy absorption densities and intrusion pressures shown in Fig. 5a and Supplementary Fig. 14.

Fig. 5b,d and Supplementary Fig. 16 demonstrate that ZIF-67 and ZIF-71 exhibit the attractive rate dependence and reusability, confirming the ZIF-8 as proof-of-concept. For ZIF-71, a very high intrusion pressure, exceeding 150 MPa, under a loading rate of ~2,000 $s^{-1}$ is expected (see Supplementary Fig. 15). However, for ZIF-7 and ZIF-9, water is permanently trapped inside (Fig. 5c). This is probably due to their smaller PLDs, as their hydrophobic apertures are narrower than the size of water molecules.[12] Therefore, ZIF-7 and ZIF-9 can only be reused after evacuating the intruded water molecules by heat treatment to regain the original porosity,[12] in contrast to ZIF-8, ZIF-67, and ZIF-71, which can be directly reused to absorb multiple impacts. Finally, the comparison between the structural analogues, ZIF-8 *vs.* ZIF-67 (Fig. 5d), and ZIF-7 *vs.* ZIF-9 (Fig. 5c), suggests that the influence of the chemical moieties is very limited.

Based on these experimental observations, four general rules emerge that can be used to design mechanical impact absorbers leveraging the high-rate water intrusion mechanism:

1. The material should be hydrophobic;
2. The material should consist of nanocages, *i.e.*, LCD > PLD;
3. The apertures connecting the nanocages should be sufficiently large to ensure reusability. Based on our experimental observations on ZIF-11 and ZIF-12 (see Supplementary Fig. 17), the PLD threshold value is ~3 Å for water intrusion systems;
4. Larger nanocages can accommodate larger water clusters and hence increase the energy absorption density at high strain rates.

In Fig. 5e and Supplementary Section S10, the 105 ZIF-like materials tabulated in Ref. 34 are tested against these design rules. Fig. 5e demonstrates that, besides the here validated ZIF-8, ZIF-67, and ZIF-71, our design rules identify an additional 17 materials as potential high-performance impact-attenuating materials *via* the high-rate water intrusion mechanism depicted in Fig. 2a. These design rules can furthermore be generalised to other porous zeolite frameworks, as validated for chabazite in Supplementary Section S2.8.

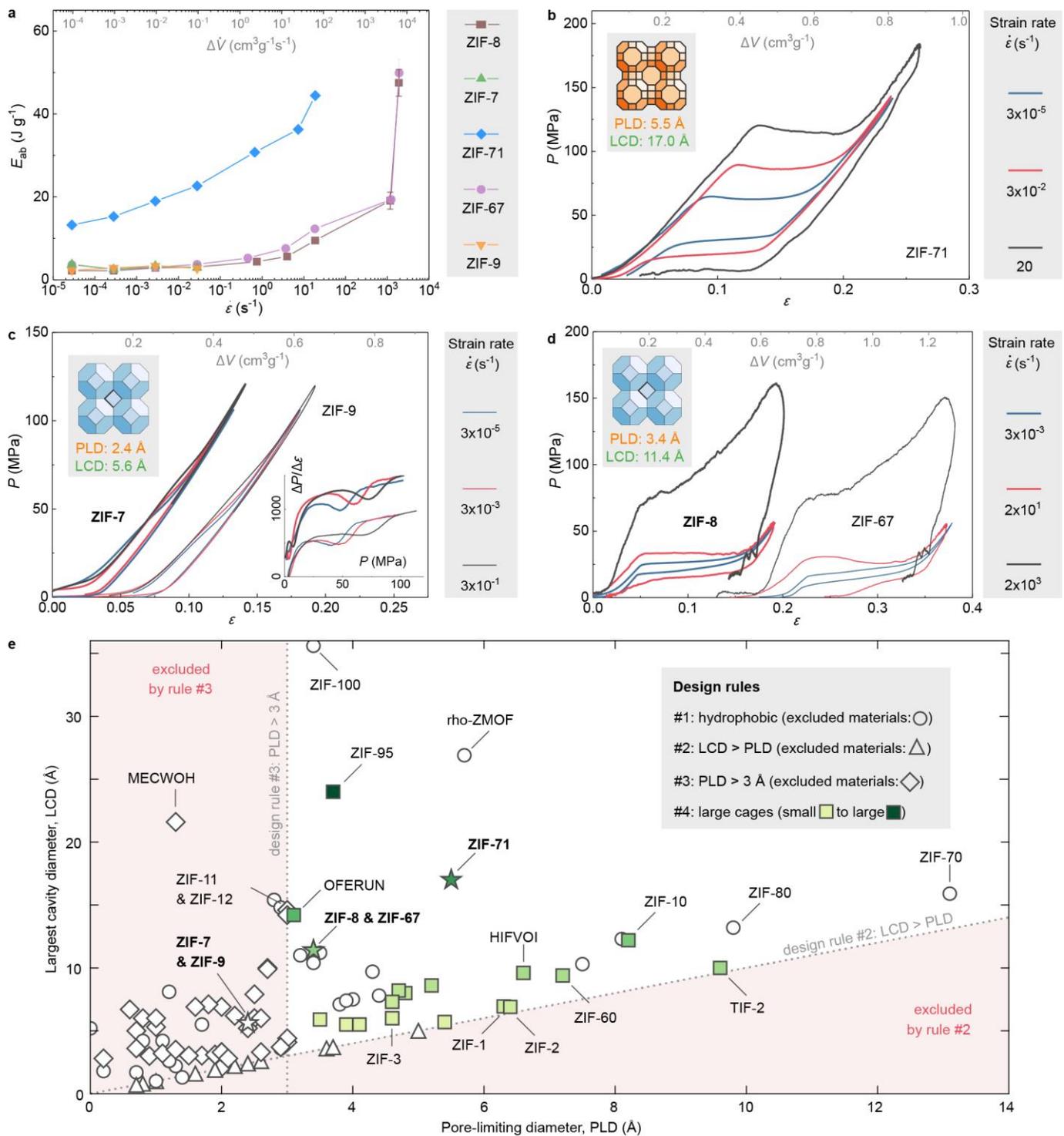

**Fig. 5 | Generalisation of the design rules. a**, Energy absorption densities of various hydrophobic cage-type ZIFs as a function of the strain rate. The error bars represent the uncertainty due to the incomplete unloading curves. **b**, Water intrusion and extrusion of ZIF-71 at three different strain rates, which correspond to a specific volume change rate $\Delta\dot{V}$ of $1\times10^{-4}$, $1\times10^{-1}$, and 80 cm$^3$ g$^{-1}$ s$^{-1}$, respectively. **c**, Water intrusion of ZIF-7 and ZIF-9 at three different strain rates, which correspond to $\Delta\dot{V}$ of $1\times10^{-4}$, $1\times10^{-2}$, and 1 cm$^3$ g$^{-1}$ s$^{-1}$, respectively. Their intrusion pressures, which can be identified from the $\Delta P/\Delta\varepsilon$ during loading process (inset), increase with the strain rate as design rule #2 is satisfied, although design rule #3 is violated. **d**, Water intrusion and extrusion of ZIF-8 and ZIF-67 at three different

strain rates, which correspond to $\Delta\dot{V}$ of $1\times10^{-2}$, 70, and $7\times10^3$ cm$^3$ g$^{-1}$ s$^{-1}$, respectively. Plots in **c** and **d** are offset horizontally for clarity. **e**, Materials selection map of the 105 ZIFs tabulated in Ref. 34 and Supplementary Section S10 according to their PLD and LCD, showing 20 promising materials fulfilling our design rules (☐ to ■); others are either not hydrophobic (as determined by their linkers, ○), not cage-type (△), or have too small a PLD (◇). Experimentally validated materials are bold-faced (☆).

**Contrasting against channel-containing frameworks**

To further test our hypothesis that the cage-type structure is essential for the rate-dependent intrusion-extrusion phenomenon of ZIFs (design rule #2), we performed water intrusion experiments for several typical channel-containing zeolites, namely ZSM-5, zeolite-$\beta$, and mordenite (see Supplementary Fig. 2). Supporting our hypothesis, Fig. 6a reveals that the rate dependence of their intrusion pressure is considerably weaker than for ZIFs. The same is true for their extrusion pressure, which is always located near the magnitude of the intrusion pressure, leading to a hysteresis area and energy absorption density that are not substantially enhanced by high loading rates (see Fig. 6d and Supplementary Section S11).

Explicitly contrasting ZSM-5 and ZIF-8 in Fig. 6c-d reveals a much higher spontaneous extrusion rate in the former. Furthermore, ZIF-8 exhibits a slight increase of the gradient with the intrusion rate (Fig. 2b), which is absent for ZSM-5 (Fig. 6b). Both observations suggest that the water flow inside channel-containing structures experiences a much lower transport resistance, in agreement with previous reports observing enhanced flow in small channels.[35-40] This suggests the advantage under high-rate mechanical impact of cage-type structures, which can be made into efficient and reusable energy absorbers thanks to the intrinsic nanosecond time scale for water organisation, over more frequently used channel-containing structures, which perform as non-linear springs with an insignificant rate dependence.

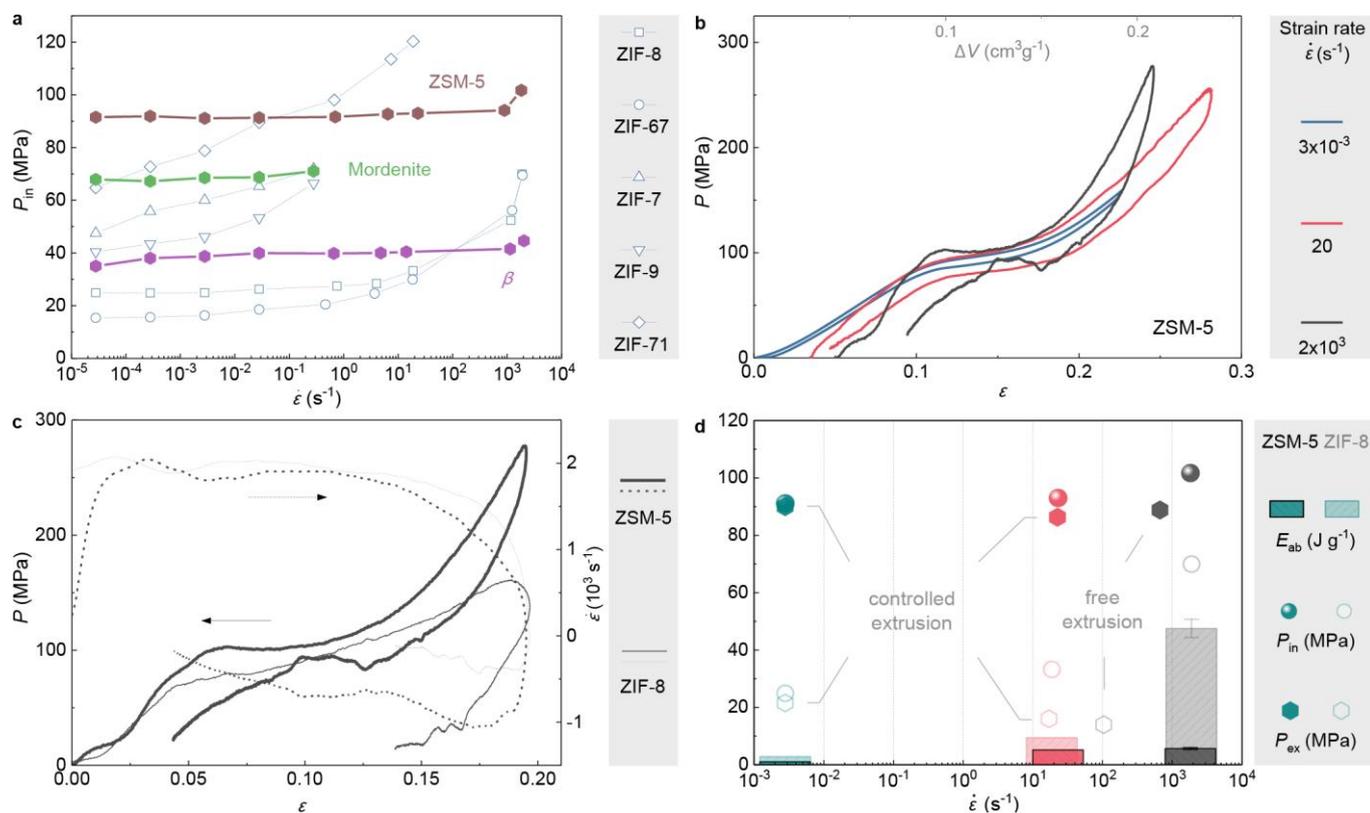

**Fig. 6 | Water intrusion and extrusion of channel-containing zeolites at different conditions. a**, Water intrusion pressure as a function of strain rate, comparing the behaviour of channel-containing zeolites against cage-type ZIFs. Due to the limited pore volume of mordenite, its intrusion pressure at higher loading rate is not available. **b**, Water intrusion and extrusion of ZSM-5 at three different strain rates, which correspond to a specific volume change rate $\Delta \dot{V}$ of $2\times10^{-3}$, 20, and $2\times10^3$ cm$^3$ g$^{-1}$ s$^{-1}$, respectively. **c**, Comparison between ZIF-8 and ZSM-5 under high-rate deformation. **d,** Intrusion pressure, extrusion pressure, and energy absorption density of ZSM-5 (dark colour and filled symbols) at the three different conditions, compared with ZIF-8 (light colour and open symbols). The error bars represent the uncertainty due to the incomplete unloading curves at high strain rates.

**Discussion**

Developing efficient impact-attenuating materials is an important societal challenge for a wide variety of applications, with current state-of-the-art energy absorption materials often showing a poor efficiency or recoverability in cyclic loading. Herein, we discovered a promising approach to mitigate mechanical impact under industrially relevant impact conditions and associated high strain rates. We revealed that the energy absorption density associated with the forced intrusion of liquid water in framework materials containing nanocages separated by hydrophobic apertures critically depends on the strain rate. Based on this concept, {ZIF+water} systems are remarkably effective as high-rate nanofluidic energy absorbers for mitigating mechanical impacts at realistic strain rates of $10^3$ s$^{-1}$.

The ZIF-8 material serves as a proof-of-concept, for which we observed a sharp rise in energy absorption density with elevating strain rates, stemming from the major enhancement of the hysteresis bound by the water intrusion-extrusion curves. Surprisingly, the fundamental process underpinning this rate-dependent water intrusion did not depend on the gate-opening mechanism that dominates its gas adsorption behaviour. Instead, we established that the unique rate-dependence is controlled by the intrinsic timescale to form critically-sized and stable hydrogen-bonded clusters in the hydrophobic cages, as they facilitate the inefficient water transport through the hydrophobic apertures separating adjacent cages. Because externally controlling the water intrusion beyond this intrinsic timescale requires substantially higher intrusion pressures, we observed a significant rise in energy absorption density at higher strain rates. We showed that this ZIF-8 proof-of-concept can be generalized to several other cage-type ZIFs, whereas it is absent for channel-containing zeolites. Based on these theoretical and experimental observations, four design rules were formulated to construct high-rate energy absorption materials, thereby identifying 17 more potential candidate materials. These generally applicable design rules shed light on the synergistic role played by the framework architecture and hydrophobicity during forced water intrusion, which are key to the design and tuning of nanoporous materials such as ZIFs towards practical and reusable impact-attenuating materials.

**References**

1. Lu, G. & Yu, T. *Energy Absorption of Structures and Materials*. (Woodhead Publishing Limited, Cambridge, 2003).

2. Gibson, L. J. & Ashby, M. F. *Cellular Solids: Structure and Properties*. 2nd edn, (Cambridge University Press, Cambridge, 1997).

3. Clough, E. C. *et al.* Elastomeric microlattice impact attenuators. *Matter* **1**, 1519-1531 (2019).

4. Fraux, G., Coudert, F. X., Boutin, A. & Fuchs, A. H. Forced intrusion of water and aqueous solutions in microporous materials: from fundamental thermodynamics to energy storage devices. *Chem. Soc. Rev.* **46**, 7421-7437 (2017).

5. Eroshenko, V., Regis, R.-C., Soulard, M. & Patarin, J. Energetics: A new field of applications for hydrophobic zeolites. *J. Am. Chem. Soc.* **123**, 8129-8130 (2001).

6. Tinti, A., Giacomello, A., Grosu, Y. & Casciola, C. M. Intrusion and extrusion of water in hydrophobic nanopores. *Proc. Natl. Acad. Sci. USA* **114**, E10266-E10273 (2017).

7. Furukawa, H., Cordova, K. E., O'Keeffe, M. & Yaghi, O. M. The chemistry and applications of metal-organic frameworks. *Science* **341**, 1230444 (2013).

8. Moghadam, P. Z. *et al.* Development of a Cambridge Structural Database subset: a collection of metal–organic frameworks for past, present, and future. *Chem. Mater.* **29**, 2618-2625 (2017).

9. Ortiz, G., Nouali, H., Marichal, C., Chaplais, G. & Patarin, J. Energetic performances of the metal–organic framework ZIF-8 obtained using high pressure water intrusion–extrusion experiments. *Phys. Chem. Chem. Phys.* **15**, 4888-4891 (2013).

10. Grosu, Y. *et al.* Stability of zeolitic imidazolate frameworks: effect of forced water intrusion and framework flexibility dynamics. *RSC Adv.* **5**, 89498-89502 (2015).

11. Ortiz, G., Nouali, H., Marichal, C., Chaplais, G. & Patarin, J. l. Energetic performances of "ZIF-71-aqueous solution" systems: A perfect shock-absorber with water. *J. Phys. Chem. C* **118**, 21316-21322 (2014).

12. Sun, Y., Li, Y. & Tan, J. C. Liquid intrusion into zeolitic imidazolate framework-7 nanocrystals: Exposing the roles of phase transition and gate opening to enable energy absorption applications. *ACS Appl. Mater. Interfaces* **10**, 41831-41838 (2018).

13. Grosu, Y. *et al.* A highly stable nonhysteretic {Cu2 (tebpz) MOF+water} molecular spring. *ChemPhysChem* **17**, 3359-3364 (2016).

14. Park, K. S. *et al.* Exceptional chemical and thermal stability of zeolitic imidazolate frameworks. *Proc. Natl. Acad. Sci. USA* **103**, 10186 (2006).

15. Banerjee, R. *et al.* High-throughput synthesis of zeolitic imidazolate frameworks and application to $CO_2$ capture. *Science* **319**, 939 (2008).

16. Ortiz, A. U., Freitas, A. P., Boutin, A., Fuchs, A. H. & Coudert, F.-X. What makes zeolitic imidazolate frameworks hydrophobic or hydrophilic? The impact of geometry and functionalization on water adsorption. *Phys. Chem. Chem. Phys.* **16**, 9940-9949 (2014).

17. Khay, I. *et al.* Assessment of the energetic performances of various ZIFs with SOD or RHO topology using high pressure water intrusion-extrusion experiments. *Dalton Trans.* **45**, 4392-4400 (2016).

18. Sun, Y., Li, Y. & Tan, J. C. Framework flexibility of ZIF-8 under liquid intrusion: discovering time-dependent mechanical response and structural relaxation. *Phys. Chem. Chem. Phys.* **20**, 10108-10113 (2018).

19. Lowe, A. *et al.* Effect of flexibility and nanotriboelectrification on the dynamic reversibility of water intrusion into nanopores: Pressure-transmitting fluid with frequency-dependent dissipation capability. *ACS Appl. Mater. Interfaces* **11**, 40842-40849 (2019).

20. Bocquet, L. Nanofluidics coming of age. *Nat. Mater.* **19**, 254-256 (2020).

21. Huang, X. C., Lin, Y. Y., Zhang, J. P. & Chen, X. M. Ligand-directed strategy for zeolite-type metal-organic frameworks: zinc(II) imidazolates with unusual zeolitic topologies. *Angew. Chem. Int. Ed.* **45**, 1557-1559 (2006).

22. Khay, I., Chaplais, G., Nouali, H., Marichal, C. & Patarin, J. Water intrusion–extrusion experiments in ZIF-8: impacts of the shape and particle size on the energetic performances. *RSC Adv.* **5**, 31514-31518 (2015).

23. Fraux, G., Boutin, A., Fuchs, A. H. & Coudert, F.-X. Structure, dynamics, and thermodynamics of intruded electrolytes in ZIF-8. *J. Phys. Chem. C* **123**, 15589-15598 (2019).

24. Sakata, Y. *et al.* Shape-memory nanopores induced in coordination frameworks by crystal downsizing. *Science* **339**, 193-196 (2013).

25. Krause, S. *et al.* The effect of crystallite size on pressure amplification in switchable porous solids. *Nat. Commun.* **9**, 1-8 (2018).

26. Bennett, T. D. & Cheetham, A. K. Amorphous metal-organic frameworks. *Acc. Chem. Res.* **47**, 1555-1562 (2014).

27. Gonzalez, M. A. & Abascal, J. L. A flexible model for water based on TIP4P/2005. *J. Chem. Phys.* **135**, 224516 (2011).

28. Ghosh, P., Kim, K. C. & Snurr, R. Q. Modeling water and ammonia adsorption in hydrophobic metal-organic frameworks: Single components and mixtures. *J. Phys. Chem. C* **118**, 1102-1110 (2014).

29. Zhang, H. & Snurr, R. Q. Computational study of water adsorption in the hydrophobic metal-organic framework ZIF-8: Adsorption mechanism and acceleration of the simulations. *J. Phys. Chem. C* **121**, 24000-24010 (2017).

30. Hobday, C. L. *et al.* Understanding the adsorption process in ZIF-8 using high pressure crystallography and computational modelling. *Nat. Commun.* **9**, 1429 (2018).

31. Moggach, S. A., Bennett, T. D. & Cheetham, A. K. The effect of pressure on ZIF-8: increasing pore size with pressure and the formation of a high-pressure phase at 1.47 GPa. *Angew. Chem. Int. Ed.* **48**, 7087-7089 (2009).

32. Durholt, J. P., Fraux, G., Coudert, F. X. & Schmid, R. Ab initio derived force fields for zeolitic imidazolate frameworks: MOF-FF for ZIFs. *J. Chem. Theory Comput.* **15**, 2420-2432 (2019).

33. Cnudde, P. *et al.* Light Olefin Diffusion during the MTO Process on H-SAPO-34: A Complex Interplay of Molecular Factors. *J. Am. Chem. Soc.* **142**, 6007-6017 (2020).

34. Phan, A. *et al.* Synthesis, structure, and carbon dioxide capture properties of zeolitic imidazolate frameworks. *Acc. Chem. Res.* **43**, 58-67 (2010).

35. Majumder, M., Chopra, N., Andrews, R. & Hinds, B. J. Enhanced flow in carbon nanotubes. *Nature* **438**, 44-44 (2005).

36. Holt, J. K. *et al.* Fast mass transport through sub-2-nanometer carbon nanotubes. *Science* **312**, 1034-1037 (2006).

37. Secchi, E. *et al.* Massive radius-dependent flow slippage in carbon nanotubes. *Nature* **537**, 210-213 (2016).

38. Radha, B. *et al.* Molecular transport through capillaries made with atomic-scale precision. *Nature* **538**, 222-225 (2016).

39. Tunuguntla, R. H. *et al.* Enhanced water permeability and tunable ion selectivity in subnanometer carbon nanotube porins. *Science* **357**, 792-796 (2017).

40. Keerthi, A. *et al.* Ballistic molecular transport through two-dimensional channels. *Nature* **558**, 420-424 (2018).

**Correspondence and requests for materials** should be addressed to Y.S., S.M.J.R., V.V.S, or J.-C.T.

**Acknowledgments**

Y.S. and J.-C.T. wish to thank the K.C. Wong Fellowship (Y.S.) and the ERC Consolidator Grant (under the grant agreement 771575 PROMOFS (J.-C.T.)) for funding the research. Y.S. also wishes to thank the University of Birmingham for startup funds. S.M.J.R., A.L, S.V.D.B., and J.W wish to thank the Fund for Scientific Research Flanders (FWO, grant nos. 12T3519N (S.M.J.R.), 11D2220N (A.L.), 11U1914N (S.V.D.B), and 1103618N (J.W.)) and the Research Board of Ghent University (BOF). Funding was also received from the European Union's Horizon 2020 Research and Innovation Programme [ERC Consolidator Grant Agreement 647755 – DYNPOR (2015-2020) (V.V.S.)]. We thank the Research Complex at Harwell (RCaH) for access to the materials characterization facilities, and Dr. Timothy Johnson at Johnson Matthey Technology Centre for providing the chabazite material. The computational resources (Stevin Supercomputer Infrastructure) and services used in this work were provided by the VSC (Flemish Supercomputer Centre), funded by Ghent University, the Research Foundation – Flanders (FWO), and the Flemish Government – department EWI.

**Author contributions**

Y.S. conceived and performed all experiments, with guidance from C.R.S. and J.-C.T. S.M.J.R. performed the force-field based MD simulations, A.L. performed the *ab initio* and umbrella sampling MD simulations, S.V. performed the GCMC and canonical MC simulations, and J.W. derived the ZIF-8 covalent force field, all under guidance of V.V.S. Y.S., S.M.J.R., J.-C.T., and V.V.S wrote the manuscript with contributions of all authors.

**Competing interests**

The authors declare no competing interests.

# Figure legends (for main text figures)

**Fig. 1 | Experimental Setup. a**, Split-Hopkinson-Pressure-Bar (SHPB) setup for high-rate experiments ($10^3$ s$^{-1}$). **b**, Hydraulic compression setup for medium-rate experiments (1-$10^2$ s$^{-1}$), with a pair of linear variable differential transformers (LVDTs) for displacement measurement. **c**, Water suspension of ZIF-8, which is sealed in stainless-steel chambers each with a pair of pistons (shown in **a-b**). **d**, Nanoporous framework structure, building blocks, and sodalite topology of ZIF-8 with indication of the 4-membered ring (4MR) and 6-membered ring (6MR) apertures. Low-rate experiments (up to 0.1 s$^{-1}$) are performed on a commercial screw driven testing machine (Instron 5582) using the stainless-steel chamber shown in **b**.

**Fig. 2 | Water intrusion and extrusion of ZIF-8 at low-rate, medium-rate, and high-rate loading conditions. a**, Schematic overview of the different processes during compression (top) and expansion (bottom), including the water intrusion/extrusion processes that give rise to distinct plateaus and indication of the PLD and LCD of the nanoporous material. **b**, Compressive stress-strain curves at three different strain rates corresponding to a specific volume change rate $\Delta \dot{V}$ of $1\times10^{-2}$, 70, and $7\times10^3$ cm$^3$ g$^{-1}$ s$^{-1}$, respectively. The unloading part of the high-rate experiment is uncontrolled (*i.e.,* without external driving force), so only part of the extrusion plateau can be recorded. **c**, Intrusion pressure, extrusion pressure, and energy absorption density under three different loading conditions, plotted as a function of the intrusion and extrusion strain rates measured by the SHPB technique (see also Supplementary Fig. 12). The error bar represents the uncertainty due to the incomplete unloading curve at high strain rate. **d**, Five consecutive high-rate experiments (~6 min interval between each cycle), showing consistent performance against multiple impacts. The small variation in their responses can be attributed to the strain rate history shown as the dashed lines in the graph. **e**, 1,000 intrusion-extrusion cycles at a strain rate of 0.03 s$^{-1}$ (or 0.1 cm$^3$ g$^{-1}$ s$^{-1}$), confirming the durability of the system. After 24 h relaxation during which the sample was kept under no mechanical pressure, another set of 1,000 cycles was recorded. The intrusion-extrusion cycles after relaxation are horizontally offset by a value of 0.12 for clarity.

**Fig. 3 | Simulated water distribution in ZIF-8 and its effect on gate opening. a,** Symmetrised water density from a 2×2×2 ZIF-8 simulation cell as obtained from canonical Monte Carlo simulations at 298 K with either 4 or 80 water molecules per unit cell in the ambient pressure (AP) phase. The axes are defined in fractional coordinates and the unit cell is sliced in five equal slabs for clarity (see **b**). The 4MR and 6MR apertures connecting the cages are denoted by (truncated) diamonds and hexagons, respectively. Other water loadings are shown in Supplementary Section S4.2. **b,** The ZIF-8 topology with the cross-sections used for the density plots in **a** as well as the 4MR and 6MR apertures. Atomistic representation of the 6MR aperture, showing in orange the pore limiting diameter (PLD) and 6MR swing angle. **c,** Probability density of the 6MR dihedral swing angle of ZIF-8 during a 5 ns MD simulation at 300 K and 0 MPa with either 0 or 76 water molecules per ZIF-8 unit cell. Experimental AP and HP (high pressure) phases are indicated in purple and green, respectively.[31] Swing angle distributions and X-ray diffraction patterns at other water loadings and increased mechanical pressures are shown in Supplementary Section S5, while the swing angle distributions obtained through *ab initio* MD simulations at various temperatures are shown in Supplementary Section S6.

**Fig. 7 | Determining the intrinsic timescale for water mobility in the ZIF-8 nanocages by non-equilibrium MD simulations. a,** A 1×1×2 ZIF-8 supercell with four inequivalent cages connected through 6MR apertures. **b,** Evolution of the number of water molecules per cage at 300 K and 0 MPa when starting from 42 water molecules in cage 1 while all other cages are initially empty. The best exponential fit of $a(1 - e^{-t/\tau})$ to the filling of cage 2 (red line) yields a time constant $\tau \sim 1$ ns (see statistical analysis in Supplementary Section S8). **c,** Visualization of water-filled ZIF-8 structures at five representative points during the simulation; a full water cluster analysis is provided in Supplementary Section S9.4. **d,** Free energy profiles associated with a water molecule transitioning from cage 1, containing a critical-sized water cluster of five additional molecules, to cage 2, initially containing in between zero (first transition) and five (sixth transition) water molecules. Results obtained through six independent sets of umbrella

sampling (US) simulations using a similar collective variable (CV, see figure inset) as in Ref. 33. Additional US simulations are reported in Supplementary Section S9.3.

**Fig. 5 | Generalisation of the design rules. a**, Energy absorption densities of various hydrophobic cage-type ZIFs as a function of the strain rate. The error bars represent the uncertainty due to the incomplete unloading curves. **b**, Water intrusion and extrusion of ZIF-71 at three different strain rates, which correspond to a specific volume change rate $\Delta \dot{V}$ of $1\times10^{-4}$, $1\times10^{-1}$, and 80 cm$^3$ g$^{-1}$ s$^{-1}$, respectively. **c**, Water intrusion of ZIF-7 and ZIF-9 at three different strain rates, which correspond to $\Delta \dot{V}$ of $1\times10^{-4}$, $1\times10^{-2}$, and 1 cm$^3$ g$^{-1}$ s$^{-1}$, respectively. Their intrusion pressures, which can be identified from the $\Delta P/\Delta \varepsilon$ during loading process (inset), increase with the strain rate as design rule #2 is satisfied, although design rule #3 is violated. **d**, Water intrusion and extrusion of ZIF-8 and ZIF-67 at three different strain rates, which correspond to $\Delta \dot{V}$ of $1\times10^{-2}$, 70, and $7\times10^3$ cm$^3$ g$^{-1}$ s$^{-1}$, respectively. Plots in **c** and **d** are offset horizontally for clarity. **e**, Materials selection map of the 105 ZIFs tabulated in Ref. 34 and Supplementary Section S10 according to their PLD and LCD, showing 20 promising materials fulfilling our design rules (□ to ■); others are either not hydrophobic (as determined by their linkers, ○), not cage-type (△), or have too small a PLD (◇). Experimentally validated materials are bold-faced (☆).

**Fig. 6 | Water intrusion and extrusion of channel-containing zeolites at different conditions. a**, Water intrusion pressure as a function of strain rate, comparing the behaviour of channel-containing zeolites against cage-type ZIFs. Due to the limited pore volume of mordenite, its intrusion pressure at higher loading rate is not available. **b**, Water intrusion and extrusion of ZSM-5 at three different strain rates, which correspond to a specific volume change rate $\Delta \dot{V}$ of $2\times10^{-3}$, 20, and $2\times10^3$ cm$^3$ g$^{-1}$ s$^{-1}$, respectively. **c**, Comparison between ZIF-8 and ZSM-5 under high-rate deformation. **d,** Intrusion pressure, extrusion pressure, and energy absorption density of ZSM-5 (dark colour and filled symbols) at the three different conditions, compared with ZIF-8 (light colour and open symbols). The error bars represent the uncertainty due to the incomplete unloading curves at high strain rates.

## Methods

**Material synthesis and characterisation.** ZIF-8 was purchased from Aldrich Sigma (Basolite® Z1200). Other ZIFs were synthesized using chemical compounds without further purification following the protocols outlined in Supplementary Section S1.2. Zeolites ZSM-5, zeolite-$β$, and mordenite were purchased from Alfa Aesar (45883, 45875, 45877 respectively). All of them were heated at 1,000 °C for 3 h and cooled in air, in order to obtain higher hydrophobicity before use. The chabazite was obtained from Johnson-Matthey (1318-02-1 22:1 CHA) and was heated at 950 °C for 3 h before use. This heat treatment procedure can increase the Si/Al ratio through dealumination, and therefore has been established as an efficient way to enhance the hydrophobicity of zeolites.[41,42] The effect of different heat treatment conditions on their water intrusion behaviours is shown in Supplementary Section S11.6. Microscopy imaging and X-ray diffraction of the synthesized samples were performed, and results are shown in Supplementary Section S1.3-14 and S2.6.

**Sample fabrication.** The obtained ZIFs and deionized water were combined and sealed in a stainless-steel chamber by precisely-fitting sealing rings. As shown in Fig. 1, the thickness of the sample is always 3 mm. In the low-rate Instron experiments and medium-rate hydraulic experiments, we adjusted the diameter of the sample to 6 mm, which includes 25 mg of ZIF material. In the SHPB experiments, we scaled the sample up to 12.7 mm in diameter, which includes 112 mg of ZIF material. As such, the pistons of the sealing chamber have the same diameter as the bars of SHPB, which means they can be impedance matched to avoid reflection of stress waves at the piston-bar interface.

Because of the relatively lower pore volume of zeolites compared to ZIFs, a higher amount of zeolites was used per sample, so that the lengths of water intrusion and extrusion plateaus are comparable to those of the ZIFs. This allows better identification of the corresponding pressures from the pressure-volume curves. Using a different amount of zeolites does not affect our observations, as shown in Supplementary Section S11.5. To fabricate a sample of 6 mm in diameter, we adopted 100 mg for ZSM-5, 50 mg for zeolite-$β$, and 50 mg for mordenite. For the larger sample, 12.7 mm in diameter, we used 448 mg for ZSM-5, 224 mg for zeolite-$β$, and 224 mg for mordenite.

**Liquid intrusion experiments.** Our liquid intrusion experimental platform is shown in Fig. 1. It is composed of a sealing chamber of the sample (Fig. 1c) and three different mechanical loading apparatus that provide appropriate driving force for water intrusion at different loading rates and allow the corresponding stress-strain measurement. The loading apparatus includes a commercial screw-driven load frame (Instron 5582) for low-rate experiments (up to 0.1 s$^{-1}$), an in-house hydraulic compression machine for medium-rate experiments (1-10$^2$ s$^{-1}$), and the Split-Hopkinson-Pressure-Bar (SHPB) setup for high-rate experiments (10$^3$ s$^{-1}$). By employing this experimental platform, pressure-volume change ($P$-$\Delta V$) curves or stress-strain curves along the water intrusion and extrusion process are obtained over a wide range of strain rates (10$^{-5}$-10$^3$ s$^{-1}$).

The efficacy of the low-rate experimental method has been reported in our recent work.[12,18] We applied a constant crosshead displacement rate, corresponding to a certain strain rate in the sample. Then, at a peak pressure at which water molecules have filled the framework porosity (*e.g.*, 56 MPa for ZIF-8), we reversed the crosshead direction to obtain the extrusion behaviour at the same displacement rate. The Instron records the force and displacement history during the loading and unloading cycles.

Medium-rate experiments were conducted on a hydraulic compression machine (Fig. 1b), consisting of a hydraulic actuator, a strain-gauge based force transducer, and a pair of linear variable differential transformers (LVDTs) for displacement measurement. An appropriate peak displacement is pre-set, at which water molecules have filled up the framework porosity. We used the same strain rate for the loading and unloading process, which is controlled by the hydraulic actuator and measured by the displacement signals from the LVDTs.

High-rate experiments were carried out on a SHPB setup driven by a gas gun. In the experiment, the impact of a striker onto the incident bar produces an incident stress wave which propagates through the sample, with a certain amount being reflected, into the transmitted bar. The wave profiles recorded by the strain gauges on the incident and transmitted bars are used to calculate the forces and displacements at the specimen-bar interfaces and hence produce stress-strain curves using standard calculations.[43] The strain rate was also recorded using the reflected wave throughout the loading and unloading process. Stress equilibrium inside the sample during the impact is checked for each experiment, by confirming that the forces at the two interfaces have the same magnitude. All bars, including the piston of the sealing chamber, are made of the same material and have the same diameter, so as to match the impedances. Tungsten is selected as the bar material, to provide sufficient impact energy to drive the water intrusion process and allowing us to obtain experimental data at higher strain rates. Pulse shapers are employed to achieve a constant strain rate during the loading process. The unloading process of the sample is uncontrolled, which means that the strain rate during the unloading process is not constant and can be different from that of the loading process. The unloading data were recorded and used to reveal the intrinsic extrusion behaviour of water from the nanoporous framework. This procedure is different from the low-rate and medium-rate experiments, where the unloading process is still displacement-controlled by the compression head and the unloading rate is set to be the same as the loading rate. Unfortunately, without any driving force, the SHPB technique does not allow us to capture a complete unloading curve, unless extremely long bars are used, which are not available. The uncontrolled free water extrusion process in high-rate experiments represents the performance of {ZIF+water} systems under realistic impact loading conditions, and the difference in the unloading setting for experiments at different strain rates does not affect our discussion on the rate-dependent energy absorption phenomenon (see Supplementary Section S2.1).

Based on the obtained $P$-$\Delta V$ or stress-strain curves, the intrusion pressure $P_{in}$ is determined as the onset of the intrusion plateau, while the extrusion pressure $P_{ex}$ is determined as the midpoint of the extrusion plateau. For the incomplete extrusion plateaus in high-rate experiments, $P_{ex}$ is taken from the last data point. The plateaus of ZIF-7, ZIF-9, and mordenite are short; therefore, their $P_{in}$ and $P_{ex}$ are determined based on the gradient of the plot. The intrusion and extrusion rates are taken from the value of the selected data point of $P_{in}$ and $P_{ex}$. The strain rate history during the intrusion and extrusion is perfectly constant in the low-rate experiments but has some oscillations in the medium-rate and high-rate experiments. Energy absorption is defined as the hysteresis area enclosed by the loading and unloading curve, which is a certain percentage of the mechanical energy stored during the loading process.

**Periodic *ab initio* simulations.** 0 K density functional theory (DFT) calculations were performed with the Vienna Ab initio Simulation Package (VASP)[44] using the projector-augmented wave (PAW) method.[45] The computational unit cell contained a total of 276 atoms, 12 of which are zinc atoms. The PBE exchange-correlation functional[46] was combined with the DFT-D3 dispersion scheme using Becke-Johnson damping.[47,48] The recommended GW PBE PAW potentials were employed for all elements and functionals (v5.4). For the zinc atoms, the 3s, 3p, 3d, and 4s electrons were included explicitly. For the carbon and nitrogen atoms, the 2s and 2p electrons were considered as valence electrons. For the hydrogen atoms, the 1s electron was treated as a valence electron. These DFT calculations were performed with a plane-wave kinetic-energy cut-off of 800 eV and using Gaussian smearing with a smearing width of 0.05 eV. Projection operators were evaluated in reciprocal space. A Γ-point k-grid was used for all volumes. The real-space FFT grid was used to describe wave vectors up to two times the maximum wave vector present in the basis set. An augmentation grid that is twice as large was used to avoid wrap-around errors in order to obtain accurate forces. The electronic (ionic) convergence criterion was set to $10^{-9}$ ($10^{-8}$) eV. The resulting energy equation of state, reported in Supplementary Section S3.1, was constructed by fixed volume relaxations in which the positions and cell shape were optimized.[49] Subsequently, the dynamical matrix was determined using 0.015 Å displacements for all atomic coordinates with respect to the equilibrium structure.

To probe the influence of temperature on the ZIF-8 swing angle, an additional set of $(N, P, \sigma_a = \mathbf{0}, T)$ *ab initio* MD simulations[50] was performed at temperatures of 100 K, 200 K, and 300 K and at 0 MPa using the CP2K software package.[51,52] In these calculations, the PBE-D3(BJ)[46-48] level of theory was used in combination with Gaussian TZVP-MOLOPT basis sets,[53] a plane wave basis set with a cut-off of 800 Ry and a relative cut-off of 60 Ry, and Goedecker-Teter-Hutter (GTH) pseudopotentials.[54] The temperature of the simulations was controlled with a Nosé-Hoover chain thermostat consisting of three beads and with a time constant of 0.1 ps.[55-58] The pressure was controlled with a Martyna-Tobias-Klein barostat with a time constant of 1 ps.[59,60] These parameters were validated before to correctly capture the flexibility of MOFs.[50] The MD time step was set to 0.5 fs. The total simulation time for the AIMD simulations comprised 11 ps, of which the first ps was discarded for equilibration.

**Force field derivation.** From the dynamical matrix determined above, a flexible and *ab initio*-based force field for the empty ZIF-8 structure was derived. The covalent part of the force field, which contains diagonal terms that describe bonds, bends, out-of-plane distances, and torsion angles, as well as cross terms was derived using the in-house QuickFF software package.[61,62] The Lennard-Jones parameters were obtained from the Dreiding force field,[63] whereas the electrostatics were modelled as Coulomb interactions between Gaussian charge distributions[64] with the atomic charges computed using the Minimal Basis Iterative

Stockholder (MBIS) partitioning method.[65] More details about the ZIF-8 force field derivation are provided in Supplementary Section S3.2.

To describe the interactions between the different adsorbed water molecules and between the water molecules and the framework, a Lennard-Jones potential and point-charge electrostatics were used. The water molecules were described with the TIP4P/2005f model.[27] This model was chosen given its agreement with experimental water adsorption isotherms,[16,28,29] although the rigid TIP4P/2005 model does underestimate the vapor pressure.[66]

**Force field based Monte Carlo simulations.** Canonical Monte Carlo (MC) and grand canonical Monte Carlo (GCMC) simulations were performed using the RASPA2 software package[67] to extract the water density plots in the ZIF-8 framework. To this end, we first performed a series of GCMC simulations at a temperature of 298 K and a water pressure that varied from 300 Pa to 6.6 kPa over the different simulations, using the Peng-Robinson equation of state to relate the water pressure and the chemical potential.[68] From these GCMC results, the saturation limit was determined and initial {ZIF-8+water} snapshots were extracted at water loadings of 4, 8, 20, 40, 60, and 80 water molecules. These snapshots were then used as the starting point for separate canonical MC simulations, in which the water loading was kept constant and at a temperature of 298 K. In these canonical MC simulations, the average density of the centres of mass of the water molecules was averaged over at least 2 million MC cycles. The framework structures for the open- and closed-gate configurations of ZIF-8 were obtained from Ref. 31 (identifiers TUDHUW and TUDJOS in the Cambridge Crystallographic Database for the AP and HP phases, respectively) and the simulations were performed in a 2×2×2 supercell. During these MC simulations, the framework and the internal coordinates of the water molecules were kept rigid. The Lennard-Jones interactions were truncated at 12 Å with analytical tail-corrections to correct for the finite cut-off. Electrostatic interactions were treated using the Ewald summation method.[69]

**Force field based molecular dynamics simulations.** Based on the GCMC determined saturation limit of 80 water molecules per unit cell, 21 initial ZIF-8 structures were generated with 0, 4 ,8, …, 80 water molecules per unit cell by varying the chemical potential during the GCMC simulations. To create an inhomogeneous water distribution, the initial structure with 80 water molecules was doubled along one of the crystal axes, and all water molecules except for those in one out of four cages were removed (see Fig. 4). For each water loading, 11 simulations were run at pressures spaced equally between 0 MPa and 100 MPa.

For each of these structures and pressures, an $(N, P, \sigma_a = 0, T)$ MD simulation[50] has been performed using our in-house developed software code Yaff[70] for a total simulation time of 3 ns (inhomogeneous water distribution) or 5 ns (homogeneous water distribution). Furthermore, for the longer simulations and the larger ZIF-8 models discussed in Supplementary Section S8, the Yaff software package was interfaced with LAMMPS to calculate the long-range interactions more efficiently.[71] During these $(N, P, \sigma_a = 0, T)$ MD simulations, the temperature was controlled to be on average 300 K using a Nosé-Hoover chain thermostat consisting of three beads and with a time constant of 0.1 ps.[55-58] The pressure was controlled with a Martyna-Tobias-Klein barostat with a time constant of 1 ps.[59,60] The integration time step was limited to 0.5 fs to ensure energy conservation when using the velocity Verlet scheme. The long-range van der Waals interactions were cut off at a radius of 12 Å, which was compensated by tail corrections. The electrostatic interactions were efficiently calculated using an Ewald summation with a real-space cut-off of 12 Å, a splitting parameter α of 0.213 Å$^{-1}$, and a reciprocal space cut-off of 0.32 Å$^{-1}$.[69] The snapshots in Fig. 4c were generated using VMD,[72] while the pore limiting diameter and largest cavity diameters were calculated with Zeo++.[73]

**Force field based umbrella sampling simulations.** To quantify the free energy associated with the hopping of a water molecule from one cage to an adjacent cage in ZIF-8, umbrella sampling (US) simulations were performed using our in-house developed software code Yaff.[70,74] A 2x2x2 ZIF-8 supercell was adopted to avoid potential, spurious interactions between periodic images of the water molecules. From the 16 cages in this supercell, 14 cages were kept empty, leaving only two adjacent cages that were potentially filled, named cage 1 and cage 2. The collective variable (CV) used in these US simulations is based on earlier diffusion work in related zeolite materials and is defined as follows.[33] First, the relative vector between the centroid of the selected water molecule that undergoes the hopping and the centre of the 6MR aperture separating cages 1 and 2 is determined. The CV is then defined as the oriented, perpendicular projection of this relative vector on the outwards normal of this 6MR aperture as observed from cage 1. Hence, CV = 0 Å corresponds to the water molecule being in the 6MR aperture, while CV < 0 Å and CV > 0 Å correspond to the molecule being in cage 1 and 2, respectively.

For each of the twenty different transitions (*vide infra*), the collective variable was divided into 47 equidistant windows centred with CV values between -11.5 Å and 11.5 Å. To restrict the simulation to each individual window, a harmonic bias potential at the centres of these equidistant windows was applied with a force constant of 25 kJ mol$^{-1}$ Å$^{-2}$. Each of these US simulations was run for 2.25 ns, including 10 ps equilibration time. After this, the free energy profile for each of the transitions was obtained from the sampling distribution in each window by the weighted histogram analysis method (WHAM).[75,76] To prevent the other water molecules from escaping from their respective cages, an additional harmonic bias potential with a force constant of 100 kJ mol$^{-1}$ Å$^{-2}$ was applied to each of those other water molecules whenever the centroid of the molecule was at least at a distance of 9 Å from the respective cage centre; this bias potential disappeared as long as the water molecules remained within a radius of 9 Å from their respective cage centres. This bias was also applied to the water molecule that undergoes the transition, but only when the molecule was sufficiently far from the 6MR aperture and hence sufficiently committed to one cage (|CV| ≥ 3 Å). In this case, the additional spherical bias, which adds to the harmonic bias defined by the umbrella sampling, does not influence the transition through the 6MR aperture that is used in the definition of the CV, but prevents the water molecule from leaving the cage through a different 6MR aperture, similar to the effect of the spherical bias on the spectator water molecules.

The above procedure was followed for twenty different transitions. In the first set of ten transitions, besides the biased water molecule, cage 1 was filled with a critical cluster of five water molecules, while cage 2 was filled with in between zero (first transition) and nine (tenth transition) water molecules. The results of the first six of these simulations are shown in Fig. 4d. In the second set of ten transitions, this procedure was repeated but with a supercritical cluster of thirty molecules instead of a critical cluster of five water molecules in cage 1 to quantify the effect of supersaturation. All results are discussed in Supplementary Section S9.3.

**Data availability.** The experimental dataset generated during the current study are available from the authors upon reasonable request. Relevant configurations for the optimizations and molecular dynamics simulations are available through Zenodo.[77] Additional computational data supporting the results of this work are available from the online GitHub repository at https://github.com/SvenRogge/supporting-info or upon request from the authors.

**Code availability.** The Yaff software used to perform the MD simulations in this manuscript is freely accessible via https://molmod.ugent.be/software/yaff. Representative input and processing scripts are available at https://github.com/SvenRogge/supporting-info.

**References**

41. Sun, Y. *et al.* Experimental study on energy dissipation characteristics of ZSM-5 zeolite/water system. *Adv. Eng. Mater.* **15**, 740-746 (2013).
42. Sun, Y. *et al.* A candidate of mechanical energy mitigation system: Dynamic and quasi-static behaviors and mechanisms of zeolite β/water system. *Mater. Des.* **66**, 545-551 (2015).
43. Gray, G. T., III. Classic split hopkinson pressure bar testing. *ASM Handbook* **8**, 462-476 (2000).
44. Kresse, G. & Furthmüller, J. Efficient iterative schemes for ab initio total-energy calculations using a plane-wave basis set. *Phys. Rev. B* **54**, 11169-11186 (1996).
45. Kresse, G. & Joubert, D. From ultrasoft pseudopotentials to the projector augmented-wave method. *Phys. Rev. B* **59**, 1758-1775 (1999).
46. Perdew, J. P., Burke, K. & Ernzerhof, M. Generalized gradient approximation made simple. *Phys. Rev. Lett.* **77**, 3865-3868 (1996).
47. Grimme, S., Antony, J., Ehrlich, S. & Krieg, H. A consistent and accurate ab initio parametrization of density functional dispersion correction (DFT-D) for the 94 elements H-Pu. *J. Chem. Phys.* **132**, 154104 (2010).
48. Grimme, S., Ehrlich, S. & Goerigk, L. Effect of the damping function in dispersion corrected density functional theory. *J. Comput. Chem.* **32**, 1456-1465 (2011).
49. Vanpoucke, D. E. P., Lejaeghere, K., Van Speybroeck, V., Waroquier, M. & Ghysels, A. Mechanical properties from periodic plane wave quantum mechanical codes: The challenge of the flexible nanoporous MIL-47(V) framework. *J. Phys. Chem. C* **119**, 23752-23766 (2015).
50. Rogge, S. M. J. *et al.* A comparison of barostats for the mechanical characterization of metal-organic frameworks. *J. Chem. Theory Comput.* **11**, 5583-5597 (2015).
51. VandeVondele, J. *et al.* Quickstep: Fast and accurate density functional calculations using a mixed Gaussian and plane waves approach. *Comput. Phys. Commun.* **167**, 103-128 (2005).
52. Hutter, J., Iannuzzi, M., Schiffmann, F. & VandeVondele, J. CP2K: atomistic simulations of condensed matter systems. *Wiley Interdiscip. Rev. Comput. Mol. Sci.* **4**, 15-25 (2014).
53. VandeVondele, J. & Hutter, J. Gaussian basis sets for accurate calculations on molecular systems in gas and condensed phases. *J. Chem. Phys.* **127**, 114105 (2007).
54. Goedecker, S., Teter, M. & Hutter, J. Separable dual-space Gaussian pseudopotentials. *Phys. Rev. B* **54**, 1703-1710 (1996).
55. Nosé, S. A molecular dynamics method for simulations in the canonical ensemble. *Mol. Phys.* **52**, 255-268 (1984).
56. Nosé, S. A unified formulation of the constant temperature molecular dynamics methods. *J. Chem. Phys.* **81**, 511-519 (1984).
57. Hoover, W. G. Canonical dynamics: Equilibrium phase-space distributions. *Phys. Rev. A* **31**, 1695-1697 (1985).
58. Martyna, G. J., Klein, M. L. & Tuckerman, M. Nosé–Hoover chains: The canonical ensemble via continuous dynamics. *J. Chem. Phys.* **97**, 2635-2643 (1992).
59. Martyna, G. J., Tobias, D. J. & Klein, M. L. Constant pressure molecular dynamics algorithms. *J. Chem. Phys.* **101**, 4177-4189 (1994).
60. Martyna, G. J., Tuckerman, M. E., Tobias, D. J. & Klein, M. L. Explicit reversible integrators for extended systems dynamics. *Mol. Phys.* **87**, 1117-1157 (1996).
61. Vanduyfhuys, L. *et al.* Extension of the QuickFF force field protocol for an improved accuracy of structural, vibrational, mechanical and thermal properties of metal-organic frameworks. *J. Comput. Chem.* **39**, 999-1011 (2018).
62. Vanduyfhuys, L. *et al.* QuickFF: A program for a quick and easy derivation of force fields for metal-organic frameworks from ab initio input. *J. Comput. Chem.* **36**, 1015-1027 (2015).
63. Mayo, S. L., Olafson, B. D. & Goddard, W. A. DREIDING: a generic force field for molecular simulations. *J. Phys. Chem.* **94**, 8897-8909 (1990).
64. Chen, J. & Martínez, T. J. QTPIE: Charge transfer with polarization current equalization. A fluctuating charge model with correct asymptotics. *Chem. Phys. Lett.* **438**, 315-320 (2007).
65. Verstraelen, T. *et al.* Minimal basis iterative stockholder: Atoms in molecules for force-field development. *J. Chem. Theory Comput.* **12**, 3894-3912 (2016).
66. Chen, J. L., Xue, B., Mahesh, K. & Siepmann, J. I. Molecular Simulations Probing the Thermophysical Properties of Homogeneously Stretched and Bubbly Water Systems. *J. Chem. Eng. Data* **64**, 3755-3771 (2019).
67. Dubbeldam, D., Calero, S., Ellis, D. E. & Snurr, R. Q. RASPA: molecular simulation software for adsorption and diffusion in flexible nanoporous materials. *Mol. Simul.* **42**, 81-101 (2015).

68. Peng, D.-Y. & Robinson, D. B. A new two-constant equation of state. *Ind. Eng. Chem. Fundam.* **15**, 59-64 (1976).
69. Ewald, P. P. Die Berechnung optischer und elektrostatischer Gitterpotentiale. *Ann. der Phys.* **369**, 253-287 (1921).
70. Yaff, yet another force field. Available online at http://molmod.ugent.be/software/.
71. Plimpton, S. Fast parallel algorithms for short-range molecular dynamics. *J. Comput. Phys.* **117**, 1-19 (1995).
72. Humphrey, W., Dalke, A. & Schulten, K. VMD: Visual molecular dynamics. *J. Mol. Graph.* **14**, 33-38 (1996).
73. Willems, T. F., Rycroft, C. H., Kazi, M., Meza, J. C. & Haranczyk, M. Algorithms and tools for high-throughput geometry-based analysis of crystalline porous materials. *Microporous Mesoporous Mat.* **149**, 134-141 (2012).
74. Torrie, G. M. & Valleau, J. P. Nonphysical sampling distributions in Monte Carlo free-energy estimation: Umbrella sampling. *J. Comput. Phys.* **23**, 187-199 (1977).
75. Kumar, S., Rosenberg, J. M., Bouzida, D., Swendsen, R. H. & Kollman, P. A. The weighted histogram analysis method for free-energy calculations on biomolecules. I. The method. *J. Comput. Chem.* **13**, 1011-1021 (1992).
76. Souaille, M. & Roux, B. Extension to the weighted histogram analysis method: combining umbrella sampling with free energy calculations. *Comput. Phys. Commun.* **135**, 40-57 (2001).
77. Rogge, S. M. J. Supporting molecular data for High rate nanofluidic energy absorption in porous zeolitic frameworks. (2021). doi: 10.5281/zenodo.4534252

Supplementary Information

for

High rate nanofluidic energy absorption in porous zeolitic frameworks

Yueting Sun[1,3,†], Sven M.J. Rogge[2,†], Aran Lamaire[2], Steven Vandenbrande[2], Jelle Wieme[2],

Clive R. Siviour[1], Veronique Van Speybroeck[2], Jin-Chong Tan[1]

[1]Department of Engineering Science, University of Oxford, Parks Road, Oxford OX1 3PJ, United Kingdom.

[2]Center for Molecular Modeling (CMM), Ghent University, Technologiepark 46, 9052 Zwijnaarde, Belgium.

[3]School of Engineering, University of Birmingham, Edgbaston, Birmingham B15 2TT, United Kingdom.

[†]These authors contributed equally to this work

**Table of Contents**

# S1 Material synthesis and characterisation

## S1.1 Molecular structure of ZIFs and zeolites investigated

As a subfamily of MOFs, ZIFs have zeolitic topologies constructed from divalent cations connected through imidazolate-based ligands. The hydrophobic ZIFs investigated in this work (Supplementary Figure 1) are of different chemical compositions and topologies, but all of them are composed of nanocages connected through relatively narrow window apertures. In contrast, most of the zeolites shown in Supplementary Figure 2 have channel-containing frameworks except for chabazite.

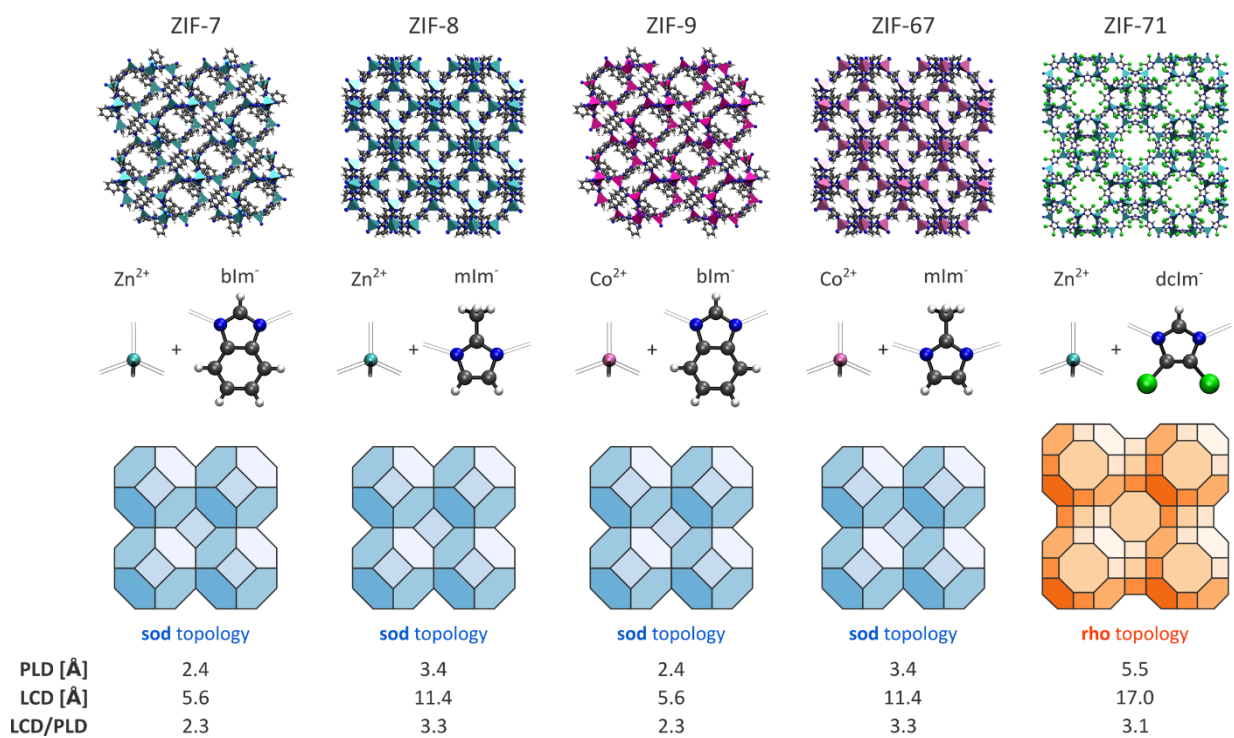

**Supplementary Figure 1 / Overview of the five ZIFs discussed in this work.** Shown are their structure, their building blocks, and their pore limiting diameter (window aperture size, PLD) and largest cavity diameter (LCD). bIm = benzimidazolate, mIm = 2-methylimidazolate, dcIm = 4,5-dichloroimidazolate.

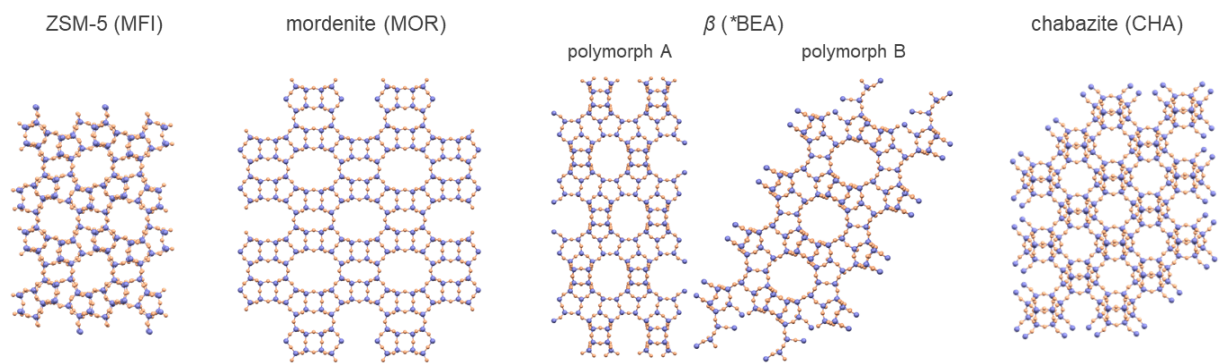

**Supplementary Figure 2 | Overview of the zeolites discussed in this work**. Shown are their framework structures (brown: oxygen; light navy: silicon). ZSM-5, mordenite, and zeolite $β$ have channel-containing frameworks while chabazite has a cage-type framework. Note that zeolite-$β$ is an intergrowth material of polymorph A and B, which are built from different stacking of the same building layer.[1]

**S1.2   Material synthesis**

ZIF-8 was purchased from Aldrich Sigma (Basolite® Z1200). The metal sources for the other ZIFs include zinc nitrate hexahydrate ($Zn(NO_3)_2 \cdot 6H_2O$, 98%, Acros Organics), cobalt nitrate hexahydrate ($Co(NO_3)_2 \cdot 6H_2O$, 98%, Acros Organics), zinc acetate dihydrate ($Zn(CH_3CO_2)_2 \cdot 2H_2O$, 98%, Acros Organics), and cobalt acetate dihydrate ($Co(CH_3CO_2)_2 \cdot 2H_2O$, 98%, Acros Organics); the organic sources include benzimidazole (HbIm, 98%, Acros Organics), 2-methylimidazole (HmIm, 99%, Acros Organics), and 4,5-dichloroimidazole (HdcIm, 98%, Alfa Aesar); the solvents include *N*,*N*-dimethylformamide (DMF, 99.5%, Fisher Chemical), methanol (MeOH, 99.8%, Fisher Chemical), toluene (analytical, Fisher Chemical), and ammonium hydroxide ($NH_3$, 28% aqueous solution, Alfa Aesar).

**ZIF-67.** In a typical synthesis of ZIF-67, 4.37 g $Co(NO_3)_2 \cdot 6H_2O$ (15 mmol) and 4.93 g HmIm (60 mmol) were dissolved in 303 mL MeOH (7.5 mol), with a molar ratio of 1 Co: 4 HmIm: 500 MeOH. The resultant solution was stirred continuously at room temperature for 1 h. The precipitates were separated from the solvent through a centrifuge (10,000 rpm for 10 min) and washed three times with MeOH. Then the product was dried at room temperature overnight.

**ZIF-71.** In a typical synthesis of ZIF-71, 0.18 g $Zn(CH_3CO_2)_2 \cdot 2H_2O$ (0.82 mmol) and 0.54 g HdcIm (3.93 mmol) were dissolved in 73 mL MeOH (1.8 mol), with a molar ratio of 1 Co: 4.8 HdcIm: 2185 MeOH. The resultant solution was kept at room temperature without stirring for 24 h. The precipitates were separated from the solvent through a centrifuge (8,000 rpm for 5 min) and then washed for three times with MeOH. Finally, the product was dried at room temperature overnight. A similar synthesis method has been reported previously.[2]

**ZIF-7.** In a typical synthesis of ZIF-7, 2.35 g $Zn(NO_3)_2 \cdot 6H_2O$ (7.9 mmol) and 5.85 g HbIm (49.5 mmol) were firstly dissolved in 750 mL DMF (9.7 mol), with a molar ratio of 1 Zn: 6.3 HbIm: 1226 DMF. The resultant solution was stirred continuously at room temperature for 2 h, followed by 22 h without stirring. The suspended precipitates were separated from the solvent through a centrifuge (12,000 rpm for 45 min) and then washed for three times with DMF. A similar synthesis method has been reported previously.[3] Finally, the product was dried at room temperature overnight, and then heated in air at 200 °C for 12 h, to obtain the guest-free Phase II of ZIF-7.[4]

**ZIF-9.** As the structural analogue of ZIF-7, ZIF-9 was synthesized in a similar way, but using $Co(NO_3)_2 \cdot 6H_2O$ instead of $Zn(NO_3)_2 \cdot 6H_2O$ as the metal source. In a typical synthesis, 0.77 g $Co(NO_3)_2 \cdot 6H_2O$ (2.64 mmol) and 7.8 g HbIm (66.02 mmol) were firstly dissolved in 125 mL DMF (1.61 mol), with a molar ratio of 1 Co: 25 HbIm: 613 DMF. The resultant solution was kept at 75 °C without stirring for 24 h. The suspended precipitates were separated from the solvent through a centrifuge (12,000 rpm for 45 min) and then washed for three times with DMF. Finally, the product was dried at room temperature overnight, and then heated in air at 200 °C for 12 h, to obtain the guest-free Phase II of ZIF-9.

**ZIF-11.** ZIF-11 was synthesized by adopting a reported method.[5] In a typical synthesis of ZIF-11, 0.45g HbIm (3.81 mmol) was dissolved in 22.73 mL MeOH (0.56 mol), followed by the addition of 19.90 mL toluene (0.19 mol) and 0.254 mL ammonium hydroxide (3.8 mmol $NH_3$) whilst stirring at room temperature. Afterwards 0.41 g $Zn(CH_3CO_2)_2 \cdot 2H_2O$ (1.88 mmol) was added. The resultant solution, which had a molar ratio of 1 Zn: 2 HbIm: 2 NH3: 300 MeOH: 100 toluene, was stirred continuously at room temperature for 3 h. The precipitates were separated from the solvent through a centrifuge (8,000 rpm for 2 min) and then washed for three times with MeOH. Finally, the product was dried at room temperature overnight and then heated in air at 200 °C for 2 h.

**ZIF-12.** As the structural analogue of ZIF-11, ZIF-12 was synthesized in the same way, but using $Co(CH_3CO_2)_2 \cdot 2H_2O$ instead of $Zn(CH_3CO_2)_2 \cdot 2H_2O$ as the metal source. In a typical synthesis, 0.45g HbIm (3.81 mmol) was dissolved in 22.73 mL MeOH (0.56 mol), followed by the addition of 19.90 mL toluene (0.19 mol) and 0.254 mL ammonium hydroxide (3.8 mmol $NH_3$) under stirring at room temperature. Afterwards 0.47 g $Co(CH_3CO_2)_2 \cdot 2H_2O$ (1.88 mmol) was added. The resultant solution, which had a molar ratio of 1 Co: 2 HbIm: 2 NH3: 300 MeOH: 100 toluene, was stirred continuously at room temperature for 3 h. The precipitates were separated from the solvent through a centrifuge (8,000 rpm for 2 min) and then washed for three times with MeOH. Finally, the product was dried at room temperature overnight and then heated in air at 200 °C for 2 h.

**ZIF-8 nanocrystals.** To investigate the effect of crystal size (Supplementary Section S2.9), nanocrystals of ZIF-8 were synthesized. In a typical synthesis, 3.0 g $Zn(NO_3)_2 \cdot 6H_2O$ (10 mmol) and 6.6 g HmIm (80 mmol) were dissolved in 200 mL MeOH (5 mol), with a molar ratio of 1 Zn: 8 HmIm: 500 MeOH. The resultant solution was stirred continuously at room temperature for 1 h. The precipitates were separated from the solvent through a centrifuge (8,000 rpm for 10 min) and washed three times with MeOH. Then the product was dried at room temperature overnight.

## S1.3 Microscopy

The morphologies of the ZIFs and zeolites were characterized by atomic force microscopy (AFM) using the neaSNOM microscope (neaspec) operating under the tapping mode. A silicon probe, Scout 350 (NuNano) was employed, which consists of a Si cantilever with a conical tip at its end (nominal radius 5 nm). The cantilever was 125 μm long, 30 μm wide, and 4.5 μm thick. Its resonance frequency was 350 kHz and its spring constant was 42 N/m. Additional scanning electron microscopy (SEM) images were collected for ZIF-8 and ZIF-71 samples; the SEM image of ZIF-8 was obtained from Zeiss Evo LS 15, while the one for ZIF-71 was obtained from Hitachi TM3030Plus.

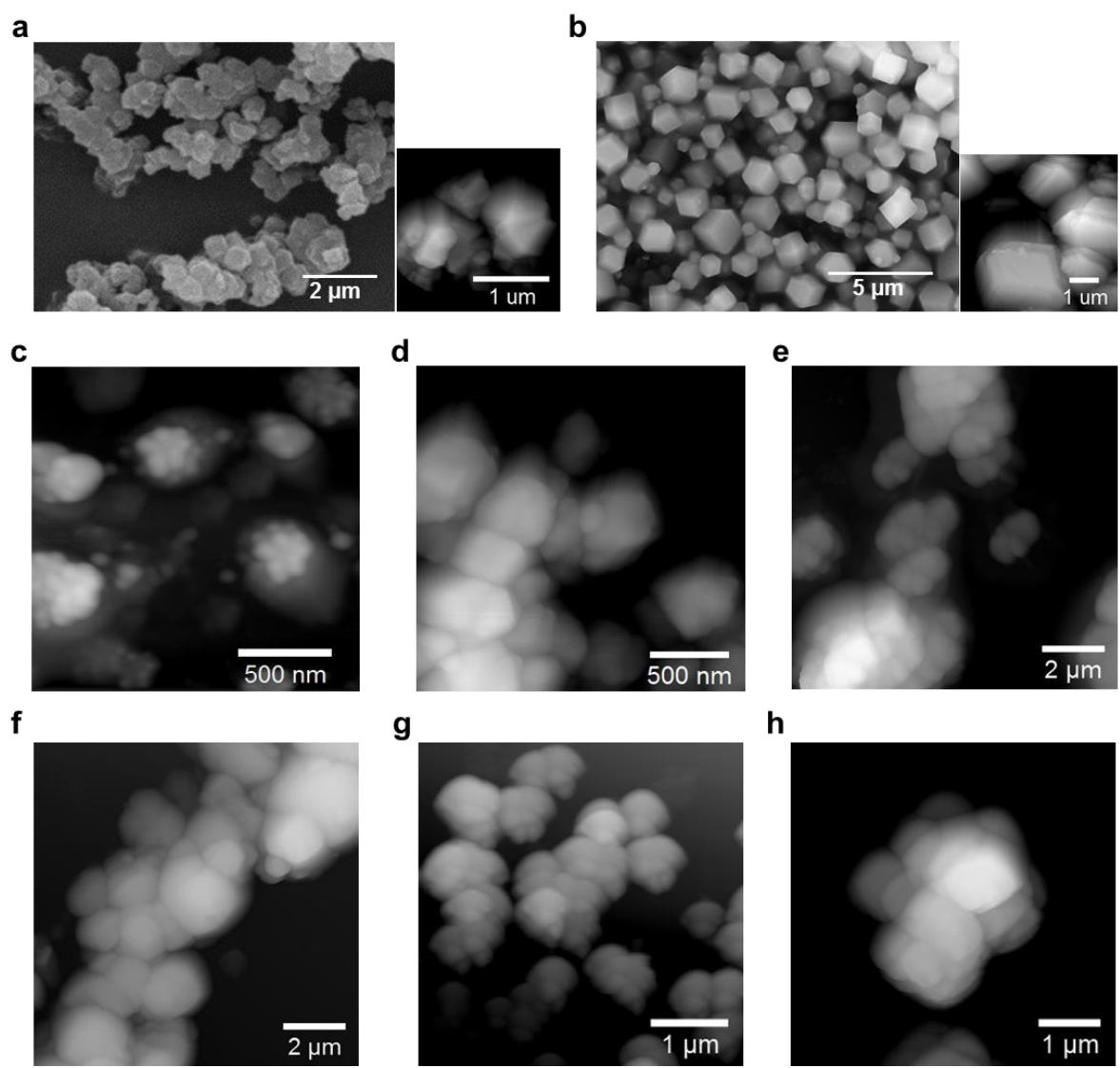

**Supplementary Figure 3 | AFM and SEM images of the ZIFs before water intrusion. a**, ZIF-8 (left: SEM, right: AFM), **b**, ZIF-71 (left: SEM, right: AFM), **c**, ZIF-7, **d**, ZIF-9, **e**, ZIF-67. **f**, ZSM-5, **g**, zeolite-*β*, **h**, mordenite. ZIF-7 and ZIF-9 are the guest-free phase II samples,[4] and the three zeolite samples are after the heating treatment at 1,000 °C.

- 8 -

## S1.4 Powder X-ray diffraction (PXRD)

Before the water intrusion experiments, the crystallinity of the powdered samples was confirmed using powder X-ray diffraction (PXRD) experiments. The PXRD patterns after water intrusion (marked with "WI" in Supplementary Figure 4 to Supplementary Figure 11) at different loading rates are included as well, showing that the molecular structures of these frameworks are intact under both quasi-static and high-rate experiments. PXRD patterns were recorded on the Rigaku Miniflex 600 using Cu Kα radiation (15 mA and 40 kV) at a scan rate of 2°/min using a $2\theta$ step-size of 0.01°.

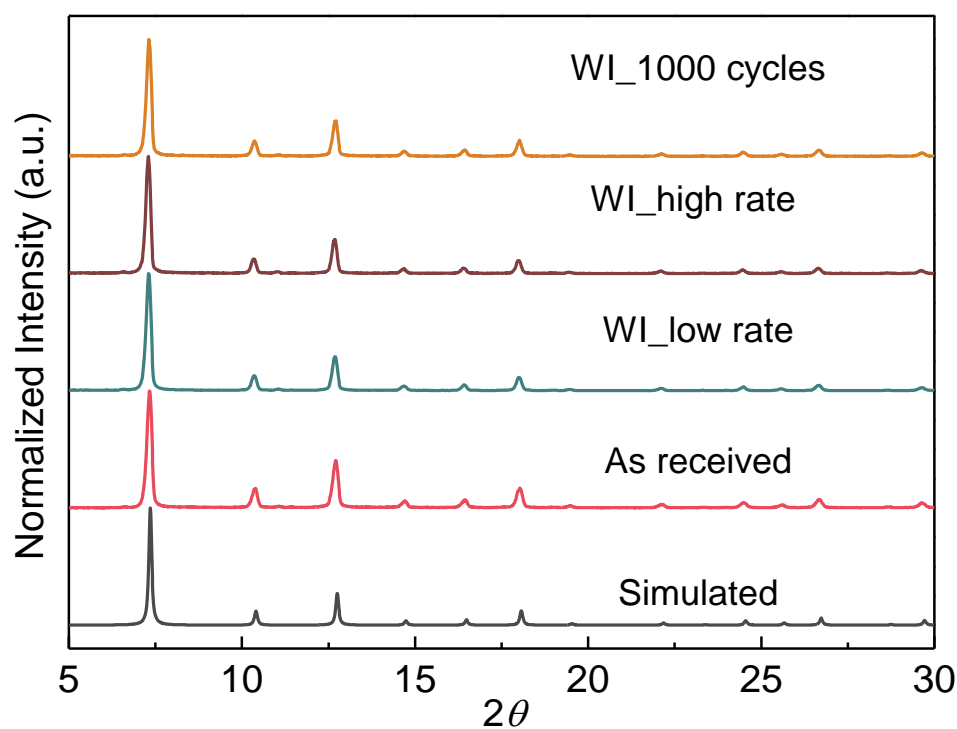

**Supplementary Figure 4 | PXRD patterns of the ZIF-8 samples.** In addition to the XRD patterns of ZIF-8 samples after low-rate and high-rate water intrusion experiments, the result after 1,000 intrusion-extrusion cycles (shown in Figure 2e of the main text) is also included. The samples maintain their intact crystalline structure after water intrusion, which is consistent with the retained level of accessible pore volume in the multicycle experiments. The XRD intensities have been normalized (a.u.) with respect to the highest peak of each pattern.

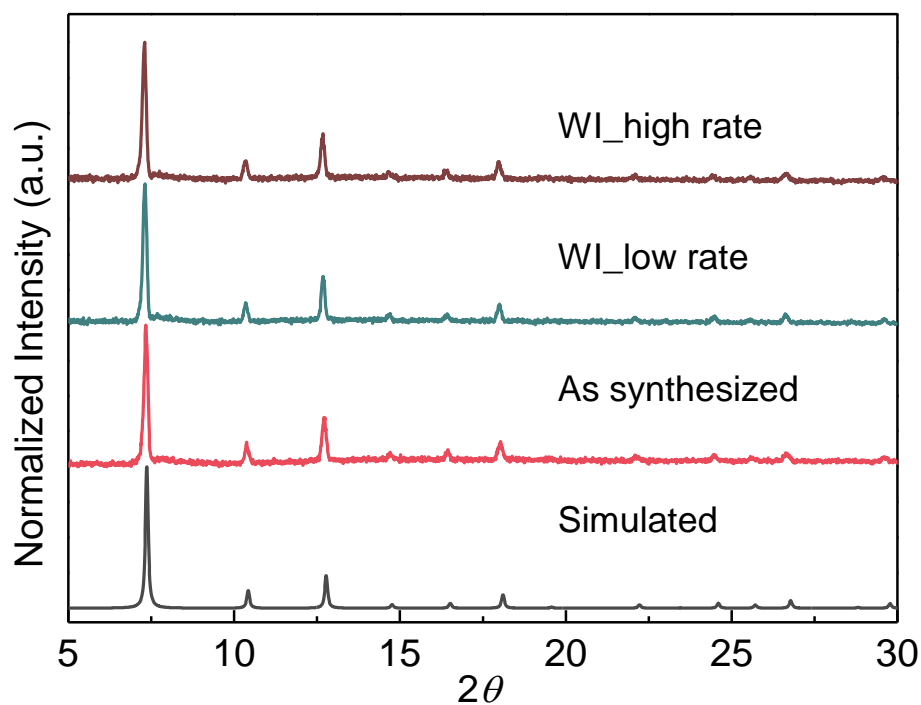

**Supplementary Figure 5 | PXRD patterns of the ZIF-67 samples.** The XRD intensities have been normalized (a.u.) with respect to the highest peak of each pattern.

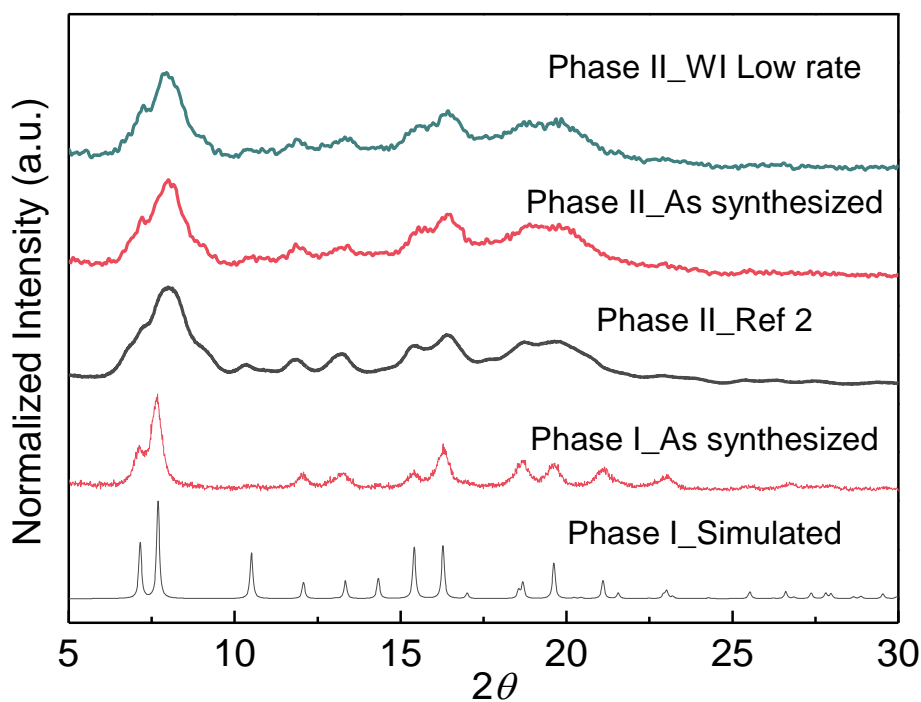

**Supplementary Figure 6 | PXRD patterns of the ZIF-7 samples.** The obtained XRD patterns agree well with the reported results in Ref. 4. The broadening of the Bragg peaks is due to the nanocrystal size of the samples. The XRD intensities have been normalized (a.u.) with respect to the highest peak of each pattern.

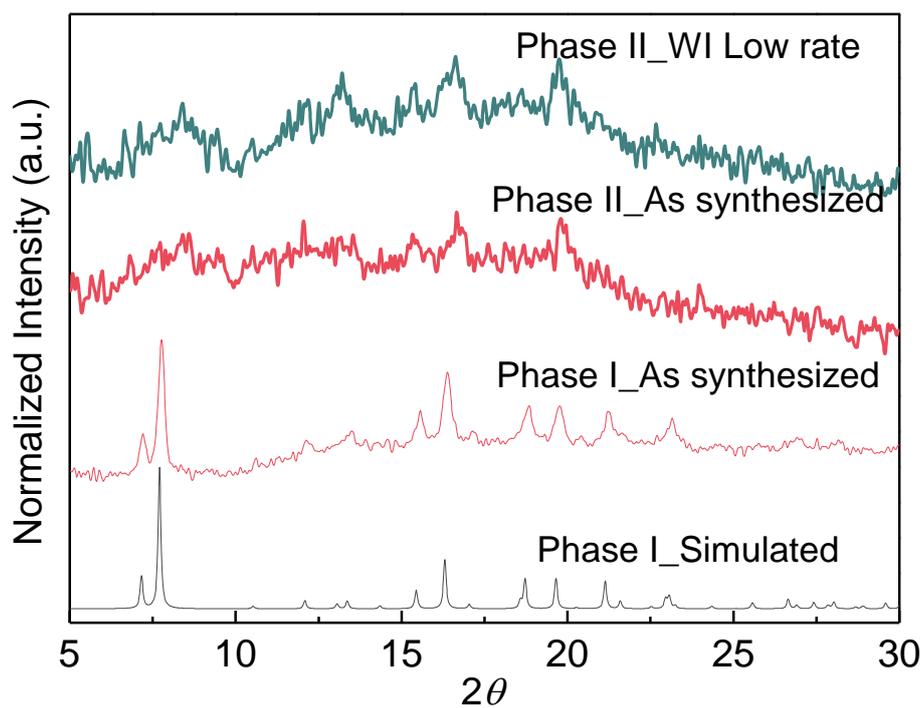

**Supplementary Figure 7 | PXRD patterns of the ZIF-9 samples.** As the structural analogue of ZIF-7, ZIF-9 has an XRD pattern similar to that of ZIF-7 (Supplementary Figure 6). The XRD intensities have been normalized (a.u.) with respect to the highest peak of each pattern.

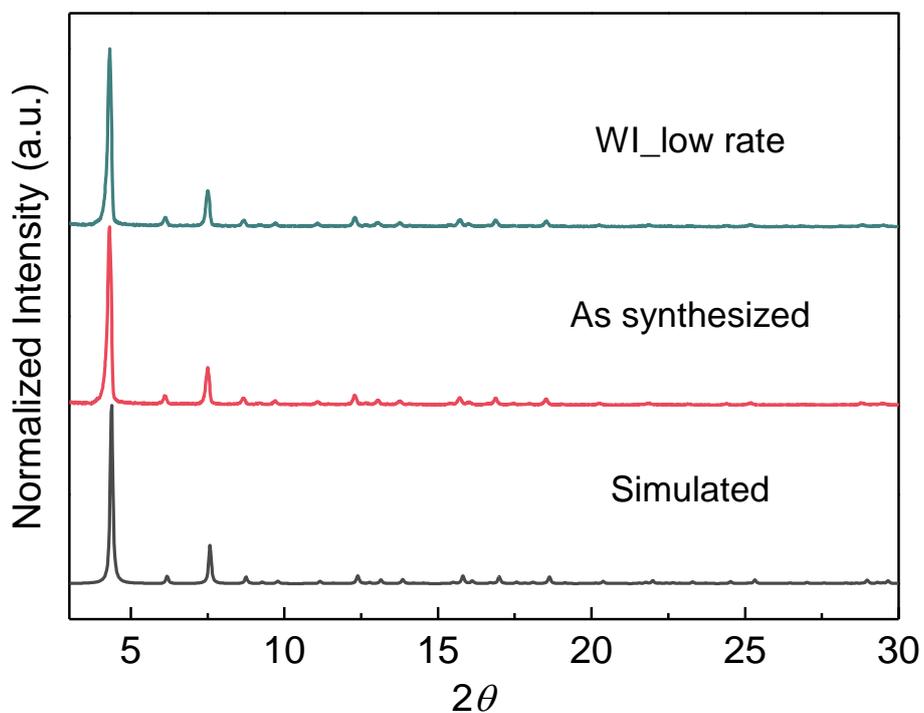

**Supplementary Figure 8 | PXRD patterns of the ZIF-71 samples.** The XRD intensities have been normalized (a.u.) with respect to the highest peak of each pattern.

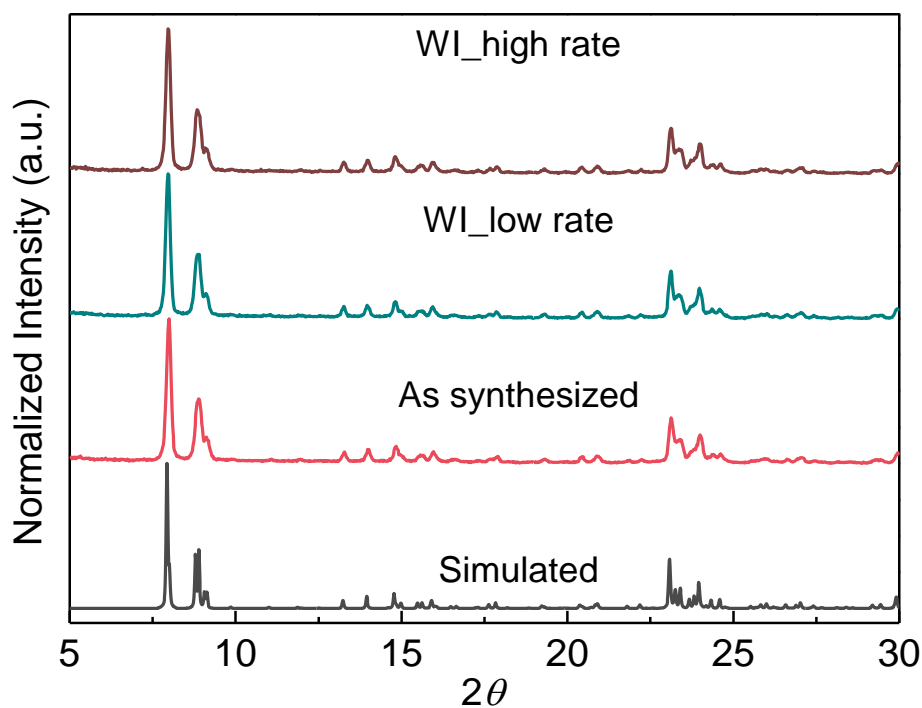

**Supplementary Figure 9 | PXRD patterns of the ZSM-5 samples.** The XRD intensities have been normalized (a.u.) with respect to the highest peak of each pattern.

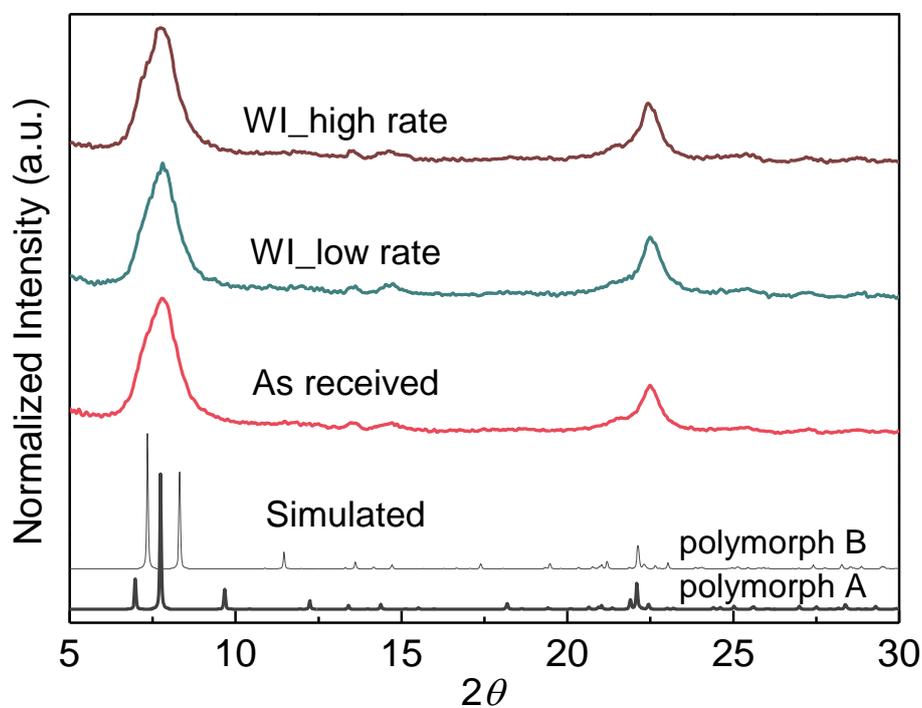

**Supplementary Figure 10 | PXRD patterns of the zeolite-*β* samples.** Zeolite-*β* is an intergrowth of polymorph A and B differing in the stacking of the 12-ring channels, and therefore has a single broad peak in the 5°-10° range which agrees well with those reported in the literature.[1] The XRD intensities have been normalized (a.u.) with respect to the highest peak of each pattern.

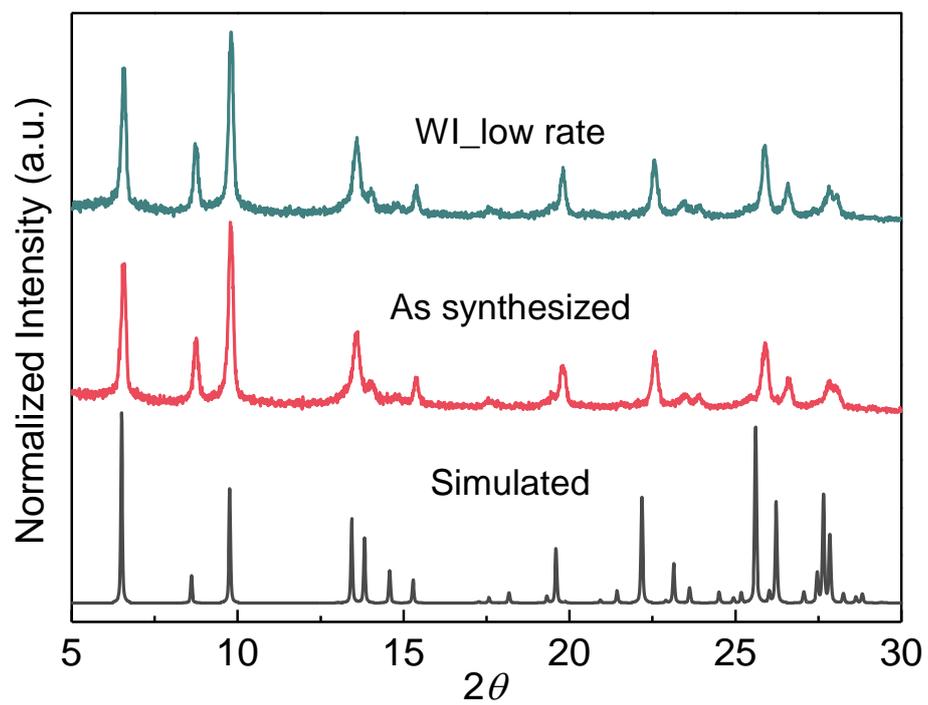

**Supplementary Figure 11 | PXRD patterns of the mordenite samples.** The XRD intensities have been normalized (a.u.) with respect to the highest peak of each pattern.

# S2 Additional experimental results of hydrophobic cage-type frameworks

## S2.1 Water intrusion and extrusion of ZIF-8

Here we show a complete profile of ZIF-8 data for Figure 2b-c of the main text including the strain rate history. The three complete intrusion/extrusion curves are plotted into Figure 2b and the extracted intrusion pressures, extrusion pressures, and energy absorption densities are plotted into Figure 2c as a function of the different strain rates during intrusion and extrusion. In the low-rate and medium-rate tests (Supplementary Figure 12a-b), the unloading process is also externally driven, and controlled at the same rate as the loading process. However, the high-rate experiment has an uncontrolled free unloading process (Supplementary Figure 12c), which is dependent on the material response. The recorded reflected waves in the SHPB experiments allow us to measure the strain rate history during the intrusion and extrusion process. As shown by the dashed line in Supplementary Figure 12c, the extrusion rate is one order of magnitude lower than the intrusion rate ($10^2$ s$^{-1}$ vs. $10^3$ s$^{-1}$).

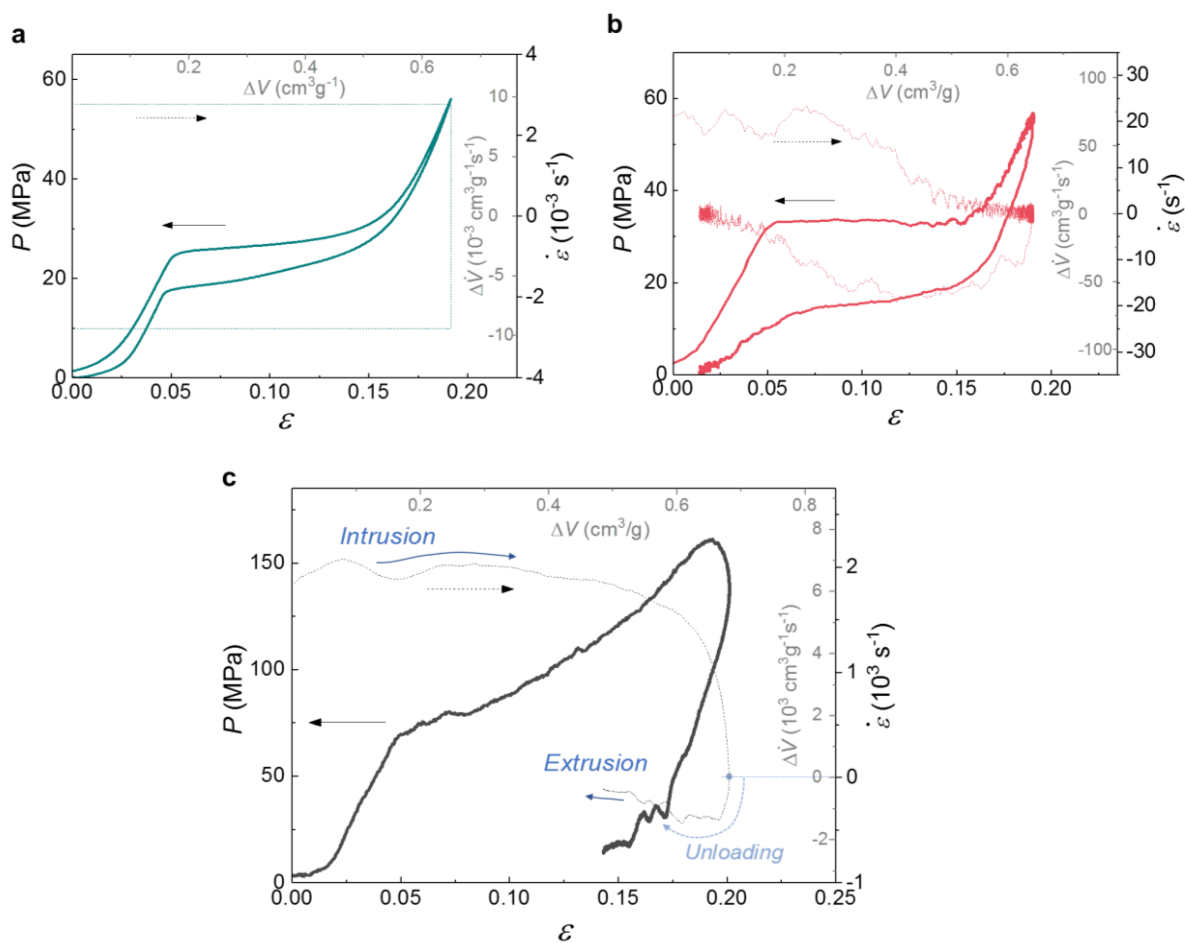

**Supplementary Figure 12 | Water intrusion and extrusion of ZIF-8**. **a**, Low-rate, **b**, medium-rate, and **c**, high-rate conditions, corresponding to the result of Figure 2b-c in the main text. Solid lines represent the pressure; dashed lines represent the corresponding strain rate.

- 14 -

The uncontrolled free water extrusion process represents the performance of {ZIF-8+water} systems under realistic impact loading conditions. In most cases of impact protection using energy absorption materials, there will be no external controls to retract the impactor or projectile from the target. This means the unloading process is determined by the intrinsic recovery of the impacted target. For this reason, we do not require the loading and unloading rates to be the same in our high-rate experiments.

There are two challenges to measuring energy absorption at an extrusion or unloading rate matching the intrusion rate of $10^3$ s$^{-1}$. Firstly, it is experimentally challenging to retract the impactor at such a rate, and secondly it is impossible to elevate the extrusion rate above the intrinsic value associated with the material. If the impactor (*i.e.*, the piston in the current experiments) were withdrawn more quickly, it would simply detach from the specimen, such that the unloading force on the impactor would drop to zero and the hysteresis would be even higher. Hence, the energy absorption is at least as high as the values reported in the main text.

## S2.2 Water extrusion of ZIF-8 revealed by high-rate experiments

To understand the factors influencing the extrusion rate of water molecules from ZIF-8, a series of high-rate experiments with different incident pulses were carried out. Three different incident pulses were designed, as shown in Supplementary Figure 13a, and their corresponding water intrusion-extrusion curves are plotted in Supplementary Figure 13b. The incident pulses are designed so that two of them (light purple and light green curves) create partial water intrusions (defined as having maximal strain $\varepsilon = 0.13$) up to a similar volumetric strain at different intrusion rates, while the other one (grey curve) results in a complete water intrusion (defined as having maximal strain $\varepsilon = 0.2$) at the same intrusion rate as the light purple curve ($\dot{\varepsilon} = 2,000$ s$^{-1}$).

The results are zoomed in for a detailed analysis of the extrusion pressure (Supplementary Figure 13c) and extrusion rate (Supplementary Figure 13d). For the two partial intrusion cases (light purple and light green curves), we found their extrusion performances to be very similar, despite their different intrusion rates. This is expected, as the extrusion of water molecules is governed by the intrinsic {ZIF-8+water} behaviour.

The comparison between the complete (grey) and partial (light purple) intrusion cases suggests that water extrusion starts at a relatively higher rate and pressure, which then gradually decreases as the extrusion progresses. This agrees well with the observation in the medium-rate test (the 20 s$^{-1}$ curve in Figure 2b of the main text), where the pressure drops gradually near the end of the extrusion: the water molecules at the end of the queue extrude at a rate as low as the moving speed of the compression head, therefore the force sensor can hardly detect any reaction force from the sample. These phenomena indicate higher mobility of water molecules when more of them are present inside the nanoporous framework, which agrees with the simulated results of hopping events at different water loadings (Supplementary Section S7.2).

Note that "complete" and "partial" water intrusion are defined only to aid the discussion in this section. These labels are essentially distinguished by their different maximal strains, independent of the completeness of water intrusion at their maximal strains. From the low- and medium-rate experimental results in Supplementary Figure 12a-b, once the strain goes up to around 0.2, the entire accessible pore volume has been occupied: the intrusion plateaus end and the system undergoes a much 'stiffer' linear reduction in volume with increasing pressure, corresponding to the elastic compression in step 3 of Fig. 2a of the main text. In high-rate experiments, due to the increased gradient of intrusion plateaus, it becomes harder to identify the actual completion of water intrusion from the high-rate stress-strain curves.

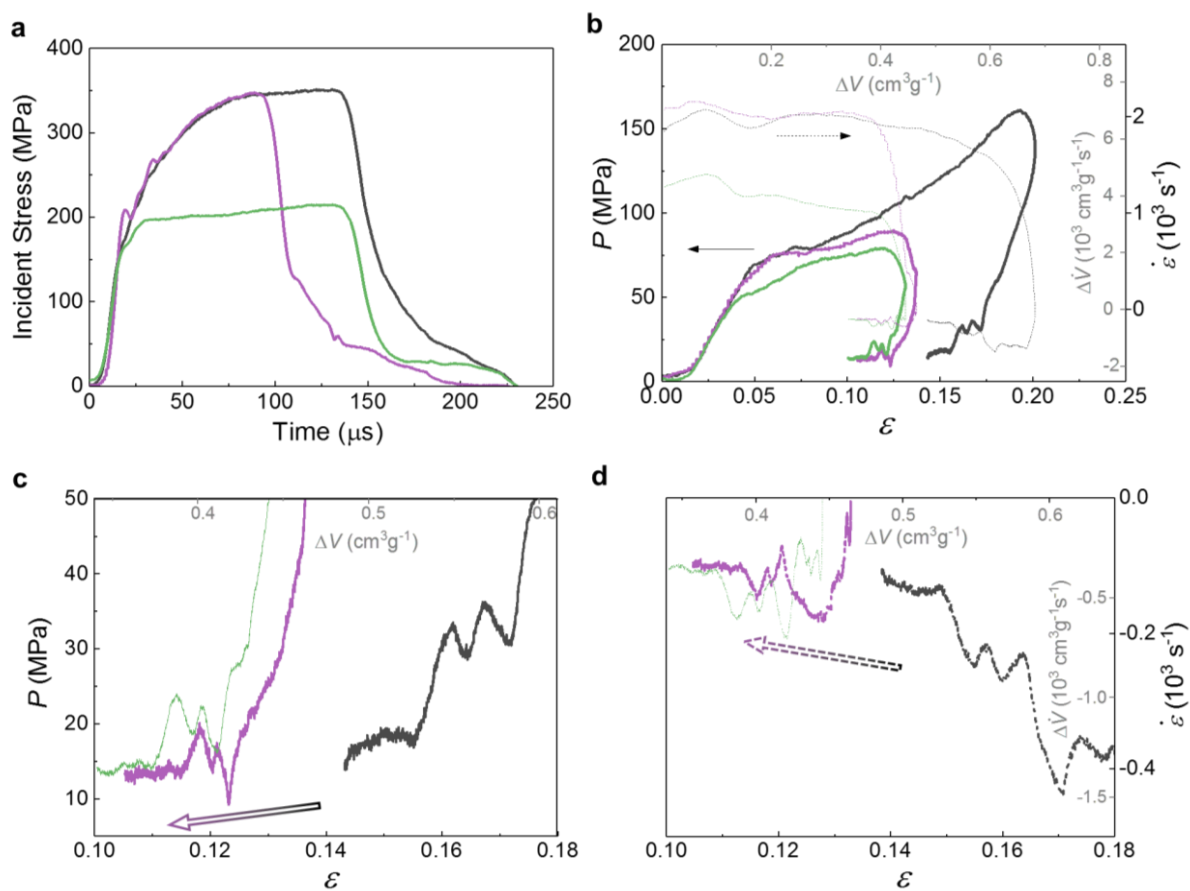

**Supplementary Figure 13 | Water intrusion and extrusion of ZIF-8 in the high-rate experiments using different incident pulses. a**, Three incident pulses of different time durations and intensities were designed as the energy input to drive the water intrusion, which was achieved by using different strikers launched at different levels of air pressures, together with different designs of pulse shapers in the SHPB experiments. **b**, Corresponding testing results. The grey and light purple curves show the complete intrusion and partial intrusion respectively at a higher intrusion rate of about $2\times10^3$ s$^{-1}$, while the light green curve shows a partial intrusion at a relatively lower intrusion rate of about $1\times10^3$ s$^{-1}$. **c**, Comparing the stress-strain curves during the water extrusion. **d**, Comparing the strain rate during the water extrusion. The arrows in **c** and **d** indicate that the extrusion pressure and extrusion rate gradually decrease as the extrusion progresses, from the complete intrusion case (grey) to the partial intrusion (light purple) case.

- 17 -

## S2.3 Water intrusion pressure of ZIFs at different strain rates

Supplementary Figure 14 shows the intrusion pressures of a group of hydrophobic ZIFs at different strain rates. Evidently, the rate-dependence of the intrusion pressure is observed for all these cage-type ZIFs. Although the intrusion pressures of ZIF-71, ZIF-7, and ZIF-9 are not available at high rates, these materials display a strong rate effect at low and medium rates, and we estimate that ZIF-71 has a very high intrusion pressure exceeding 150 MPa at a rate of ~2,000 s$^{-1}$ (see Supplementary Figure 15). The rise in intrusion pressure and fall in extrusion pressure can enlarge the hysteresis area at increasing strain rate, resulting in a greater energy absorption density, as can be seen in the result of ZIF-71 (see Figure 5b of the main text).

The comparison between the structural analogues, ZIF-8 *vs*. ZIF-67, and ZIF-7 *vs*. ZIF-9, suggests that the rate-dependence of intrusion pressure is primarily contributed by the cage-type structure, whereas the influence of chemical moieties ($ZnN_4$ *vs*. $CoN_4$) is limited. One minor difference observed is the slightly higher intrusion pressure of the zinc-based frameworks (ZIF-7 and ZIF-8) compared to the cobalt-based frameworks (ZIF-9 and ZIF-67) at low to medium strain rates.

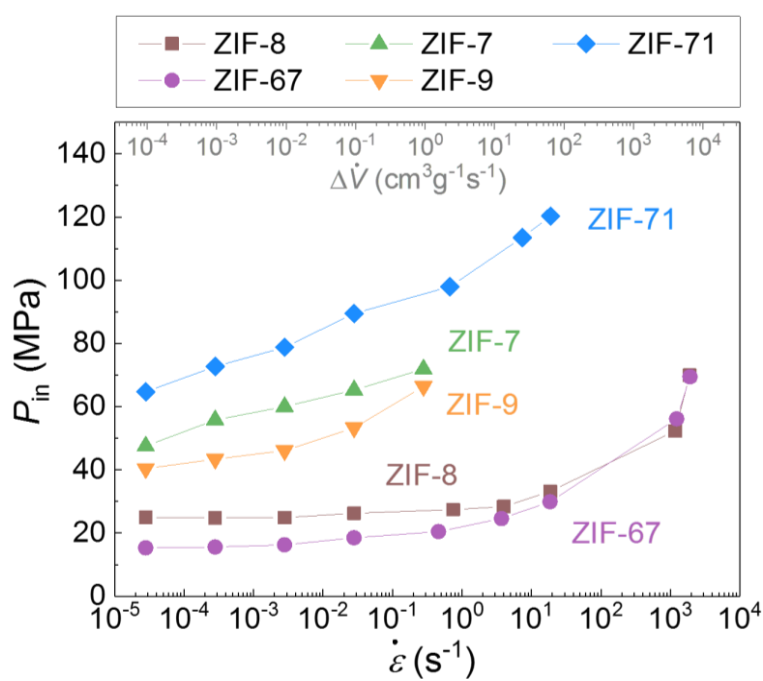

**Supplementary Figure 14 | Water intrusion pressure of a variety of ZIFs as a function of intrusion rate, showing a significant rate-dependence for all the ZIFs.** Since ZIF-7 and ZIF-9 have limited pore volume, their intrusion plateaus are difficult to recognize at higher loading rates, and therefore corresponding intrusion pressures are not available. The intrusion pressure of ZIF-71 at high strain rates is not included because it exceeds the capacity of the current SHPB setup.

**S2.4  High-rate water intrusion of ZIF-71**

Water intrusion into ZIF-71 has a very strong rate-dependence. As shown in Figure 5b of the main text, its intrusion pressure already doubles between low-rate and medium-rate conditions. Therefore, ZIF-71 is expected to have a very high intrusion pressure under high-rate conditions, which unfortunately cannot be captured using the current SHPB setup. In order to prove that it indeed has a high intrusion pressure at a high strain rate, we attempted an SHPB experiment and obtained the linear curve shown in Supplementary Figure 15. There is no intrusion plateau observed within this range, indicating that the water intrusion pressure of ZIF-71 under this condition (an average strain rate of 1865 s$^{-1}$) is at least 150 MPa. The overall gradient of the plot corresponds to the elastic compression of the {ZIF-71+water} system, which is reminiscent of the elastic part of the curves in Figure 5b of the main text.

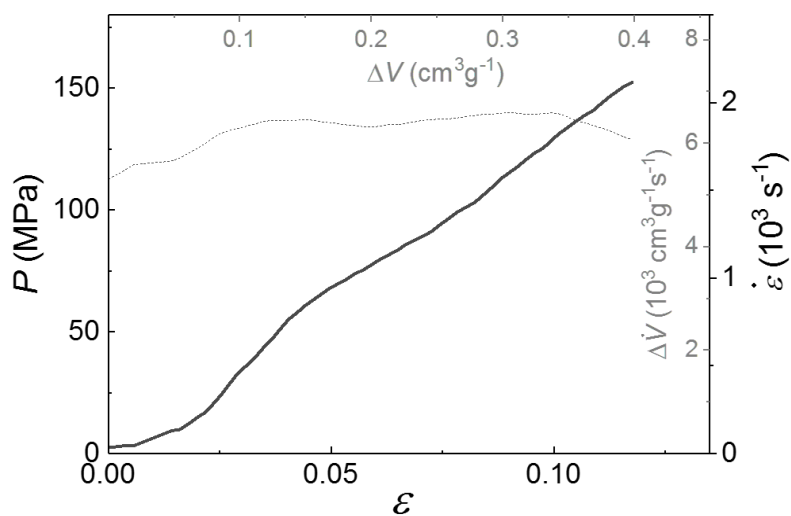

**Supplementary Figure 15 | The testing result of {ZIF-71+water} system in the high rate experiment.** The absence of an intrusion plateau indicates that the intrusion pressure exceeds 150 MPa.

## S2.5 ZIFs under multiple dynamic loading cycles

Here we show the water intrusion of ZIFs under cyclic loadings. The result of ZIF-8 has been presented in Figure 2d-e of the main text. Supplementary Figure 16a-b show that ZIF-67 and ZIF-71 also have a consistent performance during multiple dynamic loading cycles. This suggests that the intrusion of ZIF-8, ZIF-67, and ZIF-71 is recoverable, and therefore their energy absorption systems can be reused to cope with multiple impacts. ZIF-7 and ZIF-9, however, perform differently. Figure 5c of the main text shows that there is no extrusion plateau in their unloading curves. Supplementary Figure 16c-d present their second loading cycles, which confirm that no further intrusion can be observed, as all the nanopores have been occupied by the water molecules in the first loading cycle. This means that ZIF-7 and ZIF-9 cannot be reused for water intrusion. This can be explained by their very small apertures compared with the size of water molecules.[4] One way to reactivate these materials is to evacuate the water molecules by heat treatment, as reported in Ref. 4.

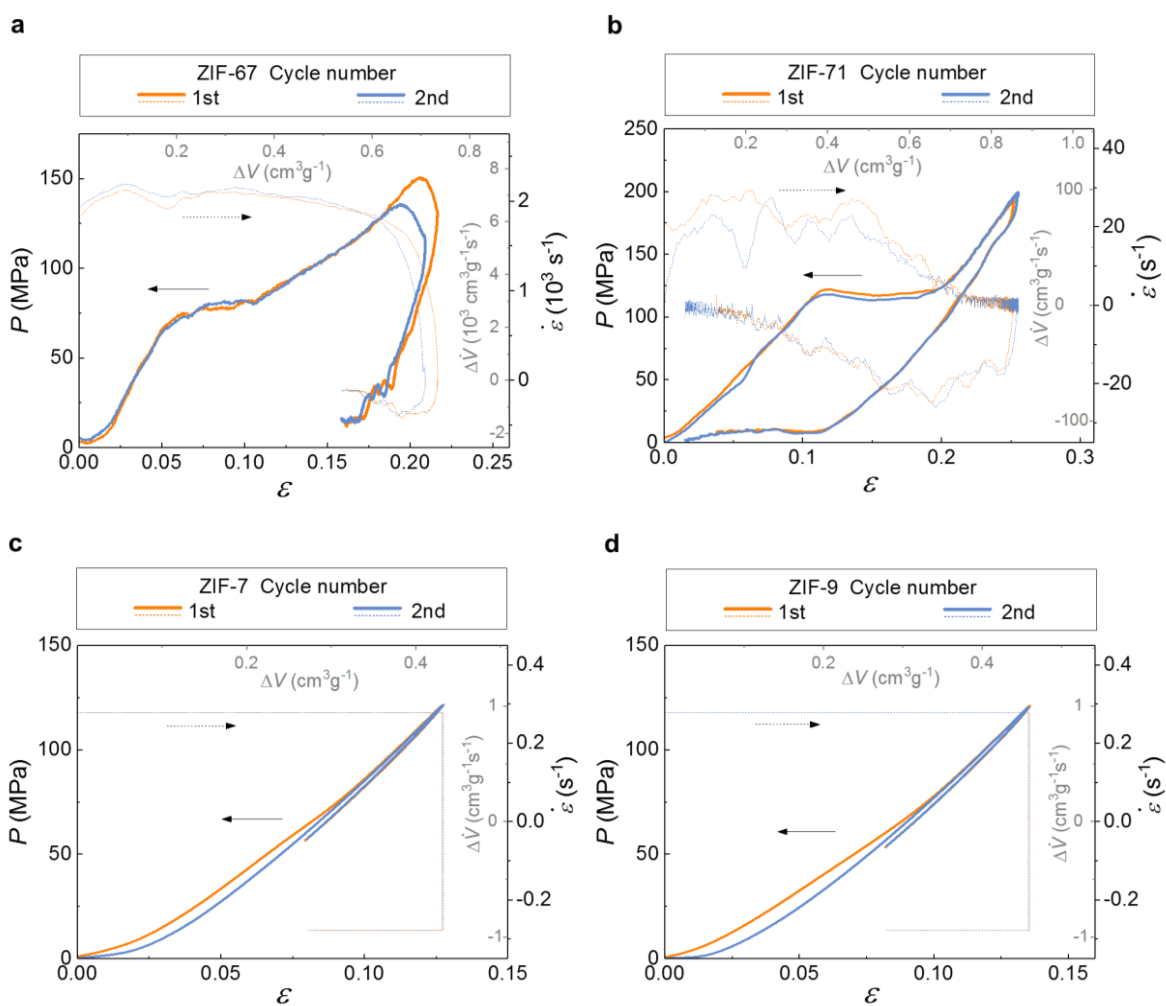

**Supplementary Figure 16 | Cyclic water intrusion and extrusion of different ZIF materials. a**, ZIF-67 at high strain rate, **b**, ZIF-71 at medium strain rate, **c**, ZIF-7 at 0.3 s$^{-1}$, and **d**, ZIF-9 at 0.3 s$^{-1}$.

- 20 -

## S2.6    Water intrusion of ZIF-11 and ZIF-12

Supplementary Figure 17 shows the cyclic water intrusion of ZIF-11 and ZIF-12 under quasi-static conditions, together with their PXRD patterns and SEM images before the experiments. We note that water intrusion into ZIF-11 has been attempted in a recent study,[6] but its water intrusion behaviour was not successfully detected, possibly due to the crystal size effect.[4]

There are no extrusion plateaus on the unloading curves, and the second loading cycles show no further intrusion, indicating that water molecules are trapped inside the framework after the intrusion process. This is similar to the non-reusable performance of ZIF-7 and ZIF-9 (Figure 5c of the main text and Supplementary Figure 16c-d). The pore limiting diameter (PLD) of ZIF-11 and ZIF-12 is 3.0 Å (see Supplementary Table 7 of Supplementary Section S10), meaning that PLD must be larger than 3.0 Å to enable the reusability.

Note that the PLD of ZIF-8 and ZIF-67 is 3.4 Å, only slightly higher than 3.0 Å, but they have a reusable performance (Figure 2d-e of the main text). Therefore, for the design rule #3 on the reusability of water intrusion energy absorbing systems, the PLD threshold value was determined to be ~3 Å.

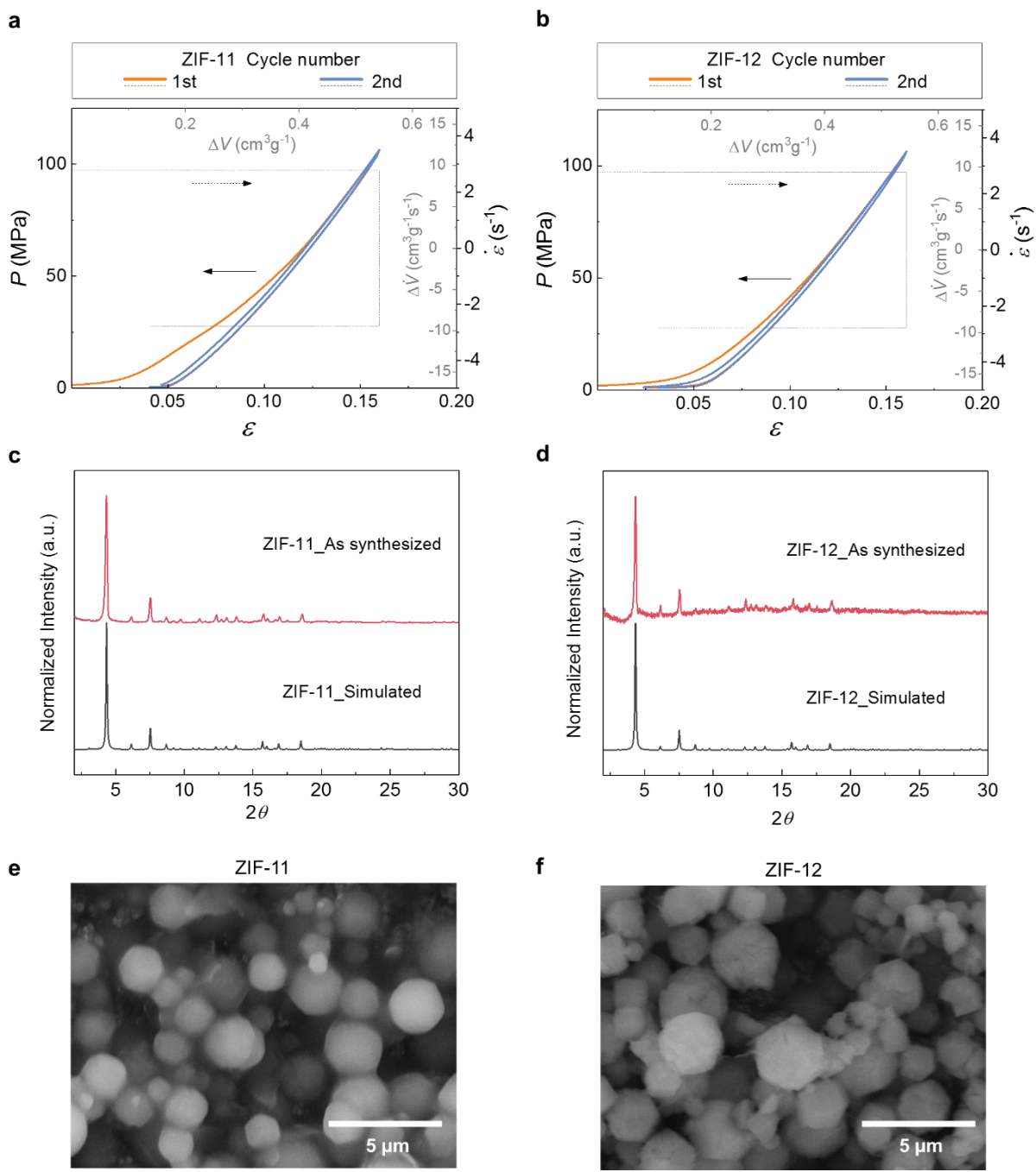

**Supplementary Figure 17 | ZIF-11 and ZIF-12 samples and their water intrusion experimental results.** Cyclic water intrusion of **a**, ZIF-11 and **b**, ZIF-12 under quasi-static loading conditions (3×10$^{-3}$ s$^{-1}$). PXRD patterns of **c**, ZIF-11 and **d**, ZIF-12 samples, which were recorded by the Rigaku Miniflex 600 using Cu Kα radiation (15 mA and 40 kV) at a scan rate of 2°/min using a 2θ step-size of 0.01°. The as-synthesized samples were obtained after the heat treatment at 200 °C for 2 h. SEM images of **e**, ZIF-11 and **f**, ZIF-12 samples obtained from Hitachi TM3030Plus.

**S2.7 Stability of ZIF-8 for water intrusion applications**

Material stability is crucial for practical applications. Figures 2d-e of the main text demonstrate that ZIF-8 exhibits consistent behaviour during the multi-cycle energy absorption tests without noticeable decay in performance. This indicates that its molecular structure remains intact, which is further supported by the X-ray diffraction patterns obtained before and after the tests (Supplementary Figure 4). To further confirm the stability of ZIF-8 for this application, we designed the following two experiments, with results shown in Supplementary Figure 18.

The first experiment is designed to confirm the stability of ZIF-8 under repeated high-rate water intrusion tests on the {ZIF-8+water} sample. We carried out a set of twenty high-rate tests on a {ZIF-8+water} sample, after which we measured the pressure-volume change response ($P$-$\Delta V$) of the tested ZIF-8 sample to evaluate its porosity based on the concept of "water porosimetry" by quasi-static water intrusion and extrusion. Water porosimetry is a technique equivalent to the more conventional mercury intrusion porosimetry but is dedicated for measuring hydrophobic porous materials.[7,8] The obtained result was then compared with that of a fresh ZIF-8 sample to identify any potential change of porosity caused by material degradation during the repeated high-rate experiment. The length of the intrusion plateau of the $P$-$\Delta V$ curves represents the total pore volume or porosity of the material. According to the results shown in Supplementary Figure 18a, there is no significant change between the observed intrusion/extrusion behaviour before and after the repeated high-rate experiments. This demonstrates that multiple high-rate water intrusion-extrusion cycles do not cause ZIF-8 to degrade, meaning that it can be stable after repeated mechanical impacts. We did not extend our high-rate tests up to hundreds of cycles due to the complexity of the experiments and the fact that such a high number of consecutive impacts is rare in real impact-attenuating applications.

It is worth noting that all experiments in this research were carried out at room temperature. The earlier reported temperature effect of quasi-static water intrusion into ZIF-8 demonstrated the possibility of structural degradation during the water intrusion, but only at high temperature (*ca*. 90 ºC).[9] The structural degradation in that work can be explained by the synergetic effect of high temperature and high pressure; by working at room temperature in this work, this synergetic effect is absent. The observed high structural-failure resistance of ZIF-8 against the impact-driven water intrusion can potentially be ascribed to the hydrophobicity of the framework and the resulting weak interactions between the framework and the intruded water molecules. This was also observed in the section 'The intrinsic water mobility timescale revealed by molecular dynamics simulations' of the main text, in which simulations demonstrate that the intruded water molecules tend to agglomerate around the cage centre and avoid the 6MR apertures.

The second experiment aims to examine the long-term water stability of ZIF-8 in this research. ZIF-8 was initially thought to be stable in water,[10] but some recent studies reported that ZIF-8 is stable in water only for a period of time and can dissolve into zinc and imidazolate ions and form new substances in the long term.[11-14] To this end, we designed an experiment to determine the possible change of ZIF-8 porosity after being immersed in water for over a week. We used the water porosimetry technique to establish a pressure-volume change ($P$-$\Delta V$) relationship for porosity evaluation. The specific procedures are as follows: we fabricated a {ZIF-8+water} sample by immersing ZIF-8 in water, and one cycle of water intrusion was conducted on the sample to ensure that water molecules were in sufficient contact with the ZIF-8 crystals. Then, the {ZIF-8+water} sample was kept inside the testing chamber (*i.e.*, with ZIF-8 in water) for 8 days, following which its $P$-$\Delta V$ curve was measured and compared with the curve obtained before the 8 days immersion. As Supplementary Figure 18b shows, no significant change of ZIF-8 porosity can be observed before and after the long-term stability test.

The good water stability of ZIF-8 in this research is likely attributed to the high mass ratio of ZIF-8 to water. A higher mass ratio of ZIF-8 to water, *i.e.*, more ZIF-8 or less water can inhibit the dissolution of ZIF-8, as the high-concentration solution can be easily saturated with the inhibitive imidazolate ligands.[11,12] In this research, we used 25 wt% as the mass ratio of ZIF-8 to water (*i.e.*, 25 mg ZIF-8 in 0.1 ml water), which is much higher than the value quoted in all of the reported studies on the water stability of ZIF-8 (ranging from 0.2 wt% to 6 wt%).[12] We used such a high mass ratio mainly because it can increase the porosity and energy absorption capacity of the {ZIF-8+water} system, but the high stability of ZIF-8 becomes another benefit.

It is worth noting, however, that this experiment has not ruled out the possibility of ZIF-8 degradation if immersed in water for months or years, which would need further investigation. Currently, there is a growing interest on the water stability of ZIFs and MOFs, and some approaches have been proposed to enhance the water stability of ZIF-8 such as surface ligand exchange.[15-17] These can be considered in the future to extend the lifecycle of the nanofluidic energy absorption systems proposed in this work.

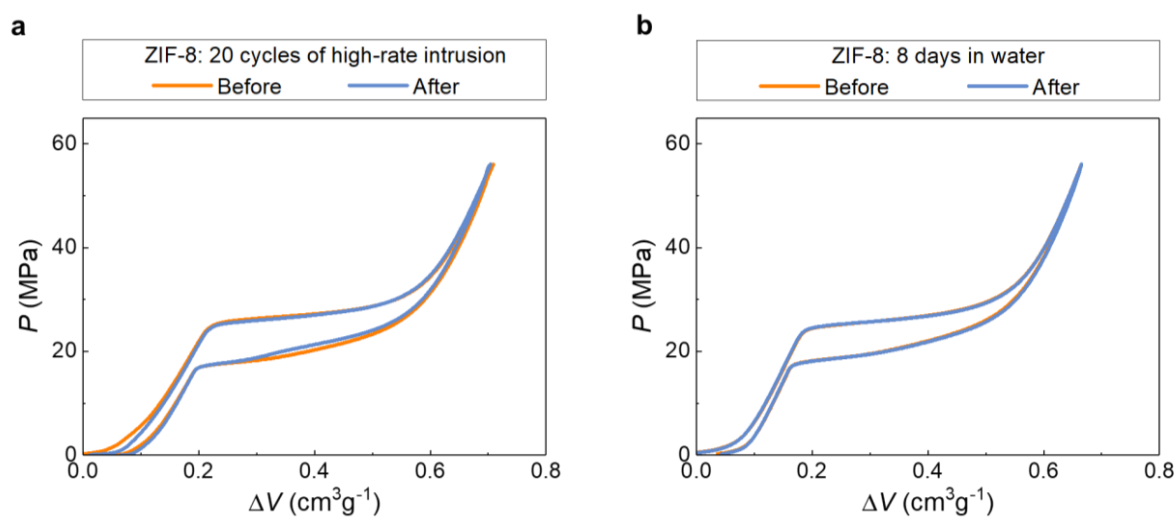

**Supplementary Figure 18 | $P$-$\Delta V$ curves of ZIF-8 before and after: a**, 20 cycles of high-rate water intrusion. **b**, 8 days in water. No appreciable change in behaviour is observed, confirming the stability of ZIF-8 with respect to multiple impact-driven water intrusion and long-term water immersion.

## S2.8 Generalisation of the design rules to hydrophobic cage-type zeolites

In order to prove that our strategy can be generalised to porous zeolitic frameworks, we tested it against chabazite, a cage-type zeolite which meets all the design rules. Chabazite consists of rhombohedral cages with 8-membered ring (8MR) apertures. The material was heated at 950 °C for 3 h to enhance the hydrophobicity (design rule #1). Its largest cavity diameter (LCD) amounts to 7.4 Å, higher than the value of the pore limiting diameter (PLD), which amounts to 3.8 Å (design rule #2). Meanwhile, its PLD is over 3 Å (design rule #3) and its LCD is also large enough (design rule #4), so the chabazite material meets all the design rules we proposed.

The water intrusion of chabazite under quasi-static conditions has been reported before,[18] showing a limited energy absorption capacity. Herein, we carried out the water intrusion experiments on chabazite at different strain rates. Supplementary Figure 19a-b shows that the strain rate strongly affects the energy absorption density during the water intrusion and extrusion, as expected from our design rules. A substantially larger energy absorption is obtained at the high strain rate, which increases from 0.6 J g$^{-1}$ under quasi-static compression to 22.3 J g$^{-1}$ at high-rate loading conditions. Furthermore, to check the reusability of the {chabazite+water} system, we carried out cyclic intrusion tests at different conditions. Supplementary Figure 19c presents two consecutive high-rate experimental results, which are highly consistent.

With these, we proved that the cage-type zeolite, by meeting all the design rules, can also have reusable and efficient energy absorption upon impact. This demonstrates that our strategy is not limited to ZIFs, but that it can be generalised to porous zeolitic frameworks. However, comparing between ZIFs and zeolites, it is worth noting that many ZIFs have a relatively larger gravimetric pore volume, meaning that they can have a relatively higher energy absorption density using the water intrusion mechanism. For example, ZIF-8 (47 J g$^{-1}$) can absorb twice the energy that can be absorbed by chabazite (22 J g$^{-1}$) at the same high-rate loading condition. The observation that cage-type zeolites can absorb mechanical energy also agrees with some reported results at low strain rates, where cage-type structures show a slightly larger hysteresis than channel-type structures under water intrusion.[19]

It is worth noting that, although the intrusion process and pressure at high strain rates cannot be easily identified in Supplementary Figure 19 from the small gradient change of the loading curves before and after the start of water intrusion in chabazite, the extrusion plateaus can still be identified from the unloading curves. The hysteresis area clearly demonstrates that the energy absorption capacity is highly rate dependent. To further confirm that the obtained high-rate stress-strain curves are underpinned by the water intrusion and extrusion process, we carried out high-rate experiments on samples with a different amount of chabazite. The results in Supplementary Figure 19d show that with the increase of sample porosity, obtained by having a higher amount of chabazite in the sample, the extrusion plateau becomes more evident, starting at a larger strain value. The extrusion plateau can be more easily identified in the high-rate experiments because the extrusion happens at a significantly lower rate than the intrusion process. Shown as the dashed line in Supplementary Figure 19d, the water extrusion from chabazite happens at $10^2$ s$^{-1}$ while the intrusion happens at $10^3$ s$^{-1}$, which are the same as the results from the experiments on ZIF-8 (see Figure 2 of the main text).

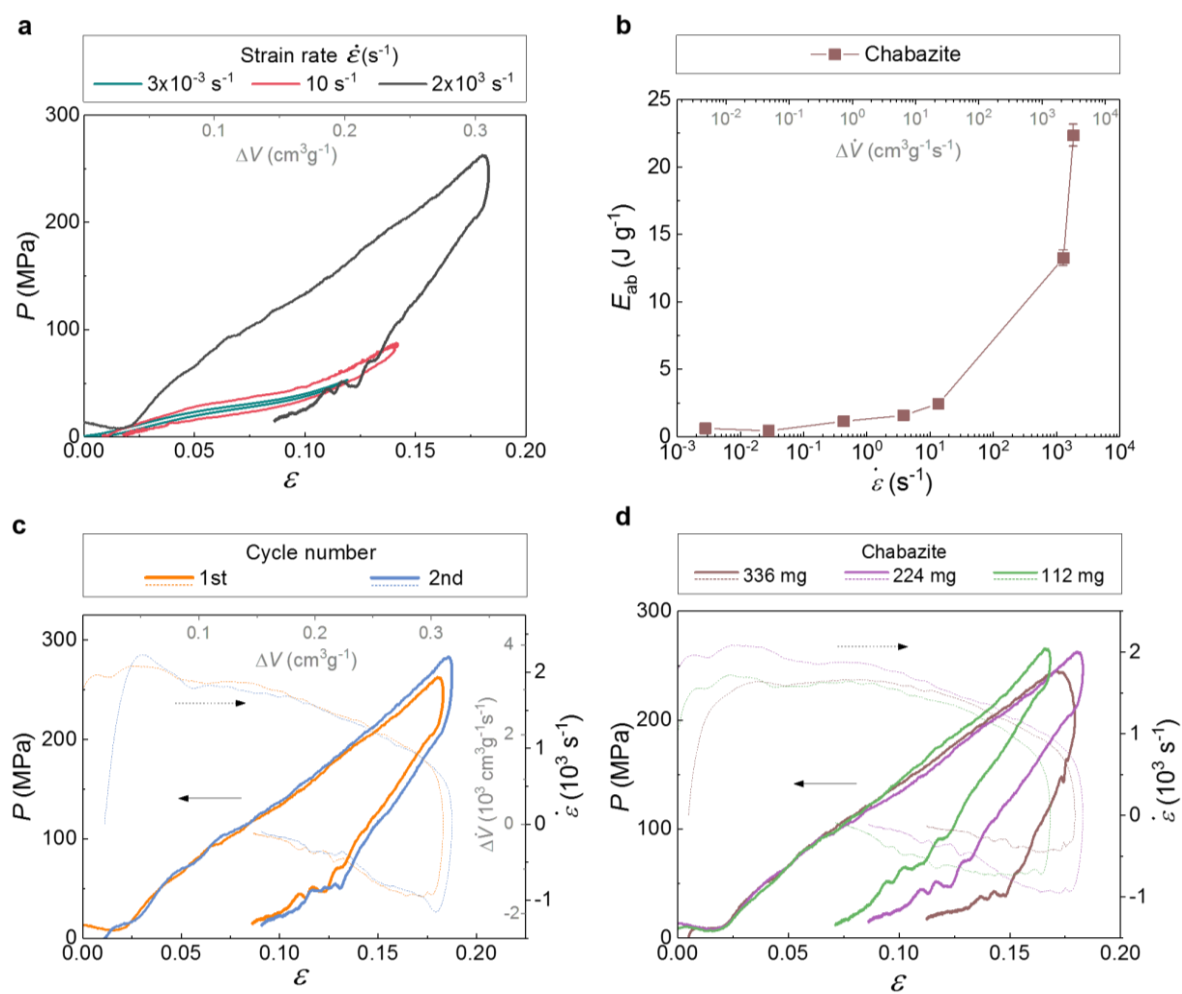

**Supplementary Figure 19 | Water intrusion of chabazite with cage structures. a**, Stress-strain curves at three different strain rates, which correspond to a specific volume change rate $\Delta\dot{V}$ of $5\times10^{-3}$, 20, and $3\times10^3$ cm$^3$ g$^{-1}$ s$^{-1}$, respectively. **b**, Energy absorption densities as a function of the strain rate. The error bars represent the uncertainty due to the incomplete unloading curves. **c**, Two consecutive high-rate experiments showing consistent performance. **d**, Influence of the zeolite mass: high-rate water intrusion and extrusion in 112 mg, 224 mg, and 336 mg of chabazite. These are the amounts of material used to make a sample of Ø12.7 mm in diameter and 3 mm in length used in the SHPB experiments.

## S2.9   Effect of crystal size on water intrusion and energy absorption

In our previous work we discovered that the crystal size of ZIF-7 plays a role in the water intrusion performance.[4] Due to the ultrasmall aperture (PLD) of ZIF-7, which is smaller than the size of water molecules, only nanocrystals of ZIF-7 can be intruded by water molecules under quasi-static conditions, while bigger crystals at micrometre scale are inaccessible to water molecules. Ref. 20 also reported that bigger ZIF-8 crystals show a slightly higher water intrusion pressure than smaller ones under quasi-static conditions, explained by the possible local defects in nanoparticles.

The effect of the crystal size on water intrusion at dynamic loading conditions has not yet been investigated. To this end, we synthesized a ZIF-8 sample with a smaller crystal size than the one in the main text. Supplementary Figure 20 shows the PXRD patterns and atomic force microscopy images of the as-synthesized ZIF-8 samples. In comparison with the larger ZIF-8 sample in this work (Supplementary Figure 20c or Supplementary Figure 3a), the AFM image shows that the new ZIF-8 sample has a significantly smaller crystal size, and therefore it also has relatively broader peaks in its PXRD pattern.

Water intrusion experiments at different strain rates were carried out and the results are shown in Supplementary Figure 21. In general, the water intrusion and extrusion behaviour is quite similar between the two ZIF-8 samples of different crystal sizes (Supplementary Figure 21a), both exhibiting a strong rate effect. Their energy absorption performances do not show a significant difference; at some loading conditions, the energy absorption density of the smaller crystals is slightly lower (Supplementary Figure 21b). A closer comparison between the two samples is presented in Supplementary Figure 21c-e, showing that the water intrusion and extrusion happen at slightly lower pressures when the crystal size becomes smaller.

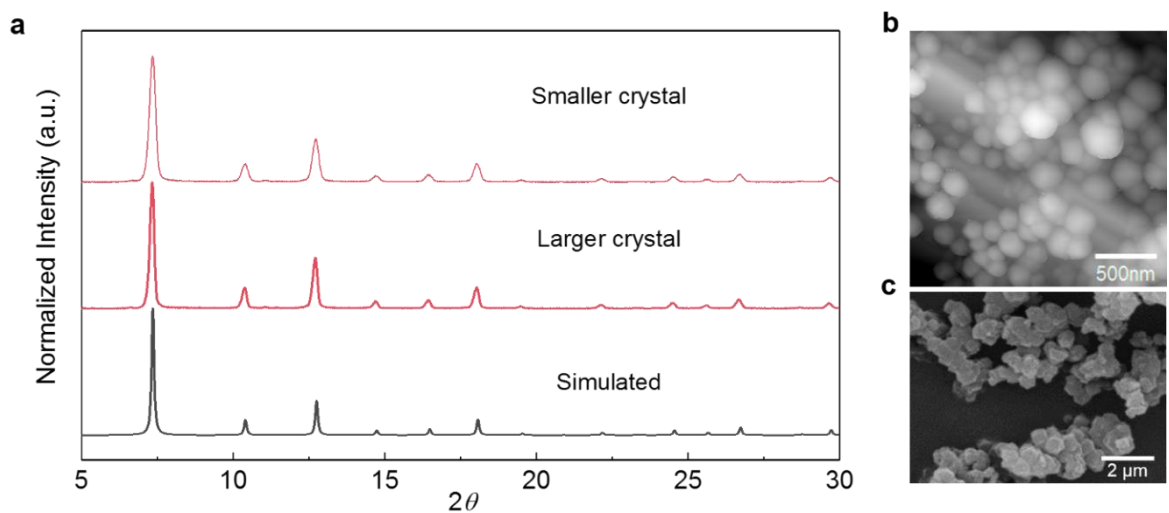

**Supplementary Figure 20 | Two ZIF-8 samples with a different crystal size. a**, PXRD patterns of the two ZIF-8 samples, **b**, AFM image of the smaller ZIF-8 sample, **c**, SEM image of the larger ZIF-8 sample (also shown in Supplementary Figure 3a).

The decrease of intrusion and extrusion pressures with the reduction in size of the crystals agrees well with the previous work on ZIF-7 and ZIF-8 under quasi-static conditions.[4,20] This is likely due to the higher structural flexibility at the crystal interfaces: near the outer surface the crystals are more flexible, directly interacting with water molecules and favouring their intrusion. Therefore, we see that the intrusion pressure increases with increasing ZIF-8 crystal size,[20] and an ultra-high intrusion pressure is required to intrude micro-sized ZIF-7 crystals which are conceived as being inaccessible.[4] It is worth noting that this phenomenon agrees with the reported discrepancy in $CO_2$ adsorption in ZIF-7 with different crystal sizes, which has been ascribed to their different degrees of structural flexibility.[21] Similarly, the increase in flexibility by crystal downsizing has also been observed by measuring the Young's moduli for the micro- and nano-sized ZIF-8 crystals.[22]

The higher structural flexibility and possible presence of disordered structures near the outer shell of the small crystals [21] may also be the reason for the observed slightly lower energy absorption density for the smaller ZIF-8 sample. The energy absorbing cage-by-cage water flow can be interrupted by the crystal interfaces so that the crystal exhibits a relatively lower energy absorption density. However, further investigation is needed to fully confirm this hypothesis and the underlying mechanisms.

With the above, we think crystal size can be a design parameter for {ZIF+water} systems, which should be kept within a reasonable range to achieve a balance between the efficacy (*i.e.*, whether water can intrude) and efficiency (*i.e.*, high energy absorption density). We do not think there would be a general quantitative threshold that can be set on all ZIFs, as it is most probably a structure-dependent threshold, varying from one material to another. For example, we observed that micro-sized ZIF-7 cannot be intruded by water molecules [4] while micro-sized ZIF-71 can be intruded (Fig. 5b of the main text); this can be potentially explained by their different structures and aperture sizes. With the knowledge obtained so far, we believe that an advisable practice to design nanofluidic energy absorption systems is to start with nanocrystals to ensure efficacy, and then consider larger crystals, as the crystal size seems to have a relatively minor effect on the absorption capacity compared to the four design rules that were formulated in this work.

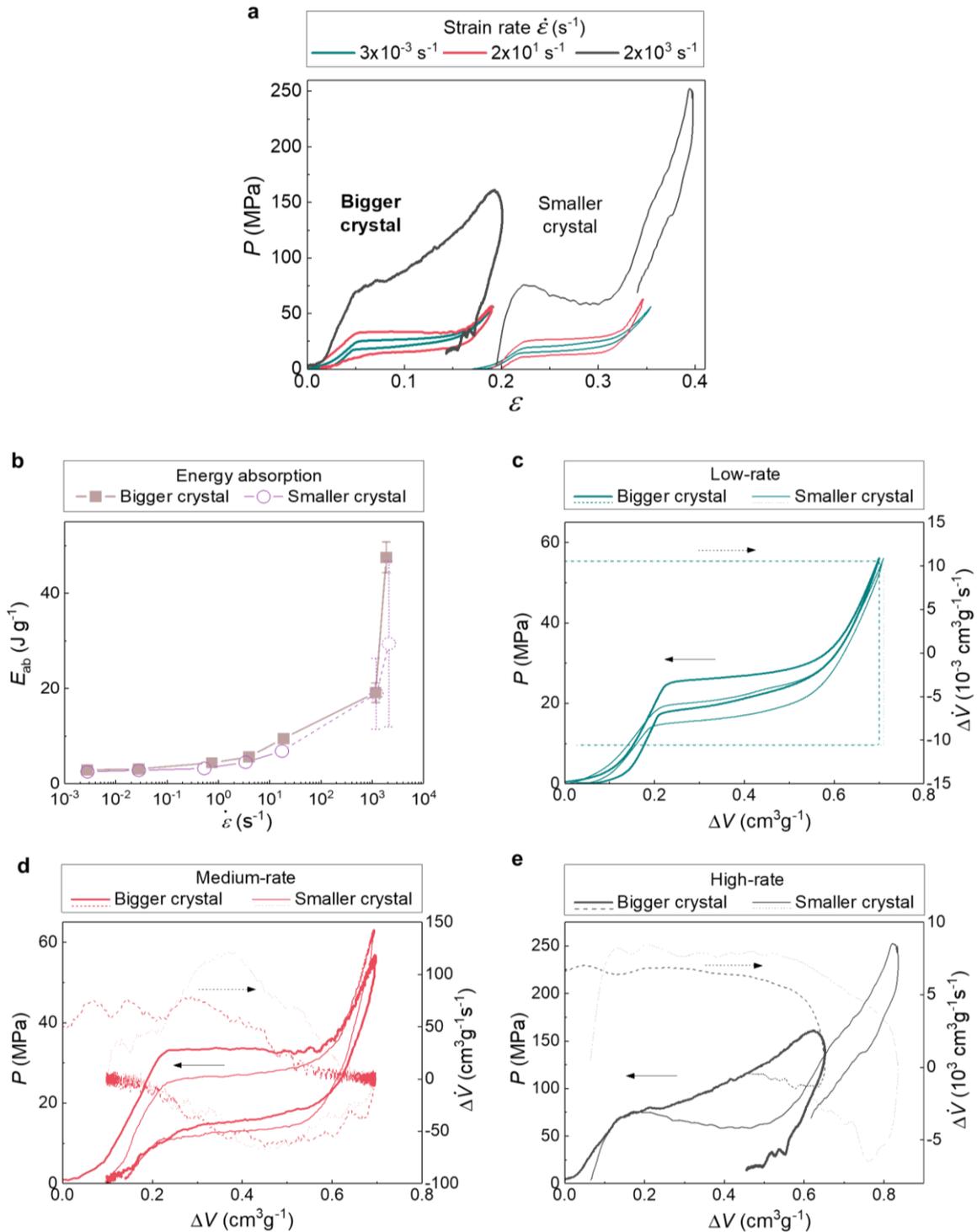

**Supplementary Figure 21 | Water intrusion of the two ZIF-8 samples with different crystal sizes.**
**a**, Stress-strain curves at three different strain rates, with the plots offset horizontally for clarity. **b**, Energy absorption densities as a function of the strain rate. The error bars represent the uncertainty due to the incomplete unloading curves. Data of the high-rate experiments (dashed) were obtained at different maximum strains for the two ZIF-8 samples. Detailed comparisons of the $P$-$\Delta V$ curves between the two samples are shown in **c** at low strain rate, **d** at medium strain rate, and **e** at high strain rate.

## S2.10 Energy absorption performance of other nanoporous materials

Currently, the materials used for nanofluidic energy absorption mainly include zeolites, MOFs, and silica. Most of the experiments reported previously have been performed under quasi-static conditions. This is the first time that stress-strain or $P$-$\Delta V$ relationships are obtained at high strain rates for the liquid intrusion of any porous material.

The performance of several microporous zeolites is discussed in this work and compared with ZIFs. For some other microporous zeolites and MOFs, their quasi-static water intrusion performance has been summarised in Ref. 19,23. These results show that most hydrophobic zeolites have a reusable spring behaviour with a very small hysteresis (< 10%), resulting in a limited energy absorption density (*ca.* 1 J g$^{-1}$) under quasi-static conditions. According to this work, at a higher strain rate, channel-type zeolites such as ZSM-5 will retain their limited energy absorption densities, while cage-type zeolites such as chabazite can obtain an enhanced energy absorption density (up to 10s J g$^{-1}$, see Supplementary Section S2.8). Compared with the ZIFs that we suggested (*i.e.*, ZIF-8, ZIF-67, and ZIF-71), the reported hydrophobic cage-type zeolites also have smaller pore volume (*ca.* 0.1 cm$^3$ g$^{-1}$) which limits their energy absorption densities. Although most {zeolite+water} systems are reusable, some have also been reported to work only for one cycle due to the formation of silanol groups during the water intrusion process.[24-26]

Mesoporous and macroporous silica can be made hydrophobic by surface treatment. We have summarized some reported quasi-static testing results in Supplementary Table 1. In contrast to microporous zeolites, the water intrusion of hydrophobic silica produces a relatively large hysteresis, and because they also have large pore volumes (accompanied by their large pore sizes), a considerable mechanical energy can be absorbed during the water intrusion process. However, their extrusion pressures are usually very low and, in many cases, result in an irreversible performance (represented by $P_{ex}$ = 0 MPa in Supplementary Table 1). The highest quasi-static energy absorption density, amounting to 11 J g$^{-1}$, is obtained by the mesoporous Fluka 100 C8. This energy absorption density can be increased up to 41 J g$^{-1}$ under impact,[27] but it cannot be reused due to the absence of water extrusion.

With the above comparison, we believe that the performance of ZIF materials represents a good combination of reusability and energy absorption efficiency to work against high-rate mechanical impact. It is worth noting that besides the forced liquid intrusion mechanism, there are also some other energy absorption mechanisms of nanoporous materials, such as the structural transformation, nanopore collapse, and chemical bond-breakage of flexible MOFs.[28-30] These mechanisms are similar to the crushing of macroscale metallic and polymeric forms,[31,32] based on a pore volume reduction process during the structure compression. As they have a very wide range of provoking pressures (from MPa to GPa) and are fundamentally different from the mechanism of this research, they are not included here.

**Supplementary Table 1 | Energy absorption capacities of selected hydrophobic silica materials under quasi-static water intrusion conditions.**

| Silica | Pore size (nm) | Pore volume (cm$^3$ g$^{-1}$) | $P_{in}$ (MPa) | $P_{ex}$ (MPa) | $E_{ab}$ (J g$^{-1}$) | Refs |
|---|---|---|---|---|---|---|
| **TMS-PhSBA-1 (2:1)** | 2.1 | 0.26 | 15 | 0 | 1.8 | 33 |
| **MCM-41** | 2.4 | 0.27 | 40 | 15 | 4.5 | 34 |
| **MSU-H** | 6.6 | 0.76 | 4.5 | 0 | 5.1 | 35 |
| **Fluka 100 C8** | 7.8 | 0.55 | 16 | 0 | 11 | 36,37 |
| **Zeoflo-TL** | 100 | 1.7 | 1.6 | 0 | 4.5 | 37 |

# S3 Force field derivation

## S3.1 *Ab initio* energy equation of state

As outlined in the Methods section of the main text, a 0 K energy equation of state for ZIF-8 was determined with VASP[38] following the procedure outlined in Ref. 39 at the PBE-D3(BJ) level of theory. The resulting data were fitted to the Rose-Vinet equation of state to obtain the equilibrium volume and bulk modulus,[40] as shown in Supplementary Figure 22. The resulting parameters are listed in Supplementary Table 2. Starting from the determined equilibrium volume, the dynamical matrix was extracted using 0.015 Å displacements for all atomic coordinates.

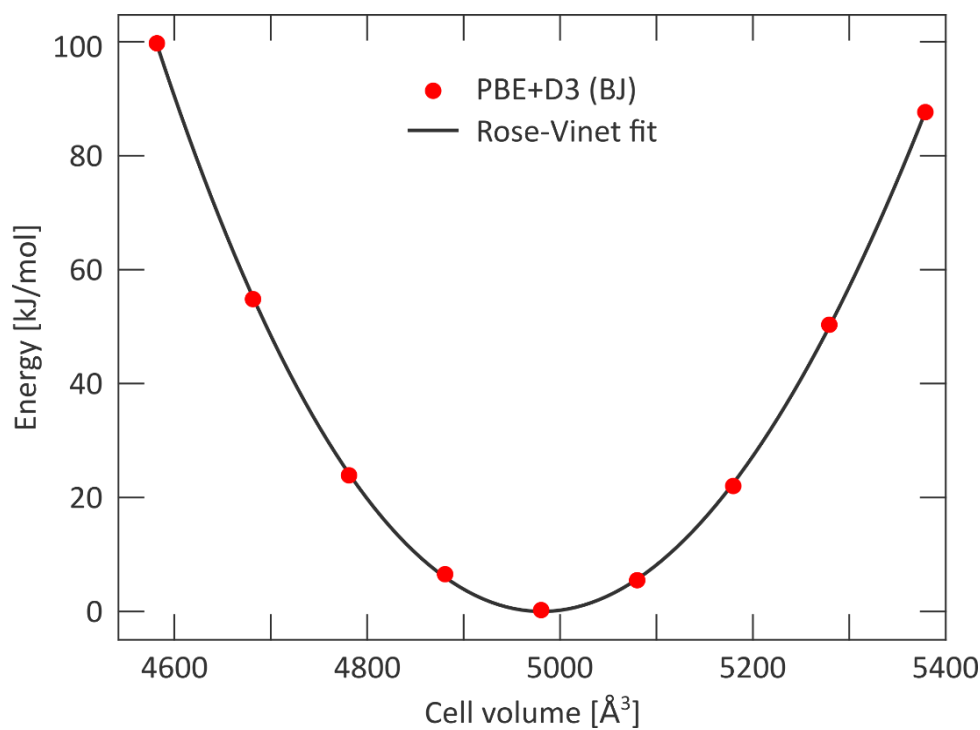

**Supplementary Figure 22 | Energy equation of state for ZIF-8.** The equation of state was obtained using the PBE+D3(BJ) level of theory with the Rose-Vinet fit.

**Supplementary Table 2 | ZIF-8 equilibrium parameters at 0 K.** The equilibrium volume, the equilibrium bulk modulus, and the pressure derivative of the bulk modulus at equilibrium. These data are extracted from the Rose-Vinet equation of state of Supplementary Figure 22.

| Equilibrium volume $V$ (Å³) | Equilibrium bulk modulus $K$ (GPa) | Pressure derivative of the bulk modulus $K'$ (–) |
|---|---|---|
| 4981 | 9.7 | 1.3 |

## S3.2 Force field derivation

The *ab initio*-based force field used to model ZIF-8 was derived from the dynamical matrix determined above using the QuickFF software package.[41,42] Within this protocol, the quantum mechanical potential energy surface (PES) is approximated by a sum of analytical functions of the nuclear coordinates that describe the covalent and noncovalent interactions. The latter is composed of electrostatic and van der Waals interactions:

$$\mathcal{V}^{FF} = \mathcal{V}_{\text{bond}} + \mathcal{V}_{\text{bend}} + \mathcal{V}_{\text{oopd}} + \mathcal{V}_{\text{torsion}} + \mathcal{V}_{\text{cross}} + \mathcal{V}_{\text{EI}} + \mathcal{V}_{\text{vdW}}.$$

The covalent interactions, which mimic the chemical bonds between the atoms, were approximated by different terms as a function of the internal coordinates (bonds, bends, out-of-plane distances, and dihedrals). The harmonic bond and bend terms are given by:

$$\mathcal{V}_{\text{bond}}^{ij} = \frac{K_{ij}}{2}(r_{ij} - r_{ij}^0)^2;$$

$$\mathcal{V}_{\text{bend}}^{ijk} = \frac{K_{ij}}{2}(\theta_{ijk} - \theta_{ijk}^0)^2.$$

Additionally, cross terms between the bonds and bends were included to improve the correspondence with the *ab initio* data. These cross terms comprise angle stretch-stretch (ASS) terms between bonds sharing a common atom and angle stretch-angle (ASA) terms, with the following mathematical expressions:

$$\mathcal{V}_{\text{ASS}}^{ijk} = K_{ijk}^{\text{ASS}}(r_{ij} - r_{ij}^0)(r_{jk} - r_{jk}^0);$$

$$\mathcal{V}_{\text{ASA}}^{ijk} = \left[K_{ijk}^{\text{ASA},1}(r_{ij} - r_{ij}^0) + K_{ijk}^{\text{ASA},2}(r_{jk} - r_{jk}^0)\right](\theta_{ijk} - \theta_{ijk}^0).$$

The out-of-plane distances (oopd) were described using a harmonic potential:

$$\mathcal{V}_{\text{oopd}}^{ijk\ell} = \frac{K_{ijk\ell}}{2}(d_{ijk\ell} - d_{ijk\ell}^0)^2.$$

This is a four-atom interaction, in which the internal coordinate is the distance between the central atom and the plane determined by its three neighbours.

The fourth covalent term is the dihedral energy term. Here, a cosine term as a function of the dihedral angle is used, which includes the multiplicity $m_\phi$ of the dihedral angle:

$$\mathcal{V}_{\text{torsion}}^{ijk\ell} = \frac{K_{ijk\ell}}{2}\left[1 - \cos\left(m_\phi(\phi_{ijk\ell} - \phi_{ijk\ell}^0)\right)\right].$$

The unknown parameters in all covalent terms (force constants, rest values, and multiplicities) were directly estimated using QuickFF.

The electrostatic interactions were modelled as Coulomb interactions between Gaussian charge distributions,[43] which allowed us to include all pairwise interactions. For ZIF-8, the atomic charges $q_i$ were derived with the Minimal Basis Iterative Stockholder (MBIS) partitioning scheme.[44]

$$\mathcal{V}_{\text{EI}} = \frac{1}{2} \sum_{\substack{i,j=1 \\ (i \neq j)}}^{N} \frac{q_i q_j}{4\pi\varepsilon_0 r_{ij}} \text{erf}\left(\frac{r_{ij}}{d_{ij}}\right).$$

Gaussian charge distributions were used with a total charge $q_i$ and radius $d_i$ centred on atom $i$. The mixed radius of the Gaussian charges were defined as $d_{ij} = \sqrt{d_i^2 + d_j^2}$.[43] The electrostatic interactions depend on the distance $r_{ij}$ between the two atoms.

The van der Waals interactions were described by the 12-6 Lennard-Jones model:

$$\mathcal{V}_{\text{vdW}}^{ij} = 4\varepsilon_{ij} \left[ \left(\frac{\sigma_{ij}}{r_{ij}}\right)^{12} - \left(\frac{\sigma_{ij}}{r_{ij}}\right)^{6} \right].$$

The two parameters $\sigma_{ij}$ and $\varepsilon_{ij}$ are the equilibrium distance and the well depth of the potential. These parameters are determined with empirical mixing rules for the interaction between atom $i$ and $j$:

$$\sigma_{ij} = \frac{\sigma_i + \sigma_j}{2} \text{ and } \varepsilon_{ij} = \sqrt{\varepsilon_i \varepsilon_j}$$

For ZIF-8, these parameters were taken from the DREIDING force field.[45] The 1-2 and 1-3 interactions were discarded to avoid a strong overestimation of the repulsion terms.

# S4 Grand canonical and canonical Monte Carlo results

## S4.1 Determining water saturation in ZIF-8

To determine the maximum number of water molecules per ZIF-8 unit cell, grand canonical Monte Carlo (GCMC) simulations were performed on the ambient-pressure (AP) phase at various water pressures. At the highest pressure, saturation is reached at about 80 water molecules per unit cell, or equivalently, 40 water molecules per cage (see Supplementary Figure 23 for a representative convergence plot).

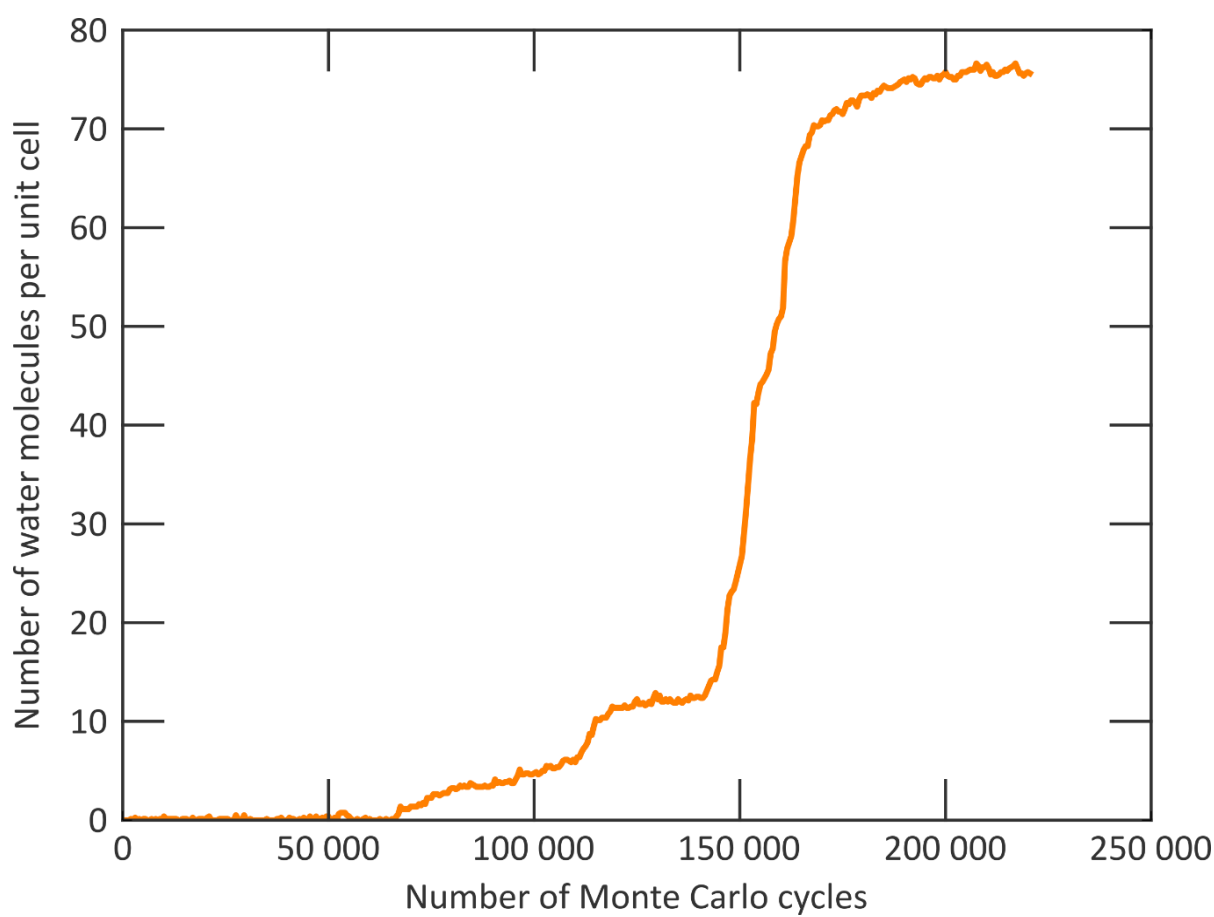

**Supplementary Figure 23 | Convergence of the number of water molecules as a function of the number of GCMC cycles.** Results are shown per conventional ZIF-8 unit cell, at a water pressure of 6 kPa and a temperature of 298 K.

## S4.2 Water distribution at different loadings

In addition to the water distribution determined from canonical MC simulations in Figure 3a of the main text, Supplementary Figure 24 to Supplementary Figure 26 provide the distributions for loadings of 4, 8, 20, 40, 60, and 80 water molecules per unit cell at 298 K, and both for the ambient pressure (AP) and high-pressure (HP) phase. These results were obtained by performing canonical MC simulations on a 2×2×2 ZIF-8 supercell at 298 K and afterwards translating the water molecules to one of the unit cells, so to speed up the convergence.

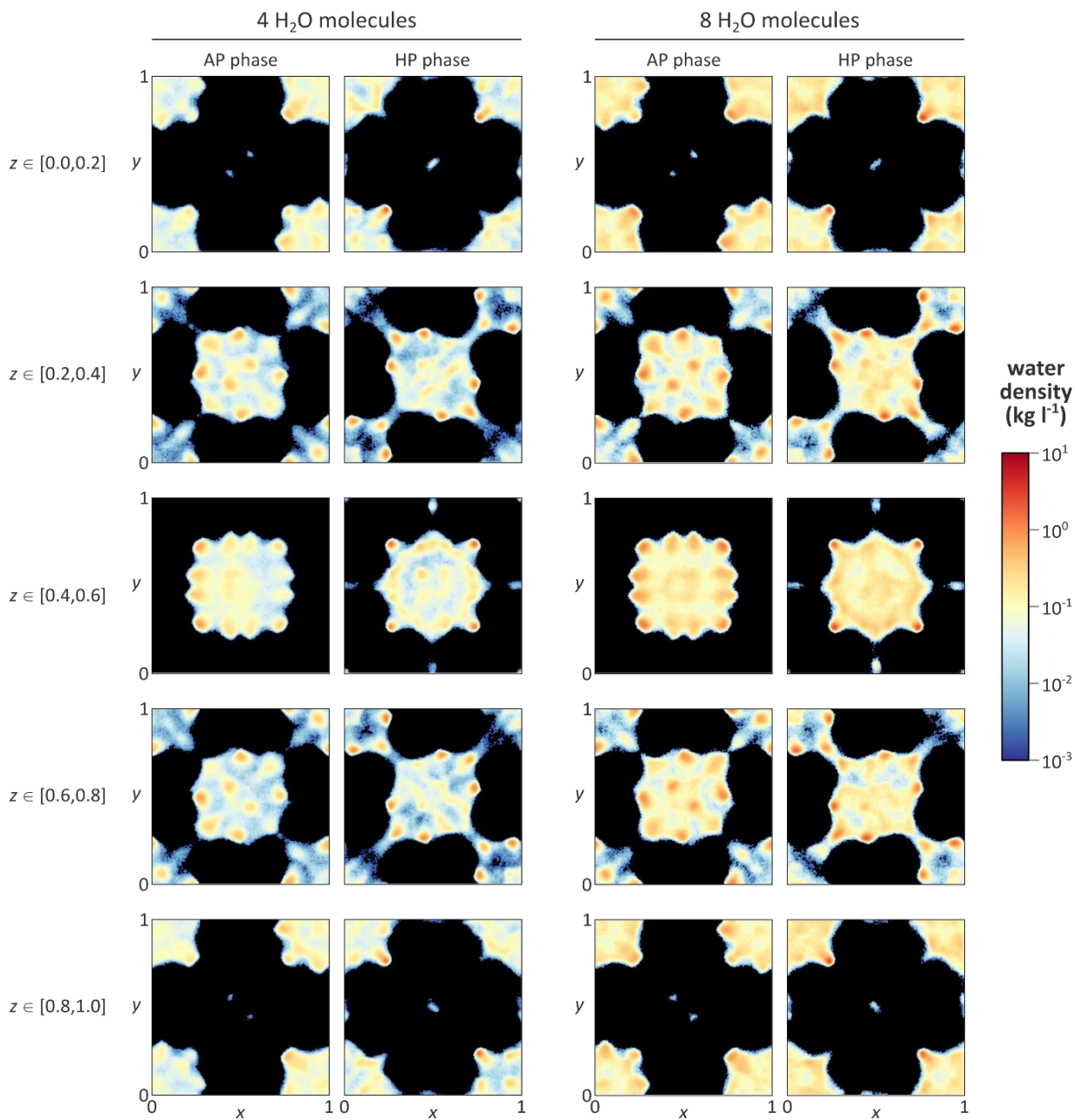

**Supplementary Figure 24 | Distribution of the water molecules in both the ambient pressure (AP) and high-pressure (HP) phase of ZIF-8 at low loading.** Water distributions are shown as a function of the fractional coordinates at loadings of either 4 or 8 water molecules per unit cell, as determined from canonical MC simulations at 300 K.

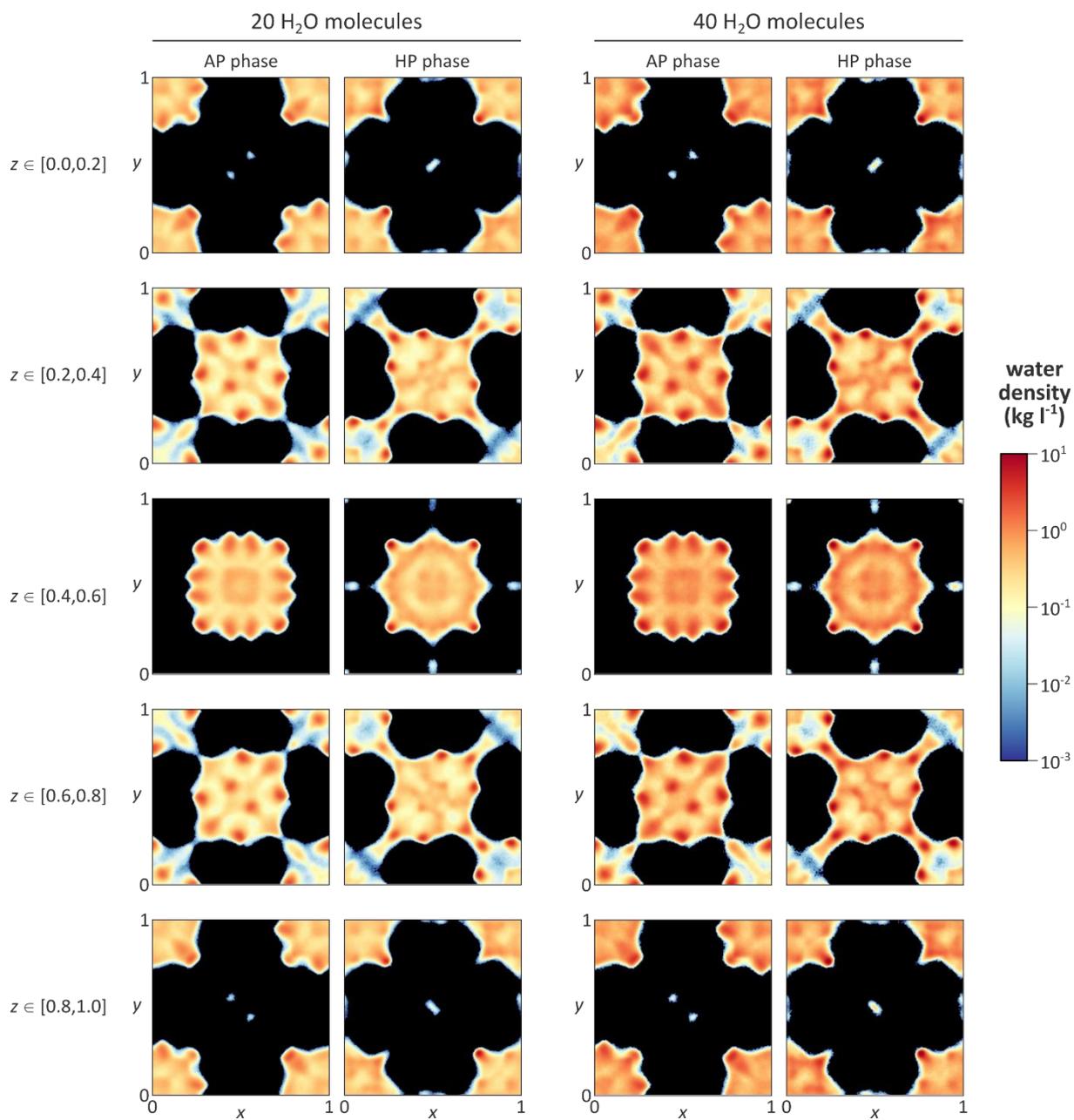

**Supplementary Figure 25 | Distribution of the water molecules in both the ambient pressure (AP) and high-pressure (HP) phase of ZIF-8 at intermediate loading**. Water distributions are shown at loadings of either 20 or 40 water molecules per unit cell, as determined from canonical MC simulations at 300 K.

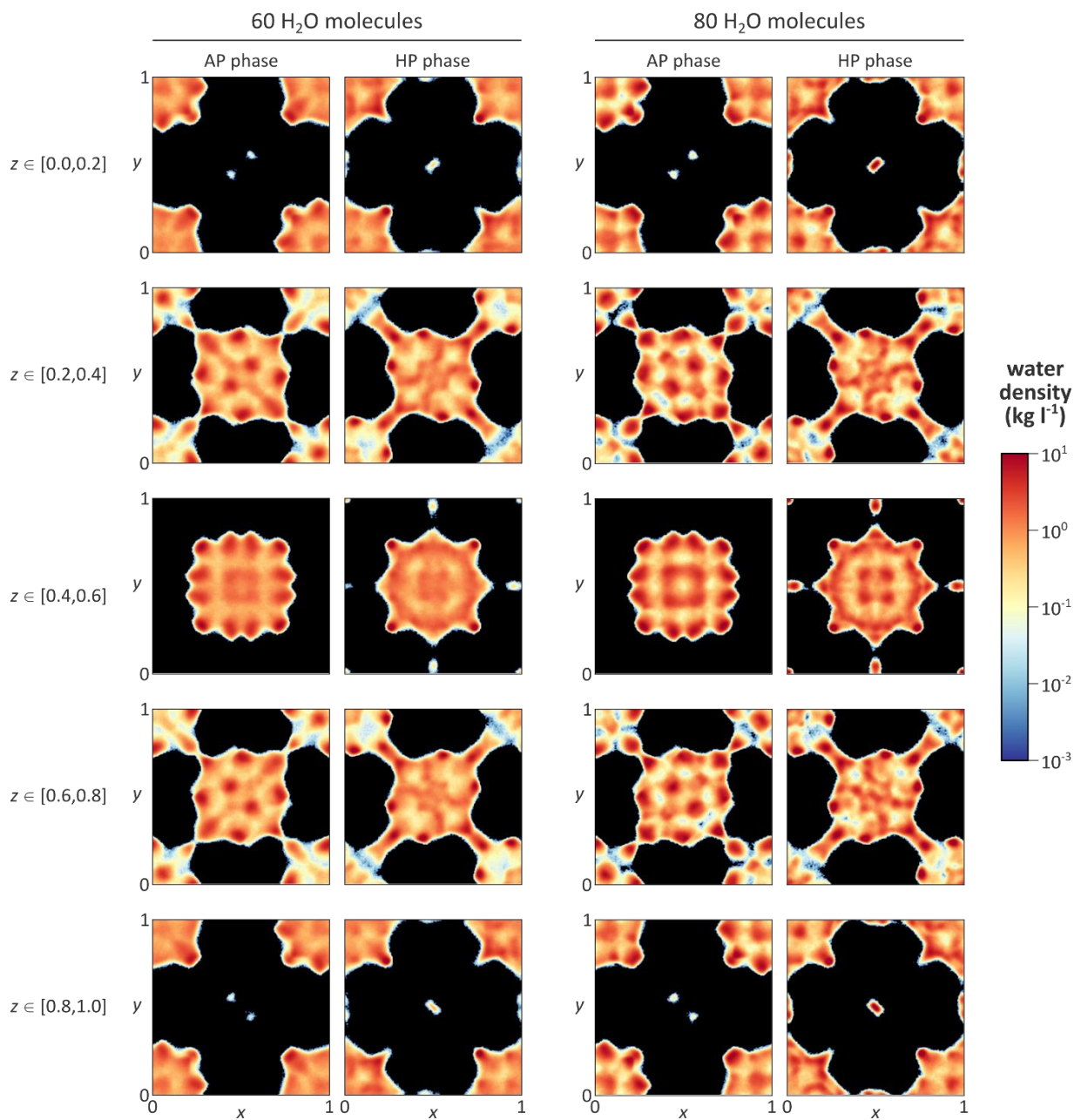

**Supplementary Figure 26 | Distribution of the water molecules in both the ambient pressure (AP) and high-pressure (HP) phase of ZIF-8 at high loading**. Water distributions are shown at loadings of either 60 or 80 water molecules per unit cell, as determined from canonical MC simulations at 300 K.

## S5 The ZIF-8 structure at different water loadings and pressures

### S5.1 Simulated powder X-ray diffraction patterns

To create the simulated powder X-ray diffraction pattern for each of the water loadings and mechanical pressures, the total $(N, P, \sigma_a = 0, T)$ simulation of 5 ns was divided into 20 equally sized subsimulations. For each subsimulation, the average structure was determined, and the PXRD of this average structure was determined using the freely available genXrdPattern program.[46] For this, the default copper Kα1 wavelength (1.54056 Å) was simulated, using a peak width of 0.572958 Å. The PXRD patterns reported in Supplementary Figure 27 to Supplementary Figure 32 are obtained by averaging over the 20 PXRD patterns of the subsimulations for each water loading and mechanical pressure separately. It is shown that with a higher water loading inside ZIF-8, the intensities of some peaks changes (*e.g.*, at 8°, 13°, 17°, 31°), which can be explained by the variation of the dihedral angles as certain reflection planes (*i.e.*, the Miller planes (110) and (211) at 2θ ~ 8° and 13°, respectively) intersect the mIm linkers of the 6MR aperture. However, importantly, the mechanical pressure has no influence on the XRD patterns.

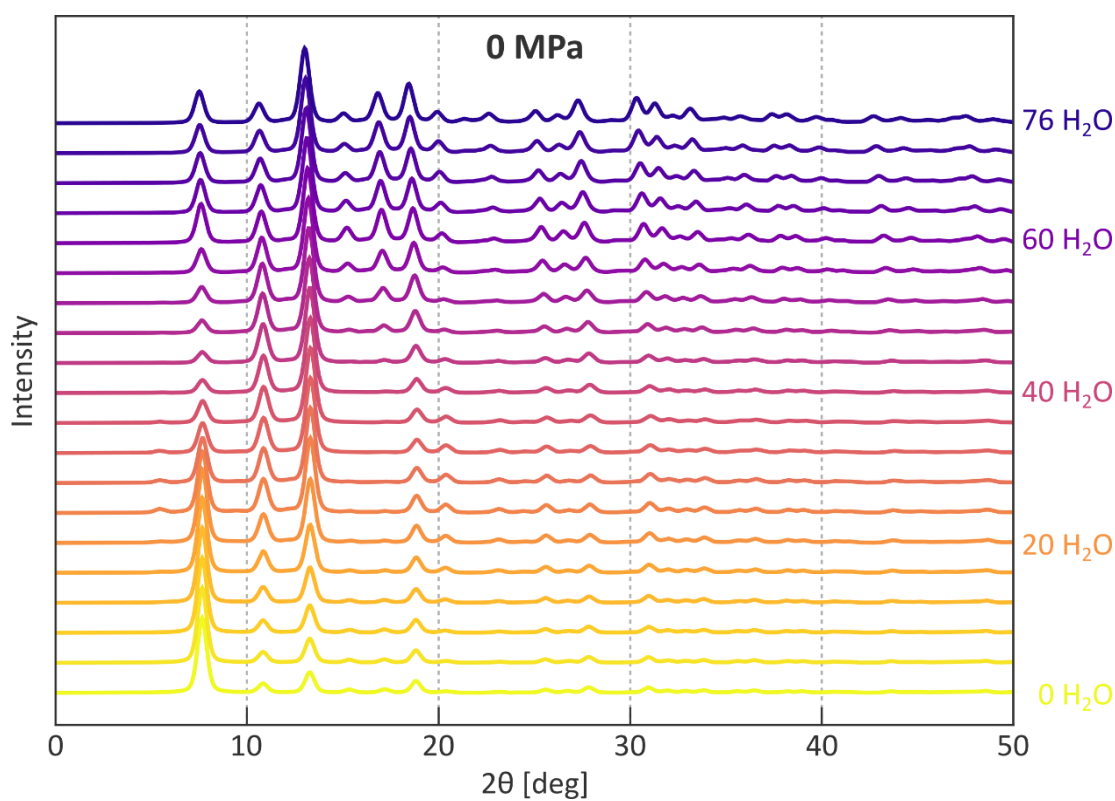

**Supplementary Figure 27 | Simulated powder X-ray diffraction patterns of ZIF-8 at different water loadings and at a mechanical pressure of 0 MPa.**

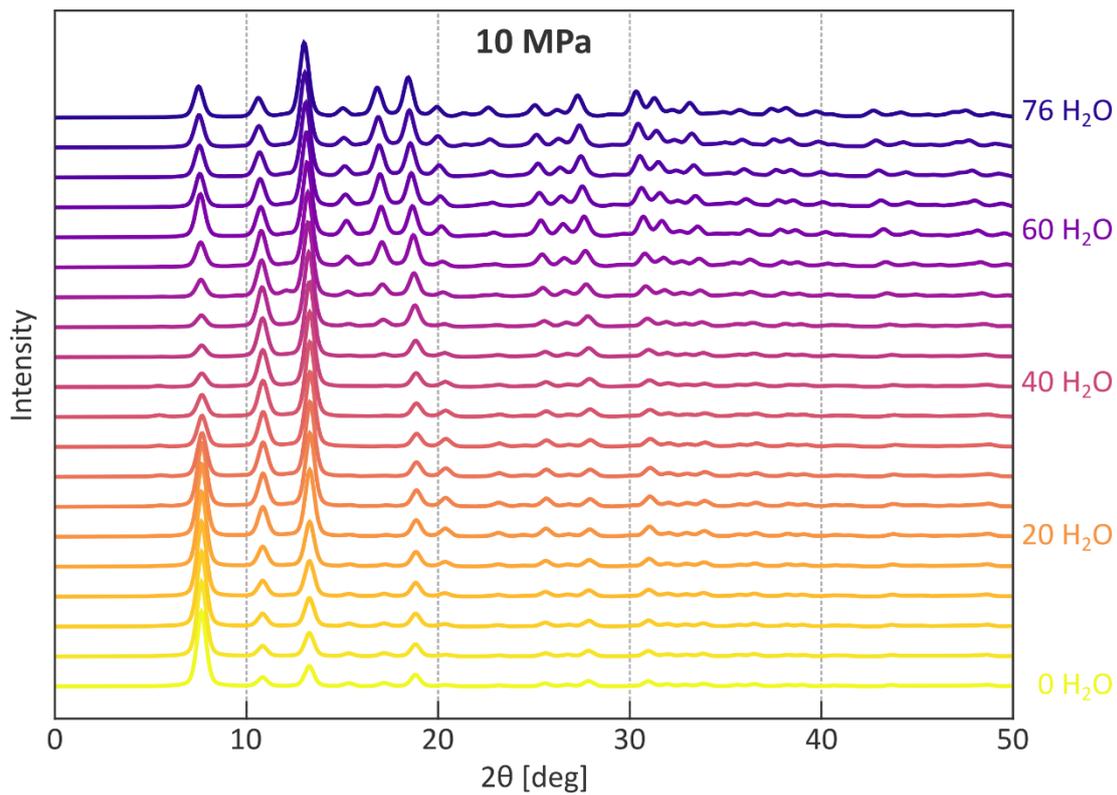

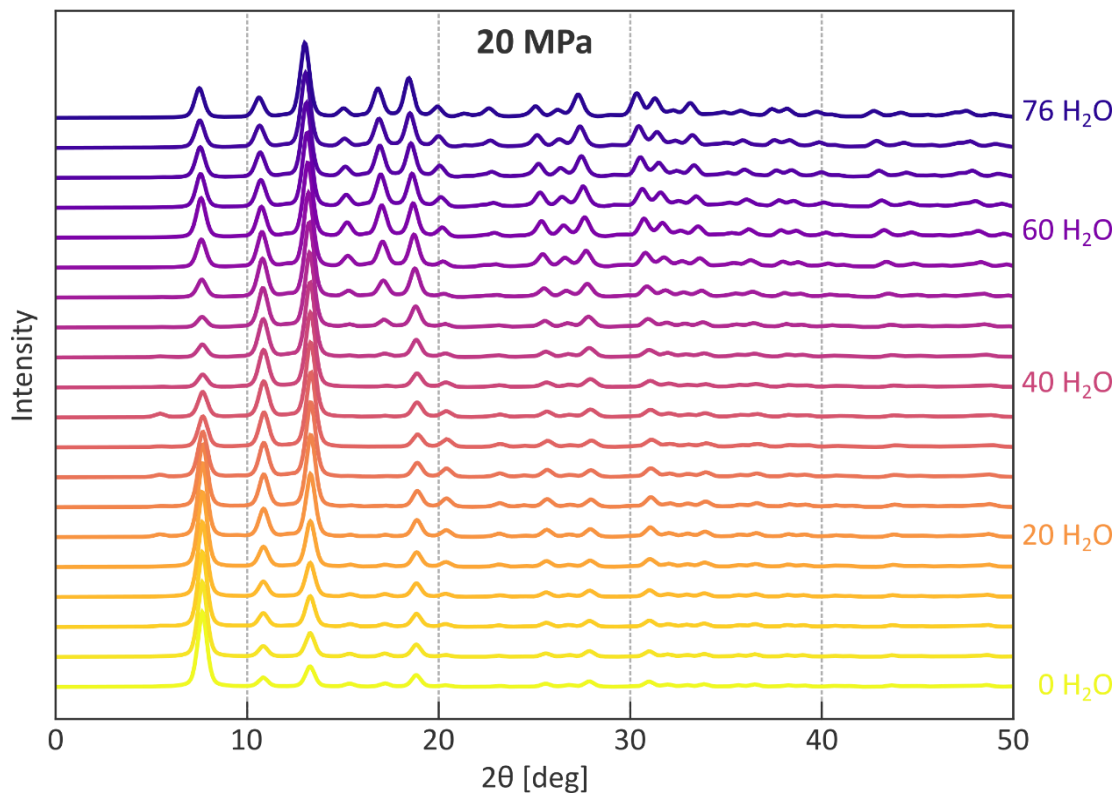

**Supplementary Figure 28 | Simulated powder X-ray diffraction patterns of ZIF-8 at different water loadings and at a mechanical pressure of either 10 MPa (top) or 20 MPa (bottom).**

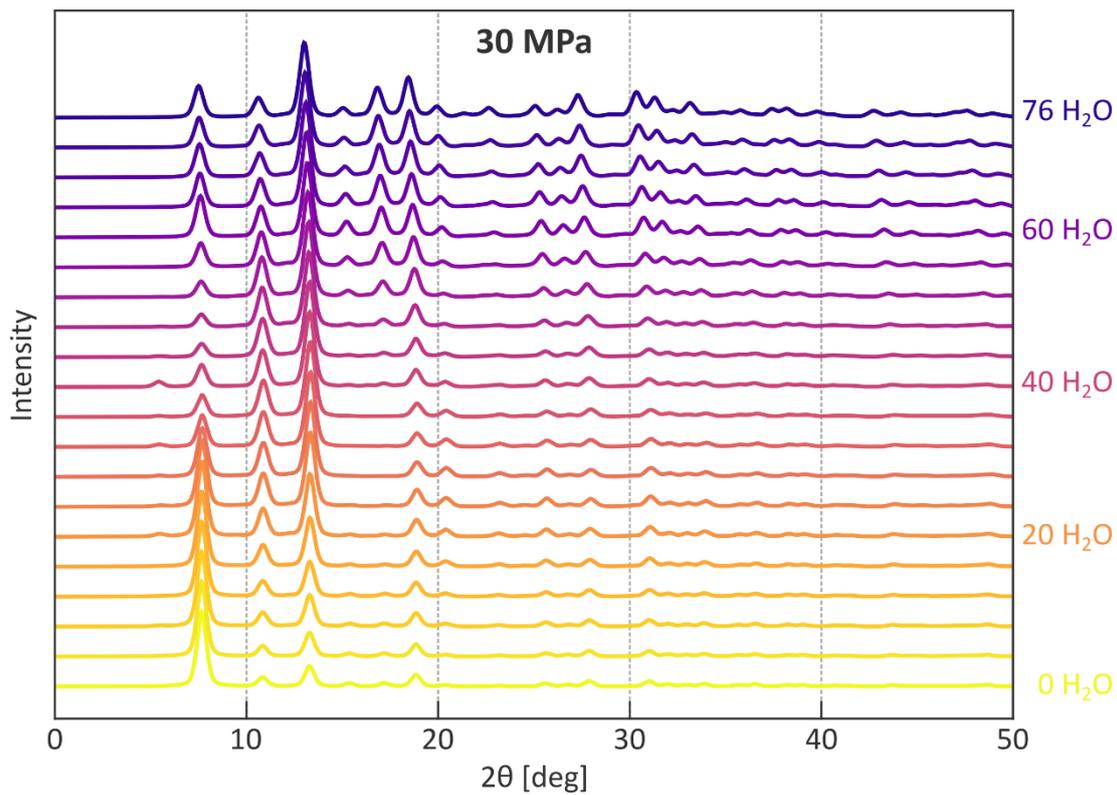

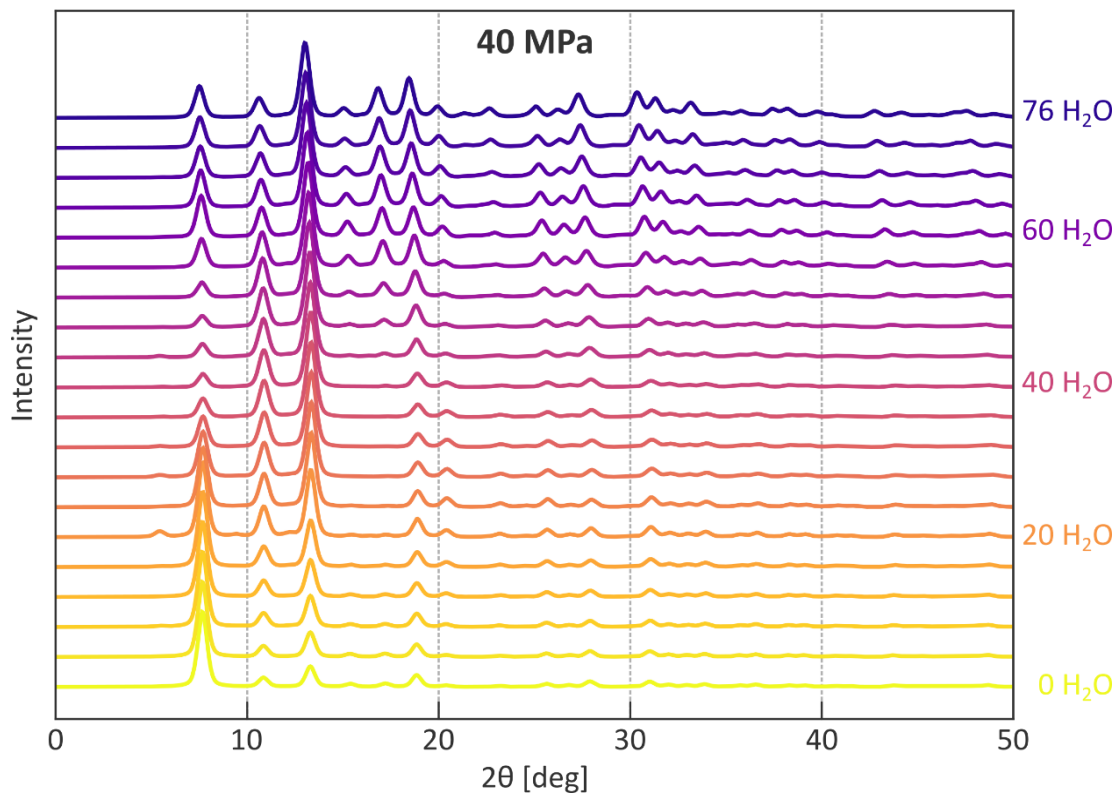

**Supplementary Figure 29 | Simulated powder X-ray diffraction patterns of ZIF-8 at different water loadings and at a mechanical pressure of either 30 MPa (top) or 40 MPa (bottom).**

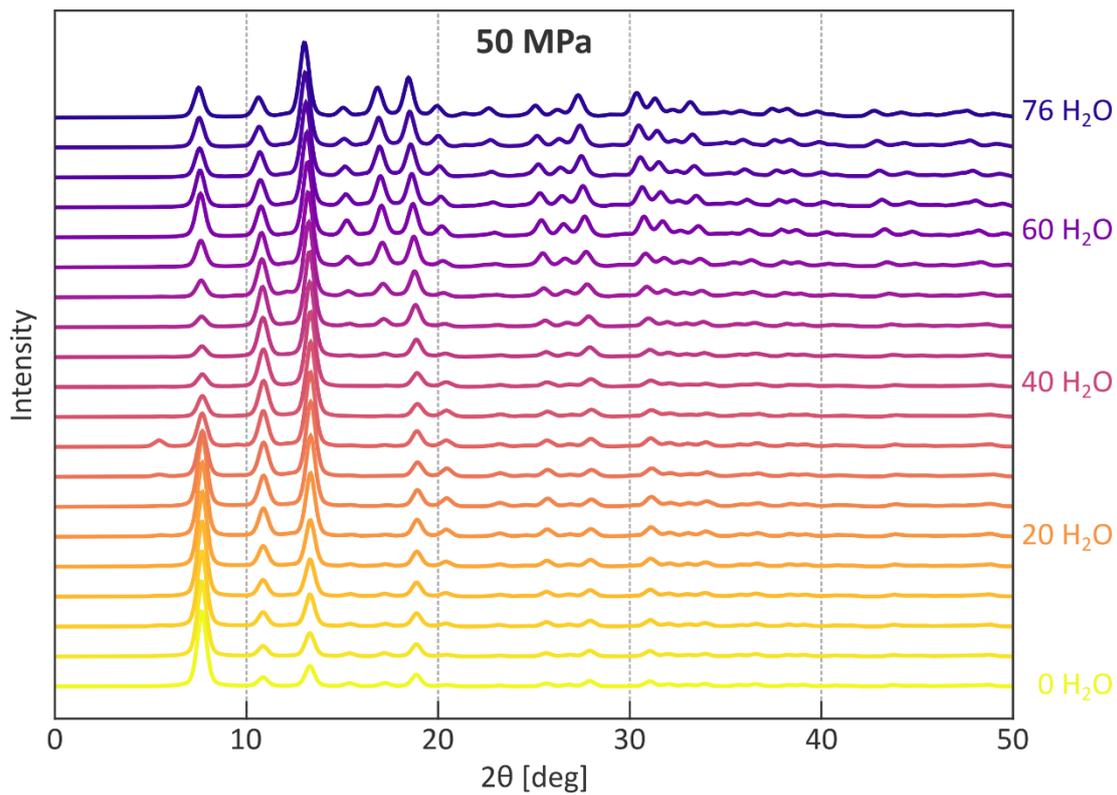
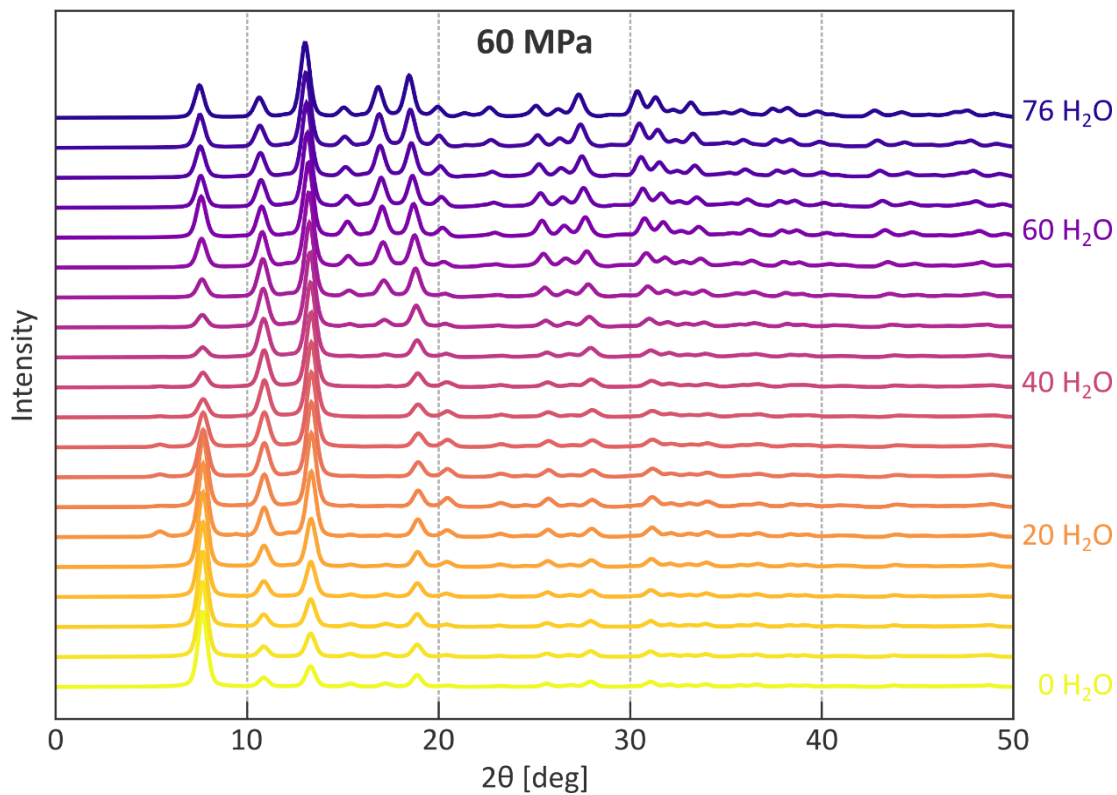

**Supplementary Figure 30 | Simulated powder X-ray diffraction patterns of ZIF-8 at different water loadings and at a mechanical pressure of either 50 MPa (top) or 60 MPa (bottom).**

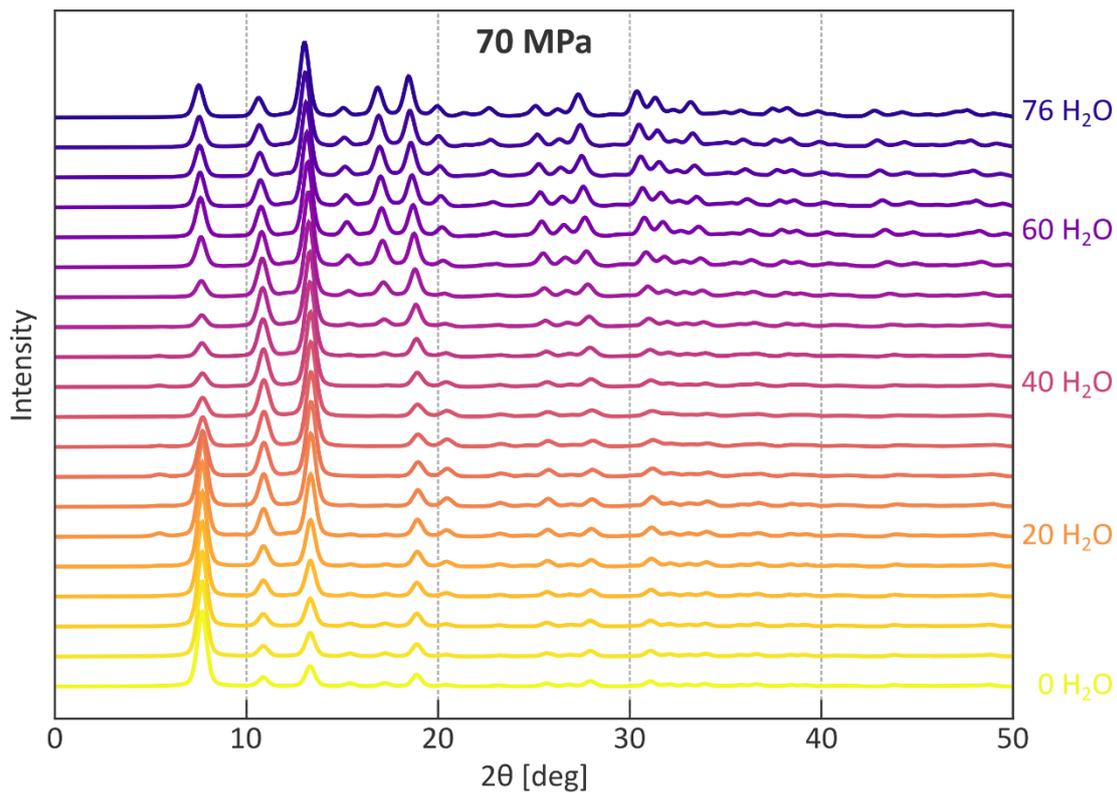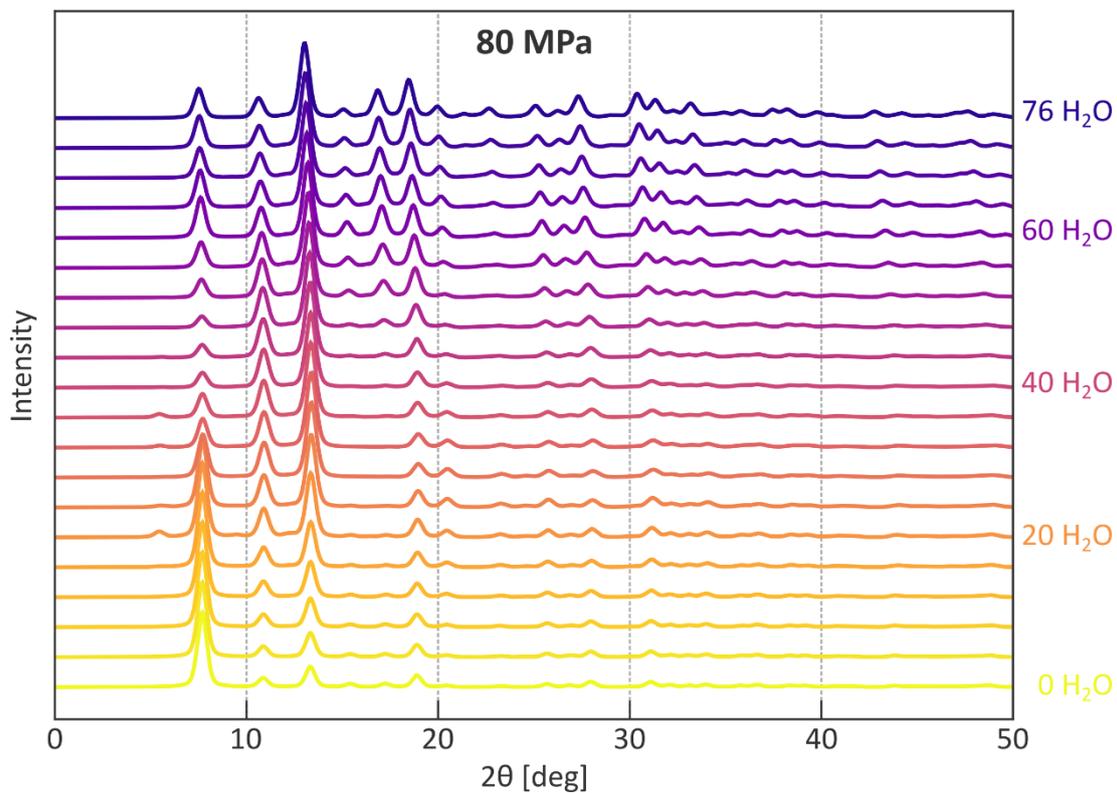

**Supplementary Figure 31 | Simulated powder X-ray diffraction patterns of ZIF-8 at different water loadings and at a mechanical pressure of either 70 MPa (top) or 80 MPa (bottom).**

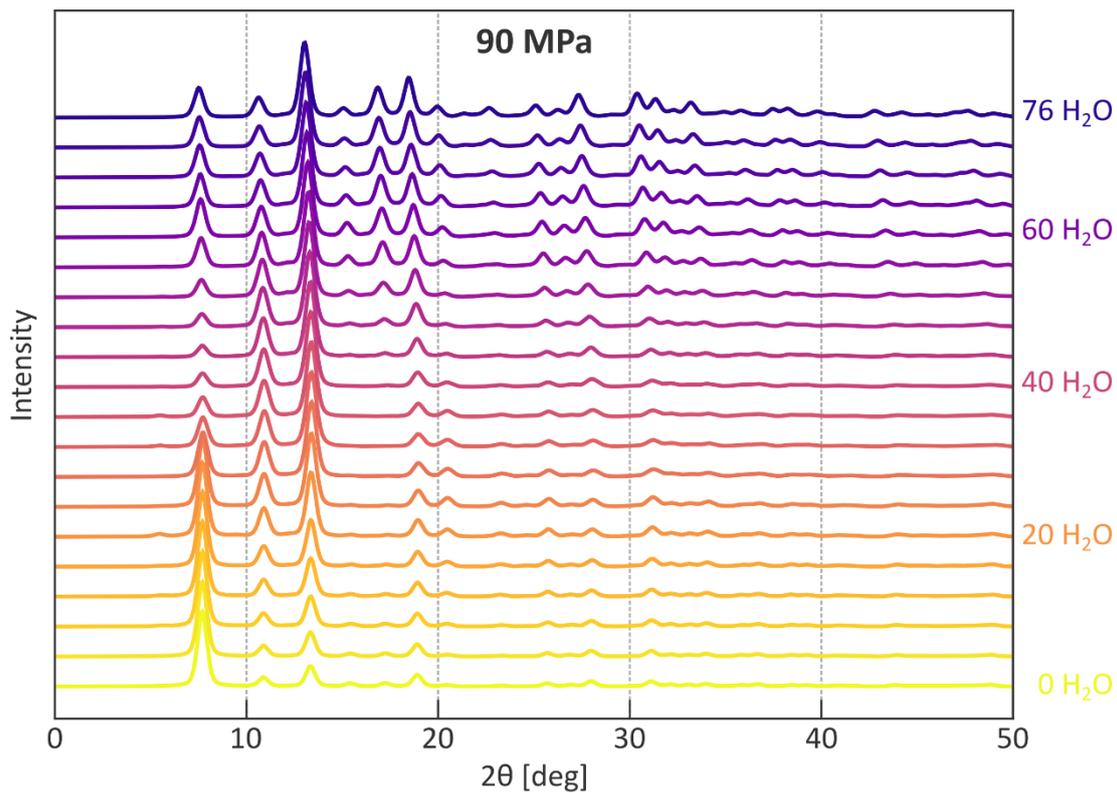

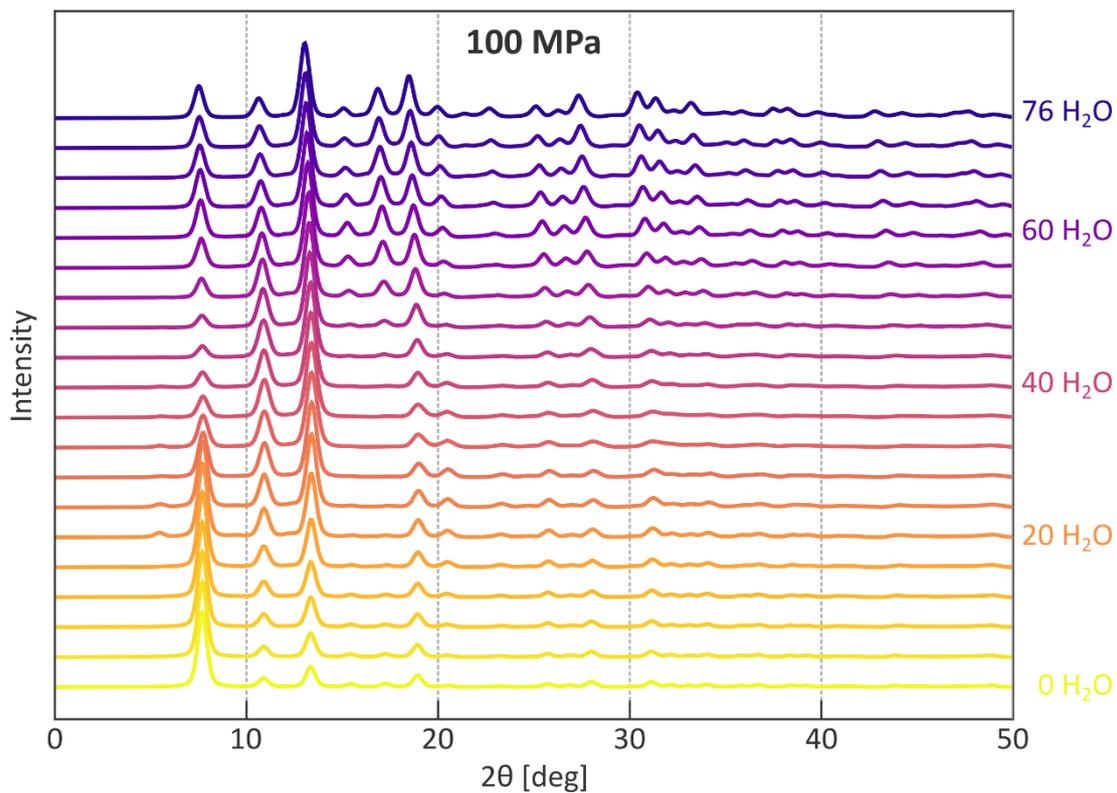

**Supplementary Figure 32 | Simulated powder X-ray diffraction patterns of ZIF-8 at different water loadings and at a mechanical pressure of either 90 MPa (top) or 100 MPa (bottom).**

## S5.2 Distribution of the swing angle

In Supplementary Figure 33 to Supplementary Figure 36, the distributions of the dihedral swing angles during the 5 ns $(N, P, \sigma_a = 0, T)$ simulation at 300 K and at various pressures are reported as a function of the water loading. They show that, irrespective of the mechanical pressure, water does not induce gate opening in ZIF-8 but rather even further closes the gate.

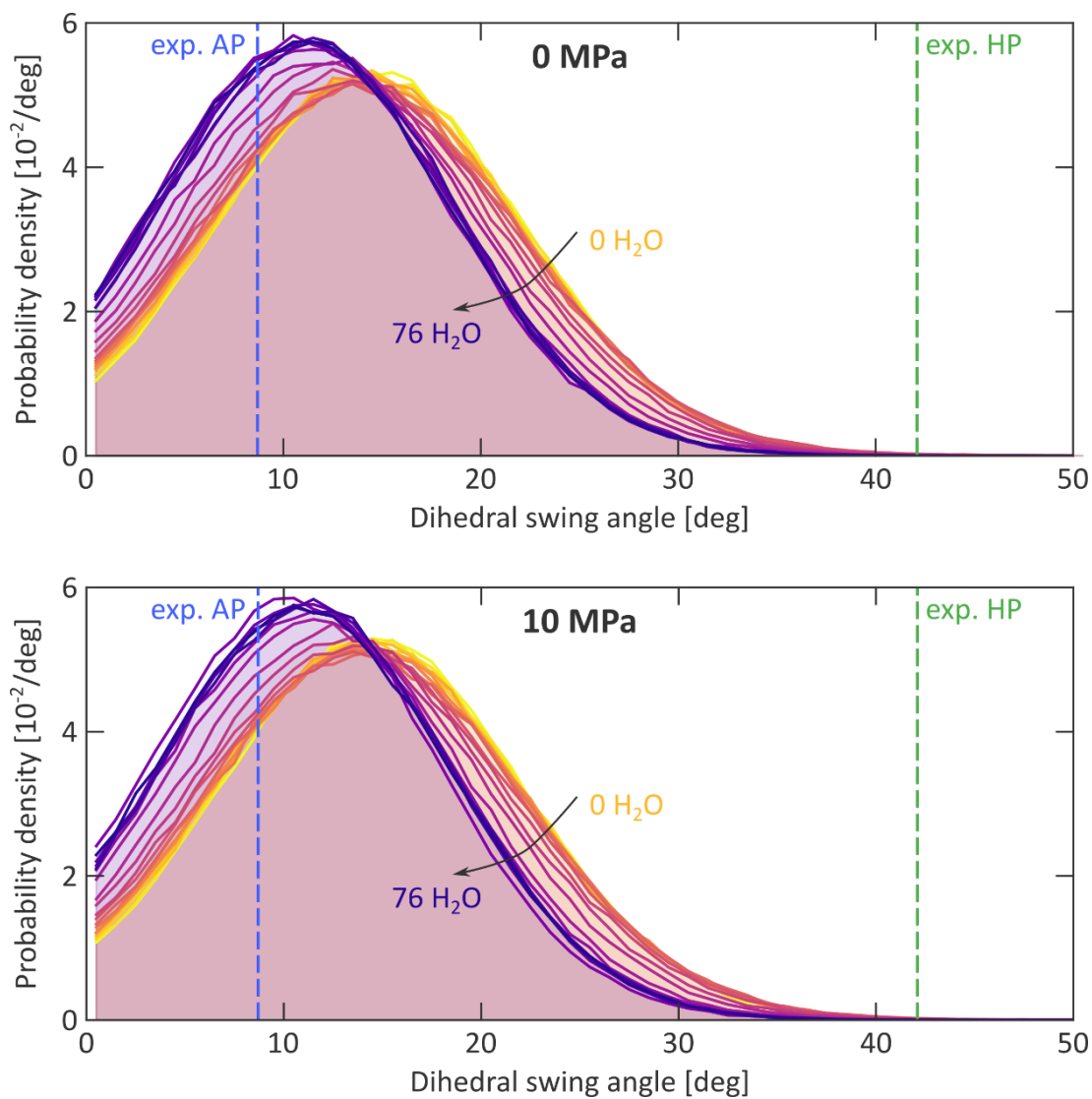

**Supplementary Figure 33 | Simulated distribution of the dihedral swing angle in ZIF-8 at 300 K and at various water loadings, at a mechanical pressure of either 0 MPa (top) or 10 MPa (bottom).** The striped lines indicate the experimental AP and HP phases.[47]

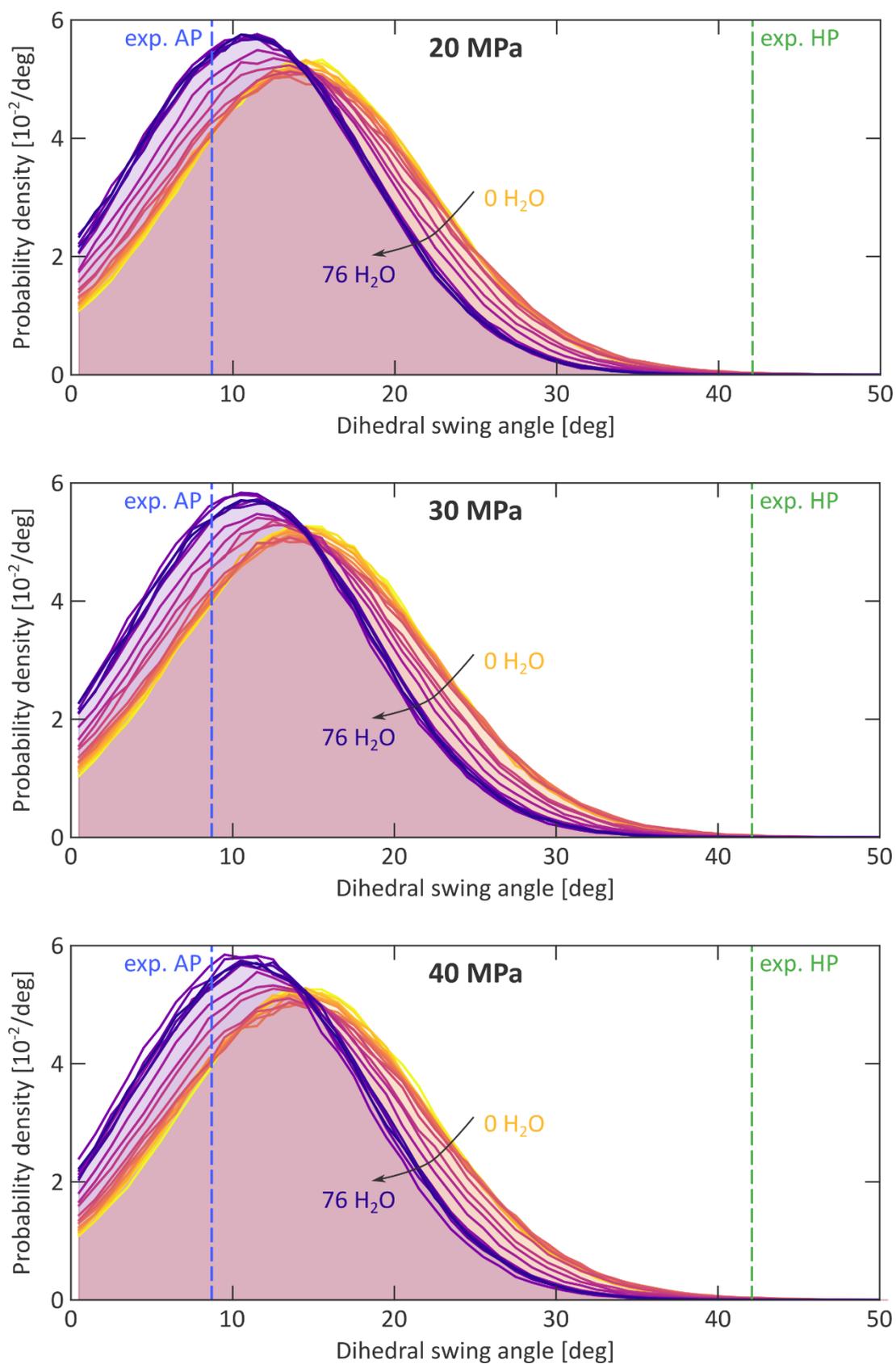

**Supplementary Figure 34 | Simulated distribution of the dihedral swing angle in ZIF-8 at 300 K and at various water loadings, at a mechanical pressure of either 20 MPa (top), 30 MPa (middle), or 40 MPa (bottom).** The striped lines indicate the experimental AP and HP phases.[47]

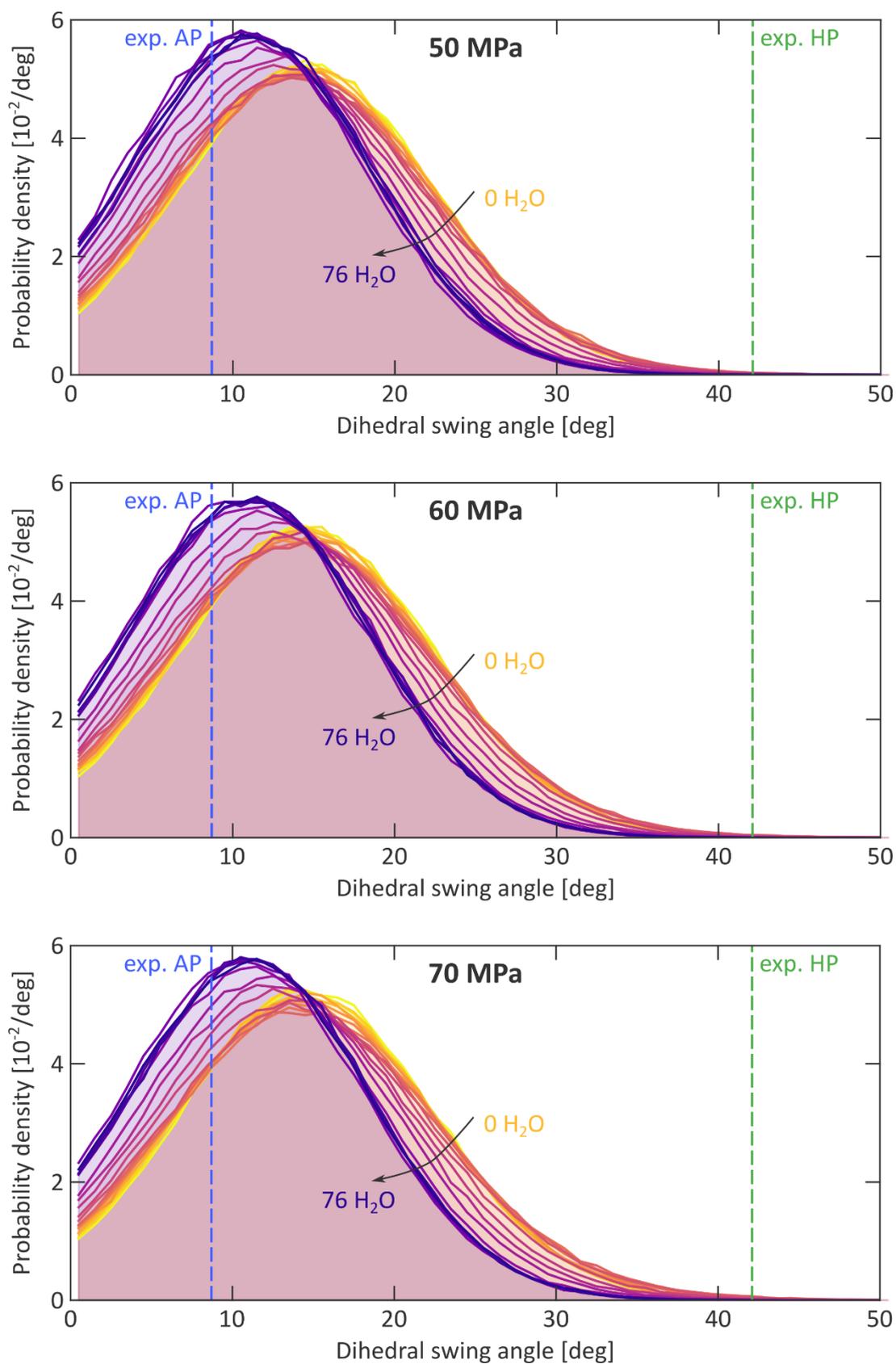

**Supplementary Figure 35 | Simulated distribution of the dihedral swing angle in ZIF-8 at 300 K and at various water loadings, at a mechanical pressure of either 50 MPa (top), 60 MPa (middle), or 70 MPa (bottom).** The striped lines indicate the experimental AP and HP phases.[47]

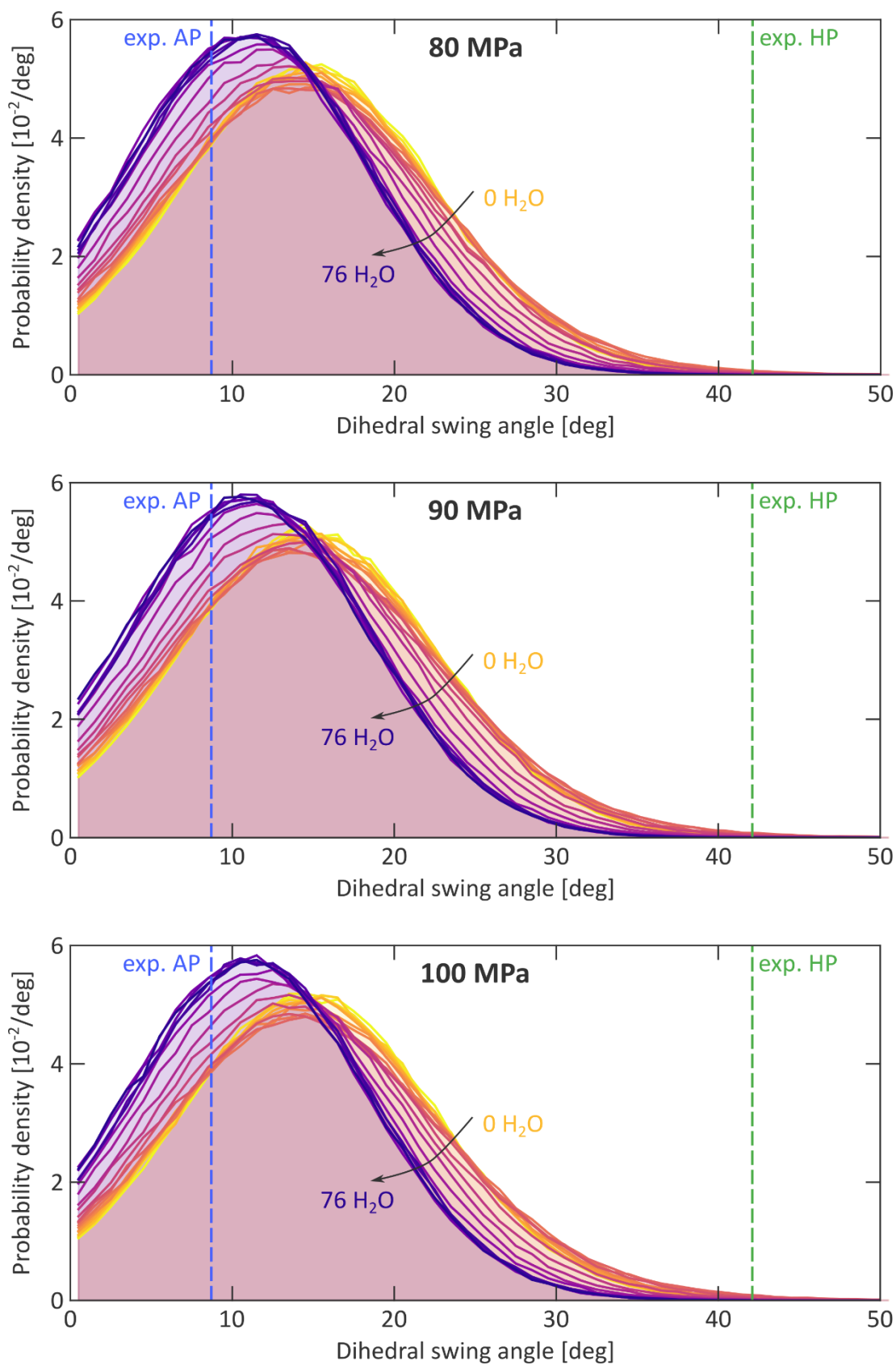

**Supplementary Figure 36 | Simulated distribution of the dihedral swing angle in ZIF-8 at 300 K and at various water loadings, at a mechanical pressure of either 80 MPa (top), 90 MPa (middle), or 100 MPa (bottom).** The striped lines indicate the experimental AP and HP phases.[47]

# S6 Swing angle distribution from *ab initio* molecular dynamics simulations

Supplementary Figure 37 reports the distribution of the dihedral swing angles in a conventional and empty ZIF-8 unit cell during a 10 ps *ab initio* $(N, P, \boldsymbol{\sigma}_a = \mathbf{0}, T)$ MD simulation at 0 MPa and at a temperature of 100 K, 200 K, and 300 K at the PBE-D3(BJ) level of theory.

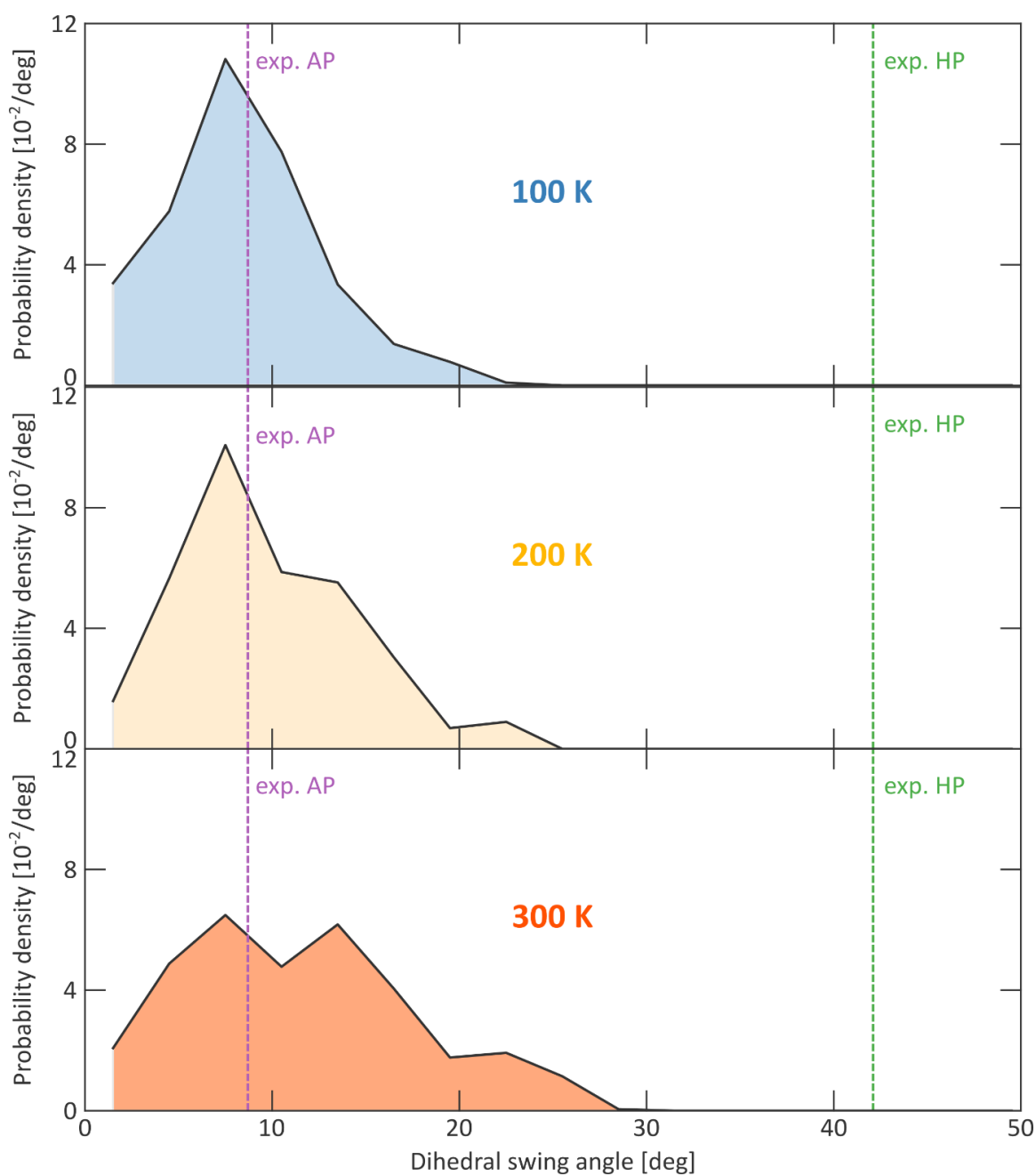

**Supplementary Figure 37 | Probability density of the dihedral swing angle of ZIF-8 at different temperatures.** Probability densities obtained during a 10 ps *ab initio* molecular dynamics simulation at 0 MPa and either 100 K, 200 K, or 300 K at the PBE-D3(BJ) level of theory. The swing angles in the experimental ambient pressure (AP) and high-pressure (HP) phases are indicated as well.[47]

# S7 The distribution and mobility of confined water in ZIF-8 in equilibrium

## S7.1 Distribution of water inside the cages

To investigate the distribution of water inside the cages of ZIF-8 at different water loadings and mechanical pressures, radial distribution functions (RDFs) were constructed. To this end, after each 5 ns $(N, P, \boldsymbol{\sigma}_a = \mathbf{0}, T)$ simulation, a virtual atom was placed in the centres of each of the 6-membered ring (6MR) cage window apertures, resulting in eight such virtual positions. Afterwards, two RDFs were constructed. The first RDF describes the pairs consisting of one of these virtual positions and the oxygens of the water molecules, while the second RDF describes each pair of virtual positions.

The RDFs reported in Supplementary Figure 38 to Supplementary Figure 48 are normalised on the cell volume and the volume of a thin spherical cell with radius $r$ and thickness $\Delta r$: $4\pi r^2 \Delta r$. In this expression, the thickness $\Delta r$ coincides with the radial spacing of 0.02 Å used in the RDF. In contrast to RDFs constructed between physical atoms, for which $r = 0$ is not possible, the oxygen atoms of the water molecules can momentarily reside at the centre of the 6MR aperture; in this case, the RDF would show a peak at $r = 0$ that would theoretically approach infinity (see for instance the 0 MPa simulation with 60 water molecules). The absence of such a peak therefore indicates that the water molecules did not visit the centre of the 6MR window during this simulation.

Finally, as the ZIF-8 cell length amounts to about 16.6 Å at 300 K and 0 MPa, these RDFs should be limited to a maximum pairwise distance of 8.3 Å – halve of the smallest in-plane distance. While in most simulations the water molecules do not visit the 6MR apertures located at $r = 0$, confirming the hydrophobicity of the apertures, we have chosen to extend the RDFs to 10 Å. Consequently, it is possible to visualise the location of the neighbouring 6MR aperture centres (the orange shade shown in each figure) and the corresponding water distribution, which shows a local minimum at these locations due to the hydrophobicity of the apertures.

**0 MPa**

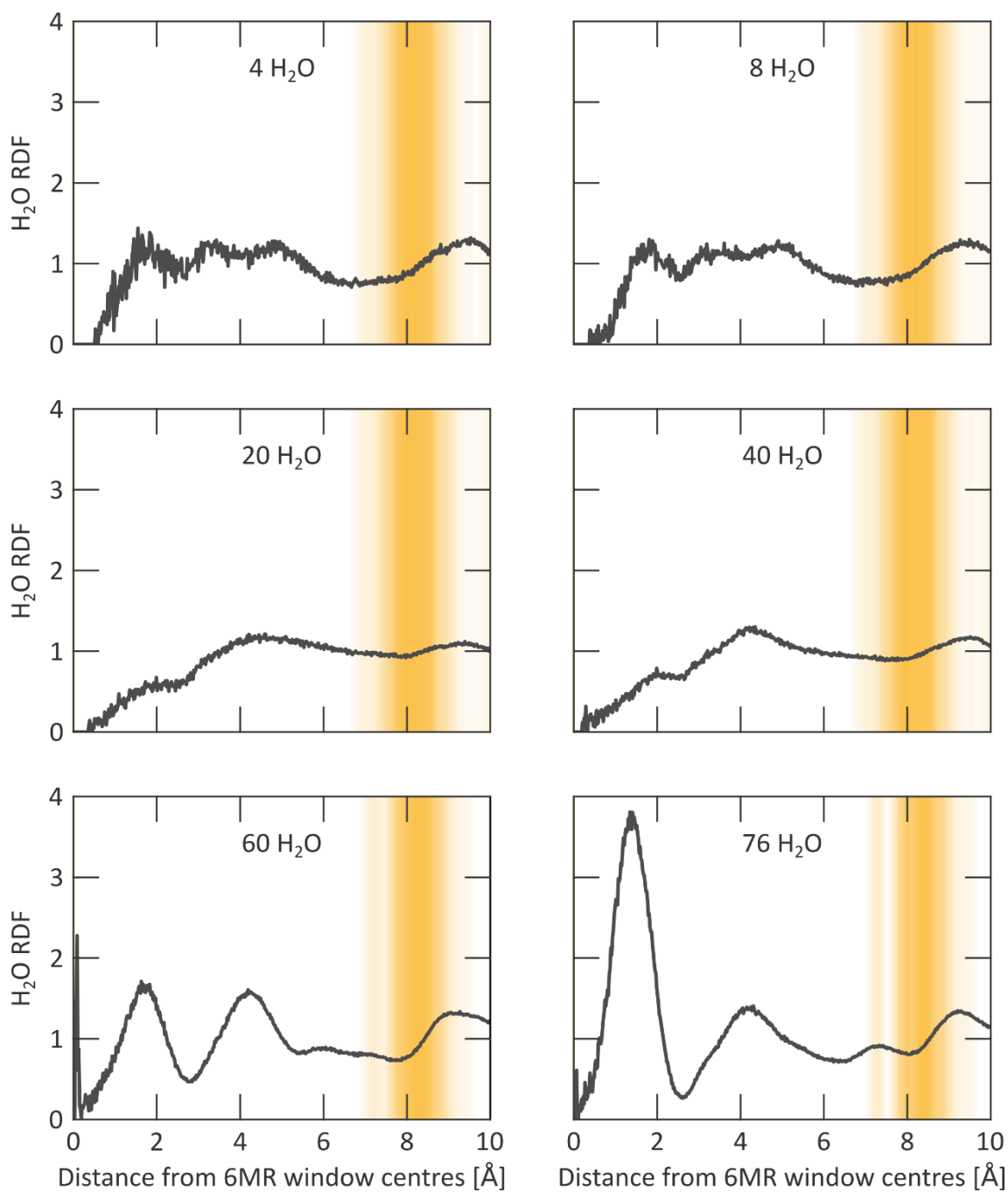

**Supplementary Figure 38 | Radial distribution functions at various water loadings at 300 K and 0 MPa.** RDFs between (i) the centres of the 6MR apertures and the oxygens of the water molecules (grey trace) and (ii) the centres of the 6MR apertures (orange shade).

**10 MPa**

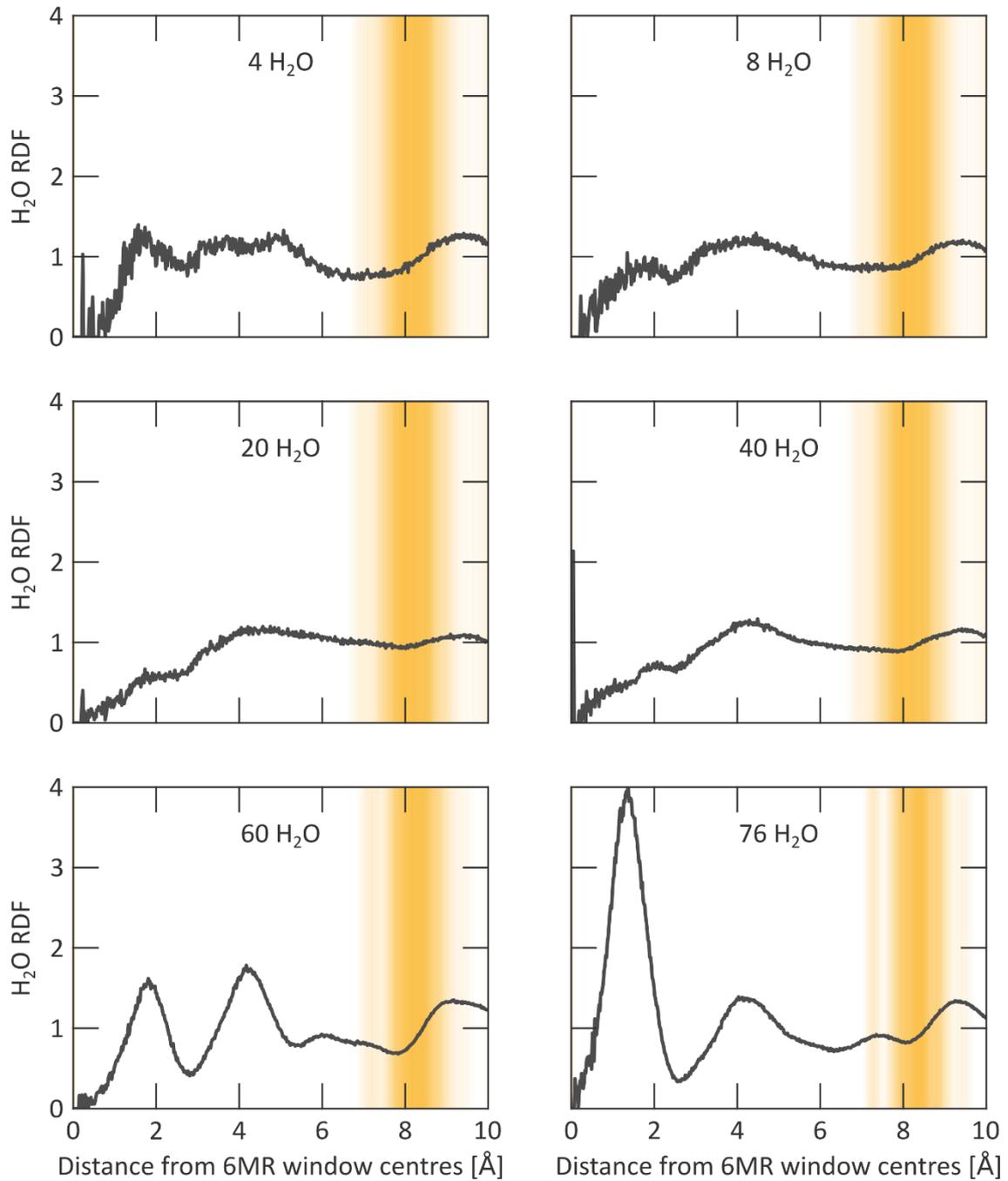

**Supplementary Figure 39 | Radial distribution functions at various water loadings at 300 K and 10 MPa.** RDFs between (i) the centres of the 6MR apertures and the oxygens of the water molecules (grey trace) and (ii) the centres of the 6MR apertures (orange shade).

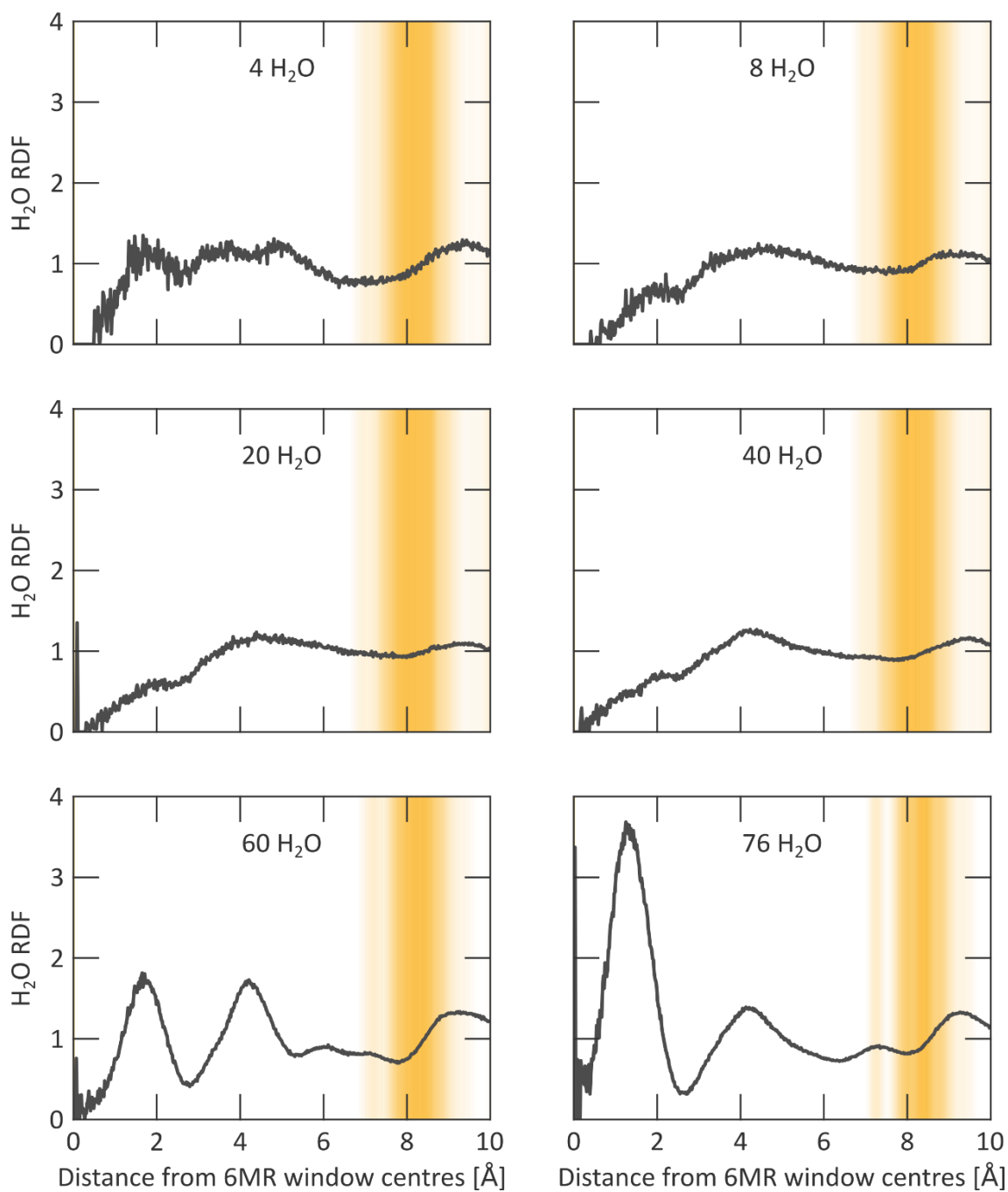

**Supplementary Figure 40 | Radial distribution functions at various water loadings at 300 K and 20 MPa.** RDFs between (i) the centres of the 6MR apertures and the oxygens of the water molecules (grey trace) and (ii) the centres of the 6MR apertures (orange shade).

**30 MPa**

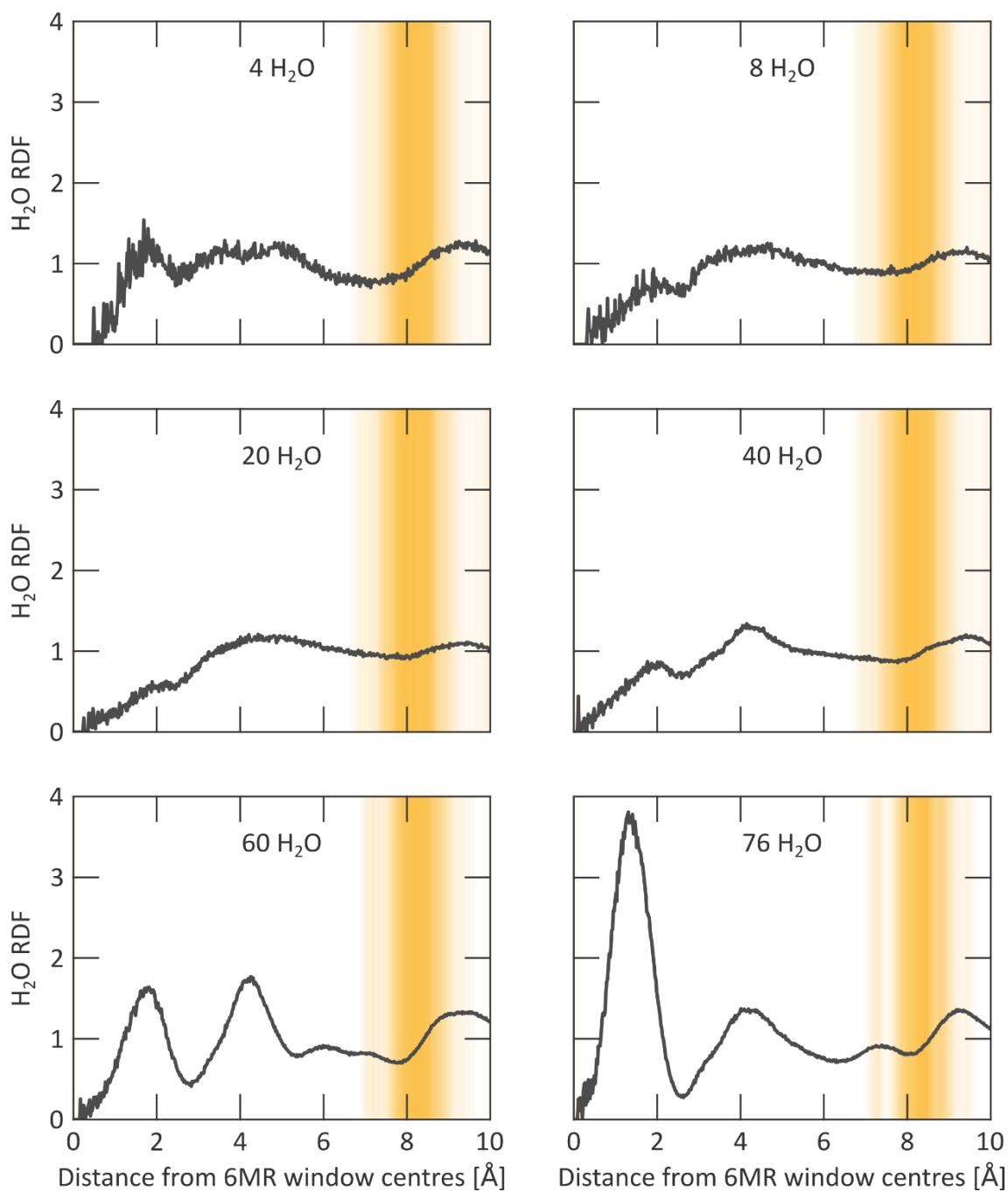

**Supplementary Figure 41 | Radial distribution functions at various water loadings at 300 K and 30 MPa.** RDFs between (i) the centres of the 6MR apertures and the oxygens of the water molecules (grey trace) and (ii) the centres of the 6MR apertures (orange shade).

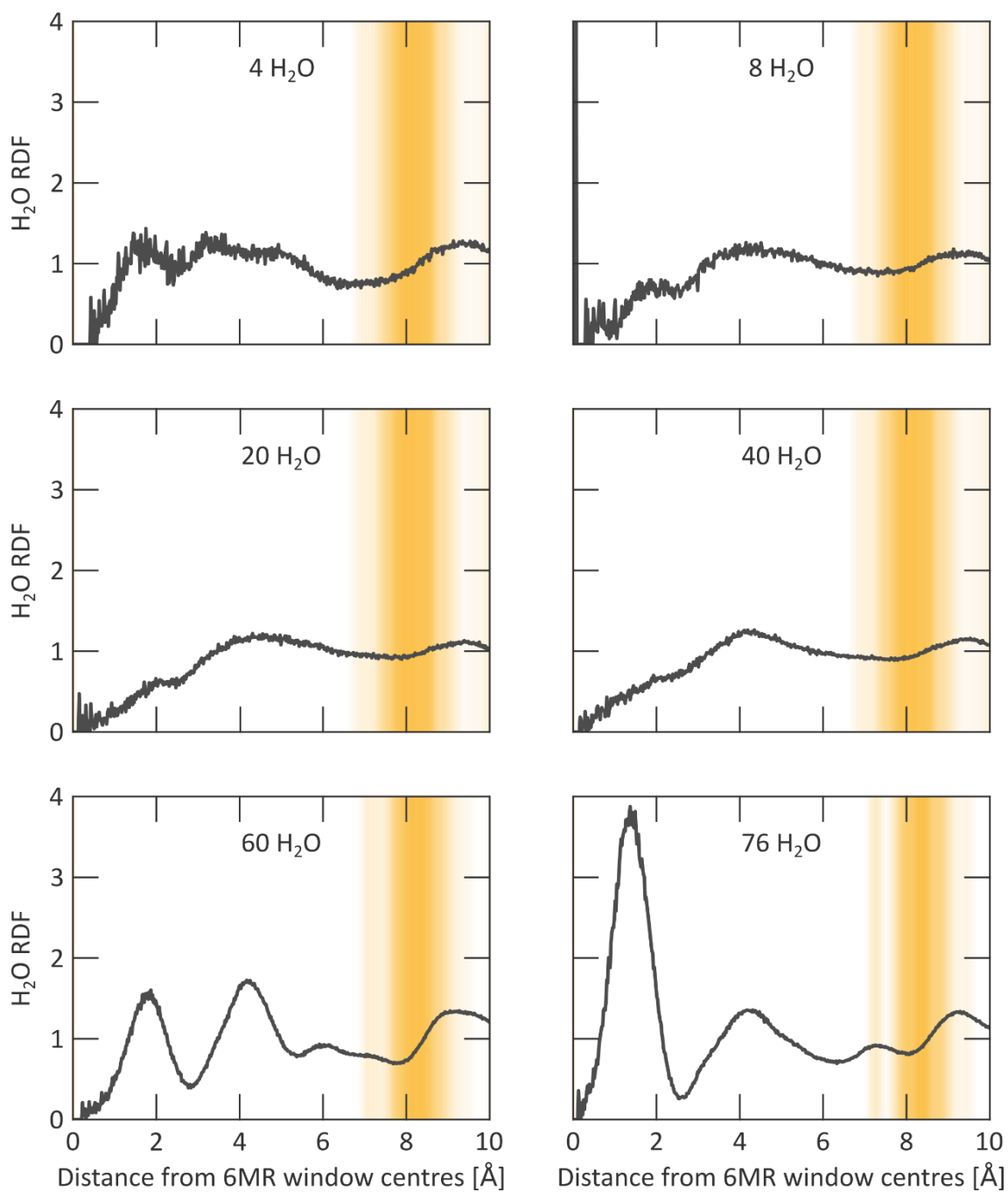

**Supplementary Figure 42 | Radial distribution functions at various water loadings at 300 K and 40 MPa.** RDFs between (i) the centres of the 6MR apertures and the oxygens of the water molecules (grey trace) and (ii) the centres of the 6MR apertures (orange shade).

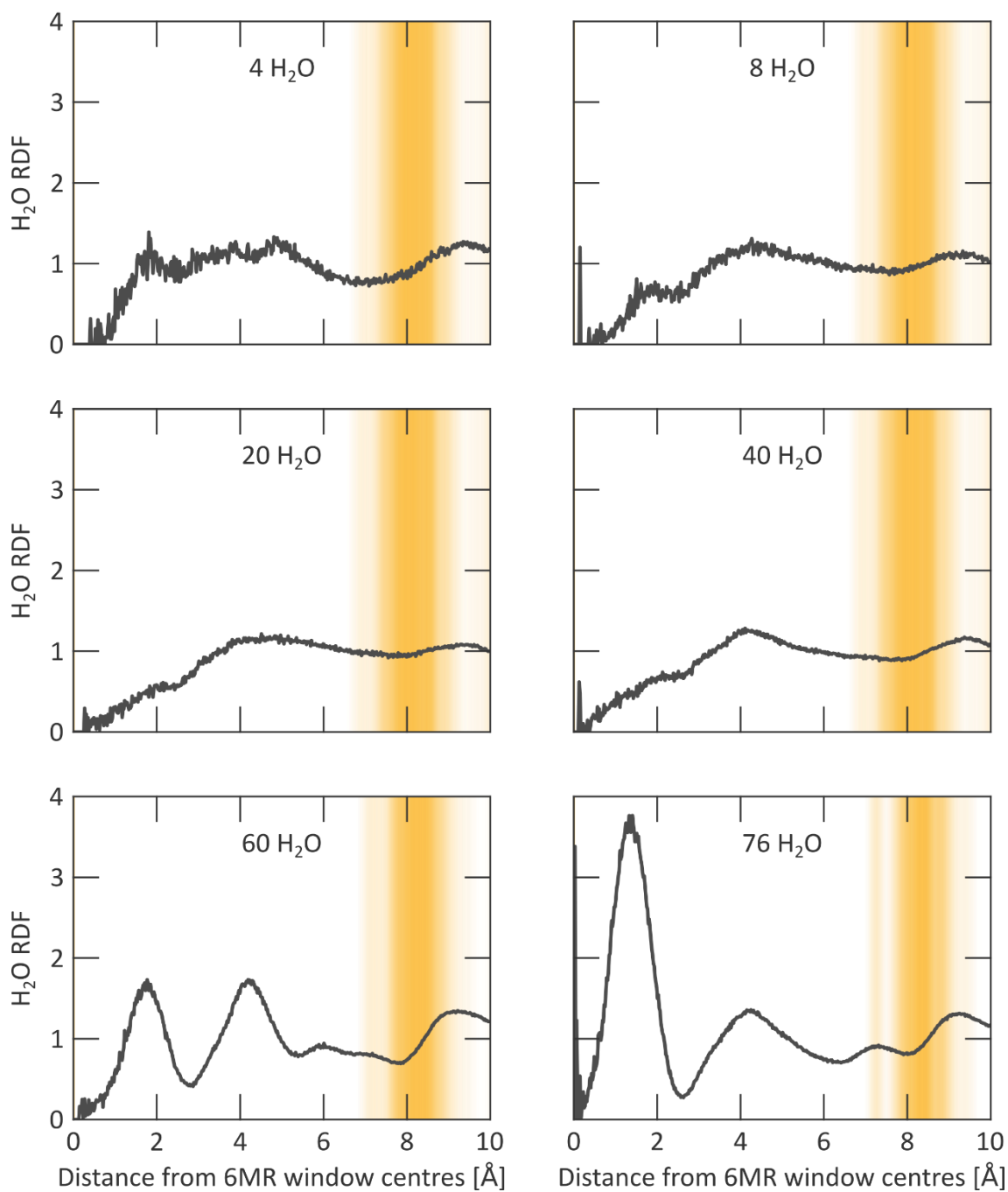

**Supplementary Figure 43 | Radial distribution functions at various water loadings at 300 K and 50 MPa.** RDFs between (i) the centres of the 6MR apertures and the oxygens of the water molecules (grey trace) and (ii) the centres of the 6MR apertures (orange shade).

**60 MPa**

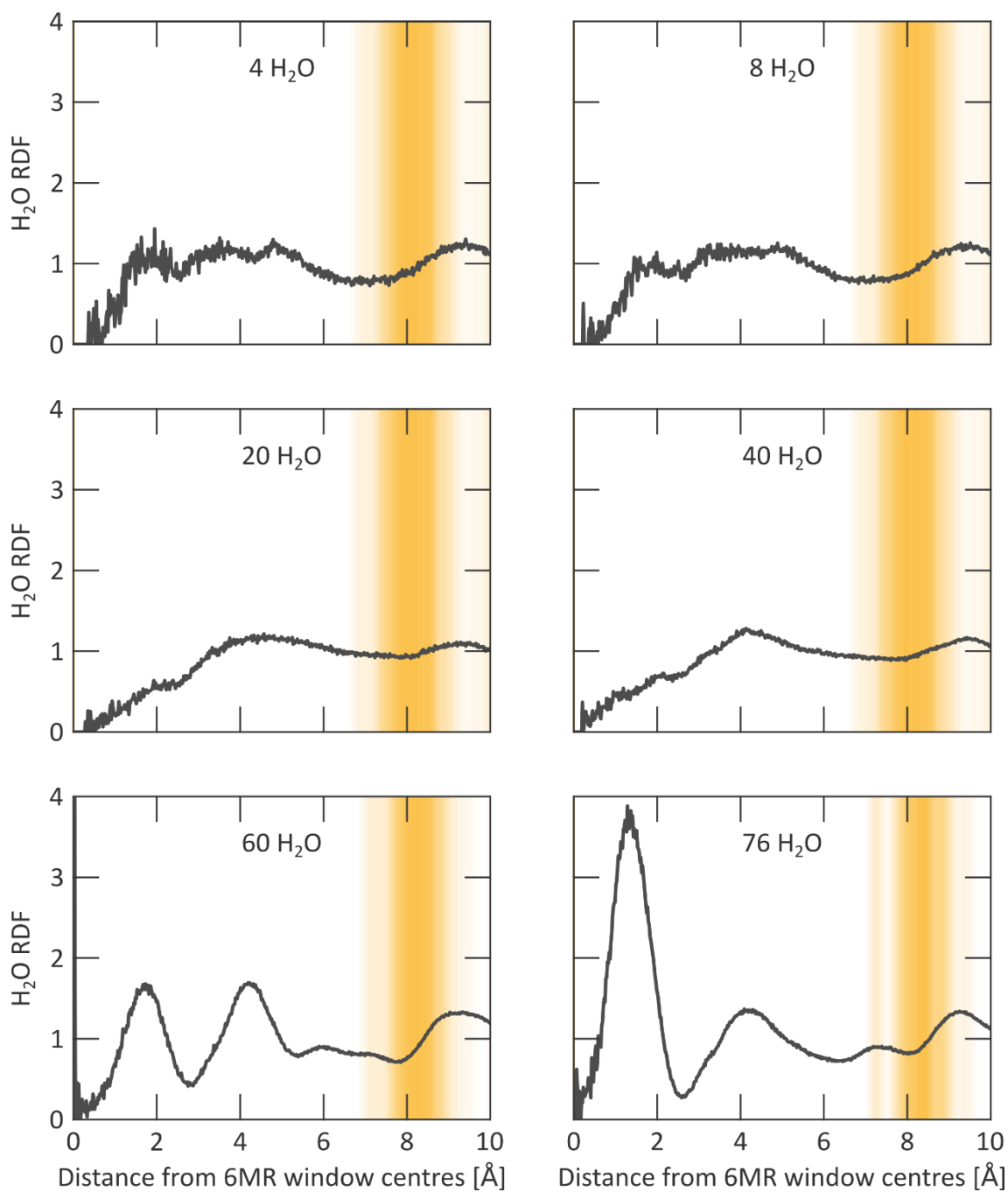

**Supplementary Figure 44 | Radial distribution functions at various water loadings at 300 K and 60 MPa.** RDFs between (i) the centres of the 6MR apertures and the oxygens of the water molecules (grey trace) and (ii) the centres of the 6MR apertures (orange shade).

**70 MPa**

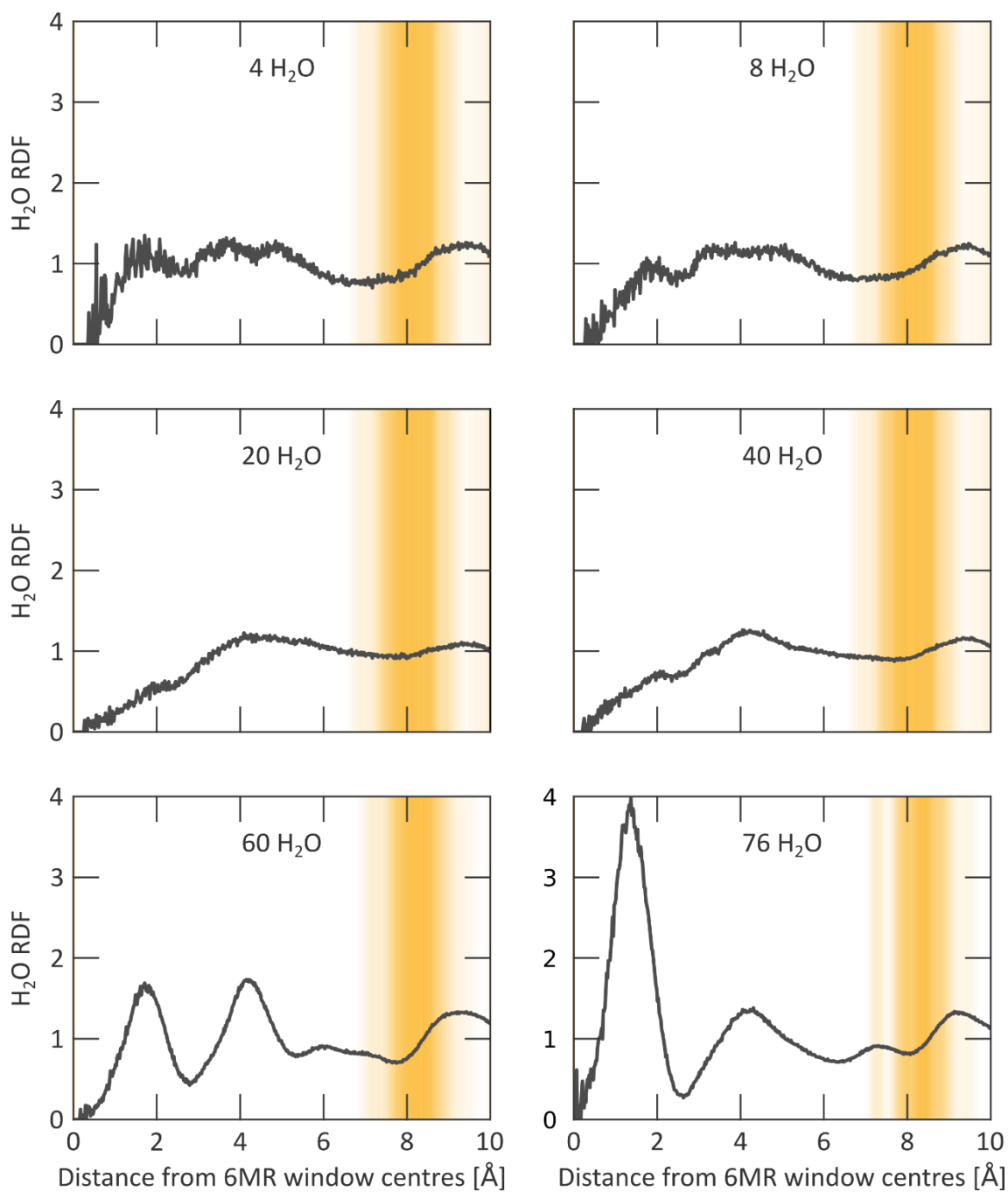

**Supplementary Figure 45 | Radial distribution functions at various water loadings at 300 K and 70 MPa.** RDFs between (i) the centres of the 6MR apertures and the oxygens of the water molecules (grey trace) and (ii) the centres of the 6MR apertures (orange shade).

**80 MPa**

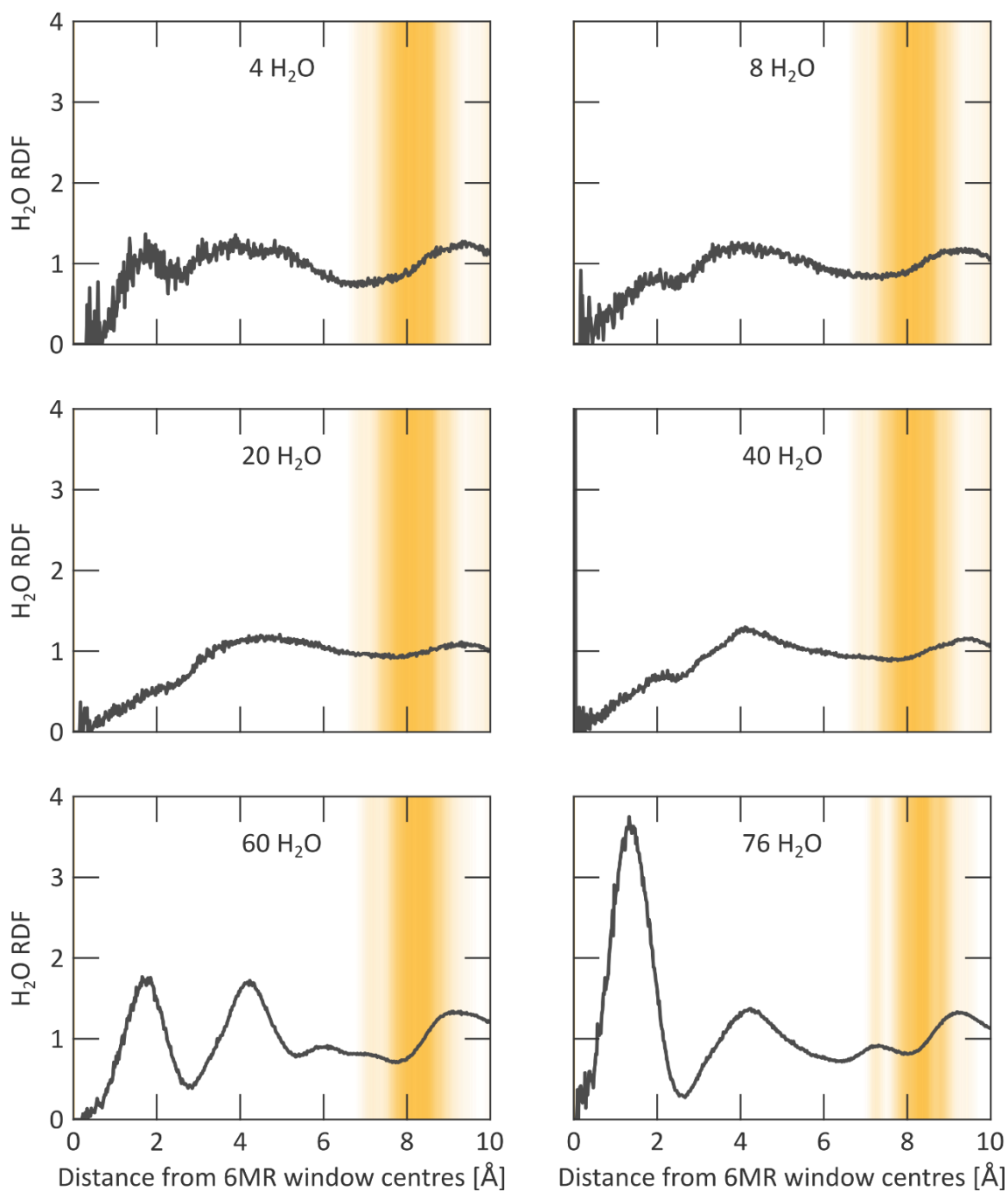

**Supplementary Figure 46 | Radial distribution functions at various water loadings at 300 K and 80 MPa.** RDFs between (i) the centres of the 6MR apertures and the oxygens of the water molecules (grey trace) and (ii) the centres of the 6MR apertures (orange shade).

**90 MPa**

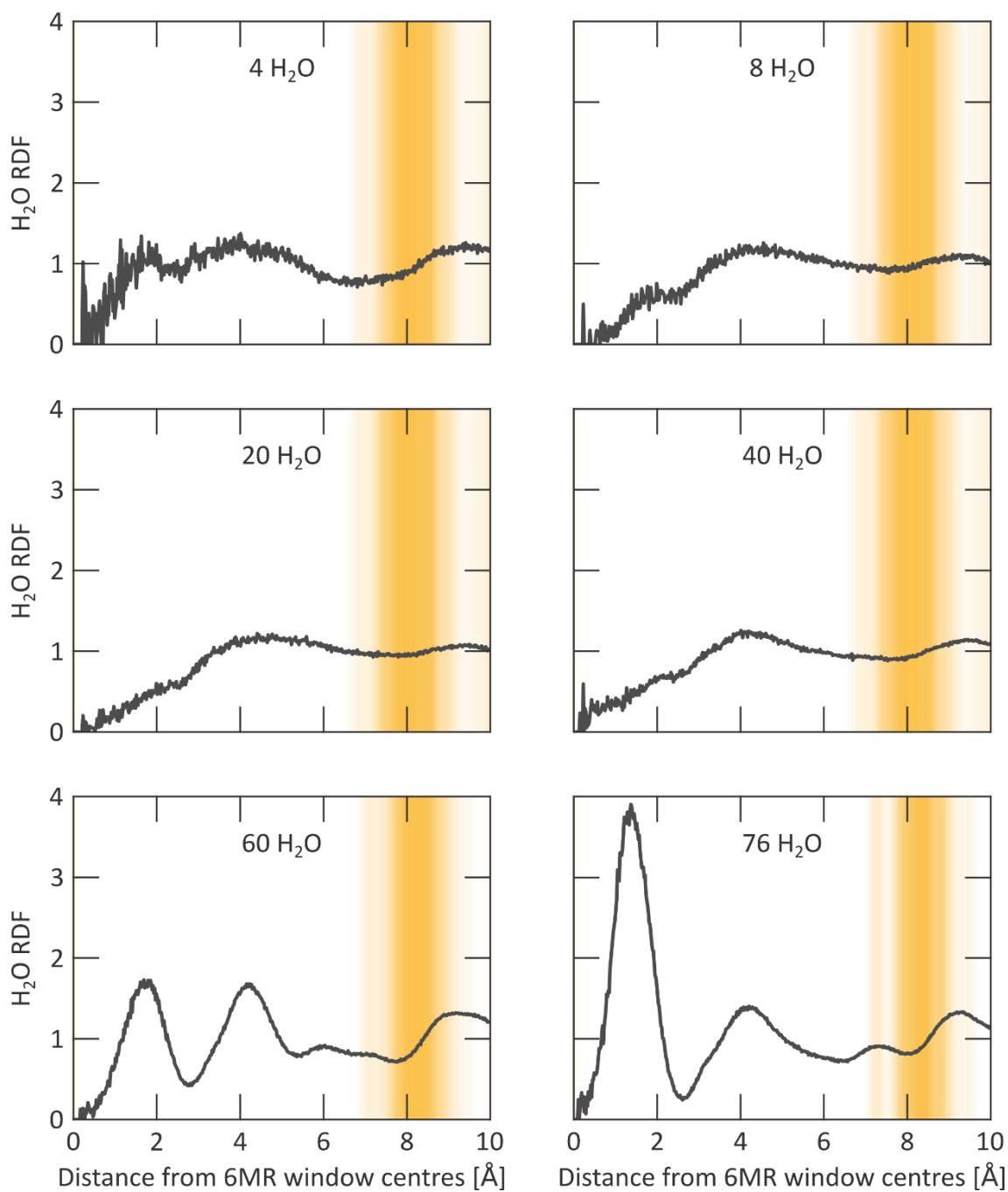

**Supplementary Figure 47 | Radial distribution functions at various water loadings at 300 K and 90 MPa.** RDFs between (i) the centres of the 6MR apertures and the oxygens of the water molecules (grey trace) and (ii) the centres of the 6MR apertures (orange shade).

- 60 -

**100 MPa**

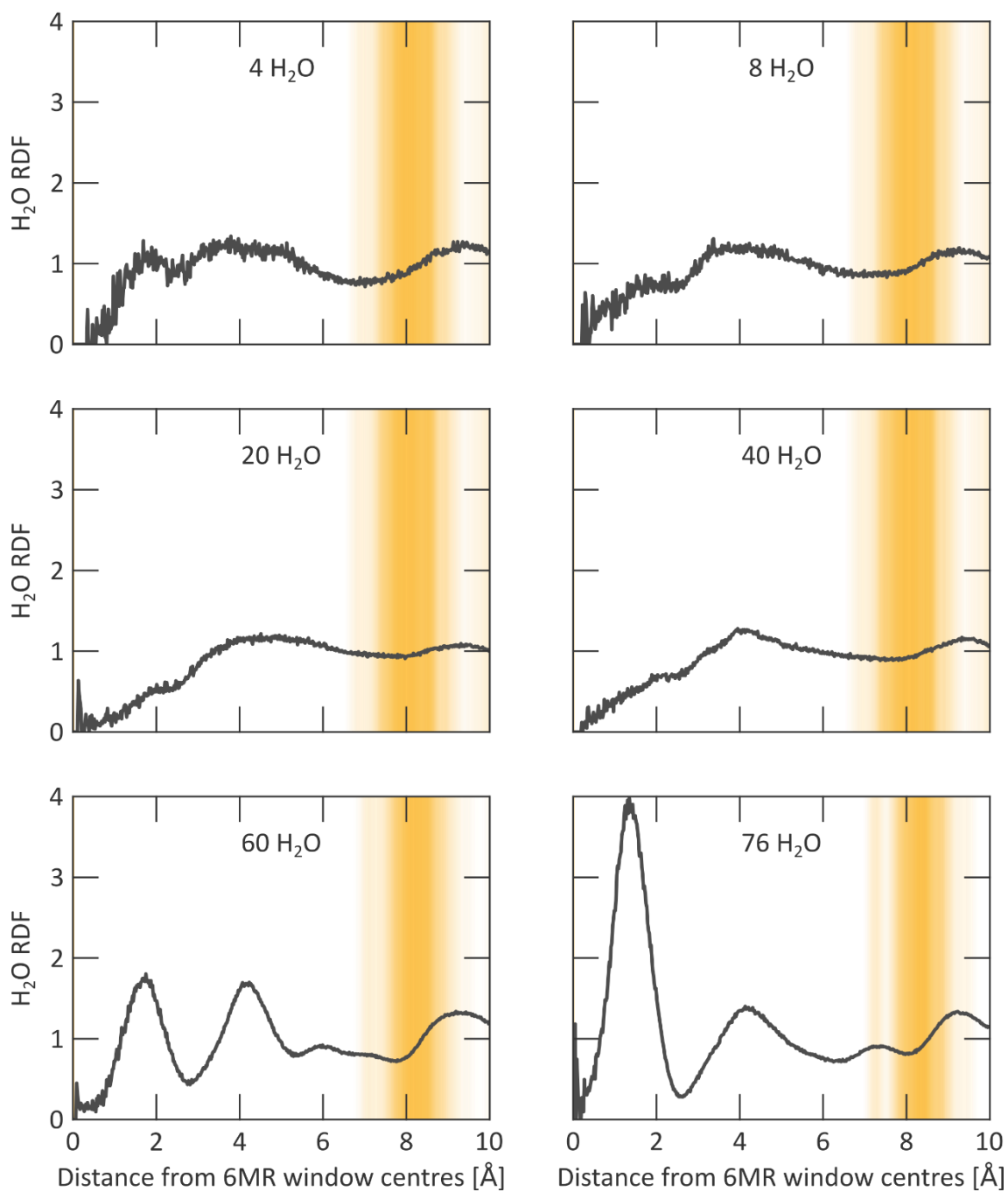

**Supplementary Figure 48 | Radial distribution functions at various water loadings at 300 K and 100 MPa.** RDFs between (i) the centres of the 6MR apertures and the oxygens of the water molecules (grey trace) and (ii) the centres of the 6MR apertures (orange shade).

## S7.2 Quantifying the hopping of water between the cages

To further validate the hydrophobicity of the 6MR apertures and its limiting effect on the diffusion of water between the ZIF-8 cages, the number of hopping events between the two cages in the conventional ZIF-8 unit cell during a 5 ns $(N, P, \boldsymbol{\sigma}_a = \boldsymbol{0}, T)$ simulation at 300 K and various mechanical pressures and water loadings are reported in Supplementary Figure 49 to Supplementary Figure 51. Due to the size of the water molecules, these events can only take place through the 6MR aperture, which acts as a 'gate' between two adjacent cages.

From these figures, it is observed that (i) in this pressure range, the mechanical pressure does not play a major role in the hopping event, and (ii) the higher the water loading, the more water hopping events are observed. This last observation is also present when normalising the number of water hopping events on the total water loading (see Supplementary Figure 50 and Supplementary Figure 51), and can be attributed to the formation of hydrogen bonds throughout the 6MR gates, which are facilitated by the higher water content. As a result, in the absence of a sizeable water gradient, a sufficiently large water cluster needs to be present in the adjacent cage for a water molecule to more easily diffuse into that cage. Even then, however, the relative number of hoppings remains relatively low (up to about 1 hopping event per water molecule and per 5 ns), confirming that this process is activated due to the hydrophobicity of the 6MR gates.

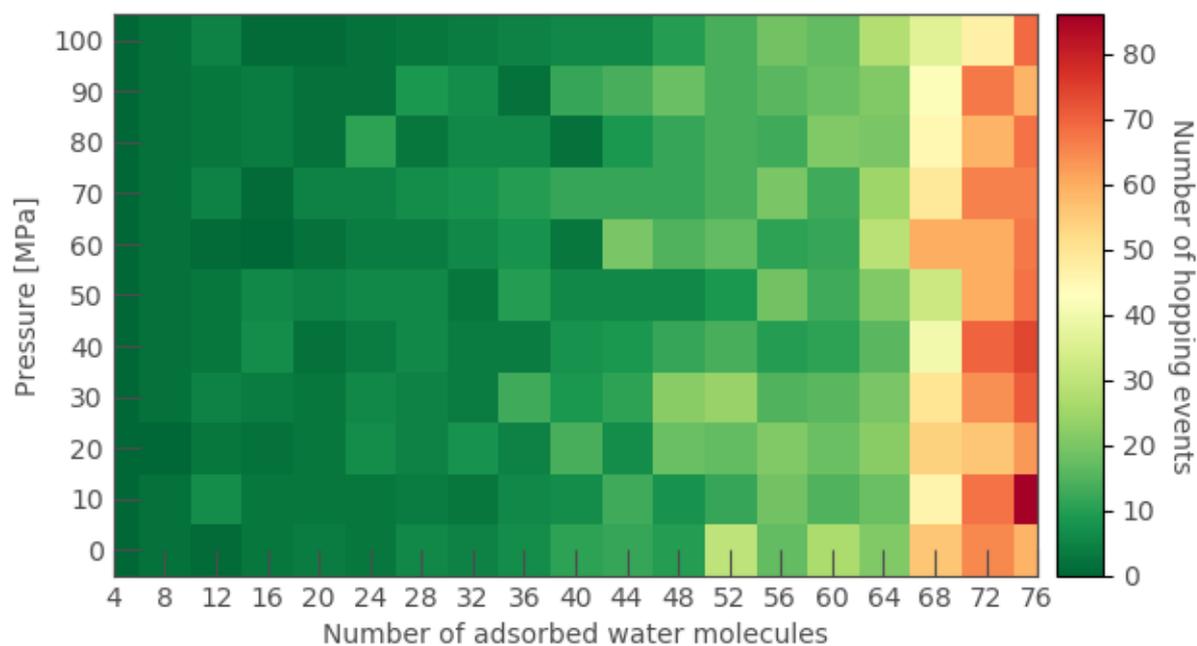

**Supplementary Figure 49 | Absolute water hopping frequency.** Number of water hopping events counted between the two cages in the conventional ZIF-8 unit cell during a 5 ns $(N, P, \boldsymbol{\sigma}_a = \boldsymbol{0}, T)$ simulation as a function of the mechanical pressure and the water loading.

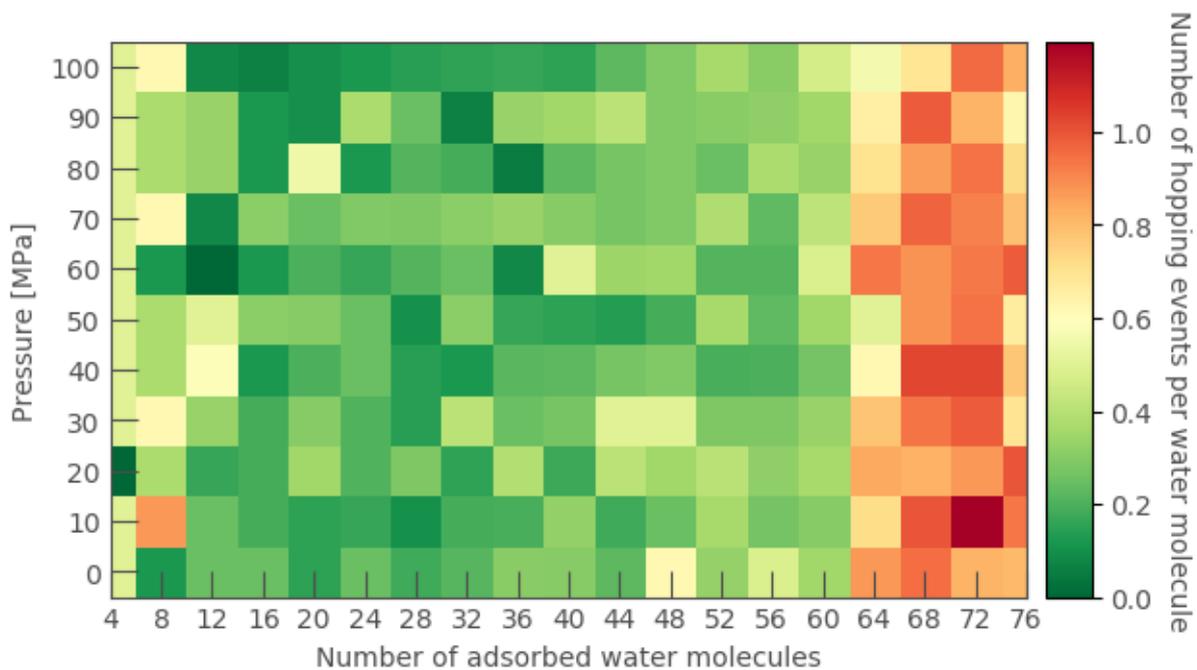

**Supplementary Figure 50 | Relative water hopping frequency.** Number of water hopping events counted between the two cages in the conventional ZIF-8 unit cell during a 5 ns $(N, P, \boldsymbol{\sigma}_a = \mathbf{0}, T)$ simulation as a function of the mechanical pressure and the water loading, normalised on the amount of water molecules present in the unit cell.

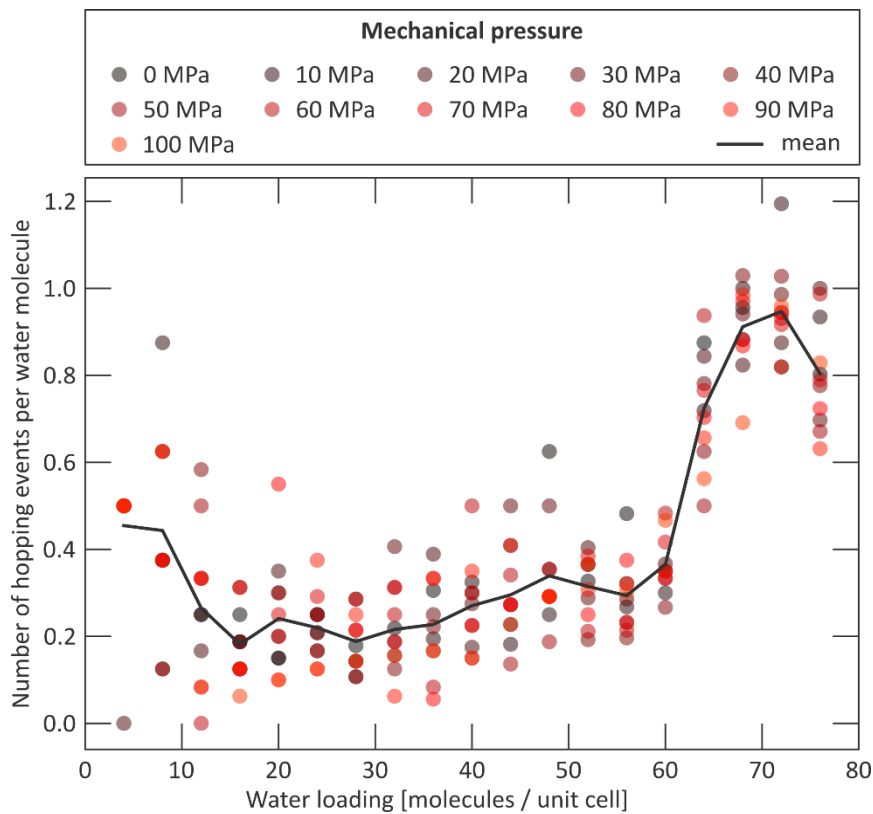

**Supplementary Figure 51 | Relative water hopping frequency** Number of water hopping events counted between the two cages in the 1×1×1 ZIF-8 unit cell during a 5 ns $(N, P, \boldsymbol{\sigma}_a = \mathbf{0}, T)$ simulation as a function of the mechanical pressure and the water loading, normalised on the amount of water molecules present in the unit cell.

## S7.3 The mobility of water inside ZIF-8

To further quantify the mobility of water inside the ZIF-8 cages, its mean squared displacement (MSD), defined as

$$\text{MSD}(t) = \frac{1}{N_{H2O}} \sum_{i=1}^{N_{H2O}} |\boldsymbol{r}_i(t) - \boldsymbol{r}_i(0)|^2$$

where $\boldsymbol{r}_i(t)$ is the position vector of the oxygen of the $i^{\text{th}}$ water molecule. From this, also the diffusion coefficient

$$D(t) = \frac{MSD(t)}{6t}$$

is calculated.

The results, shown in Supplementary Figure 52 to Supplementary Figure 62, are obtained from the 5 ns $(N, P, \boldsymbol{\sigma}_a = \boldsymbol{0}, T)$ simulations at 300 K and at various mechanical pressures and water loadings. If possible, multiple time origins were taken to speed up the convergence, ensuring that all considered windows are mutually disjoint.

Although most diffusion coefficients have not yet converged completely after 5 ns, a result of the slow mobility due to the necessity of moving through the hydrophobic 6MR apertures, it is clear that the final diffusion coefficients will be below 5 Å²/ns.

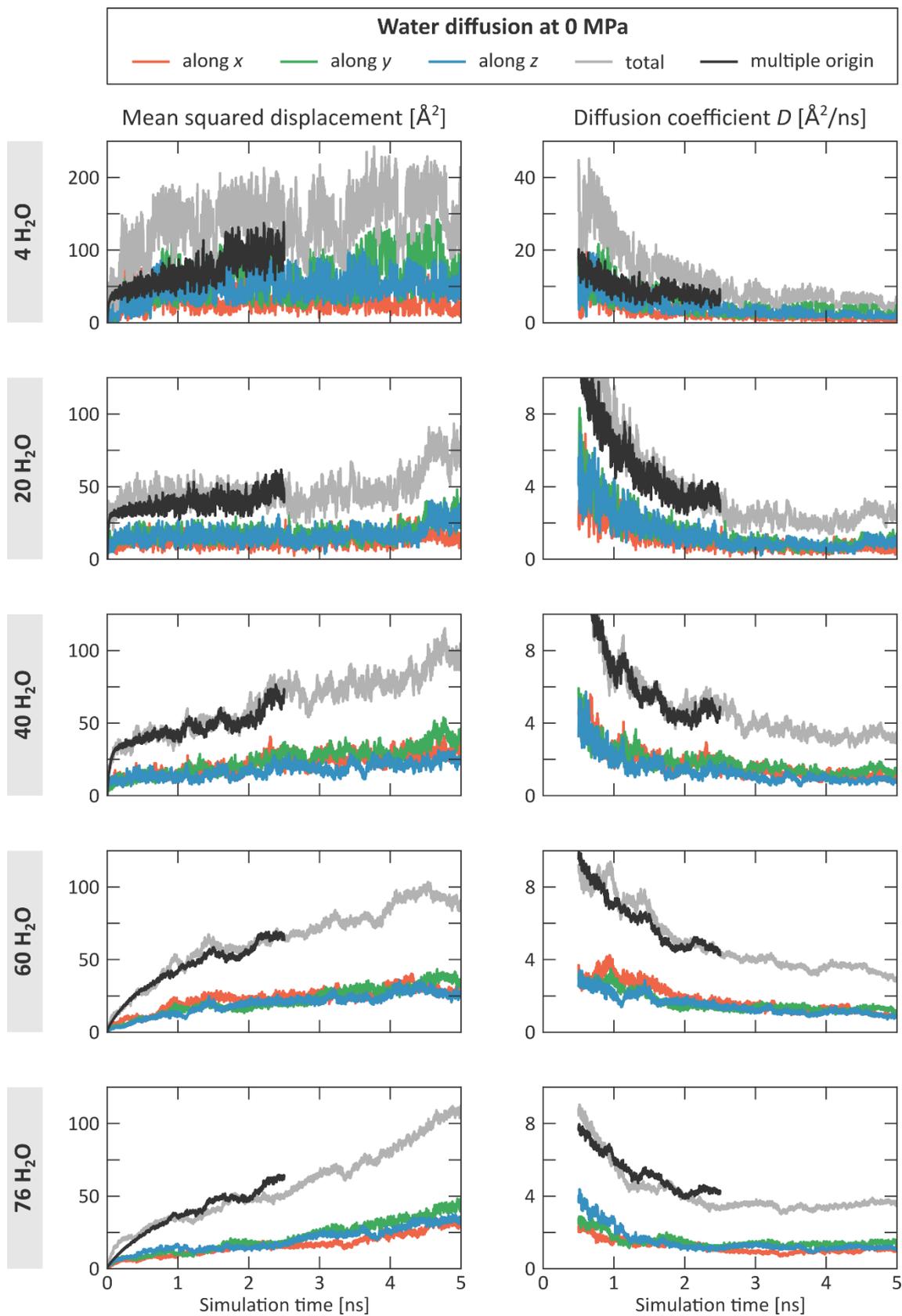

**Supplementary Figure 52 | Water mobility in ZIF-8 at 0 MPa.** Mean squared displacement and diffusion coefficient of water in ZIF-8 at different water loadings and at a mechanical pressure of 0 MPa.

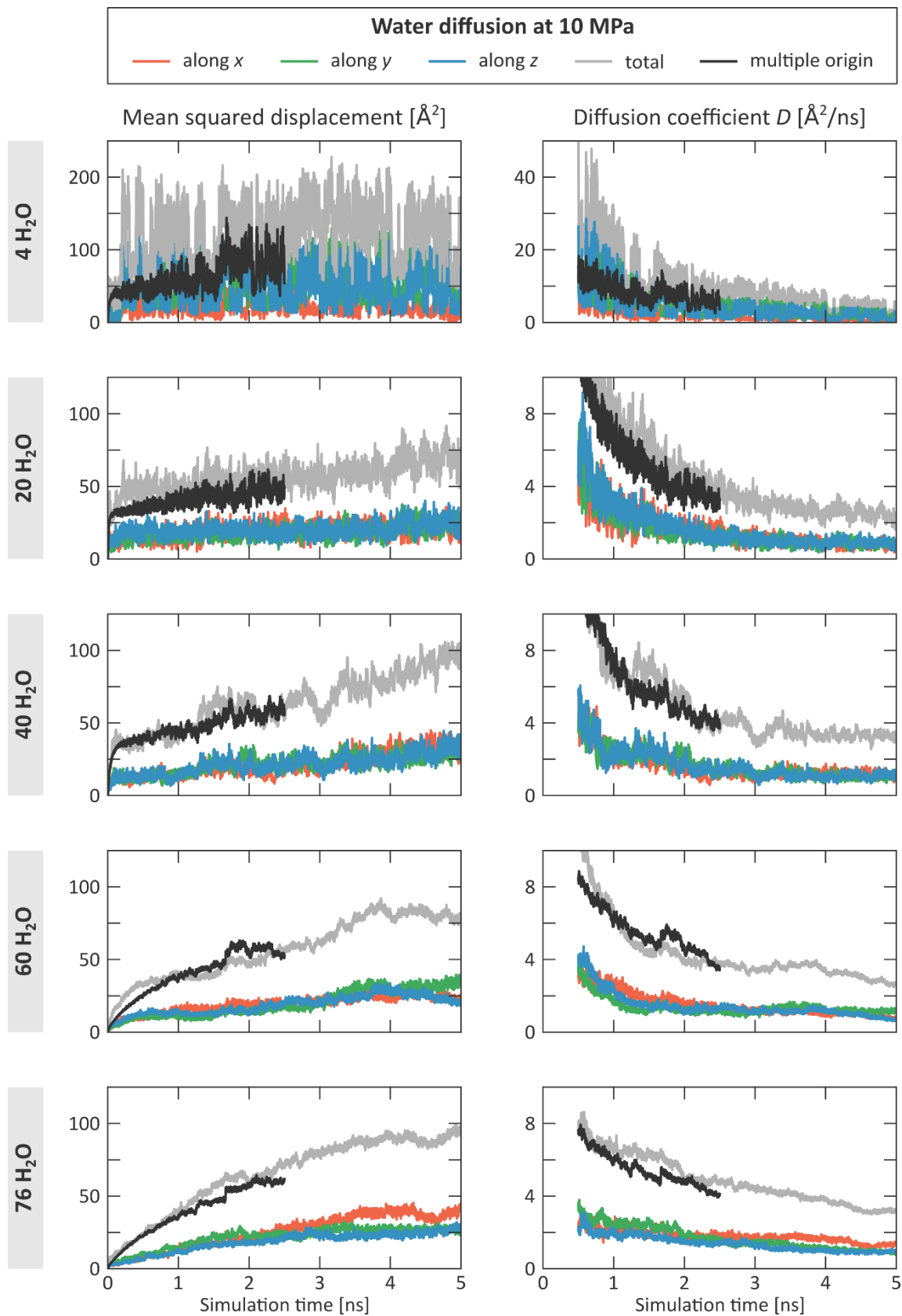

**Supplementary Figure 53 | Water mobility in ZIF-8 at 10 MPa.** Mean squared displacement and diffusion coefficient of water in ZIF-8 at different water loadings and at a mechanical pressure of 10 MPa.

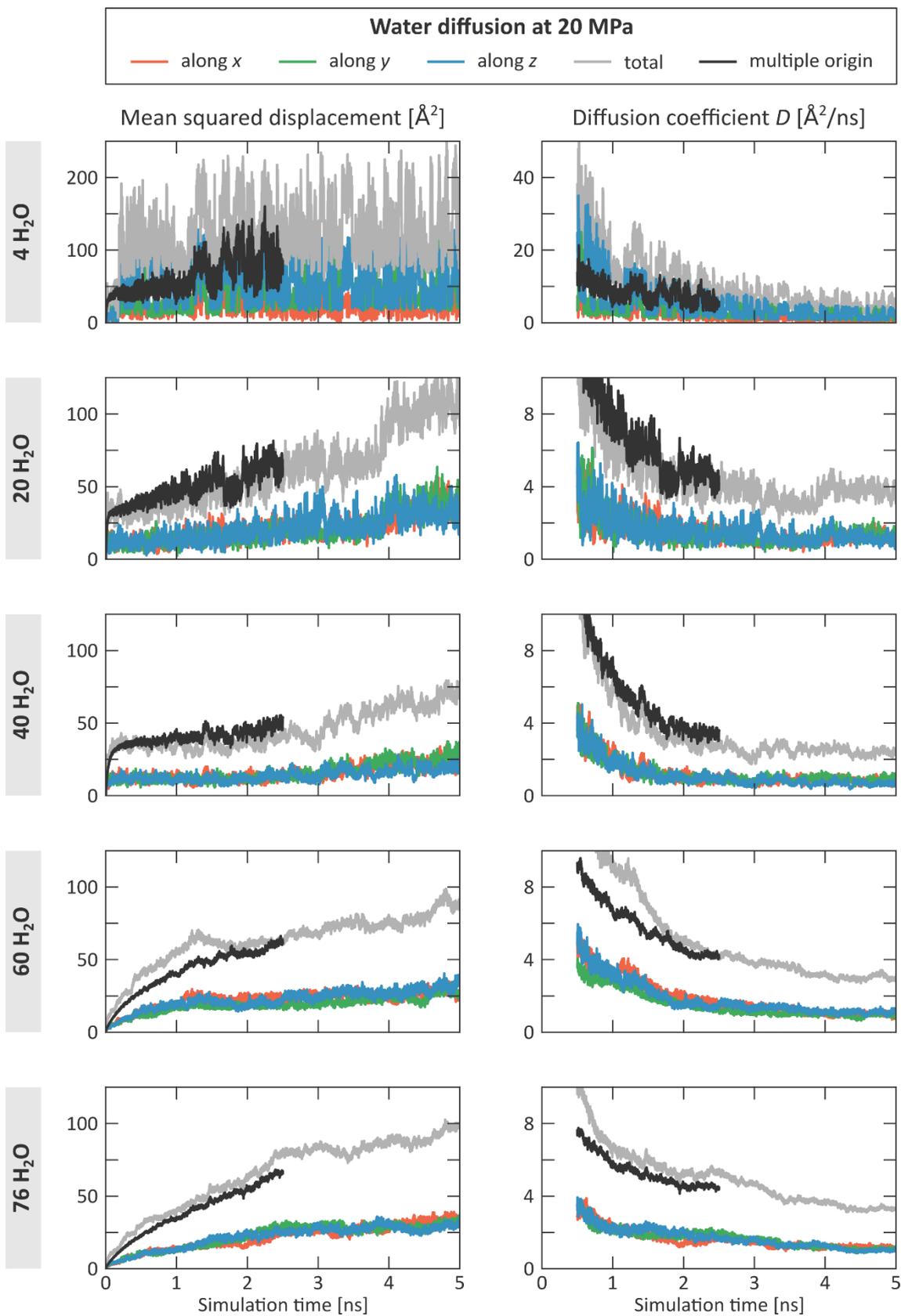

**Supplementary Figure 54 | Water mobility in ZIF-8 at 20 MPa.** Mean squared displacement and diffusion coefficient of water in ZIF-8 at different water loadings and at a mechanical pressure of 20 MPa.

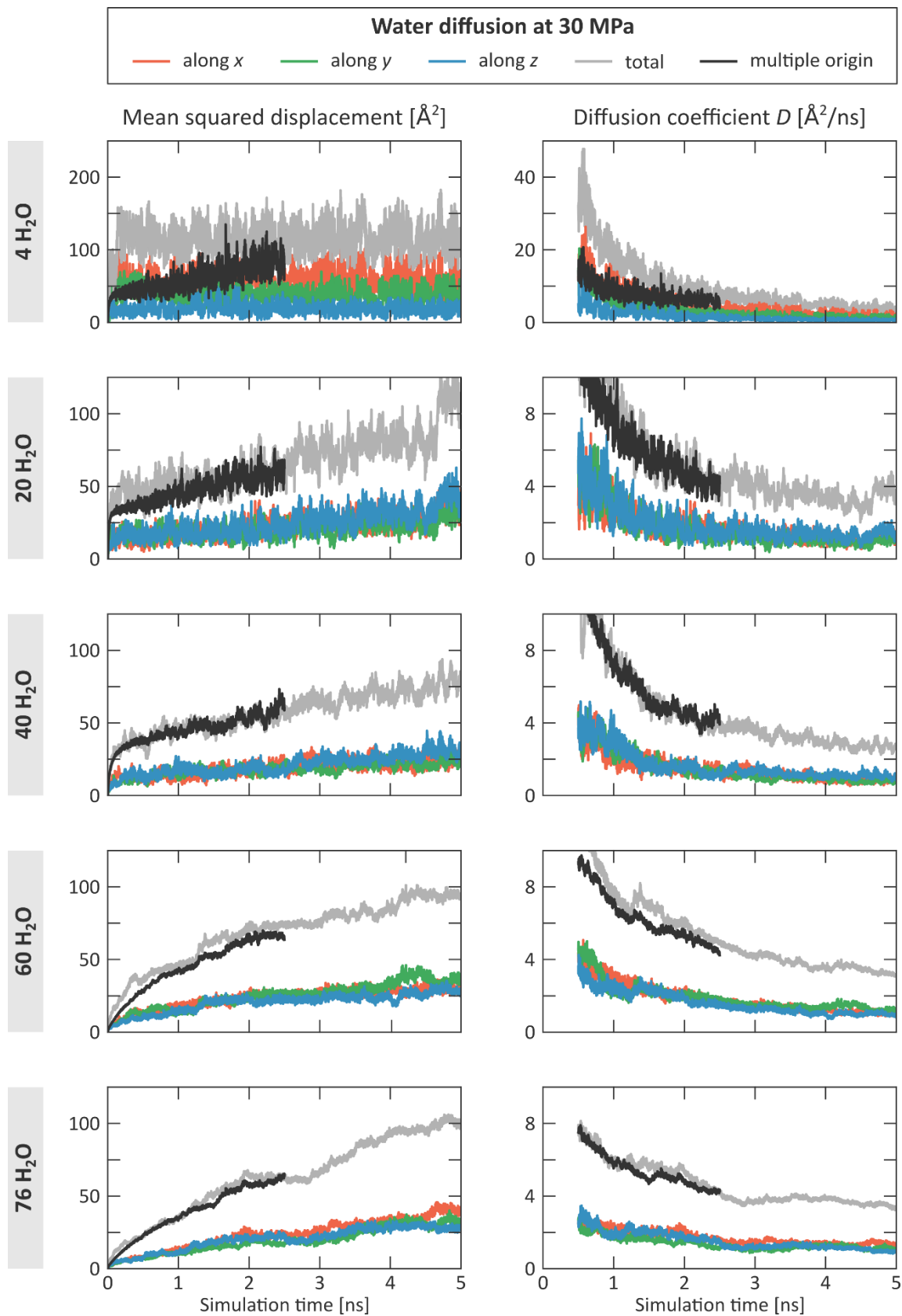

**Supplementary Figure 55 | Water mobility in ZIF-8 at 30 MPa.** Mean squared displacement and diffusion coefficient of water in ZIF-8 at different water loadings and at a mechanical pressure of 30 MPa.

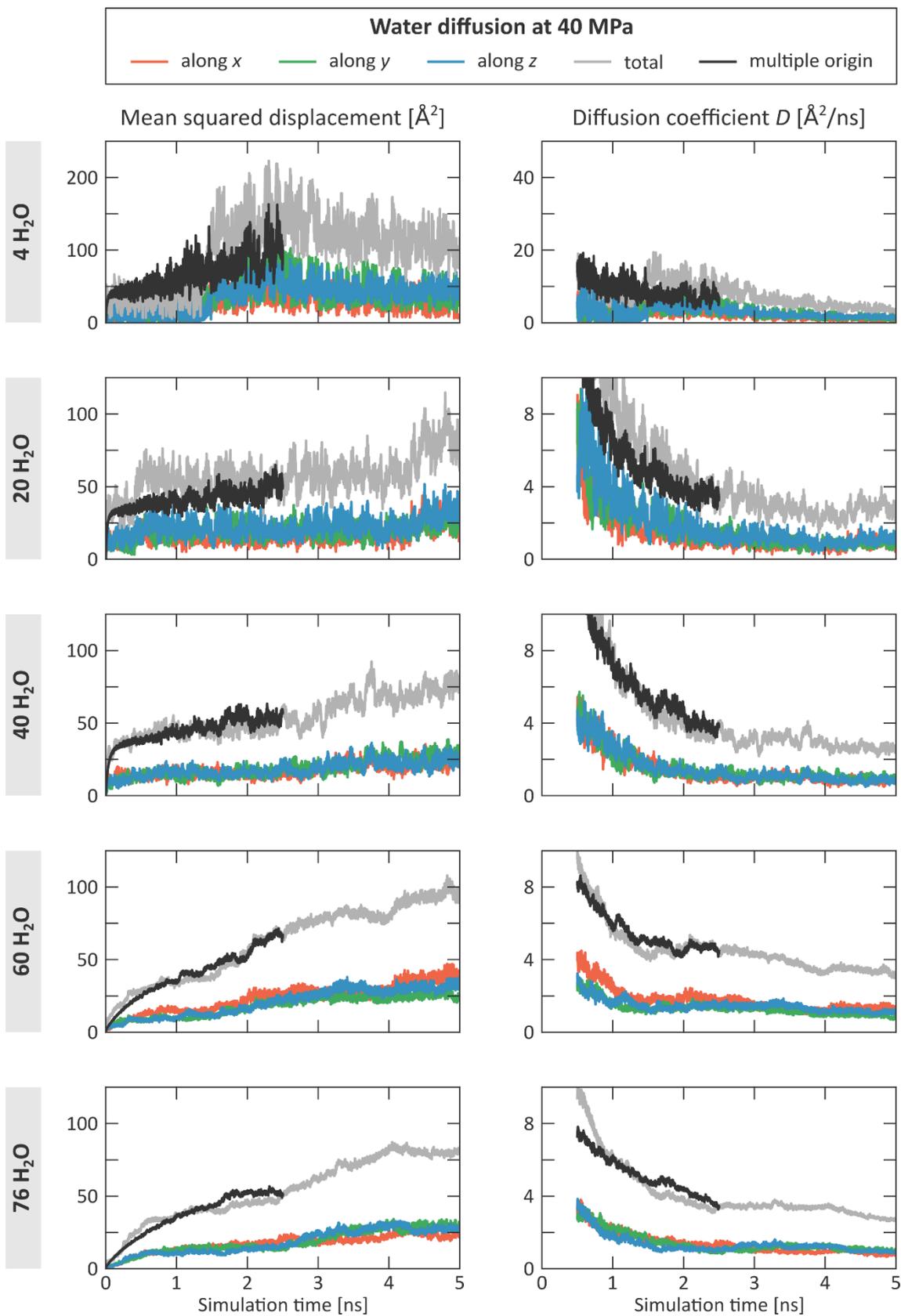

**Supplementary Figure 56 | Water mobility in ZIF-8 at 40 MPa.** Mean squared displacement and diffusion coefficient of water in ZIF-8 at different water loadings and at a mechanical pressure of 40 MPa.

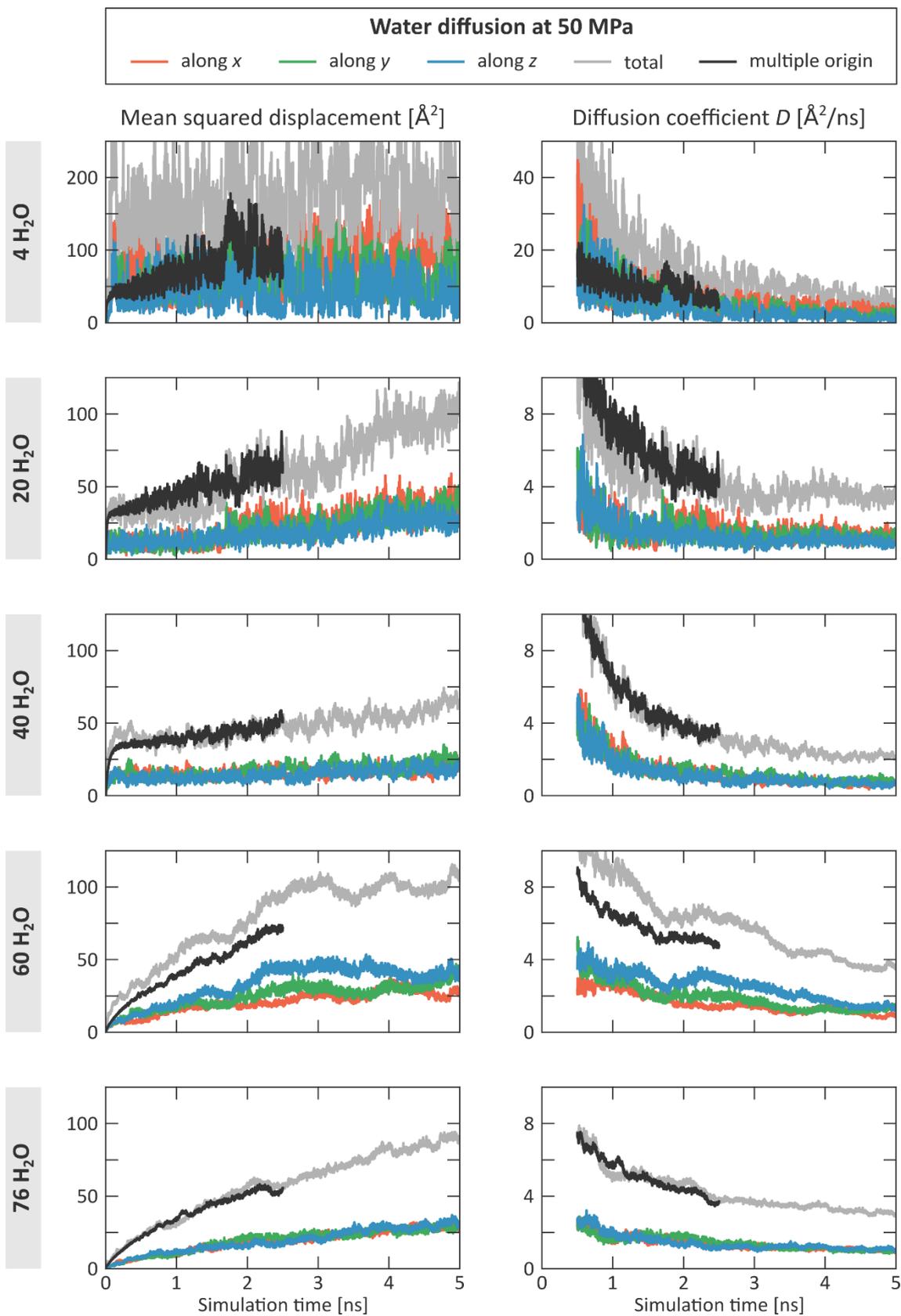

**Supplementary Figure 57 | Water mobility in ZIF-8 at 50 MPa.** Mean squared displacement and diffusion coefficient of water in ZIF-8 at different water loadings and at a mechanical pressure of 50 MPa.

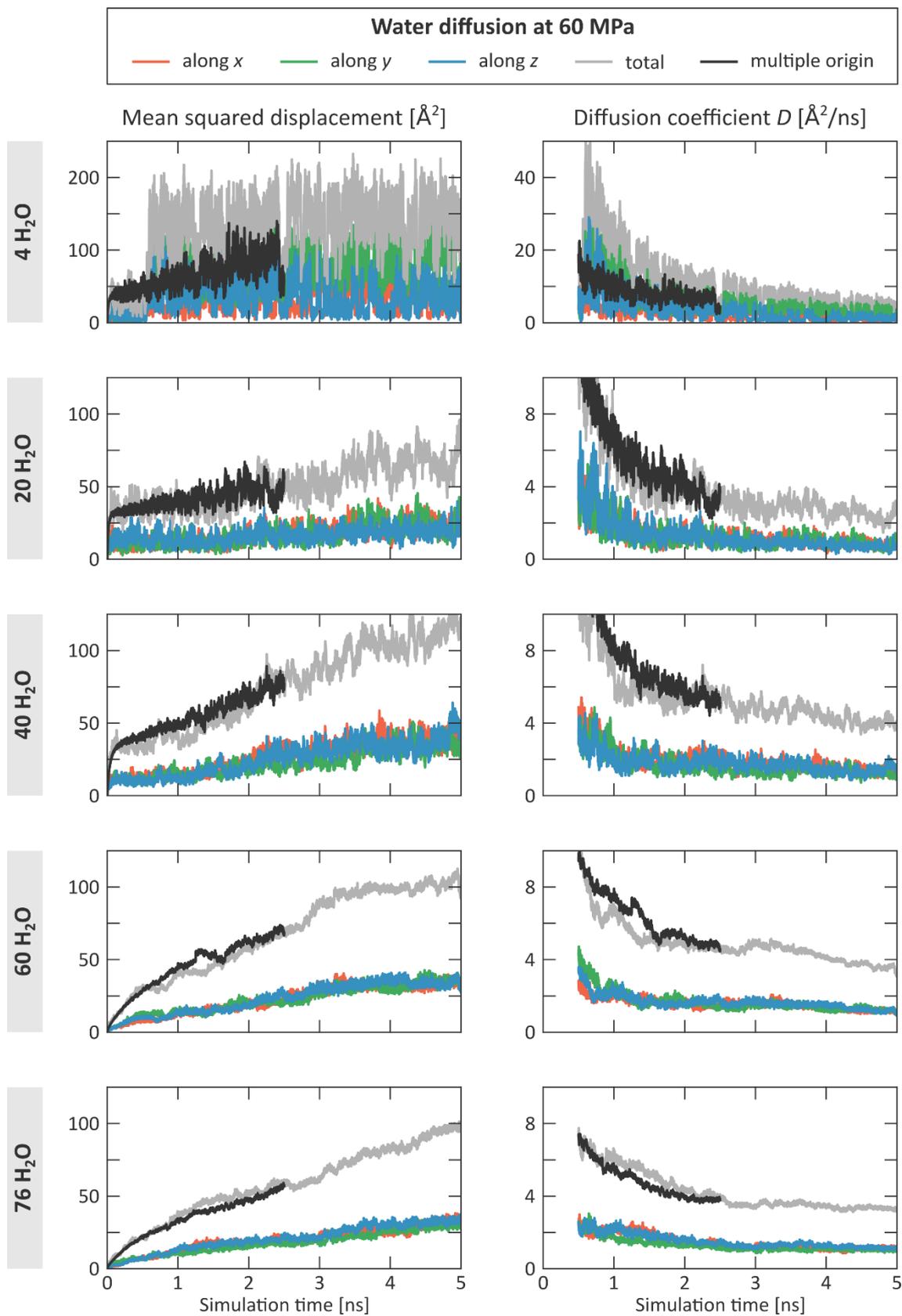

**Supplementary Figure 58 | Water mobility in ZIF-8 at 60 MPa.** Mean squared displacement and diffusion coefficient of water in ZIF-8 at different water loadings and at a mechanical pressure of 60 MPa.

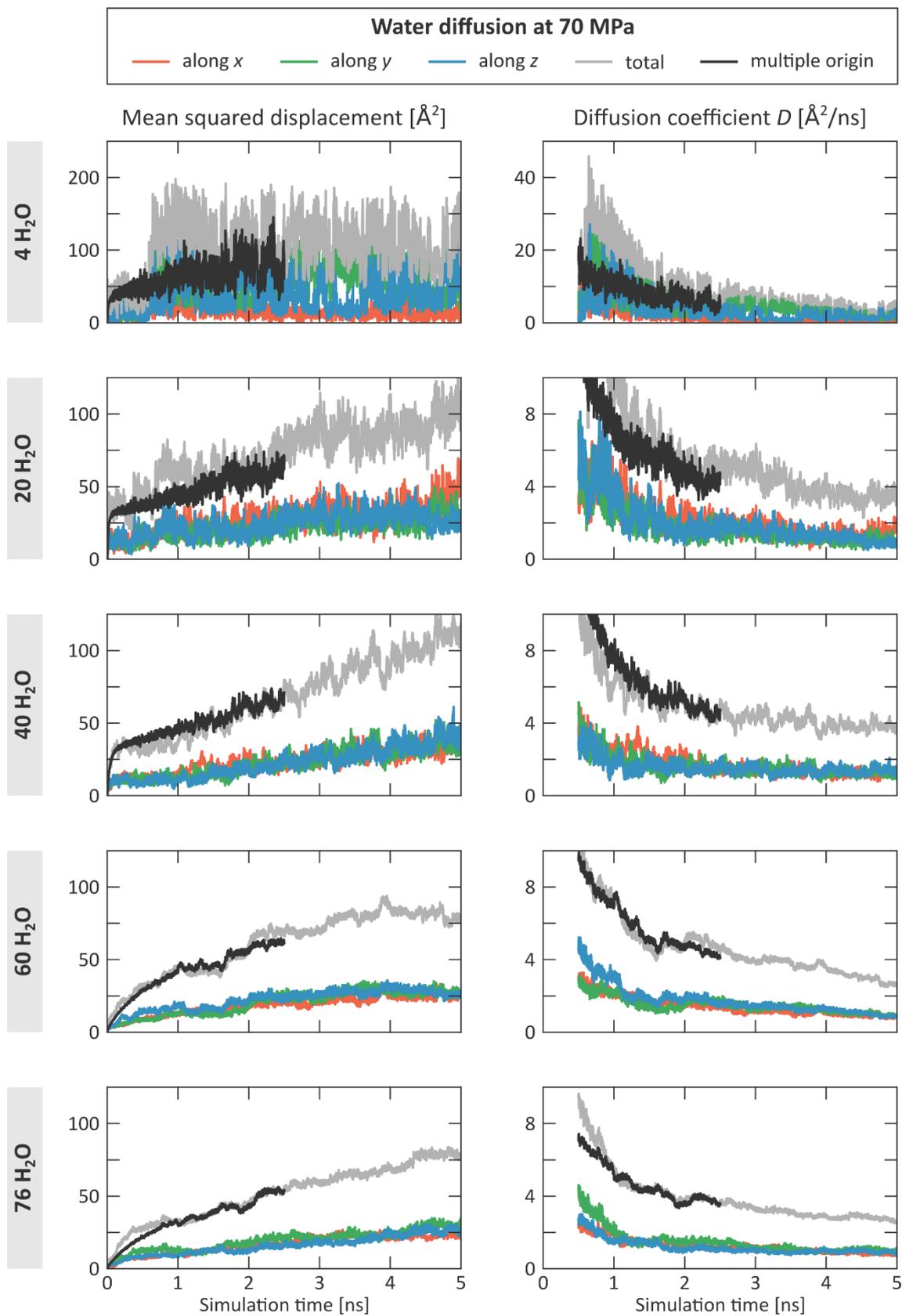

**Supplementary Figure 59 | Water mobility in ZIF-8 at 70 MPa.** Mean squared displacement and diffusion coefficient of water in ZIF-8 at different water loadings and at a mechanical pressure of 70 MPa.

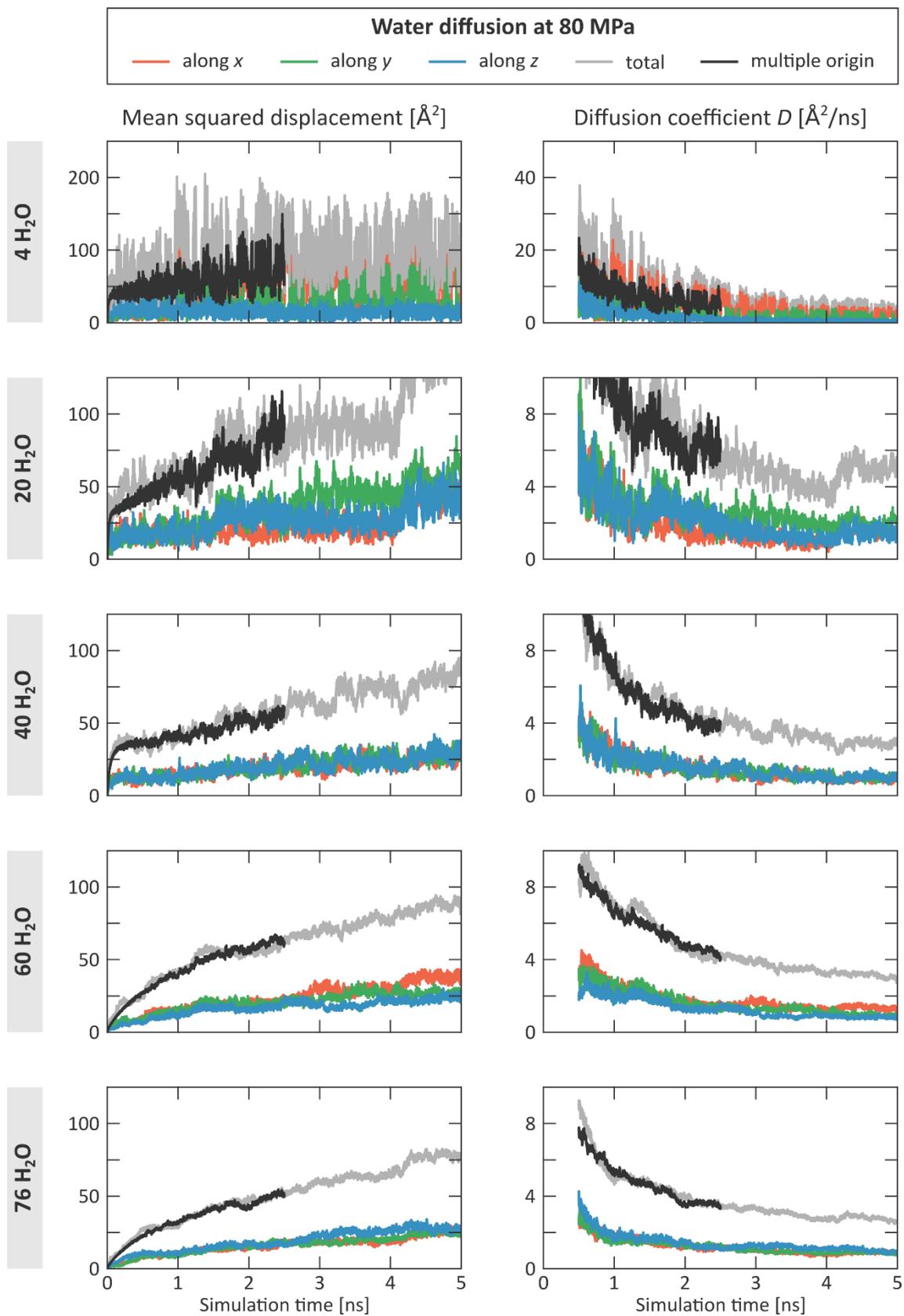

**Supplementary Figure 60 | Water mobility in ZIF-8 at 80 MPa.** Mean squared displacement and diffusion coefficient of water in ZIF-8 at different water loadings and at a mechanical pressure of 80 MPa.

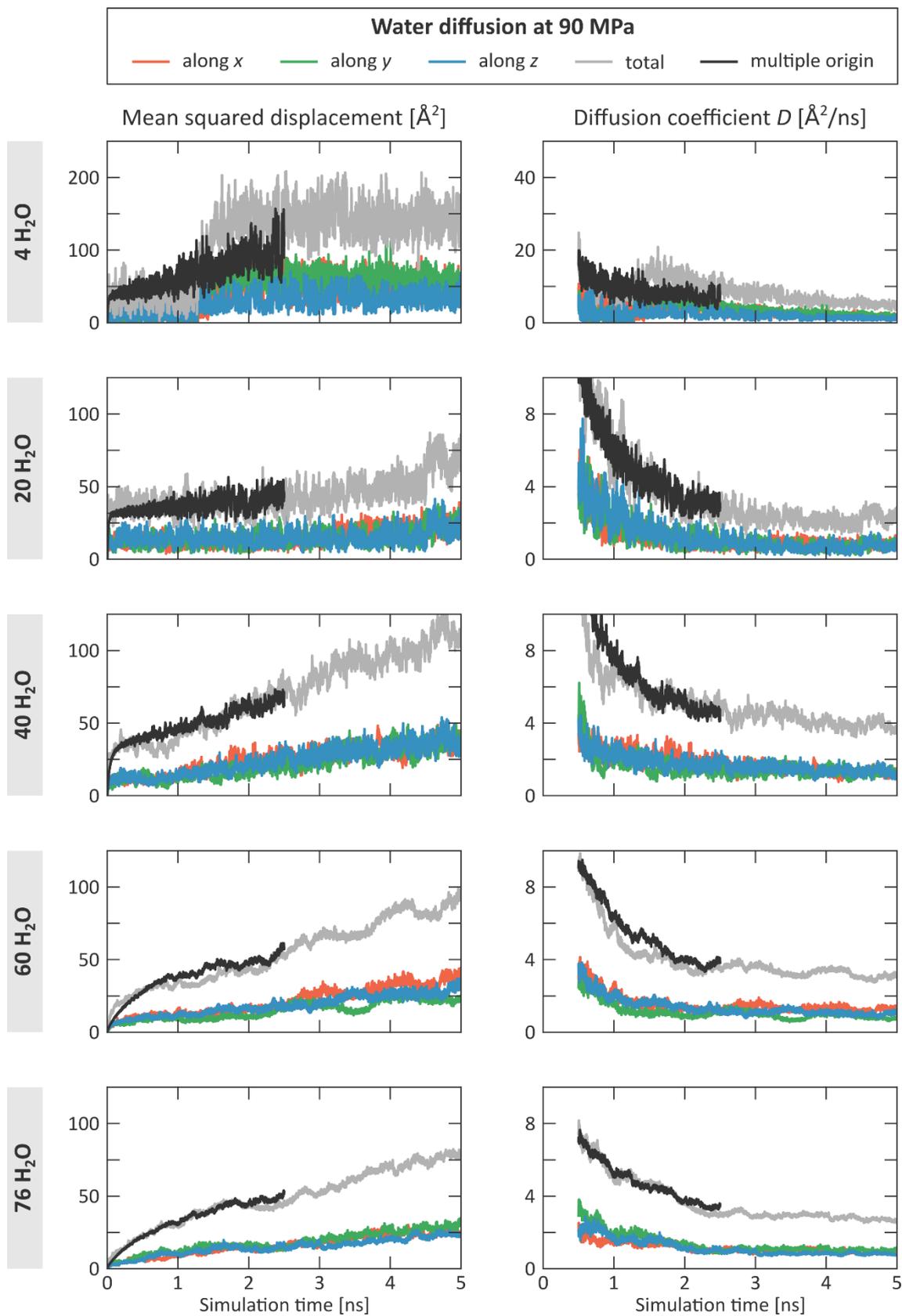

**Supplementary Figure 61 | Water mobility in ZIF-8 at 90 MPa.** Mean squared displacement and diffusion coefficient of water in ZIF-8 at different water loadings and at a mechanical pressure of 90 MPa.

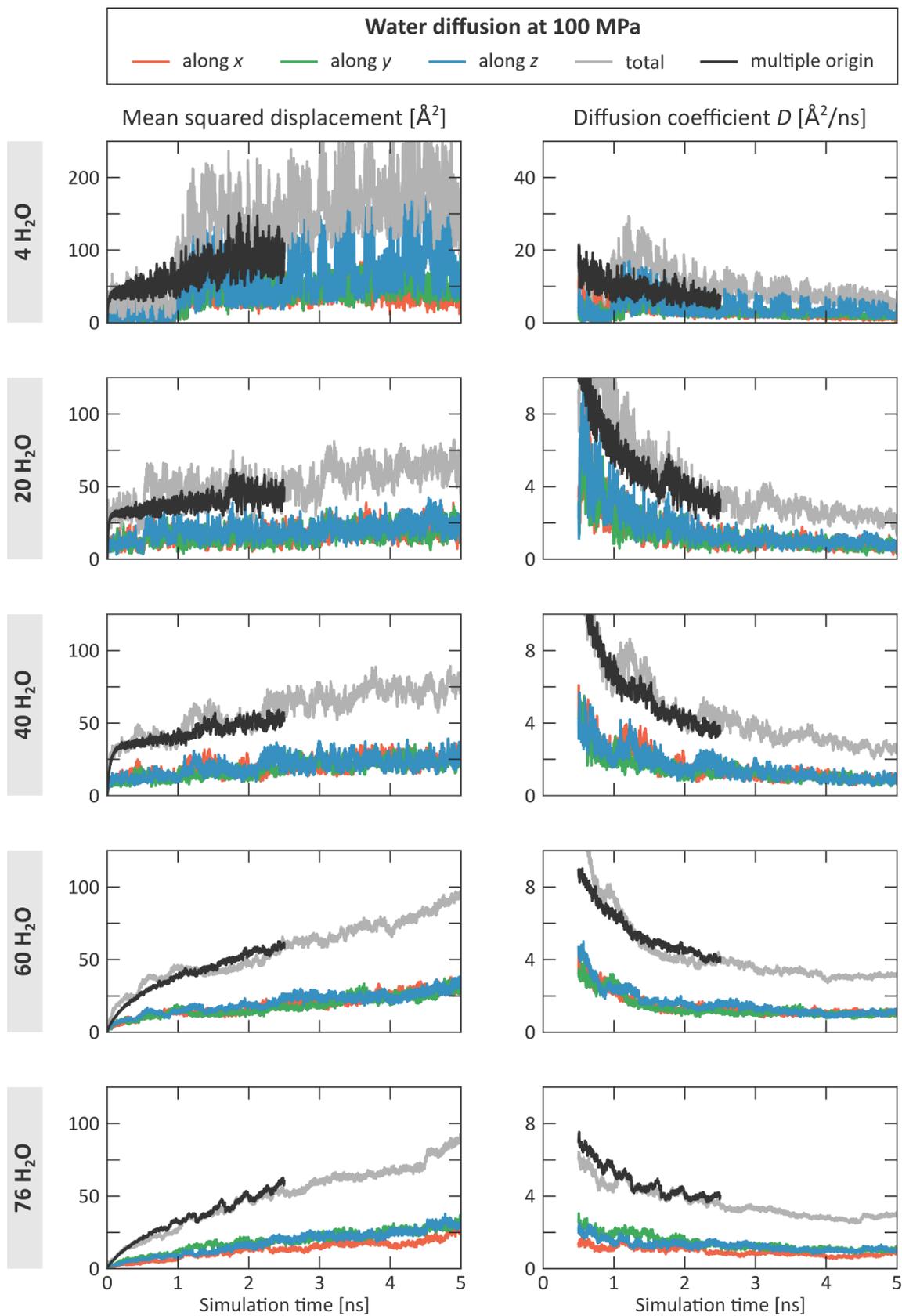

**Supplementary Figure 62 | Water mobility in ZIF-8 at 100 MPa.** Mean squared displacement and diffusion coefficient of water in ZIF-8 at different water loadings and at a mechanical pressure of 100 MPa.

# S8 The intrinsic timescale for the formation of stable water clusters

As discussed in the section 'The intrinsic water mobility timescale revealed by molecular dynamics simulations' and Figure 4 of the main text, the formation of critical-sized water clusters in the ZIF-8 cages occurs on a nanosecond timescale and is hence substantially slower than typical diffusion processes. To further confirm this intrinsic timescale and obtain additional insight in this process, the mechanism is investigated here for a series of $(N, P, \sigma_a = 0, T)$ MD simulations in ZIF-8, containing different inhomogeneous water distributions and model sizes. In Supplementary Section S8.1, ten independent 5 ns simulations on a 1×1×2 supercell of ZIF-8 are considered, starting from the same water distribution as discussed in Figure 4a of the main text, to provide a sound statistical analysis of the time needed to form stabilised water clusters. In Supplementary Section S8.2, the ZIF-8 models are substantially enlarged – up to a 2×2×12 supercell containing 13,248 frameworks atoms – and different inhomogeneous water distributions are considered. Again, for each setup, ten independent simulations are performed to obtain reliable statistics, showing that the formation of stable water clusters occurs in a layer-by-layer fashion and that the intrinsic nanosecond timescale is independent of the model size. Based on this observation, an analytical model is provided in Supplementary Section S8.3 to connect this intrinsic nanosecond timescale with the experimental total intrusion time.

## S8.1 Statistical analysis through MD simulations of a 1×1×2 ZIF-8 supercell with an inhomogeneous water distribution

To provide additional data for the statistical model, the MD simulation depicted in Figure 4 of the main text was repeated tenfold. Each of these simulations started from a 1×1×2 ZIF-8 supercell with an inhomogeneous water distribution, consisting of 42 water molecules in cage 1 and no water molecules in any of the three other cages. The velocities of these water molecules were drawn randomly from a Maxwell-Boltzmann distribution at 300 K, resulting in ten simulations with different initial conditions. The movement of these water molecules at 0 MPa and 300 K for each of these ten MD simulations was then tracked for the total simulation time of 5 ns. To enable this longer simulation time compared to the data in Figure 4b, the in-house Yaff software package[48] was interfaced with LAMMPS to calculate the long-range interactions more efficiently.[49] Afterwards, each water molecule was assigned to a specific cage based on the shortest distance between the oxygen atom of the water molecule and each of the cage centres. The results for each of these ten independent simulation are shown in Supplementary Figure 63 and Supplementary Figure 64. From these data, a series of conclusions can be drawn.

First, the formation of a stabilised water cluster can occur in three different scenarios: either only cage 2 is filled (simulation 5 and Figure 4b of the main text), either only cage 3 is filled (simulations 4, 6, 8 and 9), or either cage 2 and cage 3 are filled simultaneously (simulations 1, 2, 3, 7, and 10). Cage 4 is never filled, except for a single water molecule that briefly resides in cage 4 early on in simulation 9.

This behaviour is as expected: when the 1×1×2 ZIF-8 supercell is empty, the four cages are equivalent. At equilibrium, the water distribution would therefore be distributed homogeneously over all cages, as demonstrated by the equilibrium water distributions obtained *via* canonical Monte Carlo simulations in Supplementary Section S4.2. Especially, at equilibrium, the 42 water molecules present in our 1×1×2 supercell would adopt a distribution very similar to the left column of Supplementary Figure 25, which contains 20 water molecules per conventional unit cell. However, in the MD simulations performed here, a strong inhomogeneity is introduced in the material by initially placing 42 water molecules in cage 1 and keeping all other cages empty. This breaks the equivalency of the four cages. At the onset of these simulations, cages 2 and 3 are still equivalent as they are directly connected to the filled cage 1, but they are inequivalent with both the filled cage 1 and the empty cage 4. The inequivalency of cage 4 with respect to any of the other cages stems from the cage network in ZIF-8: cage 4 shares no 6MR aperture with cage 1, in contrast to both cage 2 and cage 3; therefore, water molecules cannot diffuse to cage 4 without first passing through either cage 2 or cage 3. This cage equivalency explains why there is no apparent preference for the cage 2 and/or cage 3 to form a critically sized water cluster during the simulation, while cage 4 is never filled beyond a single water molecule. This also shows that the intrusion process is a cage-by-cage process: a cage can only be filled if an adjacent cage is already sufficiently filled. In Supplementary Section S9, it will be shown that this requirement can be refined to require the presence of a critically sized cluster in a neighbouring cage.

Second, while the large water gradient present between cage 1 and both cage 2 and cage 3 ensures that cage 2 and cage 3 get both filled initially, the formed water cluster is not necessarily stable. For instance, in simulation 4 of Supplementary Figure 63, cage 2 fills initially but contains too few water molecules to provide sufficient stabilisation. As a result, during simulation 4, the small water cluster formed in cage 2 disappears again as water molecules move back to cage 1, against the water gradient. A similar observation was also made in the section 'The intrinsic water mobility timescale revealed by molecular dynamics simulations' of the main text, in which the initial filling of cage 3 was insufficient to provide a sufficiently stabilised cluster, eventually leading to the evacuation of cage 3 in Figure 4b. Also here, the equivalency of cage 2 and cage 3 is apparent and similar observations can be drawn in the other simulations in Supplementary Figure 63 and Supplementary Figure 64.

Finally, by fitting the number of water molecules in the different cages as a function of time using the same exponential fit as in the main text, $a(1 - e^{-t/\tau})$, the timescale $\tau$ for the formation of stable water clusters can be deduced. A time constant is extracted for each cage that is initially empty and contains at least two water molecules at the end of the simulation, in order to obtain a good fit. The so obtained time constants are visualised in Supplementary Figure 65 and largely fall within the [0.5 ns, 1.5 ns] range, further confirming the nanosecond intrinsic timescale proposed in the main text.

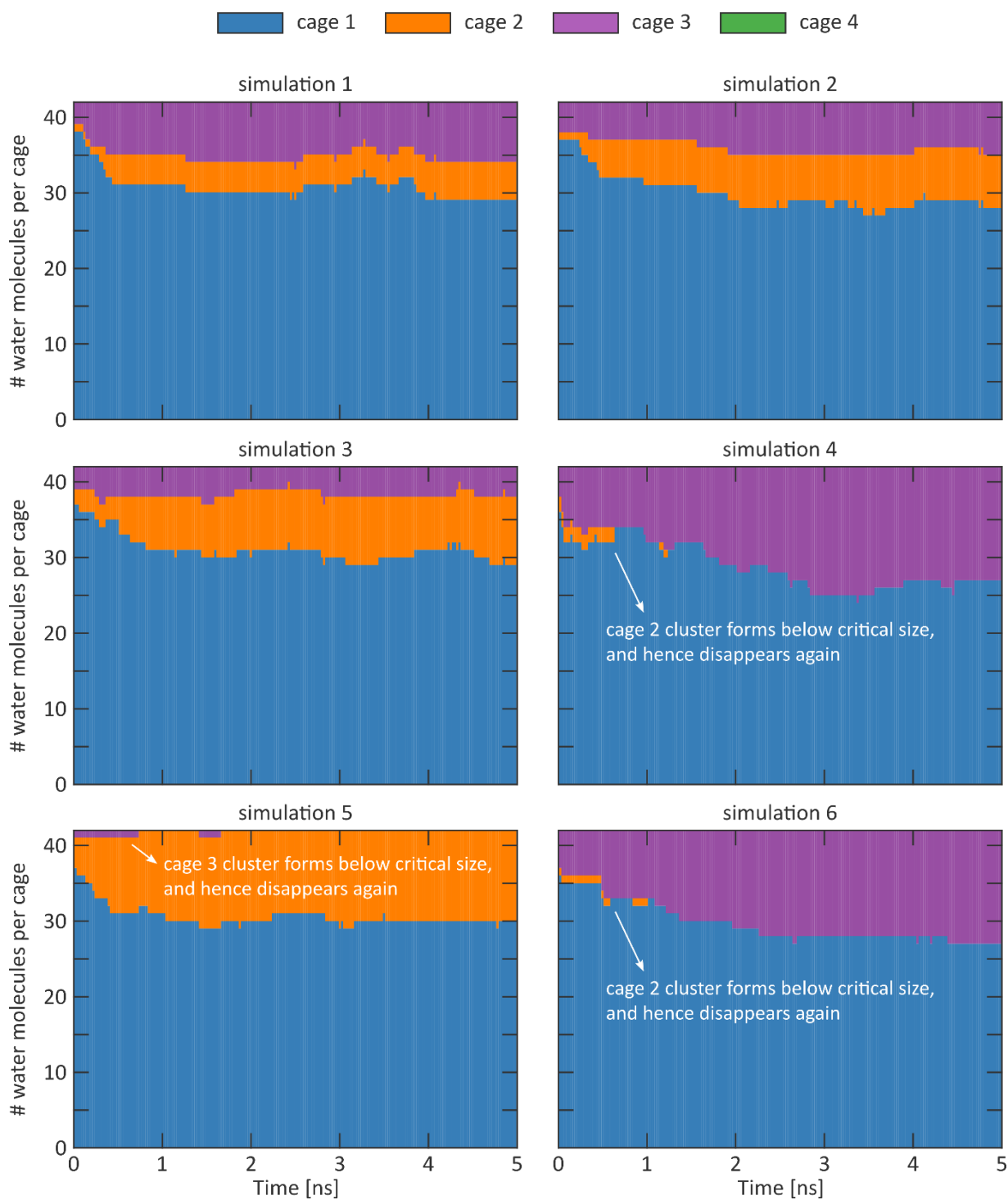

**Supplementary Figure 63 | Evolution of the number of water molecules per cage during a 5 ns MD simulation at 300 K and 0 MPa of a 1×1×2 ZIF-8 supercell with an inhomogeneous water distribution.** All simulations initially contained 42 water molecules in cage 1 while all other cages were initially empty. These simulations were initialised with different initial velocities for the water molecules, which were drawn randomly from a Maxwell-Boltzmann distribution at 300 K. The cage indexing is identical to Figure 4a of the main text. Figure continued in Supplementary Figure 64.

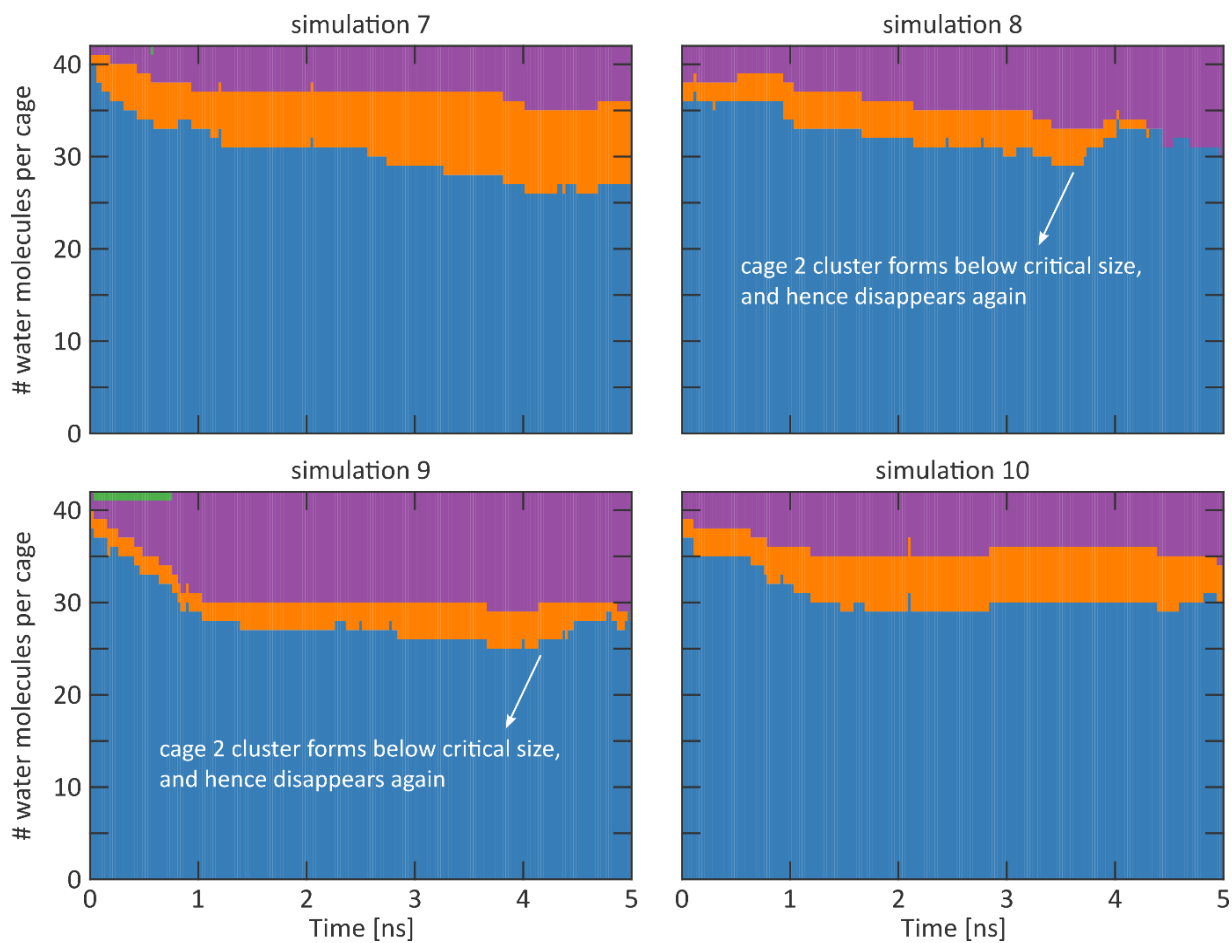

**Supplementary Figure 64 | Evolution of the number of water molecules per cage during a 5 ns MD simulation at 300 K and 0 MPa of a 1×1×2 ZIF-8 supercell with an inhomogeneous water distribution.** All simulations initially contained 42 water molecules in cage 1 while all other cages were initially empty. These simulations were initialised with different initial velocities for the water molecules, which were drawn randomly from a Maxwell-Boltzmann distribution at 300 K. The cage indexing is identical to Figure 4a of the main text. Figure continued from Supplementary Figure 63.

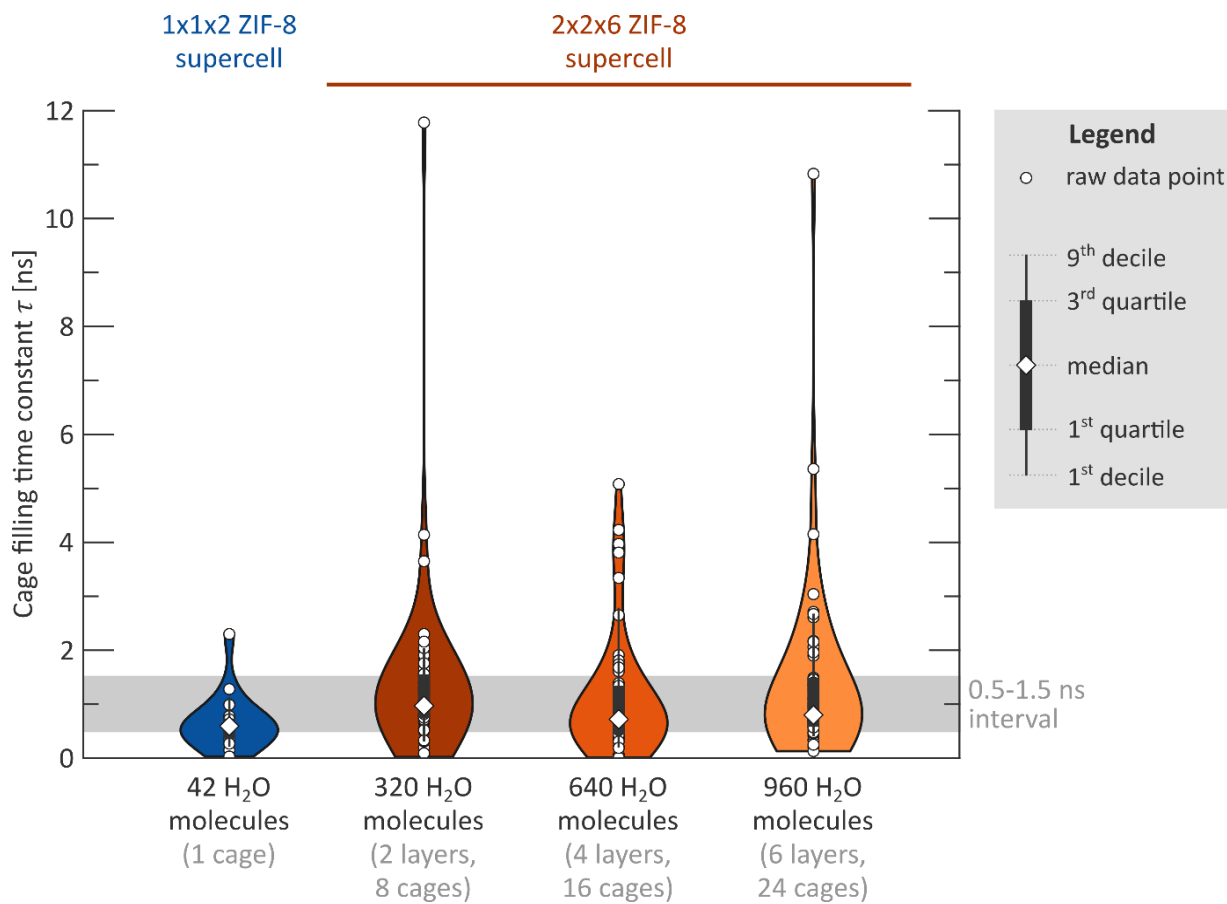

**Supplementary Figure 65 | Distribution of time constants for the formation of critically sized water clusters with different setups as discussed in Supplementary Sections S8.1 and S8.2, each repeated tenfold.** For each setup the raw data points are shown, as well as the median, the interquartile range, and the interdecile range. All simulations started from an inhomogeneous water distribution, as shown in the *x*-axis labels and further explained in the text. Time constants were derived based on an exponential fit as in Figure 4b of the main text, taking into account cages that (i) were initially empty, (ii) contained at least two water molecules at the end of the simulation, and (iii) had a fitted total cage filling lower than 42 water molecules. This last condition excludes some poor fits due to a limited amount of water molecules in this cage after the total simulation time, which would correspond to extremely large time constants. The values obtained here, for which all medians fall inside the [0.5 ns, 1.5 ns] range, will hence be a slight underestimation of the time needed to form critically sized clusters.

## S8.2 Effect of the ZIF-8 model size

To investigate whether the intrinsic nanosecond timescale for the formation of critically sized clusters is independent of the ZIF-8 model size, 2×2×6 and 2×2×12 supercells of ZIF-8 were considered, containing 6,624 and 13,248 framework atoms and having dimensions of 33.2 Å × 33.2 Å × 99.5 Å and 33.2 Å × 33.2 Å × 199 Å, respectively. Given that each conventional unit cell of ZIF-8 contains two inequivalent cages, the 2×2×6 and 2×2×12 supercells contain 48 and 96 inequivalent cages. For the 2×2×6 supercell, visualised in Supplementary Figure 66a, these 48 inequivalent cages can be grouped together in 12 layers, each containing four cages that are located at the same $z$ coordinates in equilibrium. For the 2×2×12 supercell, this procedure would lead to 24 layers, again with four cages each.

In ZIF-8, each cage is connected to eight adjacent cages through one of its 6MR apertures. When collecting the cages into layers as in Supplementary Figure 66a, the eight cages that neighbour a given cage are located in the two layers that are adjacent to the layer of the original cage, as shown in Supplementary Figure 66c. Especially, a cage is not directly connected to other cages in the same layer, nor is it directly connected with any cage that is located in layers that are not adjacent to the original layer.

The aim of this section is twofold. First, it is the intention to confirm the cage-by-cage process by which critically sized water clusters are formed, as observed earlier for the 1×1×2 ZIF-8 supercell. Given the discussion of the cage geometry above, such a process would correspond here with a layer-by-layer filling process. Second, it is the intention to verify whether the intrinsic nanosecond timescale is also retrieved for these larger systems.

To create an inhomogeneous water distribution in these models, seven different setups were simulated: three corresponding with a 2×2×6 ZIF-8 supercell and four corresponding with a 2×2×12 ZIF-8 supercell. Each of the setups consists of a slab of $2n$ adjacent layers in which all four cages per layer are completely filled (40 water molecules per cage or, equivalently, 160 water molecules per layer and $320n$ water molecules in the complete material) at the onset of the simulation, while all remaining layers only contain completely empty cages. For the 2×2×6 ZIF-8 supercells, the three different setups correspond with either two (320 water molecules), four (640 water molecules), or six (960 water molecules) adjacent layers that are completely filled, resulting in respectively ten, eight, or six adjacent layers that are initially empty. For the 2×2×12 ZIF-8 supercells, the four different setups correspond with either two (320 water molecules), four (640 water molecules), six (960 water molecules), or eight (1280 water molecules) adjacent layers that are completely filled, resulting in respectively 22, 20, 18, or 16 adjacent layers that are initially empty.

For each of the seven setups, a series of ten independent $(N, P, \boldsymbol{\sigma_a} = \boldsymbol{0}, T)$ MD simulations were started, in which the velocities of the water molecules were drawn randomly from a Maxwell-Boltzmann distribution at 300 K. The movement of these water molecules at 0 MPa and 300 K for each of these 70 MD simulations was then tracked for the total simulation time of 1, 2, or 3 ns, depending on the system size. As in Supplementary Section S8.1, these large systems were modelled by interfacing our in-house Yaff software package[48] with LAMMPS to calculate the long-range interactions more efficiently.[49] Afterwards, each water molecule was assigned to a specific cage based on the shortest distance between the oxygen atom of the water molecule and each of the cage centres.

To provide a systematic overview of the formation of stable water clusters across these 70 MD simulations, the colour scheme as shown in Supplementary Figure 66b is adopted. In this colour scheme, six distinct colours are used to distinguish between the different layers based on their initial filling:

i. If the cages inside a given layer are filled and the two adjacent layers are also completely filled, there is no empty adjacent cage to which the water molecules can move. The cages in such layers are coloured in different brown shades. In Supplementary Figure 66b, there are $2n$-2 such layers, with $n = 1$, 2 or 3 for the 2×2×6 supercell and $n = 1$, 2, 3, or 4 for the 2×2×12 supercell.

ii. If the cages inside a given layer are empty and the two adjacent layers are also completely empty, the cages cannot be directly filled. The cages in such layers are coloured in different gray shades. In Supplementary Figure 66b, there are 2×(6-$n$)-2 such layers for the 2×2×6 ZIF-8 supercell and 2× (12-$n$)-2 such layers for the 2×2×12 ZIF-8 supercell.

iii. If the cages inside a given layer are filled and one of the adjacent layers contains empty cages, the water molecules can diffuse from this layer to a nearby empty cage. Given that all $2n$ filled layers are adjacent, there are always exactly two such layers, as shown in Supplementary Figure 66b. These two layers are shown in green and blue shades; their eight cages are equivalent at the onset of the simulation similar to cage 2 and cage 3 for the 1×1×2 ZIF-8 supercell.

iv. If the cages inside a given layer are empty and one of the adjacent layers has cages that are filled, the water molecules can diffuse into this cage from this adjacent layer. Given that all empty layers are adjacent, there are always exactly two such layers, as shown in Supplementary Figure 66b. These two layers are shown in purple and orange shades; their eight cages are equivalent.

At first instance, consider the 10 MD simulations performed for the 2×2×6 ZIF-8 supercell with two layers completely filled with water, as shown in Supplementary Figure 67 and Supplementary Figure 68. These simulations show a consistent behaviour: the eight cages that were originally filled (green and blue shades) gradually empty into the nearby cages (orange and purple). While the different orange and purple cages – corresponding to cages 9, 10, 11, 12, 45, 46, 47, and 48 – are all equivalent, the exact cages that are filled depend on the initial conditions. For instance, cages 10, 12, 45, 46, 47, and 48 are filled in simulation 1, while cages 9, 11, 46, 47, and 48 are filled in simulation 2. This behaviour is conceptually the same as the filling of cages 2 and 3 in the 1×1×2 ZIF-8 supercell, and depends on the initial conditions. The further characteristics of the formation of critically sized water clusters in the 1×1×2 ZIF-8 supercell are also observed here. First, if a cage is insufficiently filled so that no sufficiently stabilised water cluster can form, the cage empties again, against the water gradient. This is for instance the case for cage 45 in simulations 2 and 4, and for cage 9 in simulation 5. Second, the cages of initially empty layers that are not adjacent to initially filled layers, indicated in gray, are never filled over the course of the simulation, similar to the earlier observation that cage 4 was never filled over the course of the 5 ns simulations on the 1×1×2 ZIF-8 supercell. This further establishes the cage-by-cage (or here layer-by-layer) character of the intrusion process; the layers shaded in gray are mere spectator layers that do not play a role on this nanosecond timescale.

These observations – the equivalency of the different cages in a given layer, the cage-by-cage process of forming stabilised water clusters, and the necessity of forming a sufficiently large water cluster – can also be made when either four layers (Supplementary Figure 69 and Supplementary Figure 70) or six layers (Supplementary Figure 71 and Supplementary Figure 72) are initially filled with water. Again, the cages shaded in gray do not play a role on this nanosecond timescale, and the filled cages that are not adjacent to empty cages – coloured in brown shades – empty more slowly since the adjacent cages are already filled with water molecules.

For each of these 30 simulations of the 2×2×6 ZIF-8 supercell, the timescale for the formation of stable water clusters has been extracted. To this end, the time constant $\tau$ and the total water cluster size $a$ are fitted according to the exponential equation $a(1 - e^{-t/\tau})$ that was also adopted for the 1×1×2 ZIF-8 supercell models. Again, cages were only considered in this analysis if (i) they were initially empty and (ii) they contained at least two water molecules at the end of the simulation. This effectively reduced the cages in this analysis to a subset of the purple- and orange-shaded cages, as they are the only ones being filled in the process. In addition, a third criterion was added: (iii) the total water cluster fitting parameter $a$ should not exceed 42 water molecules. This last criterion is added with respect to the 1×1×2 ZIF-8 supercell model given the shorter simulation times. Due to these shorter simulation times, cages that only start to fill later on in the process or do not fill appreciably during the simulation time (for instance cage 18 in simulation 10 of Supplementary Figure 70) would be poorly fit by the exponential, with a very slow time constant $\tau$ for the formation of critical-sized water clusters (up to several hundreds of nanoseconds) and an associated large total water cluster size $a$. Since this is a consequence of the limited simulation time, such cages are not retained in the analysis, which will lead to an underestimation of the intrinsic timescale for the cage filling process.

For all cages that satisfy criteria (i) to (iii), the distribution of the fitted time constants $\tau$ is shown in Supplementary Figure 65 alongside the results obtained earlier for the 1×1×2 ZIF-8 supercell. Also from this visualisation it is clear that the model size does not affect the intrinsic time scale for the formation of stable water clusters appreciably, with most results again falling inside the [0.5 ns, 1.5 ns] interval.

These 2×2×6 ZIF-8 supercell simulations therefore confirm the intrinsic nanosecond timescale obtained for the 1×1×2 ZIF-8 supercell.

To qualitatively confirm these results for even large ZIF-8 models, the results for 1 ns MD simulations of a 2×2×12 ZIF-8 supercell are shown in Supplementary Figure 73 and Supplementary Figure 74 (two layers filled initially), Supplementary Figure 75 and Supplementary Figure 76 (four layers filled initially), Supplementary Figure 77 and Supplementary Figure 78 (six layers filled initially), Supplementary Figure 79 and Supplementary Figure 80 (eight layers filled initially). These 40 simulations confirm the earlier observations:

1. Purple- and orange-shaded cages, describing initially empty layers that are adjacent to filled layers, are the only cages that get filled appreciably during the simulation time. Initially empty layers that are not adjacent to filled layers, coloured in a gray shade, are spectator layers that remain largely empty throughout the simulation. The water intrusion process is therefore a cage-by-cage process.
2. The exact cages that get filled through the simulation depend on the initial conditions, although each of these purple- and orange-shaded cages is equivalent when averaging over the ten independent simulations for each setup.
3. If a cage gets filled, one of two phenomena occur. If the cage is filled with a sufficient amount of water molecules that are sufficiently stabilised, the cage will fill further (see for instance cage 11 in in simulation 1 of Supplementary Figure 73). In contrast, if the water cluster inside the cage is insufficiently stabilised, the cage empties again back into the originally filled cage, against the water gradient (see for instance cage 9 in in simulation 1 of Supplementary Figure 73).
4. While the simulations performed for the 2×2×12 ZIF-8 supercell are too short to extract accurate time constants for the formation of critical-sized water clusters, the qualitative behaviour of the 2×2×12 ZIF-8 supercell is very similar to the 2×2×6 and 2×2×1 ZIF-8 supercells, for which an intrinsic nanosecond timescale for the formation of stable water clusters was extracted.

In conclusion, the 70 MD simulations on larger ZIF-8 models performed here demonstrate that there are no appreciable effects of the model size on the formation of stable water clusters and the associated intrinsic timescale.

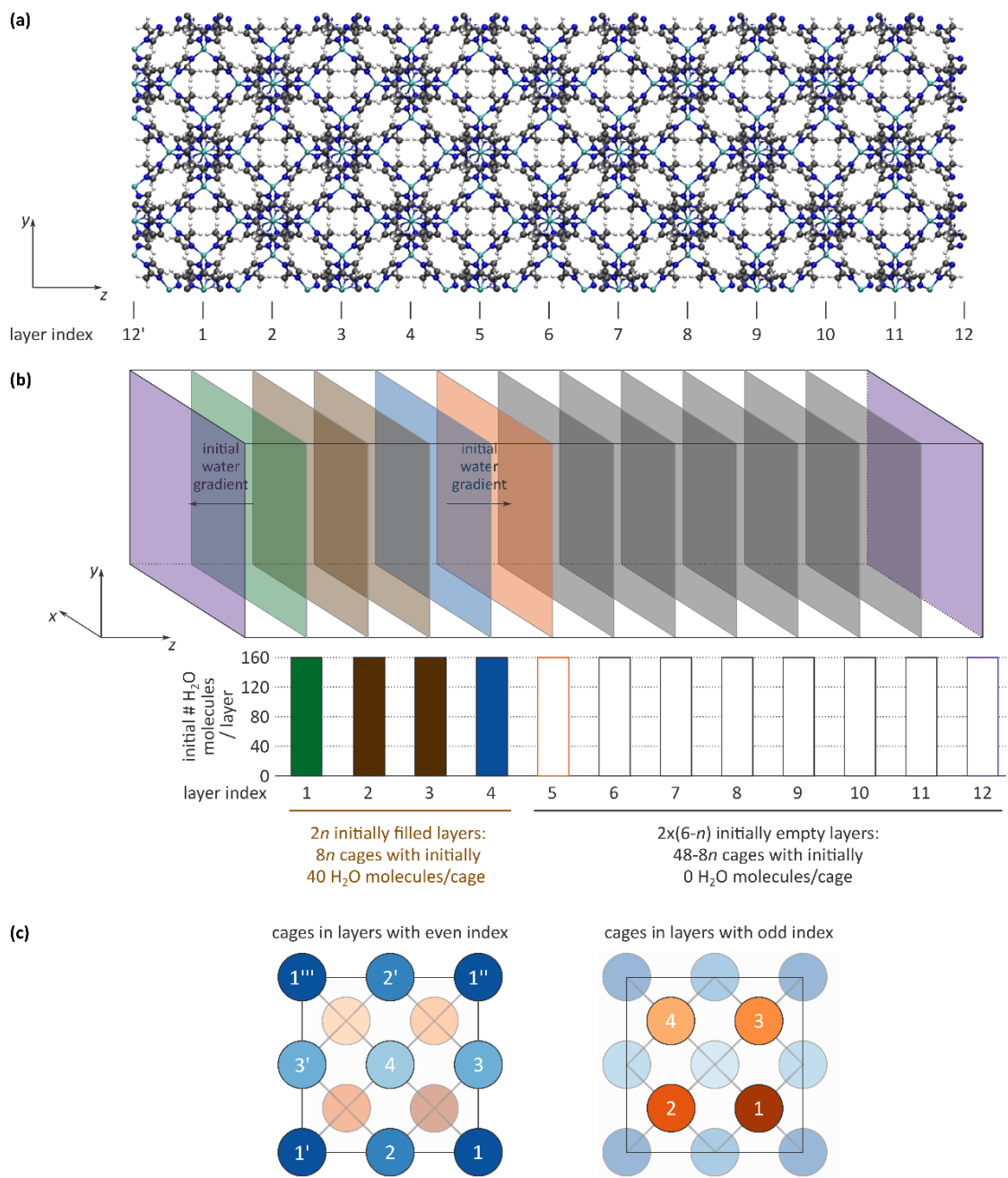

**Supplementary Figure 66 | Nomenclature and colour code for the 2×2×6 and 2×2×12 ZIF-8 supercells. a,** Atomic representation of the 2×2×6 ZIF-8 supercell showing the 12 inequivalent layers along the $z$-axis that each contain four cages. The layers with indices 12 and 12' are the same due to periodic boundary conditions. Cages of a given layer only connect to cages of one of the two adjacent layers (see panel **c**). A similar figure with 24 inequivalent layers could be made for the 2×2×12 ZIF-8 supercell. **b,** Colour code used to represent the different layers, here illustrated for the 2×2×6 ZIF-8 supercell. The cages of $2n$ of these layers are initially filled, while the remaining 2×(6-$n$) layers and 2×(12-$n$) layers of the 2×2×6 and 2×2×12 ZIF-8 supercells, respectively, are initially empty. If the cages in the layer are initially filled and are not adjacent to a layer with initially empty cages, the layer is coloured in a brown shade. If they are initially filled but also adjacent to a layer with initially empty cages, they are coloured in either a green or a blue shade. If the layer contains cages that are initially empty and is not adjacent to a layer with initially filled cages, the layer is coloured in a grey shade. Finally, if the layer contains cages that are initially empty but are adjacent to a layer with initially filled cages, the layer is coloured in a purple or an orange shade. **c,** Indexing and location of the cages in each layer, showing the cage connectivity between adjacent cages that belong to adjacent layers. Cages within a given layer that are the same due to periodic boundary conditions have the same index.

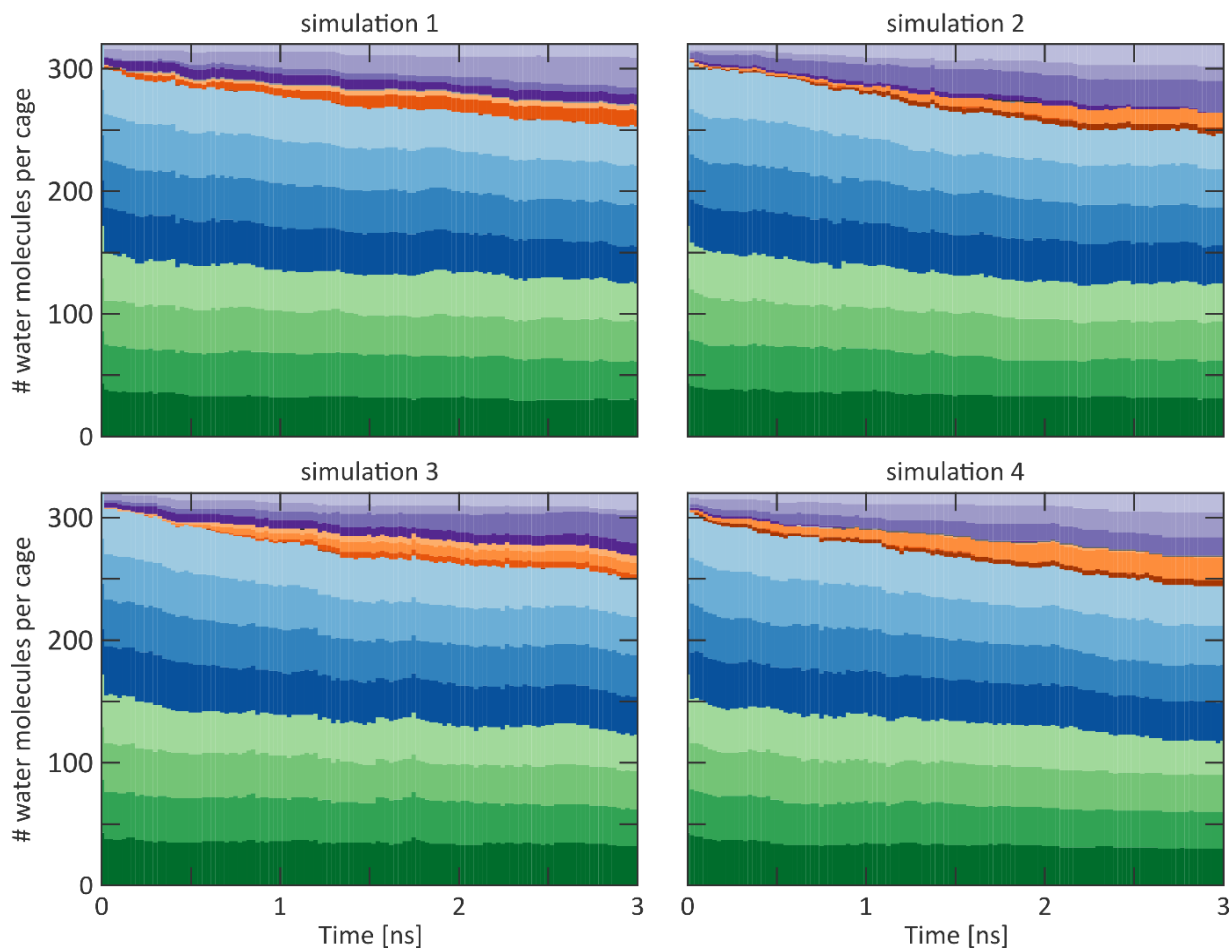

**Supplementary Figure 67 | Evolution of the number of water molecules per cage during a 3 ns MD simulation at 300 K and 0 MPa of a 2×2×6 ZIF-8 supercell with an inhomogeneous water distribution in which two layers, corresponding to a 2×2×1 ZIF-8 supercell, were initially filled.** All filled layers initially contained 40 water molecules per cage, while all other cages were initially empty, leading to a total of 320 water molecules in the system. These simulations were initialised with different initial velocities for the water molecules, which were drawn randomly from a Maxwell-Boltzmann distribution at 300 K. The cage indexing and colouring follows the scheme outlined in Supplementary Figure 66. Figure continued in Supplementary Figure 68.

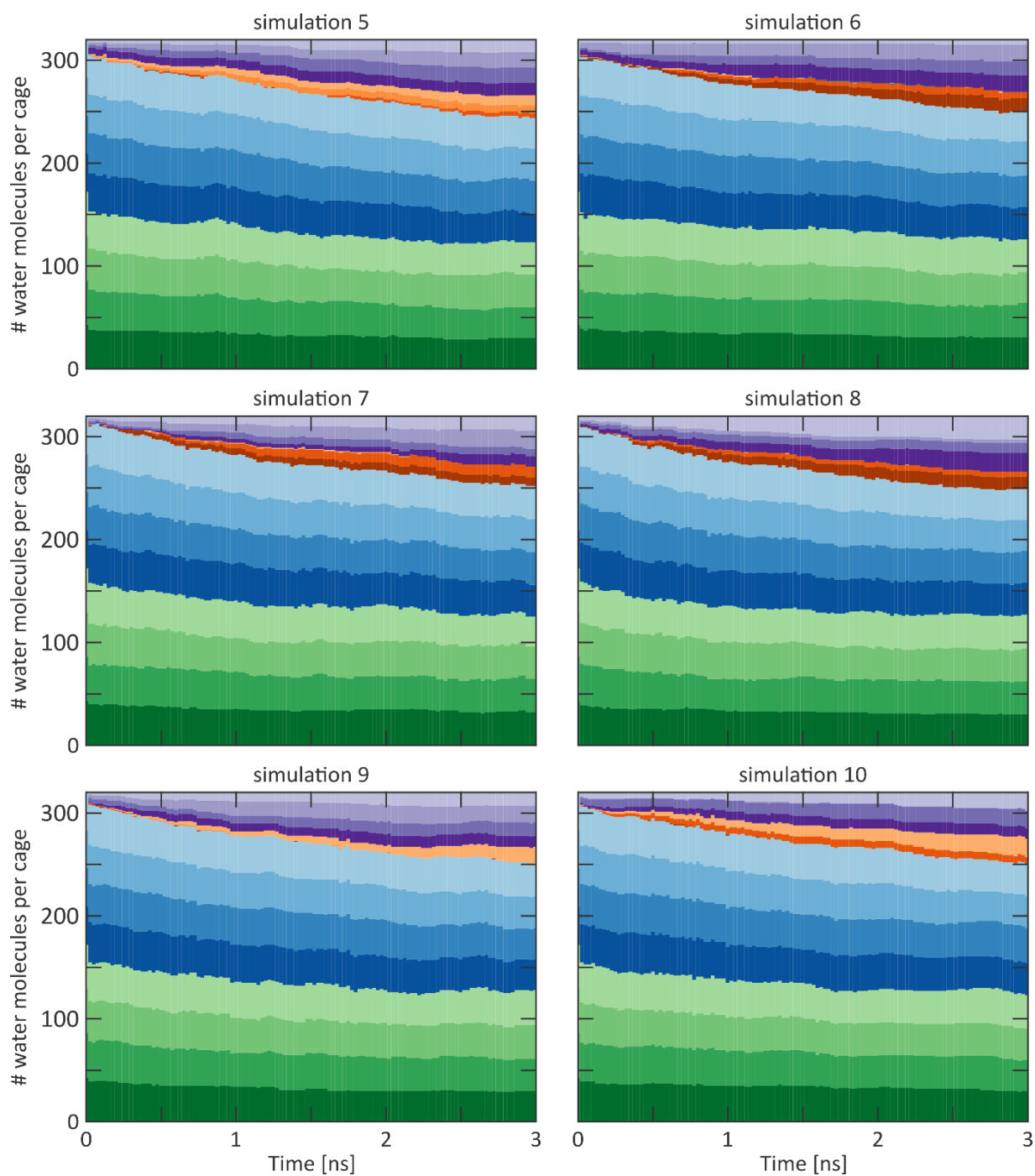

**Supplementary Figure 68 | Evolution of the number of water molecules per cage during a 3 ns MD simulation at 300 K and 0 MPa of a 2×2×6 ZIF-8 supercell with an inhomogeneous water distribution in which two layers, corresponding to a 2×2×1 ZIF-8 supercell, were initially filled.** All filled layers initially contained 40 water molecules per cage, while all other cages were initially empty, leading to a total of 320 water molecules in the system. These simulations were initialised with different initial velocities for the water molecules, which were drawn randomly from a Maxwell-Boltzmann distribution at 300 K. The cage indexing and colouring follows the scheme outlined in Supplementary Figure 66. Figure continued from Supplementary Figure 67.

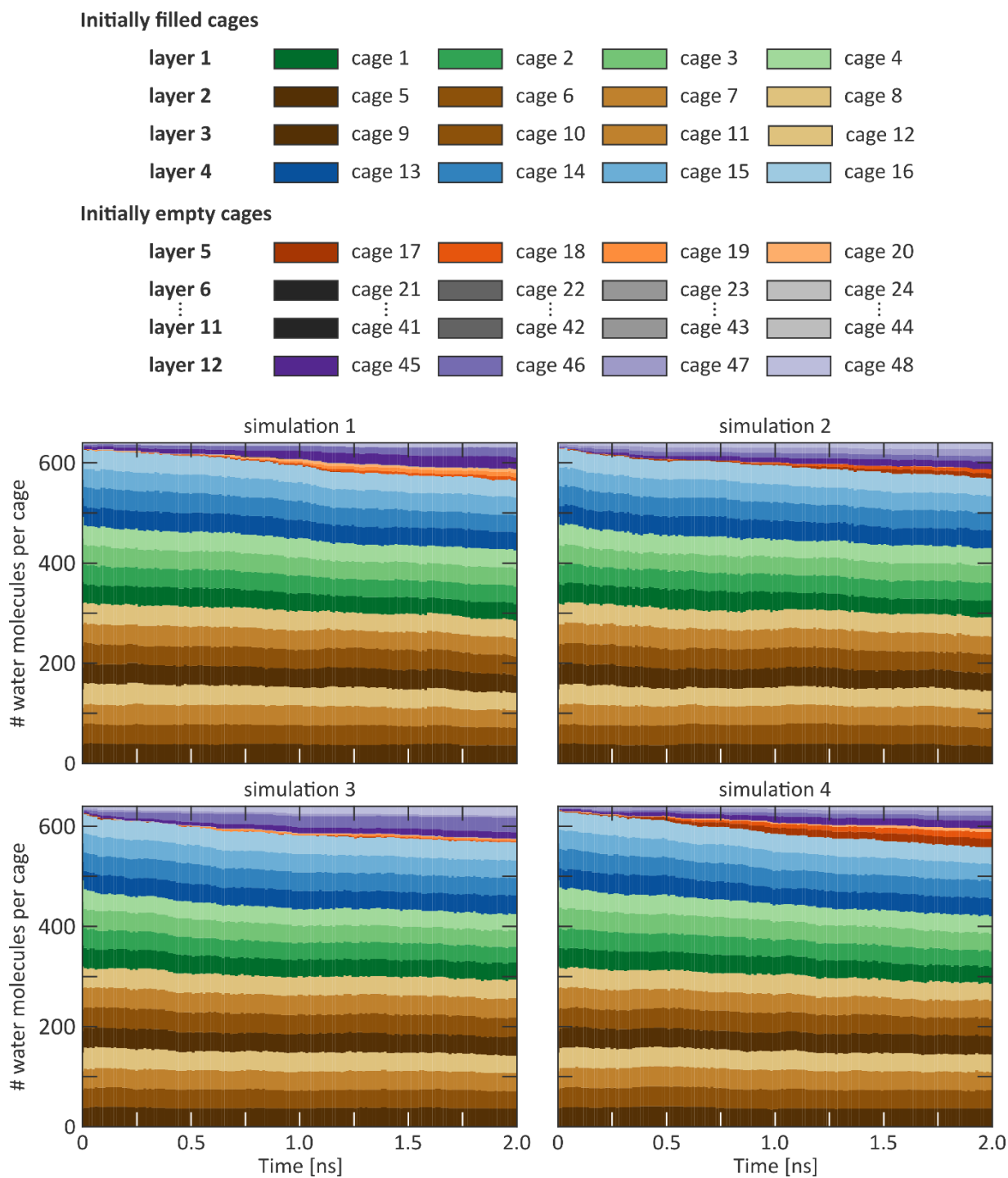

**Supplementary Figure 69 | Evolution of the number of water molecules per cage during a 2 ns MD simulation at 300 K and 0 MPa of a 2×2×6 ZIF-8 supercell with an inhomogeneous water distribution in which four layers, corresponding to a 2×2×2 ZIF-8 supercell, were initially filled.** All filled layers initially contained 40 water molecules per cage, while all other cages were initially empty, leading to a total of 640 water molecules in the system. These simulations were initialised with different initial velocities for the water molecules, which were drawn randomly from a Maxwell-Boltzmann distribution at 300 K. The cage indexing and colouring follows the scheme outlined in Supplementary Figure 66. Figure continued in Supplementary Figure 70.

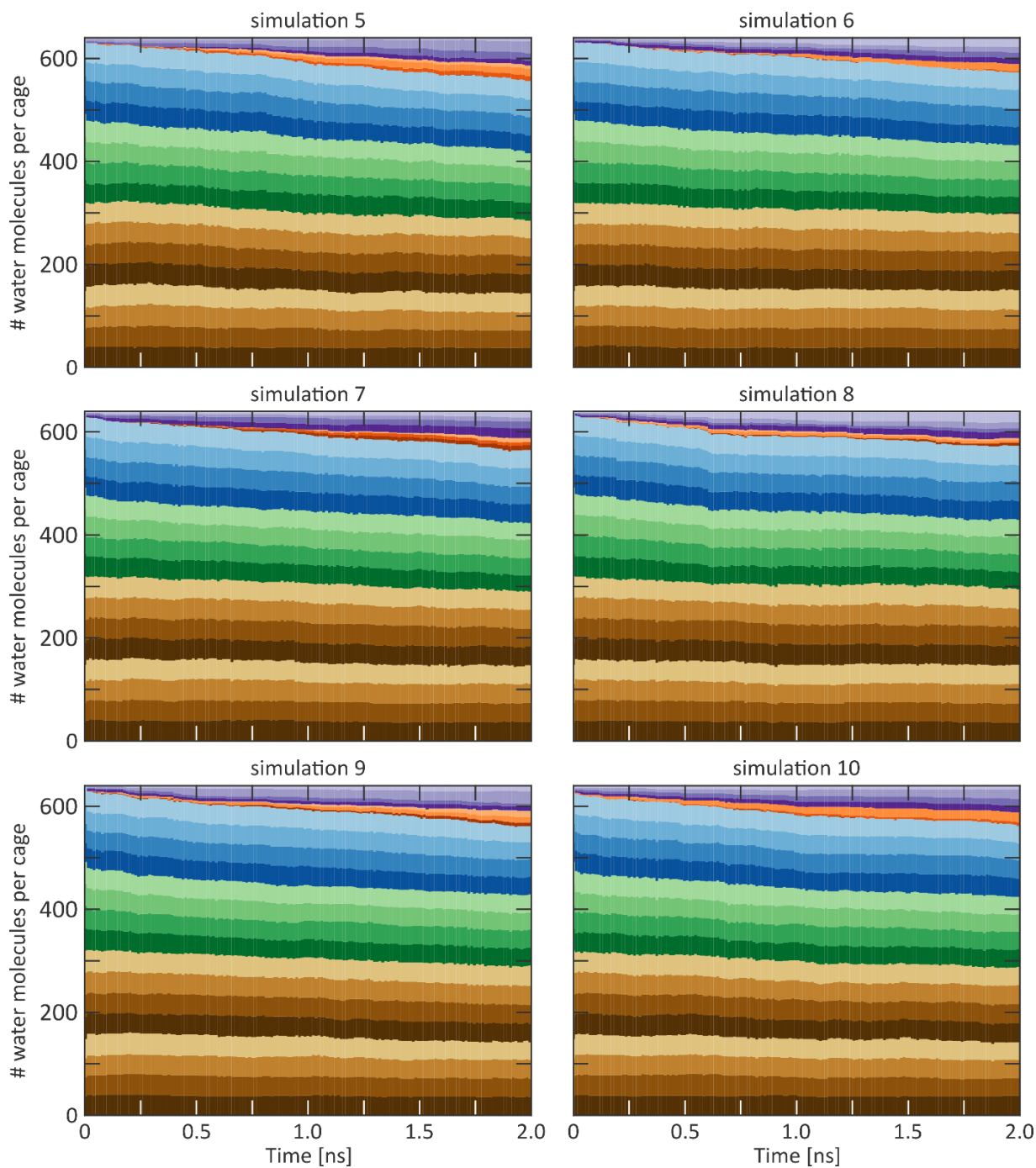

**Supplementary Figure 70 | Evolution of the number of water molecules per cage during a 2 ns MD simulation at 300 K and 0 MPa of a 2×2×6 ZIF-8 supercell with an inhomogeneous water distribution in which four layers, corresponding to a 2×2×2 ZIF-8 supercell, were initially filled.** All filled layers initially contained 40 water molecules per cage, while all other cages were initially empty, leading to a total of 640 water molecules in the system. These simulations were initialised with different initial velocities for the water molecules, which were drawn randomly from a Maxwell-Boltzmann distribution at 300 K. The cage indexing and colouring follows the scheme outlined in Supplementary Figure 66. Figure continued from Supplementary Figure 69.

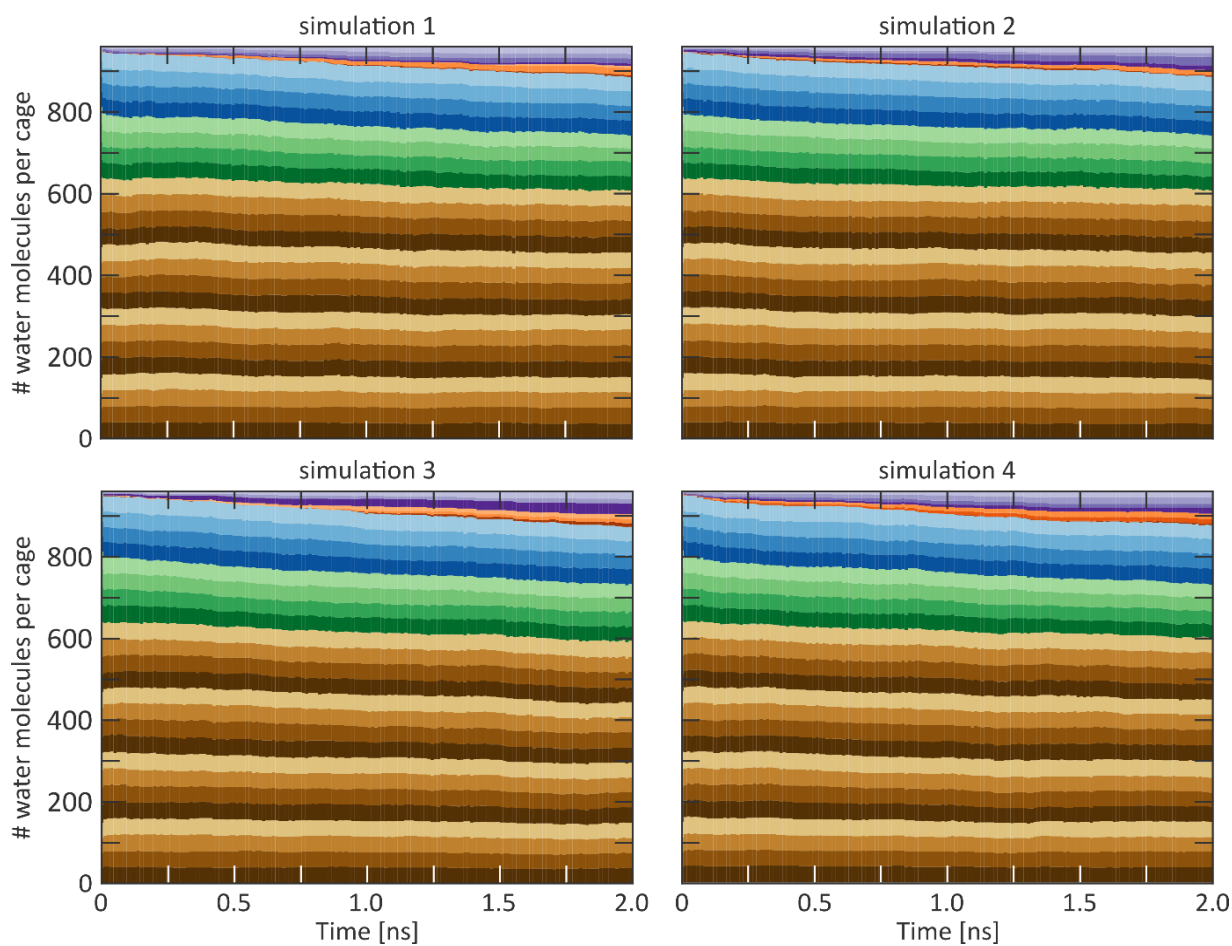

**Supplementary Figure 71 | Evolution of the number of water molecules per cage during a 2 ns MD simulation at 300 K and 0 MPa of a 2×2×6 ZIF-8 supercell with an inhomogeneous water distribution in which six layers, corresponding to a 2×2×3 ZIF-8 supercell, were initially filled.** All filled layers initially contained 40 water molecules per cage, while all other cages were initially empty, leading to a total of 960 water molecules in the system. These simulations were initialised with different initial velocities for the water molecules, which were drawn randomly from a Maxwell-Boltzmann distribution at 300 K. The cage indexing and colouring follows the scheme outlined in Supplementary Figure 66. Figure continued in Supplementary Figure 72.

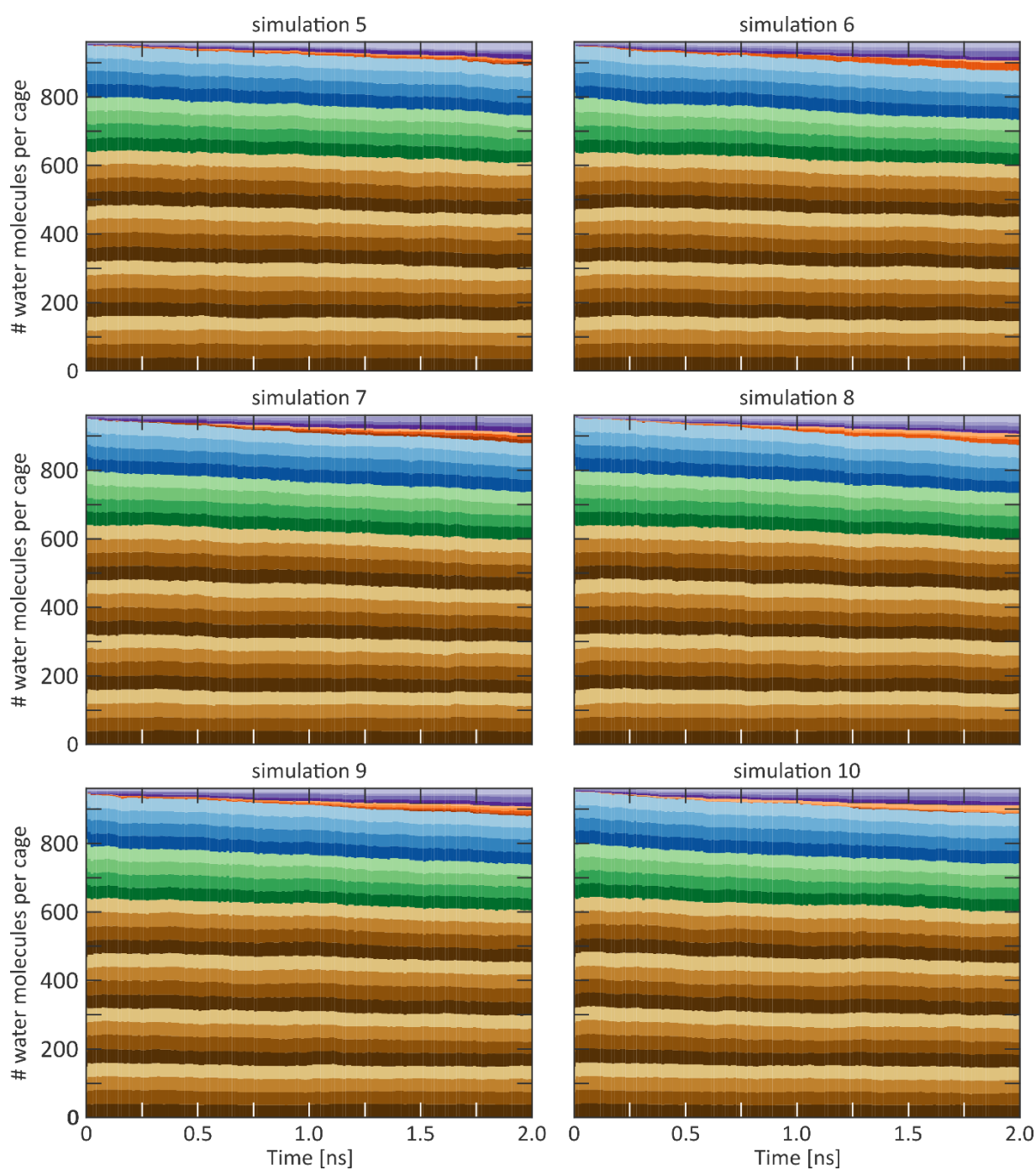

**Supplementary Figure 72 | Evolution of the number of water molecules per cage during a 2 ns MD simulation at 300 K and 0 MPa of a 2×2×6 ZIF-8 supercell with an inhomogeneous water distribution in which six layers, corresponding to a 2×2×3 ZIF-8 supercell, were initially filled.** All filled layers initially contained 40 water molecules per cage, while all other cages were initially empty, leading to a total of 960 water molecules in the system. These simulations were initialised with different initial velocities for the water molecules, which were drawn randomly from a Maxwell-Boltzmann distribution at 300 K. The cage indexing and colouring follows the scheme outlined in Supplementary Figure 66. Figure continued from Supplementary Figure 71.

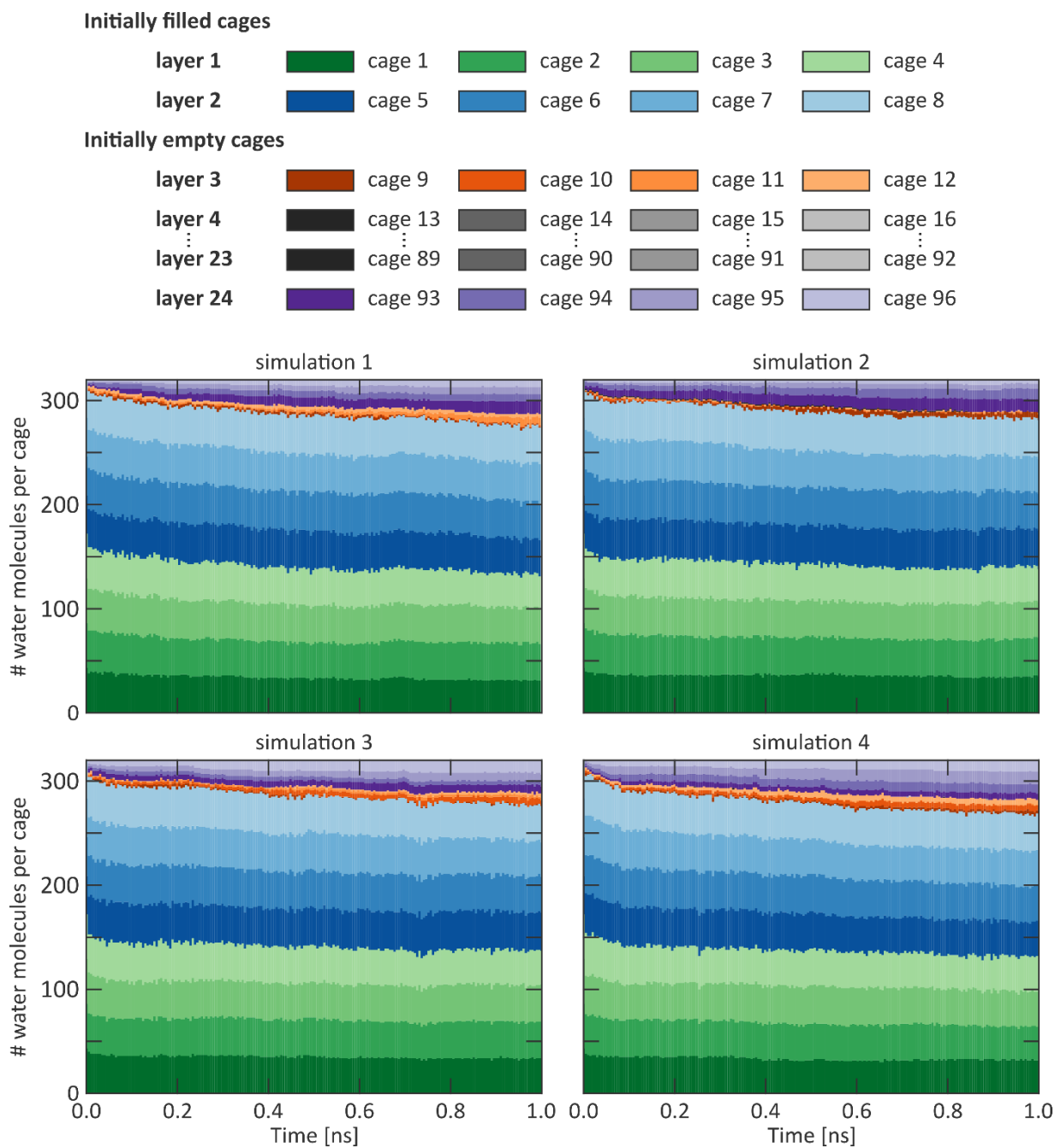

**Supplementary Figure 73 | Evolution of the number of water molecules per cage during a 1 ns MD simulation at 300 K and 0 MPa of a 2×2×12 ZIF-8 supercell with an inhomogeneous water distribution in which two layers, corresponding to a 2×2×1 ZIF-8 supercell, were initially filled.** All filled layers initially contained 40 water molecules per cage, while all other cages were initially empty, leading to a total of 320 water molecules in the system. These simulations were initialised with different initial velocities for the water molecules, which were drawn randomly from a Maxwell-Boltzmann distribution at 300 K. The cage indexing and colouring follows the scheme outlined in Supplementary Figure 66. Figure continued in Supplementary Figure 74.

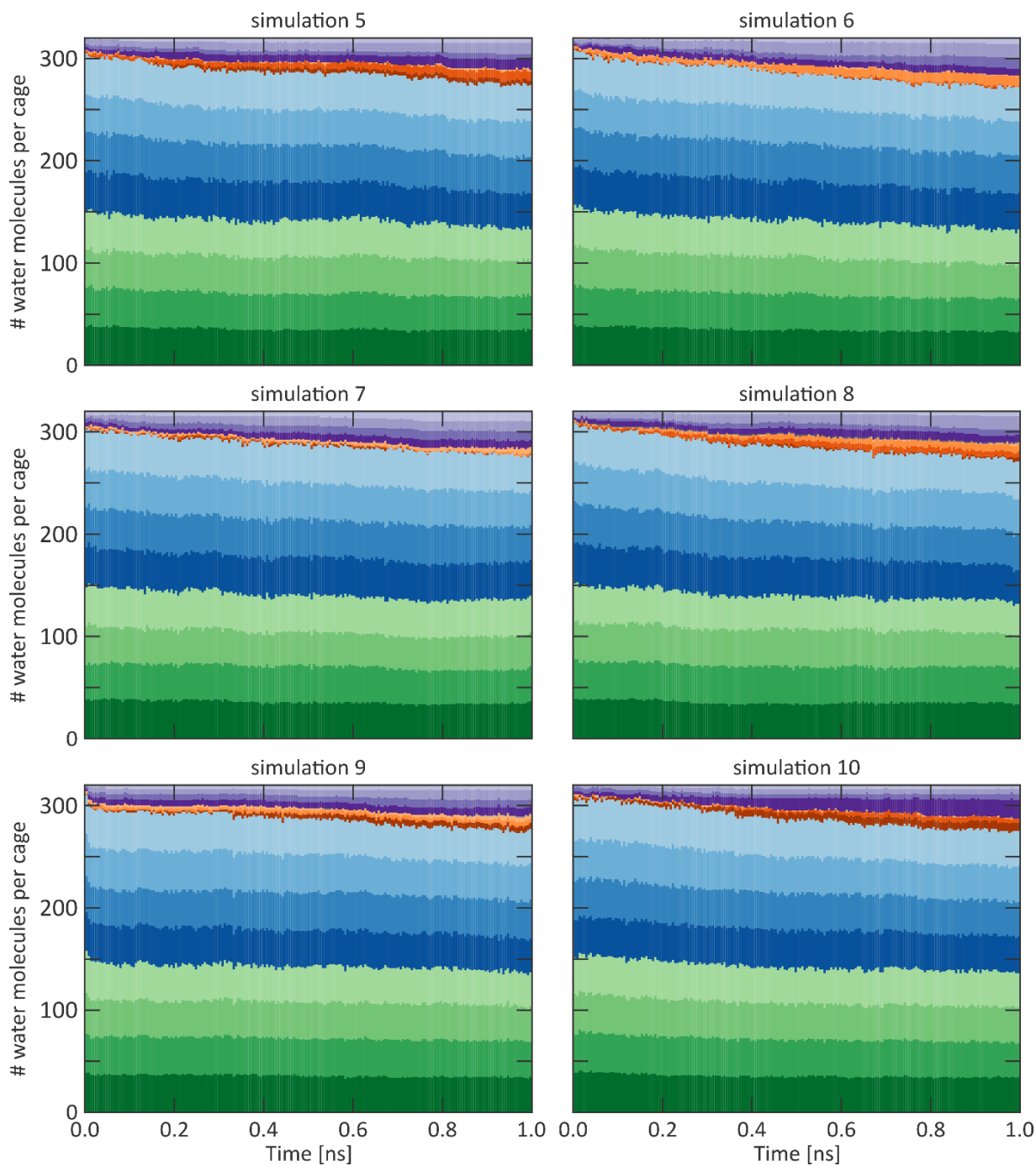

**Supplementary Figure 74 | Evolution of the number of water molecules per cage during a 1 ns MD simulation at 300 K and 0 MPa of a 2×2×12 ZIF-8 supercell with an inhomogeneous water distribution in which two layers, corresponding to a 2×2×1 ZIF-8 supercell, were initially filled.** All filled layers initially contained 40 water molecules per cage, while all other cages were initially empty, leading to a total of 320 water molecules in the system. These simulations were initialised with different initial velocities for the water molecules, which were drawn randomly from a Maxwell-Boltzmann distribution at 300 K. The cage indexing and colouring follows the scheme outlined in Supplementary Figure 66. Figure continued from Supplementary Figure 73.

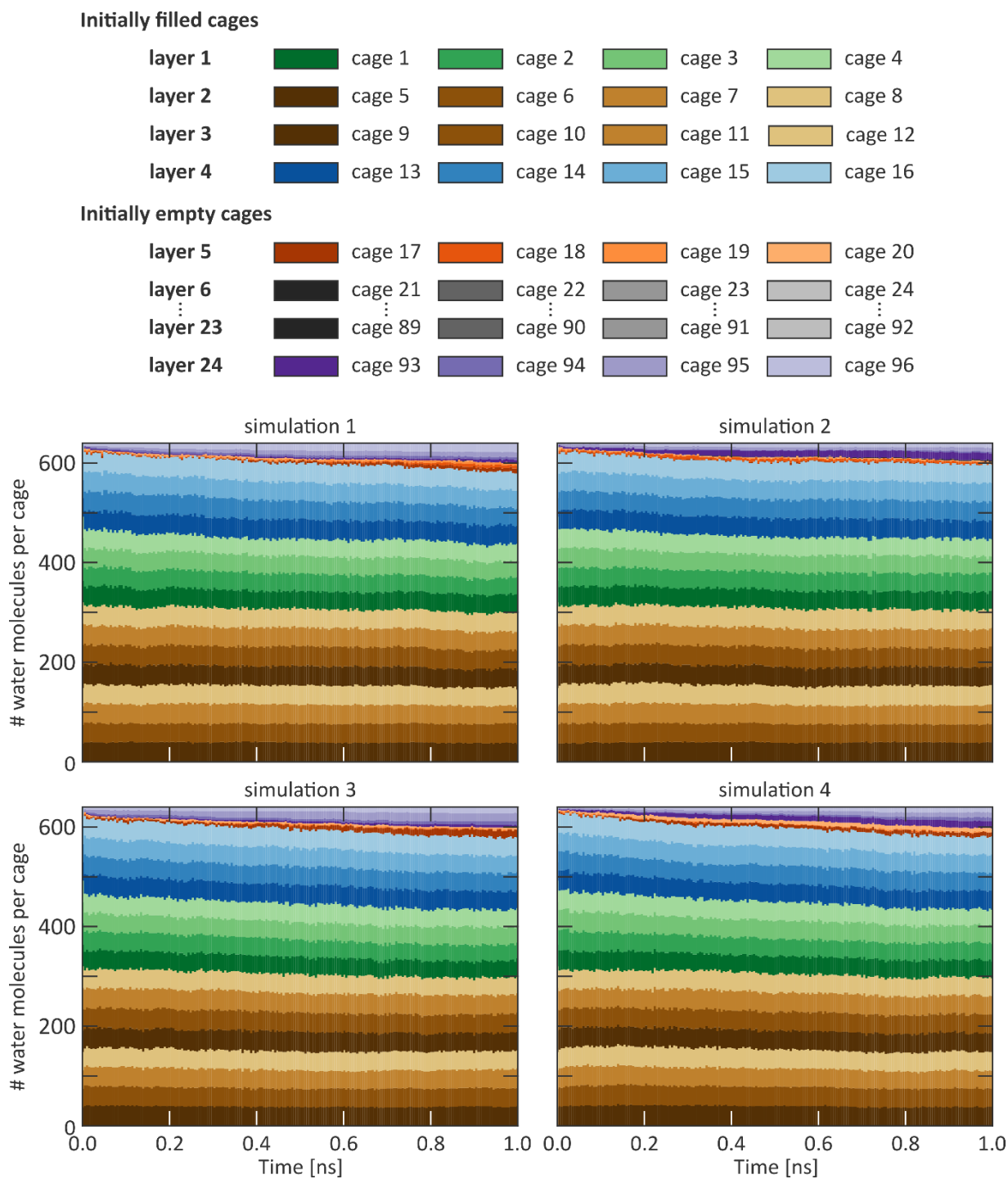

**Supplementary Figure 75 | Evolution of the number of water molecules per cage during a 1 ns MD simulation at 300 K and 0 MPa of a 2×2×12 ZIF-8 supercell with an inhomogeneous water distribution in which four layers, corresponding to a 2×2×2 ZIF-8 supercell, were initially filled.** All filled layers initially contained 40 water molecules per cage, while all other cages were initially empty, leading to a total of 640 water molecules in the system. These simulations were initialised with different initial velocities for the water molecules, which were drawn randomly from a Maxwell-Boltzmann distribution at 300 K. The cage indexing and colouring follows the scheme outlined in Supplementary Figure 66. Figure continued in Supplementary Figure 76.

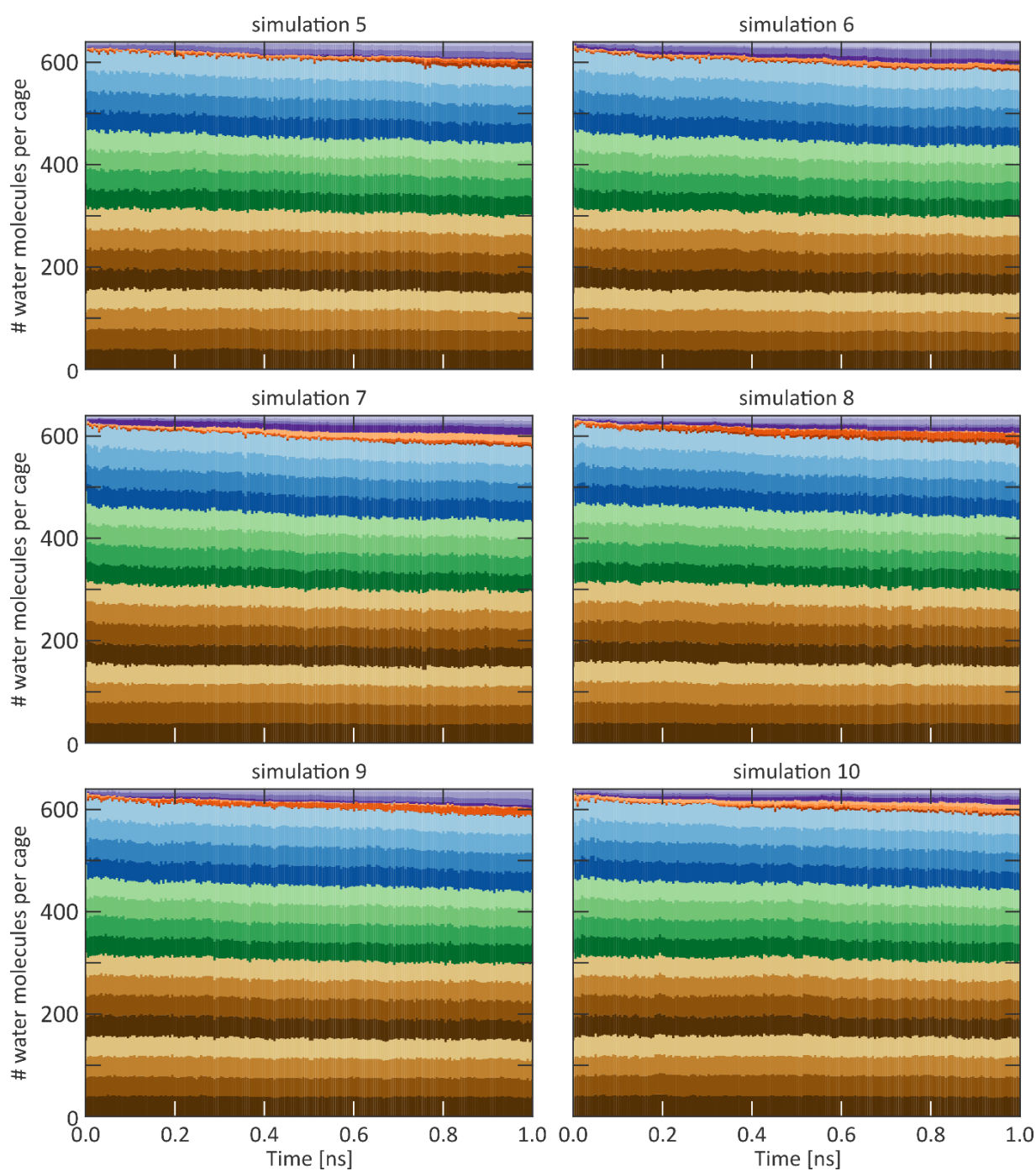

**Supplementary Figure 76 | Evolution of the number of water molecules per cage during a 1 ns MD simulation at 300 K and 0 MPa of a 2×2×12 ZIF-8 supercell with an inhomogeneous water distribution in which four layers, corresponding to a 2×2×2 ZIF-8 supercell, were initially filled.** All filled layers initially contained 40 water molecules per cage, while all other cages were initially empty, leading to a total of 640 water molecules in the system. These simulations were initialised with different initial velocities for the water molecules, which were drawn randomly from a Maxwell-Boltzmann distribution at 300 K. The cage indexing and colouring follows the scheme outlined in Supplementary Figure 66. Figure continued from Supplementary Figure 75.

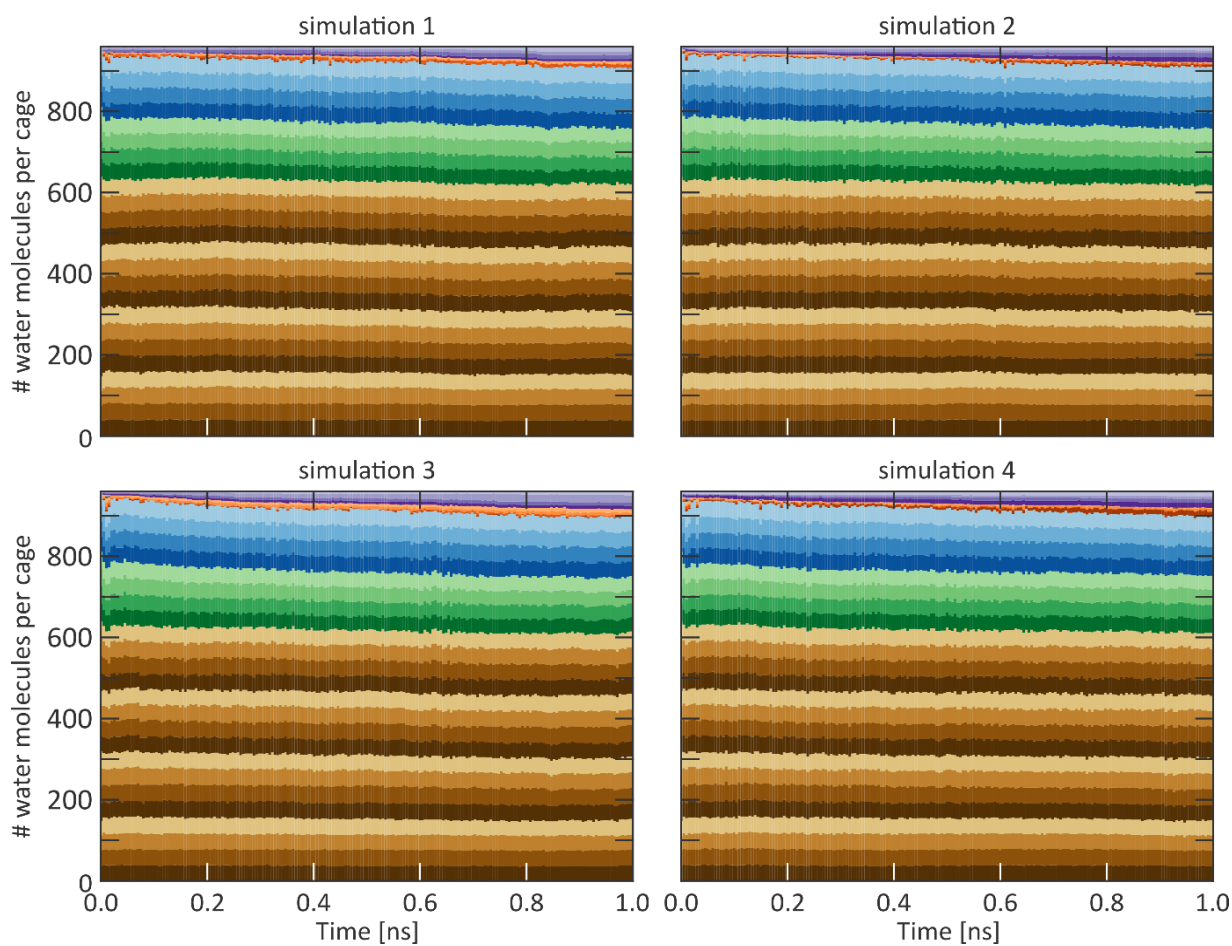

**Supplementary Figure 77 | Evolution of the number of water molecules per cage during a 1 ns MD simulation at 300 K and 0 MPa of a 2×2×12 ZIF-8 supercell with an inhomogeneous water distribution in which six layers, corresponding to a 2×2×3 ZIF-8 supercell, were initially filled.** All filled layers initially contained 40 water molecules per cage, while all other cages were initially empty, leading to a total of 960 water molecules in the system. These simulations were initialised with different initial velocities for the water molecules, which were drawn randomly from a Maxwell-Boltzmann distribution at 300 K. The cage indexing and colouring follows the scheme outlined in Supplementary Figure 66. Figure continued in Supplementary Figure 78.

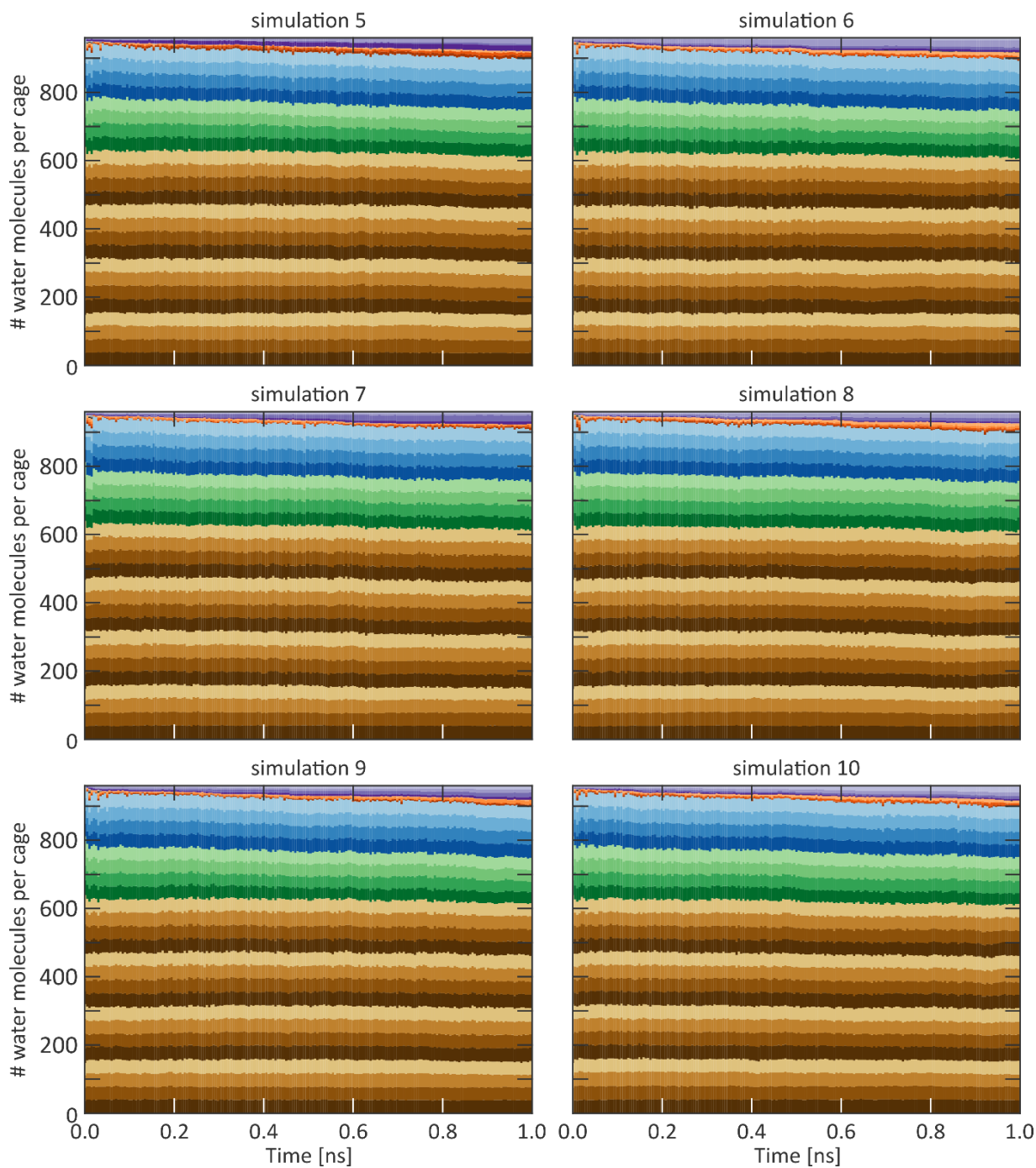

**Supplementary Figure 78 | Evolution of the number of water molecules per cage during a 1 ns MD simulation at 300 K and 0 MPa of a 2×2×12 ZIF-8 supercell with an inhomogeneous water distribution in which six layers, corresponding to a 2×2×3 ZIF-8 supercell, were initially filled.** All filled layers initially contained 40 water molecules per cage, while all other cages were initially empty, leading to a total of 960 water molecules in the system. These simulations were initialised with different initial velocities for the water molecules, which were drawn randomly from a Maxwell-Boltzmann distribution at 300 K. The cage indexing and colouring follows the scheme outlined in Supplementary Figure 66. Figure continued from Supplementary Figure 77.

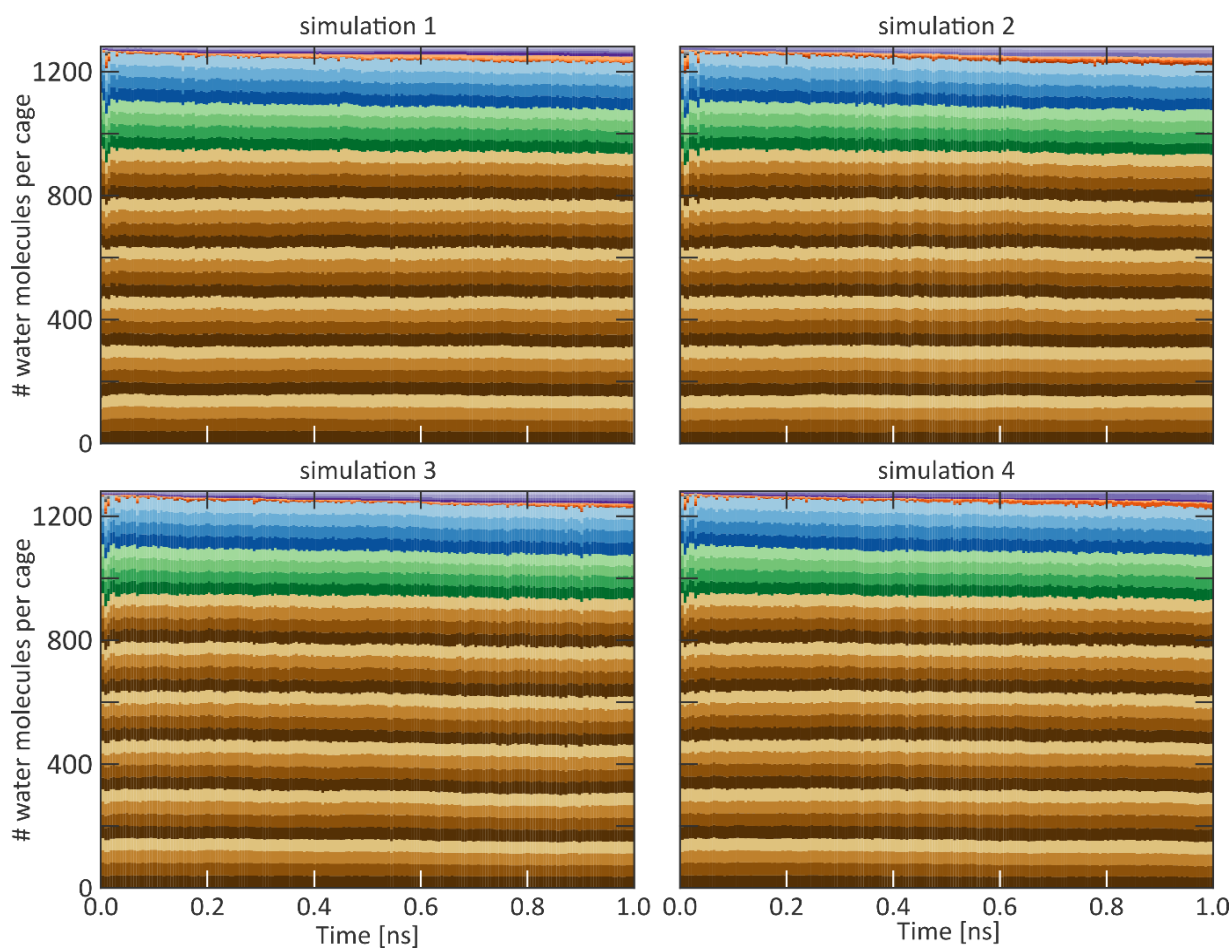

**Supplementary Figure 79 | Evolution of the number of water molecules per cage during a 1 ns MD simulation at 300 K and 0 MPa of a 2×2×12 ZIF-8 supercell with an inhomogeneous water distribution in which eight layers, corresponding to a 2×2×4 ZIF-8 supercell, were initially filled.** All filled layers initially contained 40 water molecules per cage, while all other cages were initially empty, leading to a total of 1280 water molecules in the system. These simulations were initialised with different initial velocities for the water molecules, which were drawn randomly from a Maxwell-Boltzmann distribution at 300 K. The cage indexing and colouring follows the scheme outlined in Supplementary Figure 66. Figure continued in Supplementary Figure 80.

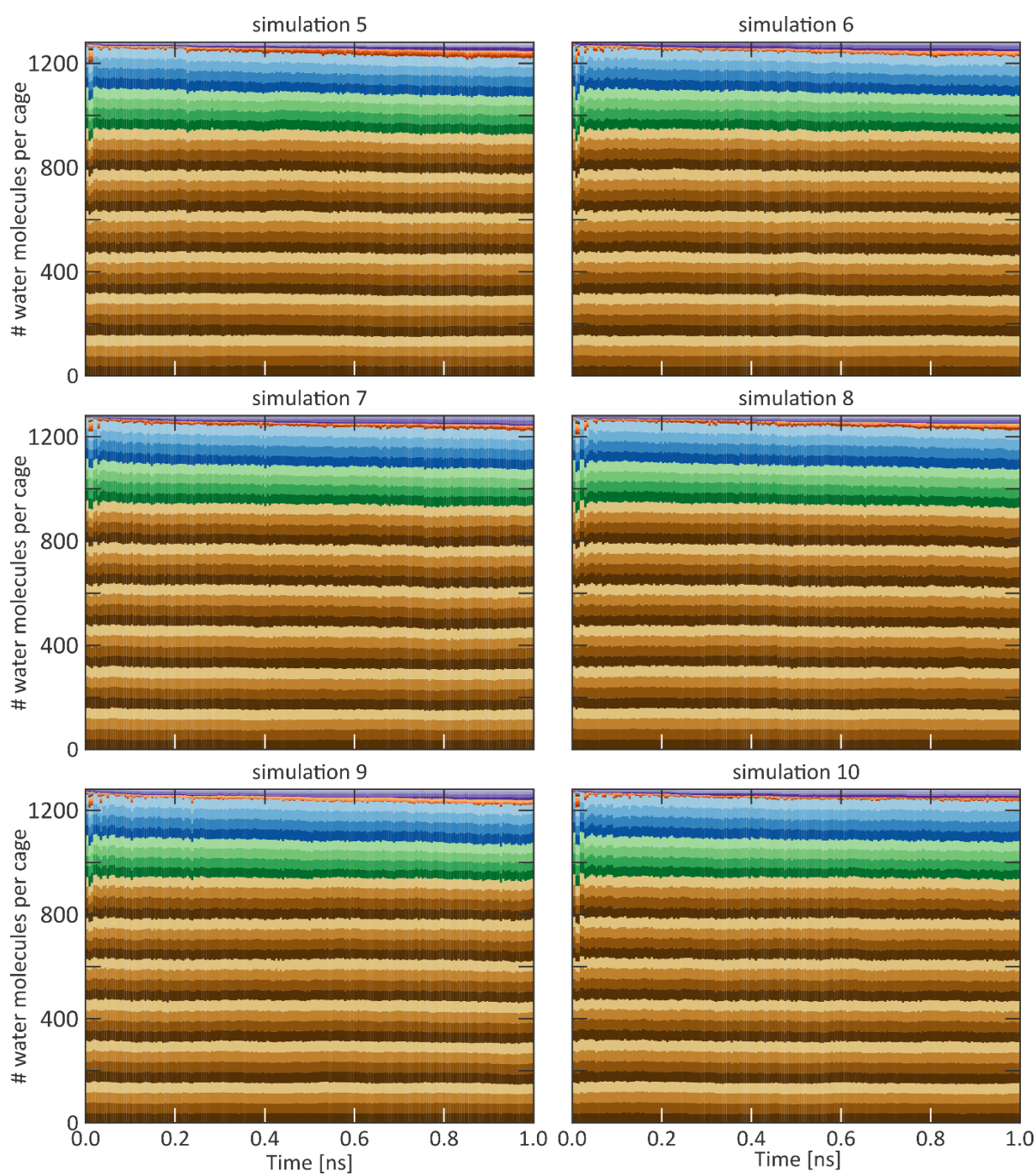

**Supplementary Figure 80 | Evolution of the number of water molecules per cage during a 1 ns MD simulation at 300 K and 0 MPa of a 2×2×12 ZIF-8 supercell with an inhomogeneous water distribution in which eight layers, corresponding to a 2×2×4 ZIF-8 supercell, were initially filled.** All filled layers initially contained 40 water molecules per cage, while all other cages were initially empty, leading to a total of 1280 water molecules in the system. These simulations were initialised with different initial velocities for the water molecules, which were drawn randomly from a Maxwell-Boltzmann distribution at 300 K. The cage indexing and colouring follows the scheme outlined in Supplementary Figure 66. Figure continued from Supplementary Figure 79.

## S8.3 Connection between the intrinsic timescale for the formation of stable water clusters and the experimental intrusion times and strain rates

Based on the discussion above, we can establish a simple analytical model to describe the intrinsic macroscopic behaviour of a ZIF-8 crystal. This model will connect the intrinsic nanosecond timescale with an intrinsic macroscopic timescale for the total intrusion of experimental ZIF-8 crystals and with an intrinsic strain rate, which can be directly compared with the experimental strain rates.

To derive such an analytical model, four assumptions need to be made:

(i) The crystal has a conventional unit cell with unit cell parameter $a$ (for ZIF-8, $a$ is approximately 1.66 nm) and can be modelled as a spherical particle with a radius $r$.

(ii) The intrusion phenomenon occurs in a cage-by-cage process, similar to the cage-by-cage process observed for the formation of critical water clusters, established in Supplementary Sections S8.1 and S8.2.

(iii) The complete filling of a cage with water molecules follows an exponential distribution, $a_{\text{filling}}(1 - e^{-t/\tau_{\text{filling}}})$, similar to the formation of a critical water cluster in such a cage (Supplementary Sections S8.1 and S8.2), and is characterised by a timescale $\tau_{\text{filling}}$. Note that $\tau_{\text{filling}}$ will be substantially larger than the intrinsic timescale for cluster formation, $\tau$, derived before. Supplementary Figure 63 and Supplementary Figure 64 indeed show that, even after 5 ns of simulation time, none of the initially empty cages are completely filled. After 5 ns, none of the originally empty cages contain more than 20 water molecules, far below the saturation limit, although some do contain a critical-sized water cluster. Therefore, we will assume that $\tau_{\text{filling}} = 10\tau$ and hence that the complete filling of a cage occurs on a timescale of a few tens of nanoseconds.

(iv) Water intrusion starts at the same time for all crystals and the complete intrusion of the crystal results in a strain difference $\Delta\varepsilon$ of the {ZIF+water} system. For the ZIF-8 crystals measured here, this strain difference amounts to about 0.1, as can be deduced from the length of the intrusion plateau in Figure 2b of the main text.

Based on assumption (i), the amount of conventional unit cells along the radius of the spherical particle can be determined as $r/a$, while the number of cages along this radius is double that, $2r/a$, given that there are two cages per conventional unit cell. Assuming the same exponential behaviour of the number of water molecules in a cage as before (assumption iii), the amount of time needed for a cage to fill to about 98% when adjacent to a saturated cage amounts to $4\tau_{\text{filling}}$, or approximately $40\tau$, with $\tau$ the intrinsic nanosecond timescale derived from MD simulations. Since water intrudes from the outer surface inwards, the cage-by-cage intrusion process (assumption ii) ensures that the total time for all cages to be completely intruded *via* the formation of critical-sized clusters, as described above, is an experimental observable that amounts to

$$T_{\text{filling}} = \frac{80r\tau}{a}$$

For a ZIF-8 particle with a crystal size of about 250 nm, such as the ones used to measure the intrusion/extrusion curves of Figure 2 in the main text, the total intrinsic intrusion time according to this model would be on the order of 12 µs, close to the intrusion times observed for the high-rate experiments (*ca*. 50 µs). For experimental intrusion times faster than this intrinsic intrusion time, the intrusion process will not take place through the formation of critical-sized clusters as these clusters have insufficient time to nucleate. As explained in Supplementary Section S9.3, critical-sized water clusters facilitate intercage water hopping through the formation of hydrogen bonds. Experimental intrusion time faster than the intrinsic intrusion time can therefore only be reached if additional energy is supplied to the system to overcome the free energy barrier for intercage hopping, explaining the increased intrusion pressure for faster intrusion times, as explained in Supplementary Section S9.3.

Based on this observation, one can also define an intrinsic strain rate. If total water intrusion of the ZIF-8 crystals occurs through the cage-by-cage nucleation of critical-sized water clusters and effectuates a strain difference $\Delta\varepsilon$, the accompanying intrinsic strain rate is, on average, given by

$$\dot{\varepsilon} = \frac{\Delta\varepsilon}{T_{\text{filling}}} = \frac{a\Delta\varepsilon}{80r\tau}$$

For the submicron ZIF-8 crystals discussed in Figure 2b of the main text, $\Delta\varepsilon$ amounts to about 0.1, resulting in an intrinsic strain rate of about $8\times10^3$ s$^{-1}$. As explained above, if a strain rate higher than this intrinsic strain rate is applied to intrude liquid water, critical-sized water clusters cannot form, and additional input pressure is necessary. For strain rates lower than this intrinsic strain rate, in contrast, the cage-by-cage process of critical-sized cluster formation as outlined above can take place.

This analytical model yields an intrinsic strain rate very close to but slightly higher than the strain rate of $2\times10^3$ s$^{-1}$ for the high-rate experiment of Figure 2b of the main text. The origin of this overestimation of the intrinsic strain rate or, equivalently, the underestimation of the intrinsic total filling time, can be ascribed to three factors. First, the time constant $\tau$ for critical water cluster nucleation and hence also $\tau_{\text{filling}}$ is underestimated given that cages in which a stable water cluster forms very slowly, beyond the 5 ns simulation limit, do not contribute to the estimation of $\tau$ as those cages are not retained in the analysis of Supplementary Sections S8.1 and S8.2. This also explains why certain cages that are adjacent to fully saturated cages do not form critical-sized water clusters during these simulations. Second, the relation $\tau_{\text{filling}} = 10\tau$ is an order of magnitude estimation. Substantially longer simulation times on larger ZIF-8 models are necessary to accurately estimate this parameter. Third, the intrusion process is assumed to start and stop at the same time for all crystals (assumption iii), which is very challenging to achieve experimentally. The current analytical model should therefore foremost be regarded as a qualitative model to explain the strain rate-dependence of the intrusion pressure.

# S9 The nucleation and structure of critical-sized water clusters inside the ZIF-8 cages

From Figure 4b in the main text and the discussion in Supplementary Section S8, it seems that the timescale for water intrusion is defined by the intrinsic timescale for critical-sized and stabilised water clusters to nucleate inside the hydrophobic ZIF-8 cages, as the latter defines the rate-limiting step of the process. In this section, the size and structure of these critical-sized water clusters will be discussed. To this end, Supplementary Section S9.1 will give a brief introduction to classical nucleation theory, in which the concept of the critical nucleus as a rate-limiting step is defined. Although classical nucleation theory is not expected to hold in the nanoconfinement of the ZIF-8 cages, various parallels can be drawn with water intrusion studied here. In Supplementary Section S9.2, the size of critical-sized water clusters will be inferred both from the MD simulations with an inhomogeneous water distribution, which were discussed in the section 'The intrinsic water mobility timescale revealed by molecular dynamics simulations' of the main text and Supplementary Section S8 and from the grand canonical Monte Carlo simulations, discussed in Supplementary Section S4. As this preliminary analysis shows that critical-sized water clusters contain about four water molecules, the free energy barrier associated with the nucleation of such a water cluster will be further quantified in Supplementary Section S9.3 using dedicated umbrella sampling (US) simulations, confirming that the nucleation of critical-sized water clusters is the rate-limiting step in this process. Finally, in Supplementary Section S9.4, the internal structure of such critical-sized water clusters will be quantified using an analysis of the hydrogen-bond network.

## S9.1 Critical-sized cluster nucleation according to classical nucleation theory

Under certain assumptions, as mentioned below, classical nucleation theory describes the growth of a stable phase β out of a metastable parent phase α.[50,51] The overall stability of phase β implies that the difference in potential difference between the two phases, $\Delta \mu = \mu^\alpha - \mu^\beta$, is positive, and that it is thermodynamically favourable to undergo the transformation from phase α to phase β. For a nucleus consisting of $N$ particles, this driving force amounts to $\Delta G_v = -N\Delta\mu$ and is proportional to the number of particles inside the nucleus. When this nucleus of phase β starts to form, however, its interface with phase α results in a repulsive interfacial term $\Delta G_s = A\gamma$, in which $A$ is the area of the surface and $\gamma$ is the surface tension. For a spherical particle with radius $r$, $\Delta G_v \propto r^3$ and $\Delta G_s \propto r^2 \propto N^{2/3}$. The combination of both effects leads to the typical free enthalpy profile as a function of the nucleus size reported in Supplementary Figure 81.

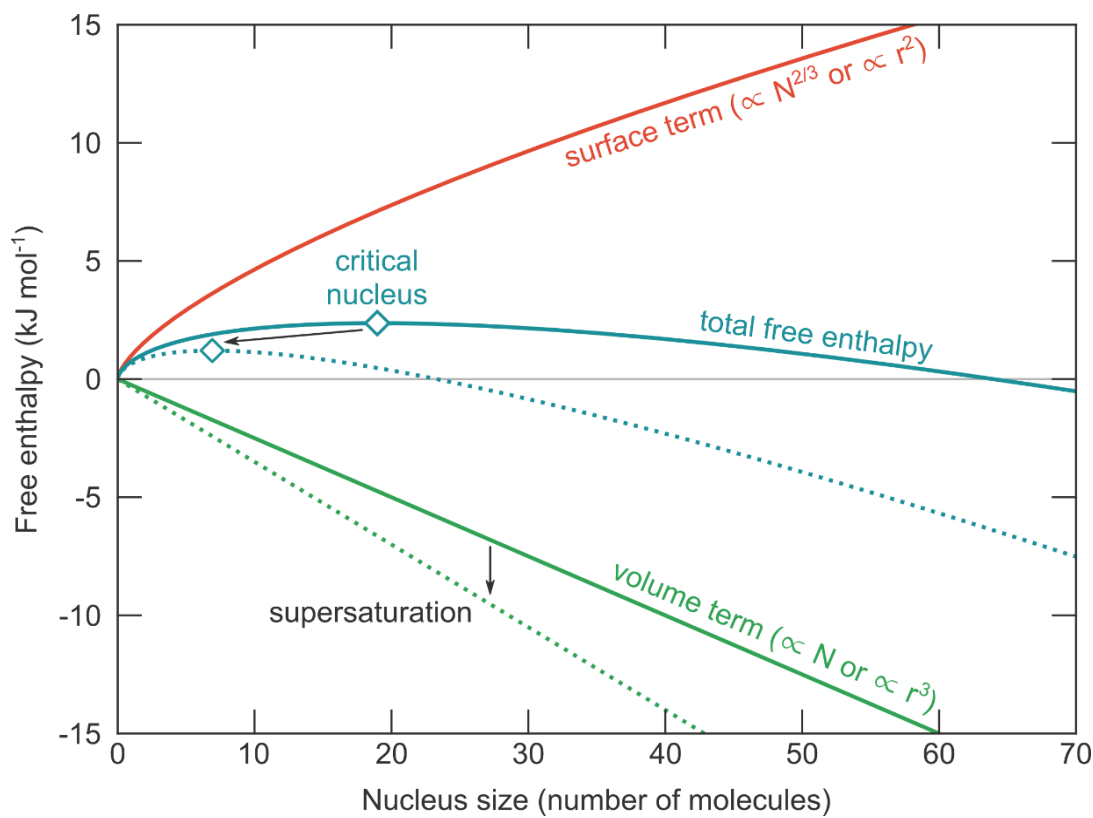

**Supplementary Figure 81 | Schematic overview of an illustrative total free enthalpy barrier and its contributions for phase nucleation according to classical nucleation theory.** The total free enthalpy is obtained as the sum of a volume and a surface contribution. In addition, the location of the critical-sized nucleus and its dependence on the degree of supersaturation are indicated.

As shown in Supplementary Figure 81, the total free enthalpy of the nucleus can be divided into two regions as a function of its size, separated by the maximum in the profile which corresponds with the critical nucleus. On average, for nuclei smaller than this critical nucleus size – so-called subcritical nuclei – the nucleus will dissolve again, while for nuclei larger than this critical nucleus size – so-called supercritical nuclei – the nucleus will grow further. The nucleus size at which the maximum free enthalpy barrier $\Delta G^{\ddagger}$ is reached is therefore a critical parameter in classical nucleation theory as it characterises the rate-limiting step for nucleation, and corresponds with the critical nucleus size. From this knowledge, one can define the nucleation rate per unit volume at a given temperature $T$ as:[50]

$$\dot{N} = jZ\rho \exp\left(-\frac{\Delta G^{\ddagger}}{k_B T}\right)$$

In this expression, $k_B$ is the Bolzmann constant, $j$ is the rate at which molecules attach to the nucleus, $\rho$ is the density of nucleation sites, and $Z$ is the so-called Zeldovich factor that defines the probability of a critical-sized nucleus to further grow into the new phase rather than dissolving again.

By altering the external conditions to further favour the chemical potential of phase β, resulting in supersaturation, the thermodynamic driving force $\Delta\mu$ for nucleation is increased. As indicated schematically in Supplementary Figure 81, this results in a decrease in both free enthalpy barrier $\Delta G^{\ddagger}$ and critical nucleus size, facilitating the growth of phase β.

Having briefly summarized classical nucleation theory, it is important to note that the extreme confinement of water in the hydrophobic ZIF-8 cages and other single-digit nanopores (pores with a diameter smaller than 10 nm) leads to significant guest-host interactions and altered guest-guest interactions compared to the bulk water phase.[52,53] This poses several nontrivial challenges to describe these systems when compared to traditional homogeneous and heterogeneous nucleation, as recently summarized in seven key challenges.[54] Specifically, as a result of this nanoconfinement, several assumptions in classical nucleation theory are violated for the {ZIF-8+water} system considered here:

1. The assumption of having a well-defined phase in extreme nanoconfinement is violated.[54] Nanoconfined solutions often show properties that are substantially different from their bulk counterparts due to the extensive host-guest interactions. Moreover, these properties are not homogeneous throughout the adsorbed phase, but rather change depending on the distance from the interface and the location inside the material.[54,55] This implies that there isn't a single and well-defined thermodynamic bulk phase present in the cage; in some cases even leading to the observation of phase separation.[54]

2. The capillarity approximation, the fundamental assumption of classical nucleation theory, assumes that the nucleus can be viewed as a (thermodynamically) large and homogeneous spherical droplet of a well-defined radius with bulk properties of the target phase β inside it and bulk properties of the original phase α outside it.[50,51] In our {ZIF-8+water} system, the water clusters remain small even at full saturation as they are sterically hindered by the cage walls and, as a result, the nuclei do not grow large enough to properly define its bulk properties in a thermodynamic sound way. This is not limited to the system studied here. Rather, it is a well-known observation that in extreme nanoconfinement, such as in MOFs, the fundamental properties of water are substantially different from the thermodynamic properties in bulk water.[55,56]

3. Even if one would be able to soundly define a thermodynamic phase in the ZIF-8 cages, phase transitions in nanoconfined systems are strongly distorted.[54] Several phenomenological corrections have been introduced to account for this, but understanding phase transitions in single-digit nanopores remains one of the recently defined key challenges for this class of materials.[54] Faucher *et al.* recognise that the thermodynamics and phase behaviour of confined water represent a significant knowledge gap that needs to be understood with new theory and modelling approaches supported by comprehensive experimental data.[54]

4. There is not a unique, well-defined nucleation barrier during the intrusion process. For water to transport from one ZIF-8 cage to a neighbouring ZIF-8 cage, it has to overcome the free energy barrier associated with the hydrophobic 6MR aperture. Once this barrier has been overcome, this water molecule adheres almost directly to the existing water cluster. As a result, as every additional water molecule needs to diffuse through the 6MR aperture before potentially adhering to the existing nucleus, each of these addition steps is characterised by its own free energy barrier, as demonstrated in Supplementary Section S9.4. This violates the assumption of nucleation as a one-step process.[50]

Despite the fact that these essential assumptions for classical nucleation theory are not satisfied here, the concepts of a critical nucleus size, the free energy barrier of the rate-limiting step for nucleation, and oversaturation have natural analogies for the {ZIF-8+water} system, as discussed in Supplementary Sections S9.2, S9.3, and S9.4.

## S9.2 Critical-sized clusters derived from MD simulations with an inhomogeneous water distribution and GCMC simulations

From the discussion of classical nucleation theory in Supplementary Section S9.1, a subcritical water cluster inside a given hydrophobic cage can be defined as a water cluster that is insufficiently stabilised, so that it dissolves again. Consider for instance the ten simulations performed on the 1×1×2 ZIF-8 supercell, which are visualised in Supplementary Figure 63 and Supplementary Figure 64, and the simulation discussed in Figure 4b of the main text. In those simulations, six subcritical nuclei can be defined: the nuclei formed in cage 3 of Figure 4b and simulation 5 in Supplementary Figure 63, and the nuclei formed in cage 2 of simulations 4, 6, 8, and 9 of Supplementary Figure 63 and Supplementary Figure 64. In all those simulations, the originally formed water cluster is insufficiently stabilised and spontaneously dissolves, against the water gradient, hence corresponding to subcritical water clusters. One observes that these subcritical water clusters contain at most four water molecules, which establishes a first lower limit for the size of a critical-sized water cluster. The same analysis also holds for the MD simulations with inhomogeneous water distributions performed for other ZIF-8 model sizes, as discussed in Supplementary Section S8.2.

A similar approach can be followed to estimate the critical cluster size from the grand canonical Monte Carlo simulations discussed in Supplementary Section S4.1. To this end, consider the GCMC simulation at a water pressure of 6 kPa and a temperature of 298 K, as discussed previously in Supplementary Figure 23. This simulation was performed on a 2×2×2 ZIF-8 supercell, containing 16 cages, to speed up convergence, and afterwards averaged and normalized per conventional unit cell. In Supplementary Figure 82, this same information is present in the top panel, additionally showing the filling of the different invidual cages in the bottom panel.

From the bottom panel of Supplementary Figure 82, it is clear that each of the pore filling processes consists of three steps. Initially, the cage is empty, and small fluctuations up to about three water molecules per cage take place. These clusters are subcritical, as they quickly dissolve again. After a variable number of Monte Carlo cycles, different for each of the cages, a steep increase in the number of water molecules is observed as a function of the MC cycles, quickly reaching saturation around 40 water molecules per cage. Finally, once saturation is reached, again small fluctuations are present, but this time around the saturation limit. This occurs for each cage separately, confirming the statement that water intrusion is a cage-by-cage process.

The behaviour found in Supplementary Figure 82 is completely in line with what would be expected from a hydrophobic material and the concept of critical-sized cluster formation:

1. The initial pore filling is not favourable due to the hydrophobicity of the material. Water molecules that are adsorbed, have the tendency to desorb again rather rapidly. This corresponds to the concept of a subcritical nucleus.
2. This changes once a critical amount of water molecules are present in the cage, which occurs here around four to five water molecules per cage. At that moment, trying to adsorb additional water molecules becomes favourable thanks to the extensive hydrogen bonds with the already existing cluster.
3. Once the cage is completely filled, there is a small chance of water molecules being desorbed from the cluster again, but this desorption is quickly counteracted by a subsequent adsorption. As a result, once saturated, the water loading around each of the cages fluctuates around the saturation limit.

Based on this explanation, one would therefore expect a critical nucleus size of about four to five water molecules, in line with the estimate based on the MD simulations with an inhomogeneous water distribution.

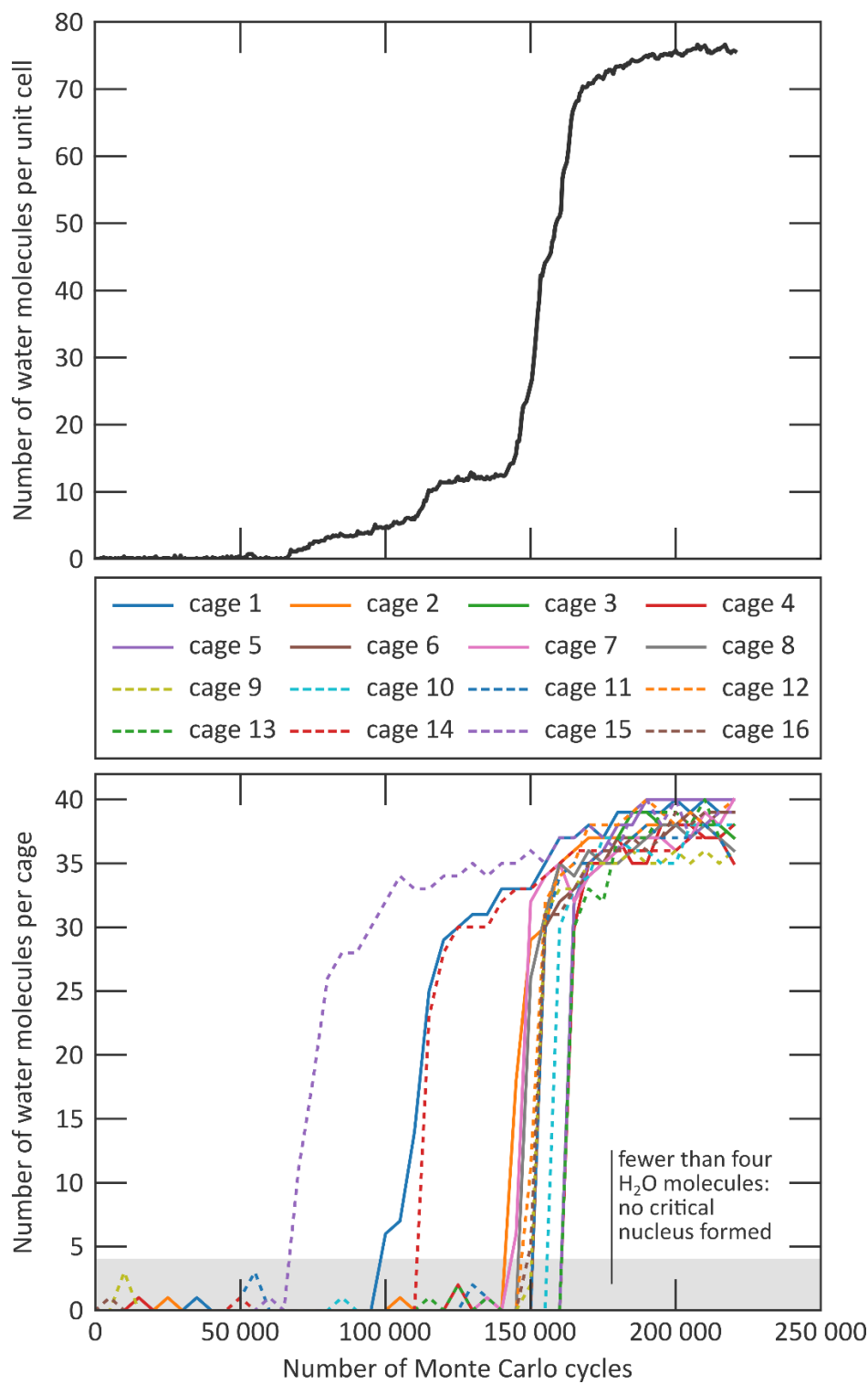

**Supplementary Figure 82 | Filling of the ZIF-8 cages during a grand canonical Monte Carlo simulation: (top) averaged over a 2×2×2 supercell, (bottom) for each of the 16 cages separately.** Results obtained at a water pressure of 6 kPa and a temperature of 298 K. The top figure corresponds with Supplementary Figure 23.

## S9.3 Umbrella sampling simulations to determine the critical cluster size and the associated free energy barrier

To quantify the size of the critical water nucleus, the associated barrier for nucleation, and the effect of oversaturation, a series of umbrella sampling (US) simulations have been carried out using our in-house software code Yaff and a 2×2×2 ZIF-8 supercell.[48,57] In all these US simulations, the collective variable (CV) as visualised in Supplementary Figure 83 has been used to describe and steer the transition of a water molecule from cage 1 to an adjacent cage 2 through the 6MR window that separates them, while the other cages were all empty. To define this CV, first define the relative position of the centroid of the water molecule that undergoes the transition with respect to the centre of the 6MR window through which the transition occurs:

$$\boldsymbol{r}_{\text{rel}} = \boldsymbol{r}_{\text{centroid},H_2O} - \boldsymbol{r}_{\text{centre,6MR}}$$

Second, let $\boldsymbol{n}_{\text{6MR}}$ be the normal of the 6MR aperture, defined such that it points from the original cage 1 to cage 2. Similar to our earlier work on diffusion in related zeolite systems,[58] the CV is then defined as the orthogonal projection of the earlier defined relative position of the water molecule onto this normal:

$$\text{CV} = \boldsymbol{r}_{\text{rel}} \cdot \boldsymbol{n}_{\text{6MR}}$$

The CV defined here has the dimension of ångström and separates cage 1 (CV < 0 Å) from cage 2 (CV > 0 Å) through the 6MR aperture (CV = 0 Å). To define the average free energy of both cages, the free energy has been averaged over all CV values smaller than -2.5 Å (cage 1) or larger than 2.5 Å (cage 2), thereby excluding the [-2.5 Å, 2.5 Å] region where the hydrophobic effect of the 6MR aperture is present, which will result in a diffusion free energy barrier (*vide infra*).

Using this CV, twenty different sets of US simulations have been performed. The first ten sets of US simulations describe the gradual filling of cage 2 when a critical-sized cluster is present in cage 1. Therefore, cage 1 is filled with five water molecules (in addition to the water molecule that undergoes the transition), based on the estimation of the size of a critical-sized cluster in Supplementary Section S9.2. The filling of cage 2 varies between the different sets of US simulations, starting at zero water molecules for the first transition to nine water molecules for the tenth transition, excluding the water molecule that undergoes the transition.

The second ten sets of US simulations describe the gradual filling of cage 2 when a supercritical-sized cluster is present in cage 1. It is conceptually identical to the first ten sets of US simulations, except that cage 1 is initially filled with a supercritical cluster of thirty water molecules rather than a critical cluster of five water molecules as before.

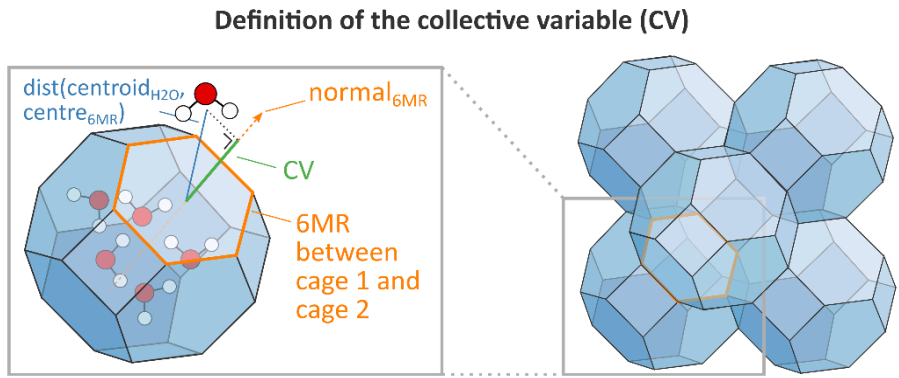

**Supplementary Figure 83 | Definition of the collective variable (CV) used during the US simulations.** Inset shows a cluster of five water molecules present in cage 1.

For each of the twenty different transitions, the collective variable was divided into 47 equidistant windows centred at CV values between -11.5 Å and 11.5 Å, yielding in total 940 distinct US simulations. To restrict the simulation to each individual window $i$, a harmonic bias potential as a function of the CV was applied with a force constant $K_1 = 25$ kJ mol$^{-1}$ Å$^{-2}$ and centred around the centres of these equidistant windows, defined by $CV_i$:

$$V_{\text{bias}}^i = \frac{1}{2} K_1 (CV - CV_i)^2$$

The combination of this force constant with the definition of the different US windows ensures that there is substantial overlap between neighbouring windows, as shown in Supplementary Figure 86 for the case of the fifth transition of Supplementary Figure 84. This is a requirement to construct the total free energy profile from the sampling probability distributions within each separate US window.

To prevent the other water molecules from escaping from their respective cages, an additional harmonic bias potential with a force constant $K_2 = 100$ kJ mol$^{-1}$ Å$^{-2}$ was applied to each of those other water molecules if the centroid of the molecule was at least at a distance $r_0 = 9$ Å from the respective cage centre $r_{\text{centre, cage }j}$; this bias potential disappeared as long as the water molecules remained within a radius of 9 Å from their respective cage centres:

$$V_{\text{sphere}}^j = \frac{1}{2} K_2 \left( \left| \boldsymbol{r}_{\text{centroid,H}_2\text{O}} - \boldsymbol{r}_{\text{centre,cage}j} \right| - r_0 \right)^2 \mathbb{H} \left( \left| \boldsymbol{r}_{\text{centroid,H}_2\text{O}} - \boldsymbol{r}_{\text{centre,cage}j} \right| - r_0 \right)$$

In this expression, $\mathbb{H}$ denotes the heaviside step function. This "spherical" bias was also applied to the water molecule that undergoes the transition, in addition to the harmonic bias defined by the umbrella sampling, but only when the molecule was sufficiently far from the 6MR aperture and hence sufficiently committed to either cage 1 or cage 2 ($|CV| \geq 3$ Å). In this case, the additional spherical bias, which adds to the harmonic bias defined by the umbrella sampling, does not influence the transition through the 6MR aperture that is used in the definition of the CV, but only prevents the water molecule from leaving the cage through a different 6MR aperture, similar to the effect of the spherical bias on the spectator water molecules.

Each of these 940 US simulations was run for 2.25 ns, including 10 ps equilibration time. After this, the free energy profile for each of the transitions was obtained from the sampling distribution in each window by the weighted histogram analysis method (WHAM).[59,60] The resulting free energy profiles are visualised in Supplementary Figure 84 and Supplementary Figure 85 for the case of a critical and a supercritical cluster in cage 1, respectively.

In Supplementary Figure 84, the diffusion free energy profiles are shown when a critical cluster of five water molecules is present in cage 1, with key free energy differences tabulated in Supplementary Table 3. For CV < 0 Å, this means a total of six water molecules are present in cage 1, including the water molecule that undergoes the transition, while for CV > 0 Å only the critical cluster of five water molecules remains in cage 1 as the water molecule that undergoes the transition is in cage 2. A maximum in free energy is always present at CV = 0 Å, corresponding with the hydrophobic 6MR aperture. For the first transition, when cage 2 is initially empty, a large difference in mean free energy of the two cages is observed, amounting to 14.6 kJ mol$^{-1}$. This large difference reflects that the water molecule that leaves cage 1, with a critical water cluster that stabilises the presence of the water molecule through hydrogen bonds, enters cage 2 in which no other water molecules are present with which it can potentially hydrogen bond. As a result, the forward free energy barrier is large (23.4 kJ mol$^{-1}$), while the reverse free energy barrier is rather low (8.7 kJ mol$^{-1}$). This indicates that it is much more likely that the water molecule will spontaneously return from cage 2 to cage 1, against the water gradient, than that it would make the transition to cage 2 in the first place. This is qualitatively the same behaviour as observed for subcritical water clusters in our MD and GCMC simulations. These same observations – a mean free energy of cage 2 that is substantially larger than the free energy of cage 1 and a large forward free energy barrier – can also be made for the second and the third transition, when respectively one and two water molecules are present in cage 2 before the additional water molecule makes the transition. This indicates that also clusters with either two or three water molecules are subcritical. In all these cases, the reverse transition is more likely than the forward transition.

Supplementary Figure 84 demonstrates that this picture changes from the fourth transition onwards. In the fourth transition, the water molecule that makes the transition can hydrogen bond with the three water molecules that are already present in cage 2, thereby forming a cluster of four water molecules. In this case, the difference in mean free energies of the two cages is small (2.1 kJ mol$^{-1}$) and the forward and reverse barriers are similar and relatively small compared to the first transition (16.3 and 18.3 kJ mol$^{-1}$), confirming the formation of a stable hydrogen-bonded water cluster in cage 2. These observations – a small difference in mean free energies (< 5 kJ mol$^{-1}$) and a relatively small forward free energy barrier (< 20 kJ mol$^{-1}$) can be consistently drawn from the fourth transition onwards.

With these observations, one can explain how increasing the strain rate above a certain threshold – the intrinsic strain rate defined in Supplementary Section S8.3 – leads to a substantial increase in intrusion pressure and energy absorption capacity, as observed experimentally. If the strain rate is sufficiently small, water clusters can nucleate spontaneously in cages adjacent to filled cages. To this end, water molecules have to temporarily overcome the free energy barrier for cage hopping located at the 6MR window, which is the rate-limiting step in this process, through thermal fluctuations and steered by the water gradient. Once a critical-sized cluster has nucleated, it facilitates the hopping of subsequent water molecules into this cage. In contrast, if the strain rate is too high, water clusters do not have sufficient time to organise in cages adjacent to filled cages and hence do not form such critical-sized water clusters. In this high-rate regime, additional energy needs to be provided to the system for the water molecules to cross the 6MR aperture and overcome the higher diffusion barrier. This additional energy is then dissipated once the water molecules have crossed the 6MR aperture. This additional energy is provided through the additional work exerted on the system, which implies that a higher intrusion pressure is needed in this case, as observed experimentally.

The above observations indicate that the concept of a critical cluster – here a water cluster of four water molecules – and the barriers associated with the formation of this critical cluster as the rate-limiting steps are similar to these concepts within classical nucleation theory (see Supplementary Section S9.1), although the presence of multiple roughly equivalent barriers in this nucleation process contradicts the assumption of a one-step process in classical nucleation theory. A final parallel with classical nucleation theory can be drawn by exploring the concept of supersaturation. Here, supersaturation will be modelled by considering a supercritical instead of a critical water cluster in cage 1.

In Supplementary Figure 85, the free energy profiles are shown for the first ten transitions of a water molecule from cage 1, with a supercritical cluster of an additional thirty water molecules, towards cage 2, which initially contains in between zero (first transition) and nine (tenth transition) water molecules. The tabulated free energy differences in Supplementary Table 4 show the same qualitative behaviour as for the critical water cluster discussed before. The first few transitions are accompanied by a substantial forward free energy barrier and exhibit a large difference in mean free energy between the two cages, whereas for further transitions the forward free energy barrier decreases and the mean free energy difference between the cages largely disappears. We observed before in Supplementary Figure 84 that, if a critical cluster is present in cage 1, a critical cluster in cage 2 is formed when four molecules are present in this cage, as for the fourth transition onwards a small difference in mean free energies ($< 5$ kJ mol$^{-1}$) and a relatively small forward free energy barrier ($< 20$ kJ mol$^{-1}$) were observed consistently. In the case of supersaturation, shown in Supplementary Figure 85, applying the same criteria would lead to the definition of a critical cluster already at the third transition, when three molecules are present in cage 2. The only small deviation from these criteria occurs in transition 7, when a free energy difference of 5.1 kJ mol$^{-1}$ is observed, which falls inside the numerical accuracy. Hence, supersaturation as defined here seems to decrease the size of the critical water cluster from four to three water molecules, in line with the effect of supersaturation in classical nucleation theory (see Supplementary Figure 81). In contrast to classical nucleation theory, however, no appreciable decrease in nucleation free energy barrier is observed at supersaturation, although this can be explained through the limited validity of classical nucleation theory for this process and the limited amount of supersaturation that can be reached, given that at most about 40 water molecules can enter a single cage.

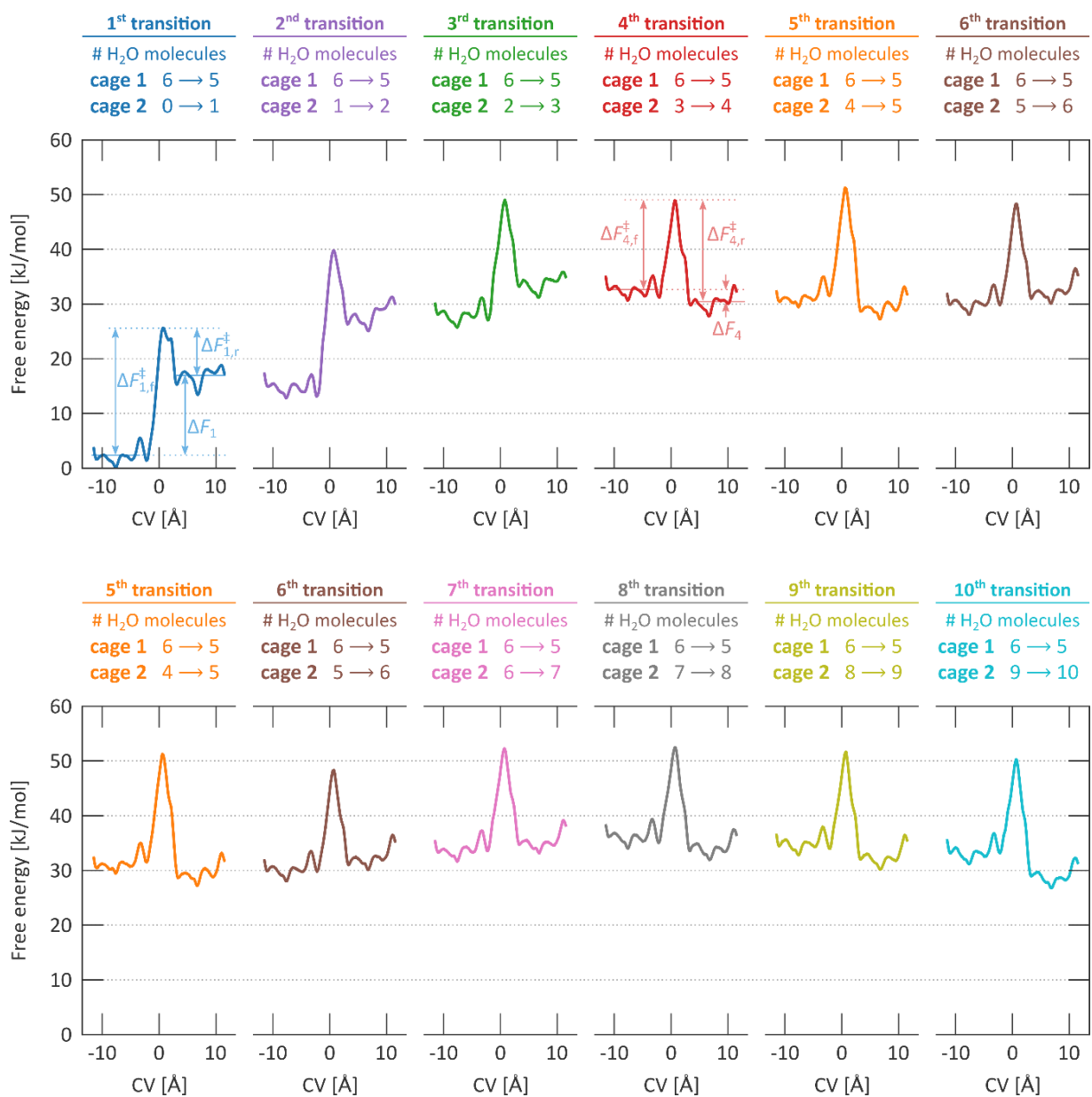

**Supplementary Figure 84 | Critical nucleus size and nucleation free energy barriers determined *via* ten sets of US simulations describing the transition of a selected water molecule from a critical cluster of an additional five water molecules present in cage 1 (CV < 0 Å) to cage 2 (CV > 0 Å), which is filled with in between 0 (first transition) and 9 (tenth transition) water molecules.** Visualised are the free energy profiles for each of the ten transitions, with the forward ($\Delta F_{i,f}^{\ddagger}$) and reverse barrier ($\Delta F_{i,r}^{\ddagger}$) and the difference in mean free energy between the two cages ($\Delta F_i$) tabulated in Supplementary Table 3.

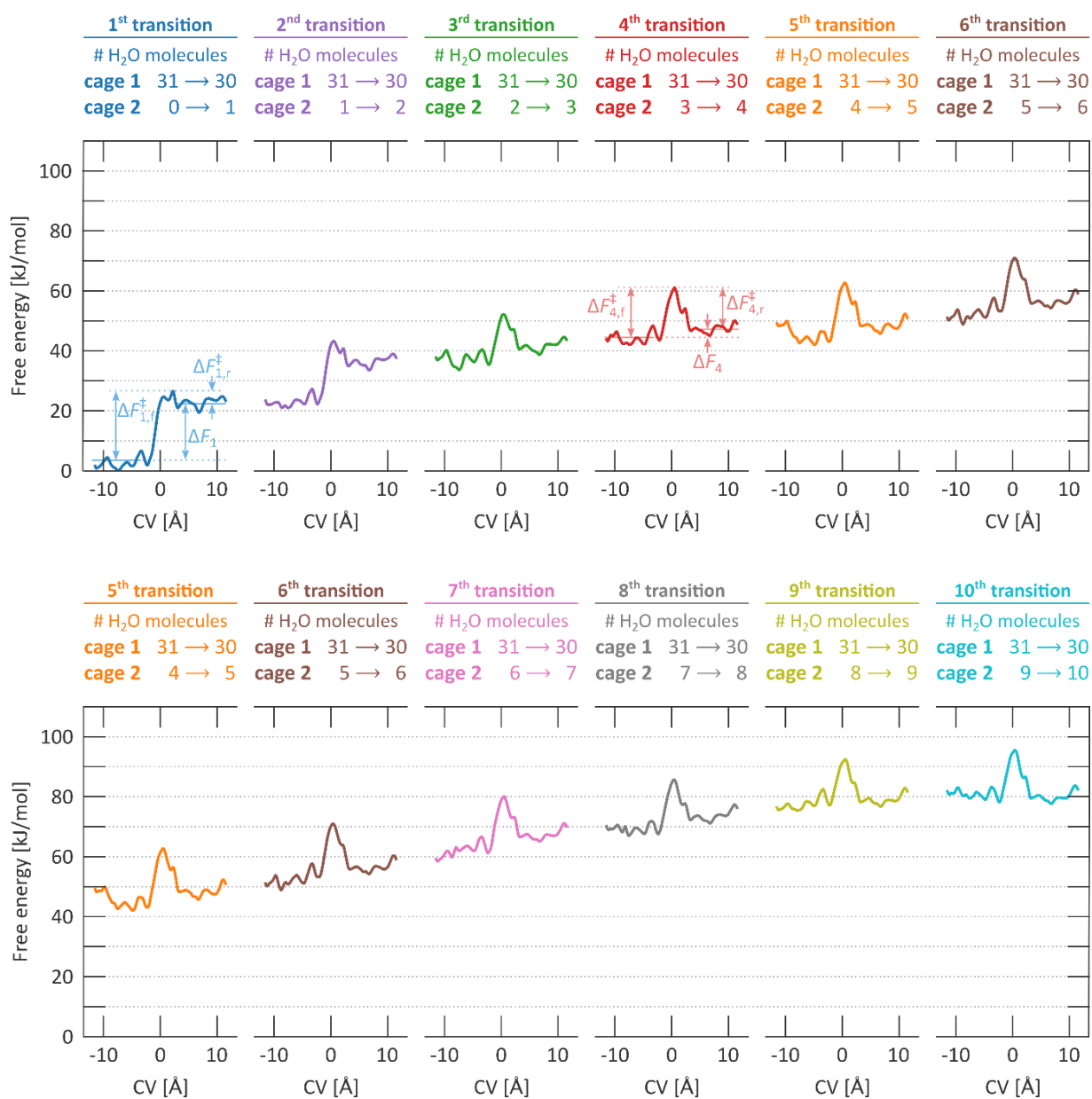

**Supplementary Figure 85 | Critical nucleus size and nucleation free energy barriers determined *via* ten sets of US simulations describing the transition of a selected water molecule from a super-critical cluster of an additional thirty water molecules present in cage 1 (CV < 0 Å) to cage 2 (CV > 0 Å), which is filled with in between 0 (first transition) and 9 (tenth transition) water molecules.** Visualised are the resulting free energy profiles for each of the ten transitions, with the forward ($\Delta F_{i,f}^{\ddagger}$) and reverse barrier ($\Delta F_{i,r}^{\ddagger}$) and the difference in mean free energy between the two cages ($\Delta F_i$) tabulated in Supplementary Table 4.

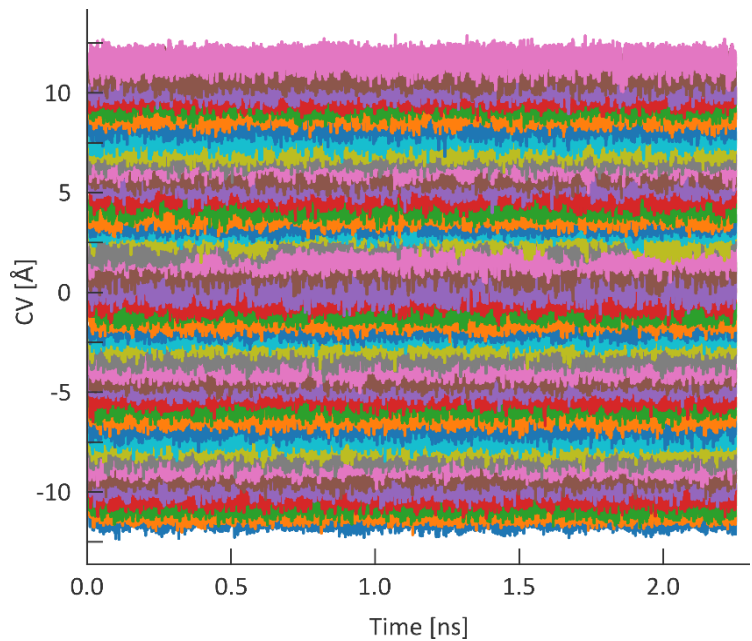

**Supplementary Figure 86 | Overview of the values of the CV attained during the 47 different US simulations for the fifth transition of Supplementary Figure 84.** The overlap between each pair of adjacent windows ensures that the overall free energy profile can be reconstructed *via* WHAM.

**Supplementary Table 3 | Free energy differences as extracted from the US simulations with a critical cluster of five water molecules in the first cage, as shown in Supplementary Figure 84.** The mean free energies of the two cages are defined by averaging the free energy over either CV < -2.5 Å (cage 1) or CV > 2.5 Å (cage 2). From transition 4 onwards, the free energy differences between the cages is consistently below 5 kJ mol$^{-1}$ and the forward free energy barrier is below 20 kJ mol$^{-1}$.

| Transition # | Mean free energy cage 1 (kJ mol$^{-1}$) | Mean free energy cage 2 (kJ mol$^{-1}$) | Free energy difference between the cages $\Delta F_i$ (kJ mol$^{-1}$) | Forward free energy barrier $\Delta F_{i,f}^{\ddagger}$ (kJ mol$^{-1}$) | Reverse free energy barrier $\Delta F_{i,r}^{\ddagger}$ (kJ mol$^{-1}$) |
|---|---|---|---|---|---|
| 1 | 2.2 | 16.9 | 14.6 | 23.4 | 8.7 |
| 2 | 2.0 | 15.3 | 13.3 | 25.0 | 11.7 |
| 3 | 2.4 | 8.2 | 5.7 | 20.9 | 15.2 |
| 4 | 4.9 | 2.8 | -2.1 | 16.3 | 18.3 |
| 5 | 4.2 | 2.6 | -1.6 | 19.9 | 21.5 |
| 6 | 2.3 | 4.5 | 2.2 | 18.0 | 15.8 |
| 7 | 2.3 | 3.8 | 1.5 | 18.4 | 16.9 |
| 8 | 4.4 | 2.4 | -1.9 | 16.2 | 18.1 |
| 9 | 4.8 | 2.5 | -2.3 | 16.7 | 19.0 |
| 10 | 6.6 | 2.4 | -4.2 | 16.9 | 21.1 |

**Supplementary Table 4 | Free energy differences as extracted from the US simulations with a supercritical cluster of thirty water molecules in the first cage, as shown in Supplementary Figure 85.** The mean free energies of the two cages are defined by averaging the free energy over either CV < -2.5 Å (cage 1) or CV > 2.5 Å (cage 2).

| Transition # | Mean free energy cage 1 (kJ mol⁻¹) | Mean free energy cage 2 (kJ mol⁻¹) | Free energy difference between the cages $\Delta F_i$ (kJ mol⁻¹) | Forward free energy barrier $\Delta F^{\ddagger}_{i,f}$ (kJ mol⁻¹) | Reverse free energy barrier $\Delta F^{\ddagger}_{i,r}$ (kJ mol⁻¹) |
|---|---|---|---|---|---|
| 1 | 2.6 | 22.8 | 20.2 | 24.0 | 3.7 |
| 2 | 1.9 | 15.4 | 13.5 | 20.3 | 6.8 |
| 3 | 3.7 | 8.0 | 4.2 | 14.8 | 10.6 |
| 4 | 2.2 | 5.5 | 3.2 | 16.8 | 13.6 |
| 5 | 3.5 | 6.6 | 3.1 | 17.3 | 14.2 |
| 6 | 3.6 | 7.7 | 4.1 | 18.5 | 14.4 |
| 7 | 3.8 | 8.9 | 5.1 | 17.8 | 12.7 |
| 8 | 2.3 | 6.7 | 4.4 | 16.4 | 12.1 |
| 9 | 2.2 | 3.7 | 1.5 | 14.9 | 13.4 |
| 10 | 3.3 | 2.5 | -0.8 | 14.6 | 15.4 |

## S9.4 Water organisation inside the critical clusters

To obtain further insight in the local structure of the water clusters formed during the cage hopping process, the organisation of the water molecules of the MD simulation of a 1×1×2 ZIF-8 simulation cell with an inhomogenous water distribution discussed in Figure 4b of the main text is analysed.

To this end, at each of the twenty snapshots shown in Figure 4b, all water molecules were selected, and hydrogen bonds between each pair of molecules were identified when the distance between the hydrogen atom of the hydrogen-bond donor molecule and the oxygen atom of the hydrogen-bond acceptor molecule was below 2.5 Å and when the two atoms were not covalently bound. In this way, a hydrogen-bonded graph is created that describes the connectivity of the different water molecules inside the hydrophobic cages. Second, in line with the concept of a water cluster, this total graph was divided into subgraphs by requiring that all water molecules inside a given subgraph would still be connected with all other water molecules inside that given subgraph if a single hydrogen bond were to be cut. Third, for each of the so created subgraphs, each combination of three to ten hydrogen-bonded water molecules was selected and compared with the 36 local water clusters containing at least three water molecules identified in Ref. 61, which act as a reference set. If the connectivity of the selected combination of water molecules is the same as the connectivity of the reference water cluster, this reference water cluster was said to be present in this specific subgraph. This process was repeated for each subgraph and for each reference water cluster of Ref. 61. Since a water molecule can form part of multiple reference clusters, the number of identified local clusters can be larger than the number of water molecules in the system. Finally, the identified local clusters are combined into superclusters, which are defined as the maximum combination of local clusters that form part of the same hydrogen-bonded network. In practice, this always clusters the local water clusters in cage 1 and cage 2, but may also combine these two superclusters to a single supercluster that extends over the two cages if the two superclusters are connected *via* a hydrogen bond through the 6MR aperture.

When applying this procedure to the simulation shown in Figure 4b of the main text, the result as shown in Supplementary Figure 87 is obtained. Of the 36 local water clusters with at least three water molecules described in Ref. 61, fourteen are identified in the simulation. All identified local clusters, shown in the top part of the figure, contain in between three and nine water molecules. In all cases, all local clusters in cage 1 are part of the same supercluster, and all local clusters in cage 2 are part of the same supercluster. Both superclusters are intermittently connected through a hydrogen bond at various instances throughout the simulation, allowing water molecules to move from cage 1 to cage 2 through the 6MR aperture.

In the main text, point (ii) in the simulation, at 0.45 ns, was identified as a crucial point. At that point, two water molecules were present in cage 3, which were insufficient to stabilise a water cluster, while six water molecules were present in cage 2, forming a critical water cluster that could grow further.

From Supplementary Figure 87, the local cluster in cage 2 at this point is identified with local cluster #17, demonstrating that all water molecules in cage 2 are indeed connected through hydrogen bonds. In contrast, no local cluster is identified in cage 3, neither at snapshot (ii) nor at any other point during the simulation, explaining why the few water molecules in cage 3 are insufficiently stabilised.

While point (ii) was indicated as a critical point in the main text, Supplementary Figure 87 shows that a first critical cluster in cage 2 is formed already after 0.15 ns. At that moment, a local cluster #3 connects all four water molecules in this cage. Both at 0.15 ns and 0.45 ns, the local cluster in cage 2 is connected to the supercluster in cage 1 *via* a single hydrogen bond through the 6MR aperture, so that additional water molecules can move from cage 1 to cage 2 and join the cluster in cage 2. Therefore, at 0.15 ns, it seems that the nucleation of such a local cluster of four water molecules is already sufficient to promote the further growth of the water cluster in this cage; local cluster #3 in this simulation can therefore be regarded as a critical cluster.

In between point (ii) and point (iii), Figure 4b of the main text indicates that the cluster in cage 2 persists but does not grow. This is also clear from Supplementary Figure 87: between these two points, there is no hydrogen bond between the supercluster in cage 1 and the supercluster in cage 2. Even more, at point (iii), the water molecules in cage 2 do not form a closely-packed cluster, but are rather extended, as shown in Figure 4c of the main text. It is only after point (iii) that local clusters form again in cage 2 and hydrogen bond with the supercluster in cage 1, resulting in the growth of the water cluster of cage 2 (see Figure 4b of the main text ).

From this systematic analysis, it is clear that the critical water nucleus defined in Supplementary Section S9.3, containing about four water molecules, is a prerequisite to further grow the water cluster in cage 2. This observation further confirms that the process with which water moves through the hydrophobic cages of ZIF-8 crucially depends on the formation of such water clusters, which can be systematically defined using the procedure discussed above.

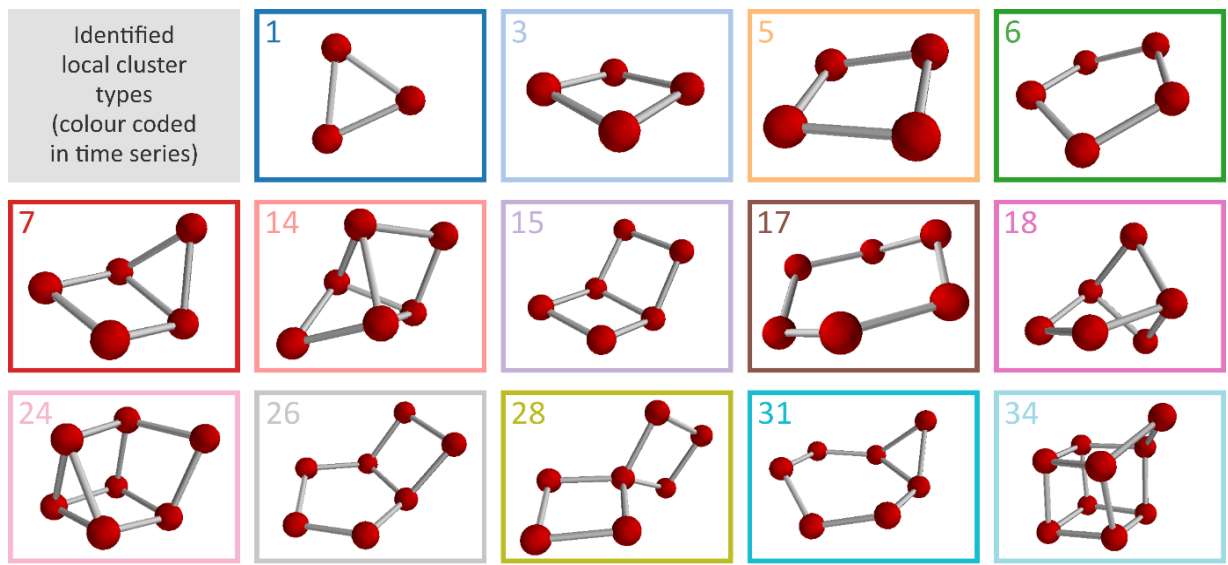
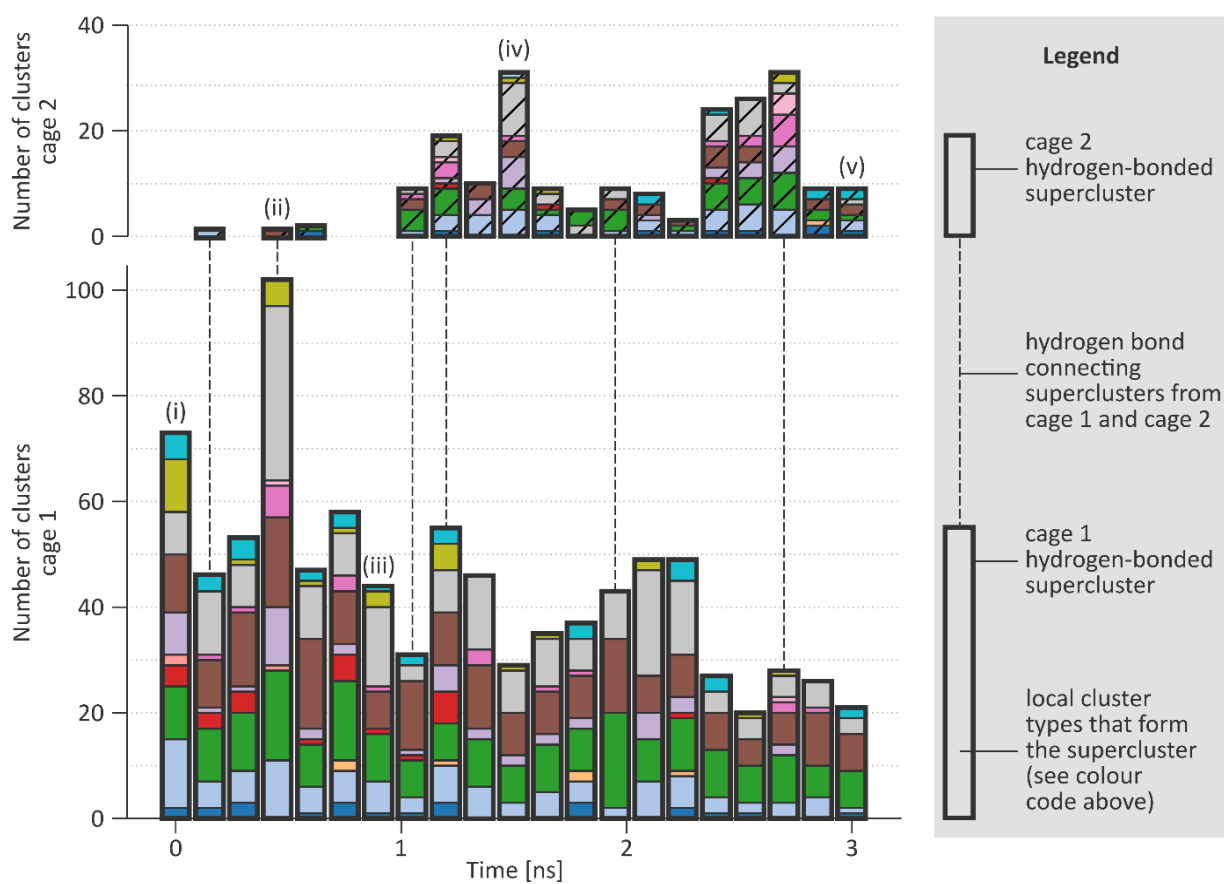

**Supplementary Figure 87 | Analysis of the hydrogen-bond network and organisation of local water clusters and superclusters for the MD simulation with inhomogenous water distribution of Figure 4b in the main text.** Number and organisation of local clusters as a function of the simulation time. Local clusters are hydrogen-bonded networks that can be identified with the organised clusters defined in Ref. 61; the identified clusters are numbered according to this reference and are colour coded according to the top part of the figure. Superclusters are defined as the maximum combination of local clusters that form part of the same hydrogen-bonded network. In each frame, all clusters inside a given cage are part of the same supercluster, intermittently those two superclusters are connected via hydrogen bonds through the 6MR aperture separating cages 1 and 2. A water molecule can form part of multiple local clusters; the total number of local clusters can hence exceed the total number of water molecules. Labels (i) to (v) correspond to the labels discussed in Figure 4c of the main text.

- 119 -

# S10 Classification of ZIFs as effective impact absorbers under high-rate water intrusion according to the formulated design rules

To demonstrate the general applicability of the approach formulated in the main text, we have classified the 105 ZIF-like materials tabulated in Ref. 62 according to the four design rules. This leads to four classes of materials: those that fail the first design rule and are not completely hydrophobic (Supplementary Table 5), those that are hydrophobic but are not cage-type materials (Supplementary Table 6), those that are formed by hydrophobic nanocages but have an insufficient aperture size such that they are not recoverable under water intrusion (Supplementary Table 7), and finally those materials that are predicted to be good candidates for energy absorption by leveraging the high-rate water intrusion mechanism (Supplementary Table 8). These results are visualized in Figure 5e of the main text.

**Supplementary Table 5 | Materials that violate design rule #1.** CCDC code, name, topology, pore-limiting diameter (PLD), and largest cavity diameter (LCD) for the 30 ZIF-like materials tabulated in Ref. 62 that are at least partially hydrophilic (as determined by their linkers) and hence violate the first design rule. They are indicated with open circles in Figure 5e of the main text.

| CCDC code | Name | Topology | PLD (Å) | LCD (Å) | Refs |
|---|---|---|---|---|---|
| GIZJOP | ZIF-62 | **cag** | 1.4 | 1.3 | 63 |
| MIHHOB | ZIF-23 | **dia** | 1.1 | 4.2 | 64 |
| VIGHID | N/A | **dia** | 0.7 | 1.7 | 65 |
| XASGON | N/A | **dia** | 0.2 | 1.8 | 66 |
| MOXKEQ | BIF-6 | **fes** | 1.3 | 2.2 | 67 |
| GITVOV | ZIF-73 | **frl** | 1.0 | 1.0 | 63 |
| GITWIQ | ZIF-77 | **frl** | 2.9 | 3.6 | 63 |
| GITVUB | ZIF-74 | **gis** | 1.2 | 2.6 | 63 |
| GITWAI | ZIF-75 | **gis** | 1.2 | 2.62 | 63 |
| GITTUZ | ZIF-68 | **gme** | 7.5 | 10.3 | 63 |
| GITVAH | ZIF-69 | **gme** | 4.4 | 7.8 | 63 |
| GITVEL | ZIF-70 | **gme** | 13.1 | 15.9 | 63 |
| ZIF-78 | ZIF-78 | **gme** | 3.8 | 7.1 | 68 |
| ZIF-79 | ZIF-79 | **gme** | 4.0 | 7.5 | 68 |
| ZIF-80 | ZIF-80 | **gme** | 9.8 | 13.2 | 68 |
| ZIF-81 | ZIF-81 | **gme** | 3.9 | 7.4 | 68 |
| ZIF-82 | ZIF-82 | **gme** | 8.1 | 12.3 | 68 |
| MIHHAN | ZIF-20 | **lta** | 2.8 | 15.4 | 64 |
| MIHHER | ZIF-21 | **lta** | 2.8 | 15.4 | 64 |
| MIHHIV | ZIF-22 | **lta** | 2.9 | 14.8 | 64 |
| EGEHOO | usf-ZMOF | **med** | 4.3 | 9.7 | 69 |
| NOFQEF | ZIF-100 | **moz** | 3.4 | 35.6 | 70 |
| TEFWIL | rho-ZMOF | **rho** | 5.7 | 26.9 | 71 |
| WOJGEI | ZIF-90 | **sod** | 3.5 | 11.2 | 72 |
| GITTIN | ZIF-65 | **sod** | 3.4 | 10.4 | 63 |
| TEFWOR | sod-ZMOF | **sod** | 1.2 | 8.1 | 71 |
| ZIF-91 | ZIF-91 | **sod** | 3.2 | 11 | 72 |
| ZIF-92 | ZIF-92 | **sod** | 0.0 | 5.2 | 72 |
| MOXKOA | BIF-8 | **srs-c-b** | 0.8 | 4.2 | 67 |
| MOXKIU | BIF-7 | **ths-c-b** | 1.7 | 5.5 | 67 |

**Supplementary Table 6 | Materials that violate design rule #2.** CCDC code, name, topology, pore-limiting diameter (PLD), and largest cavity diameter (LCD) for the 14 ZIF-like materials tabulated in Ref. 62 that are hydrophobic but are not cage-type (the LCD is not larger than the PLD) and hence violate the second design rule. They are indicated with open triangles in Figure 5e of the main text.

| CCDC code | Name | Topology | PLD (Å) | LCD (Å) | Refs |
| --- | --- | --- | --- | --- | --- |
| MECWIB | ZIF-14 | **ana** | 2.2 | 2.2 | 63, 73 |
| EQOCES01 | N/A | **cag** | 2.4 | 2.4 | 74 |
| NAFGOR | N/A | **cag** | 1.0 | 1.0 | 74 |
| VEJYUF01 | N/A | **cag** | 0.8 | 0.8 | 75 |
| VEJYEP01 | N/A | **crb** | 2.2 | 2.2 | 75 |
| LEMVOP | N/A | **crs** | 1.6 | 1.6 | 76 |
| MOXKUG | BIF-2Li | **dia-c-b** | 2.4 | 2.4 | 67 |
| MUCLIG | BIF-2Cu | **dia-c-b** | 2.6 | 2.6 | 67 |
| GIZJUV | ZIF-72 | **lcs** | 1.9 | 1.9 | 63 |
| GITWEM | ZIF-76 | **lta** | 1.9 | 1.9 | 63 |
| HICGEG | N/A | **zec** | 5.0 | 5.0 | 75 |
| GITTAF | ZIF-61 | **zni** | 0.7 | 0.7 | 63 |
| IMIDZB | N/A | **zni** | 3.6 | 3.6 | 77 |
| IMZYCO01 | N/A | **zni** | 3.7 | 3.7 | 78 |

**Supplementary Table 7 | Materials that violate design rule #3.** CCDC code, name, topology, pore-limiting diameter (PLD), and largest cavity diameter (LCD) for the 41 ZIF-like materials tabulated in Ref. 62 that are formed by hydrophobic nanocages but have an insufficient aperture size such that they are not recoverable under water intrusion (PLD $\leq$ 3 Å) and hence violate the third design rule. They are indicated with open diamonds in Figure 5e of the main text.

| CCDC code | Name | Topology | PLD (Å) | LCD (Å) | Refs |
| --- | --- | --- | --- | --- | --- |
| VEJYUF | ZIF-4 | **cag** | 2.0 | 2.1 | 10 |
| QOSYAZ | TIF-4 | **cag** | 2.0 | 6.9 | 65 |
| EQOCOC | N/A | **coi** | 2.5 | 6.0 | 78 |
| IMZYCO | N/A | **coi** | 2.5 | 6.0 | 79 |
| GITTEJ | ZIF-64 | **crb** | 2.5 | 7.9 | 63 |
| NAFGOR01 | N/A | **crb** | 0.9 | 3.0 | 74 |
| ZIMMEN | N/A | **dia** | 0.2 | 2.8 | 80 |
| BAYPUN | N/A | **dia-c** | 1.0 | 5.3 | 81 |
| BAYQAU | N/A | **dia-c** | 0.8 | 6.0 | 81 |
| BAYQAU01 | N/A | **dia-c** | 2.0 | 3.3 | 78 |
| BAYQAU02 | N/A | **dia-c** | 2.0 | 3.3 | 82 |
| VEJZAM | ZIF-5 | **gar** | 1.7 | 3.03 | 10 |
| EQOCOC01 | ZIF-6 | **gis** | 1.5 | 3.03 | 10 |
| QOSYIH | TIF-5Zn | **gis** | 1.0 | 6.0 | 83 |
| QOSYED | TIF-5Co | **gis** | 0.7 | 5.0 | 83 |
| DAYVIJ | N/A | **mab** | 1.1 | 3.2 | 84 |
| CUIMDZ03 | N/A | **mog** | 1.3 | 3.5 | 85 |
| IMIDFE | N/A | **mog** | 1.9 | 3.1 | 86 |
| IMIDFE01 | N/A | **mog** | 1.9 | 3.1 | 87 |
| IMIDZA | N/A | **mog** | 2.0 | 3.3 | 77 |
| EQOBUH | N/A | **neb** | 0.6 | 6.7 | 78 |
| EQOCES | N/A | **neb** | 1.8 | 7.1 | 78 |
| EQOCIW | N/A | **neb** | 1.6 | 6.9 | 78 |
| QOSXUS | TIF-3 | **pcb** | 2.2 | 6.2 | 83 |
| CAGLIF | N/A | **qtz** | 2.6 | 6.0 | 88 |
| MECWOH | N/A | **rho** | 1.3 | 21.6 | 73, 89 |
| VEJZOA | ZIF-11 | **rho** | 3.0 | 14.6 | 10 |
| VEJZUG | ZIF-12 | **rho** | 3.0 | 14.6 | 10 |
| AKUGES | N/A | **sod** | 2.4 | 5.2 | 89, 90 |
| MECWEX | N/A | **sod** | 3.0 | 14.2 | 73 |
| VEJZEQ | ZIF-9 | **sod** | 2.4 | 5.6 | 90 |
| VELVIS | ZIF-7 | **sod** | 2.4 | 5.6 | 90 |
| MUCLOM | BIF-3Li | **sod-b** | 2.7 | 10.0 | 67 |
| MOXJOZ | BIF-3Cu | **sod-b** | 2.7 | 9.9 | 67 |
| QOGPIM | TIF-1Zn | **zea** | 3.0 | 4.1 | 91 |
| TIF-1Co | TIF-1Co | **zea** | 3.0 | 4.1 | 91 |
| MOXJEP | BIF-1Li | **zni-b** | 3.0 | 4.4 | 67 |
| MOXJIT | BIF-1Cu | **zni-b** | 2.9 | 3.7 | 67 |
| BETHUE | N/A | N/A | 2.1 | 2.8 | 89, 92 |
| MOXJUF | BIF-4 | N/A | 0.7 | 3.6 | 67 |
| MOXKAM | BIF-5 | N/A | 2.6 | 3.3 | 67 |

**Supplementary Table 8 | Materials that satisfy all design rules.** CCDC code, name, topology, pore-limiting diameter (PLD), and largest cavity diameter (LCD) for the 20 ZIF-like materials tabulated in Ref. 62 that satisfy all four design rules and are hence expected to form effective and reusable materials to absorb energy through the here identified high-rate water intrusion mechanism. They are indicated with filled squares in Figure 5e of the main text.

| CCDC code | Name | Topology | PLD (Å) | LCD (Å) | Refs |
|---|---|---|---|---|---|
| LODCUC | N/A | **crb** | 4.8 | 8.0 | 93 |
| VEJYEP | ZIF-1 | **crb** | 6.3 | 6.94 | 10 |
| VEJYIT | ZIF-2 | **crb** | 6.4 | 6.9 | 10 |
| VEJYIT01 | N/A | **crb** | 5.4 | 5.7 | 75 |
| HIFVOI | N/A | **dft** | 6.6 | 9.6 | 75 |
| VEJYOZ | ZIF-3 | **dft** | 4.6 | 6.0 | 10 |
| HIFVUO | N/A | **gis** | 5.2 | 8.6 | 75 |
| GITSUY | ZIF-60 | **mer** | 7.2 | 9.4 | 63 |
| VEJZIU | ZIF-10 | **mer** | 8.2 | 12.2 | 63 |
| AFIXAO | N/A | **nog** | 4.1 | 5.5 | 94 |
| AFIXAO01s | N/A | **nog** | 3.9 | 5.5 | 78 |
| AFIXES | N/A | **nog** | 3.5 | 5.9 | 94 |
| HIFWAV | N/A | **nog** | 4.7 | 8.2 | 75 |
| NOFQAB | ZIF-95 | **poz** | 3.7 | 24.0 | 70 |
| GITVIP | ZIF-71 | **rho** | 5.5 | 17.0 | 63 |
| CUIMDZ01 | N/A | **sod** | 4.6 | 7.3 | 85 |
| GITTOT | ZIF-67 | **sod** | 3.4 | 11.4 | 63 |
| OFERUN | N/A | **sod** | 3.1 | 14.2 | 95 |
| VELVOY | ZIF-8 | **sod** | 3.4 | 11.4 | 90 |
| QOSXOM | TIF-2 | **zeb** | 9.6 | 10.0 | 83 |

# S11 Additional experimental results of hydrophobic channel-containing frameworks

## S11.1 Water intrusion and extrusion of ZSM-5

Here we show a complete profile of ZSM-5 data for Figure 6b-d of the main text including the strain rate history. In the low-rate and medium-rate tests (Supplementary Figure 88a-b), the unloading process is externally driven, while the high-rate experiment has an uncontrolled free unloading process (Supplementary Figure 88c). The strain rate history shown as the dashed line in Supplementary Figure 88c suggests that the extrusion rate is relatively high, especially compared with that of ZIF-8 (Supplementary Figure 12c).

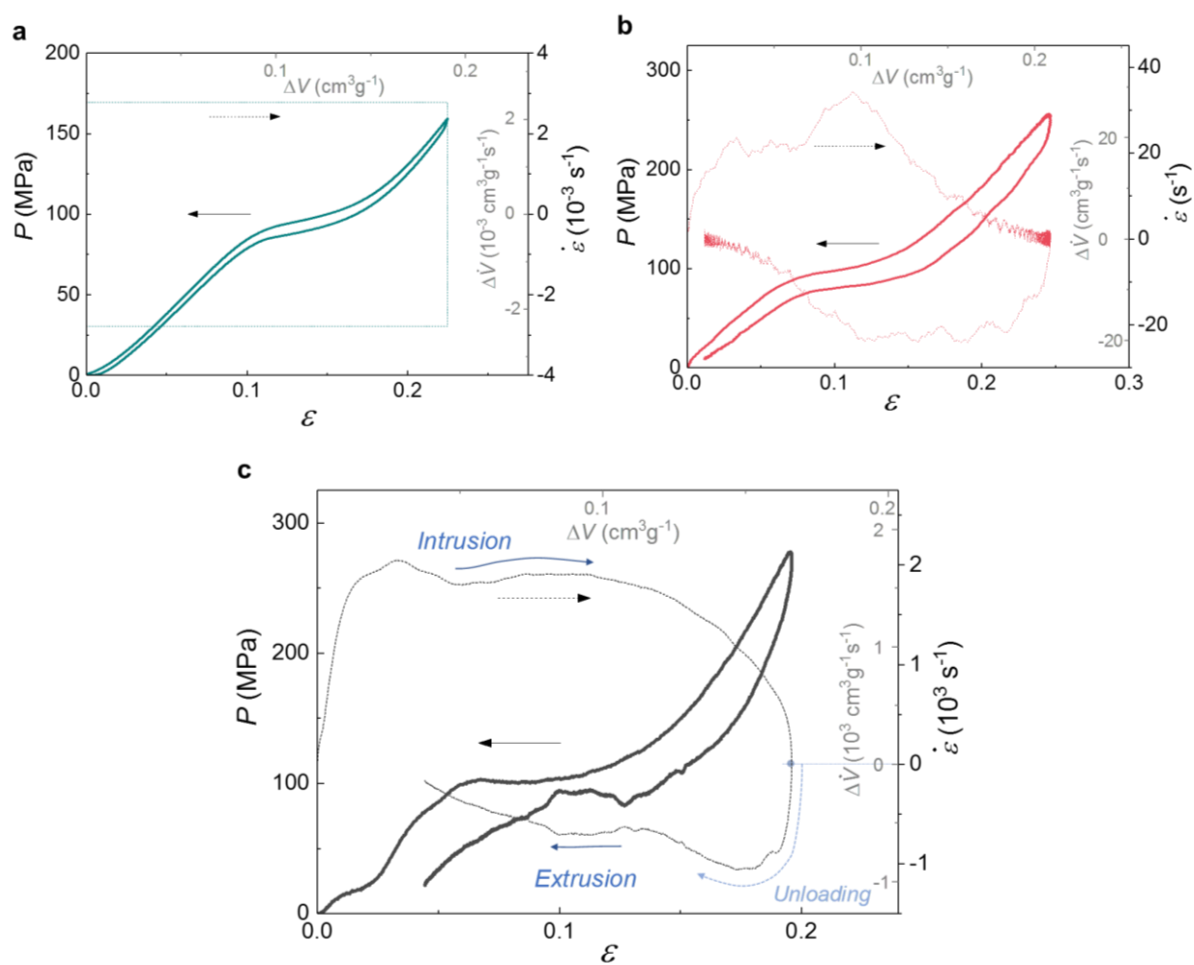

**Supplementary Figure 88 | Water intrusion and extrusion of ZSM-5. a,** low-rate, **b,** medium-rate, and **c,** high-rate loading conditions, corresponding to the results of Figure 6b-d in the main text.

## S11.2 Water intrusion and extrusion of zeolite-β and mordenite

The performance of zeolite-β and mordenite at different strain rates are shown in Supplementary Figure 89. They have a non-linear spring-like performance which is similar to ZSM-5. They have limited energy absorption capacity and are not sensitive to the loading rate.

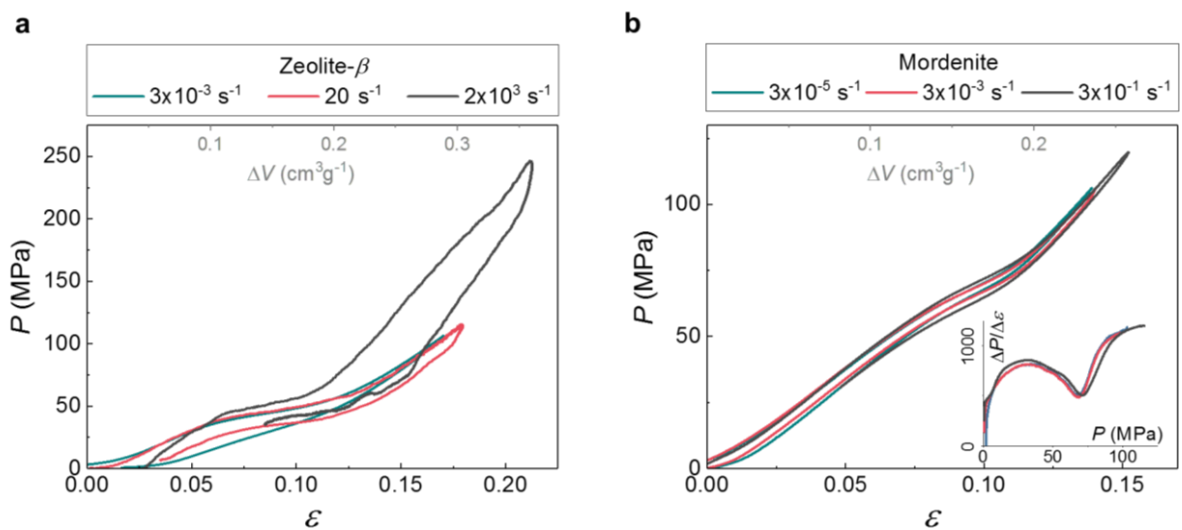

**Supplementary Figure 89 | Water intrusion and extrusion of zeolite-β and mordenite. a**, Water intrusion and extrusion of zeolite-β at three different strain rates, which correspond to a specific volume change rate $\Delta \dot{V}$ of $5\times10^{-3}$, 20, and $3\times10^{3}$ cm$^3$g$^{-1}$s$^{-1}$, respectively. **b,** Water intrusion and extrusion of mordenite at three different strain rates, which correspond to a specific volume change rate $\Delta \dot{V}$ of $5\times10^{-5}$, $5\times10^{-3}$, and $5\times10^{-1}$ cm$^3$ g$^{-1}$ s$^{-1}$, respectively. The inset shows the gradient $\Delta P/\Delta \varepsilon$ during loading, from which the intrusion plateaus and pressures can be identified.

## S11.3 Energy absorption density of zeolites at different strain rates

Supplementary Figure 90 reveals that for channel-containing zeolites, the rate dependence of their energy absorption density is considerably weaker than the one detected for cage-type ZIFs. There is no significant change in their intrusion and extrusion behaviours at different strain rates, as shown in Figure 6b of the main text and Supplementary Figure 89.

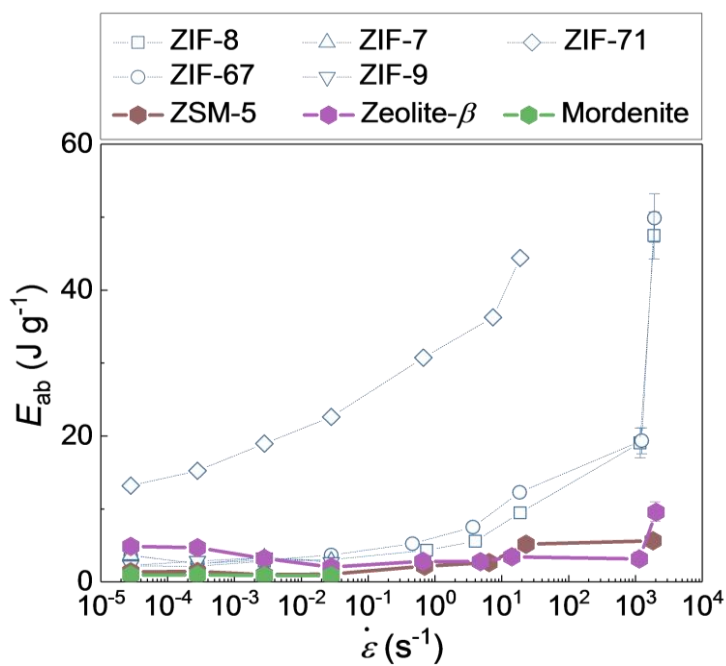

**Supplementary Figure 90 | Energy absorption density as a function of strain rate for the channel-containing zeolites.** These results are compared to the behaviour of cage-type ZIFs. Due to the limited pore volume of mordenite, its energy absorption density at high strain rates is not available. The error bars represent the uncertainty due to the incomplete unloading curves at high strain rates.

## S11.4 Zeolites under multiple dynamic loading cycles

Here we show the water intrusion of zeolites under cyclic loading. All zeolites show consistent performance against multiple dynamic loading cycles. Therefore, they can work as a non-linear liquid spring, which is reusable but unfortunately cannot absorb much mechanical energy.

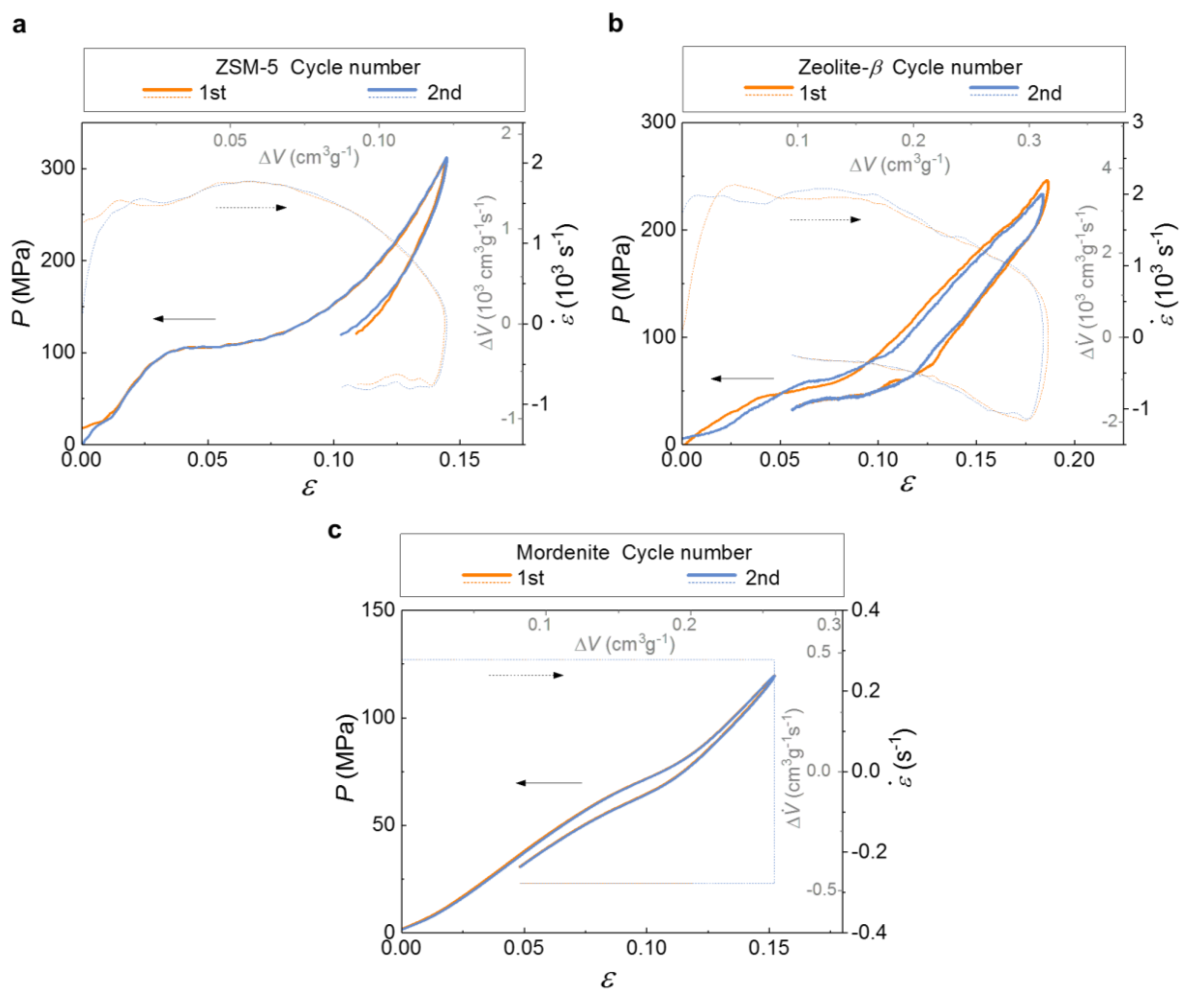

**Supplementary Figure 91 | Cyclic water intrusion and extrusion of the channel-containing zeolites.**
**a**, ZSM-5 at high strain rate, **b**, zeolite-$\beta$ at high strain rate, and **c**, mordenite at 0.3 s$^{-1}$.

## S11.5  Effect of sample mass of zeolites in high-rate experiments

In this work, we used a higher amount of zeolites per sample compared to ZIFs, so that the lengths of water intrusion and extrusion plateaus can be made comparable to those of the ZIFs and corresponding pressures can be better recognized. One may have the concern whether the mass of zeolites available for water intrusion will affect the observed rate-dependence, especially under high-rate loading conditions. Therefore, we carried out a group of experiments comparing samples with different masses of zeolites. Supplementary Figure 92 provides their high-rate experimental results at the strain rate of ~2,000 s$^{-1}$, which shows that the mass of zeolites has no significant influence on the intrusion pressure and extrusion performance. This should be attributed to the overall weak rate-dependence of water intrusion and extrusion inside these channel-containing zeolite structures.

As expected, the length of the intrusion plateaus indicates the available pore volume of a sample, which is proportional to the amount of zeolites. Note that the sample length is kept at 3 mm in these experiments despite the different amount of zeolites inside the samples.

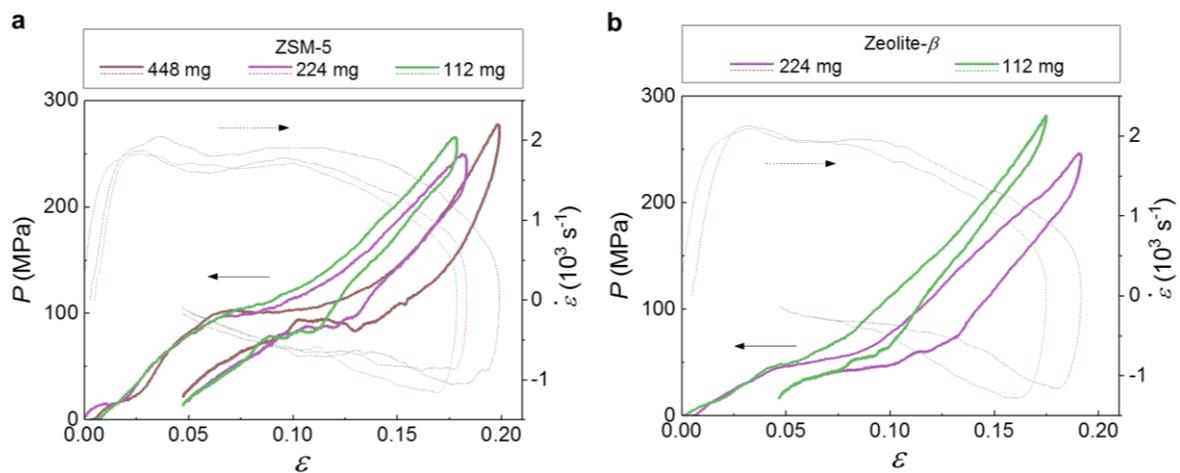

**Supplementary Figure 92 | Influence of the zeolite mass on the high-rate intrusion and extrusion behaviour of water into the zeolite. a,** High-rate water intrusion and extrusion in 112 mg, 224 mg, and 336 mg of ZSM-5. **b,** High-rate water intrusion and extrusion in 112 mg and 224 mg of zeolite-$\beta$. These are the amount of materials to make a sample of Ø12.7 mm in diameter used in the SHPB experiments. For ZIFs, 112 mg is used for each sample.

## S11.6 Effect of heat treatment of zeolites

In this work, zeolites were heated at 1,000 °C for 3 h before use. Heat treatment increases the Si/Al ratio and hydrophobicity through dealumination of the zeolites. This method has been used and reported before on ZSM-5 and zeolite-$\beta$,[96,97] but the water intrusion behaviour of mordenite is presented here for the first time. Supplementary Figure 93 presents the water intrusion and extrusion curves of the three zeolites heated at different temperatures. It is shown that the intrusion pressure increases with the heating temperature, and more nanopore volume becomes accessible after the treatment.

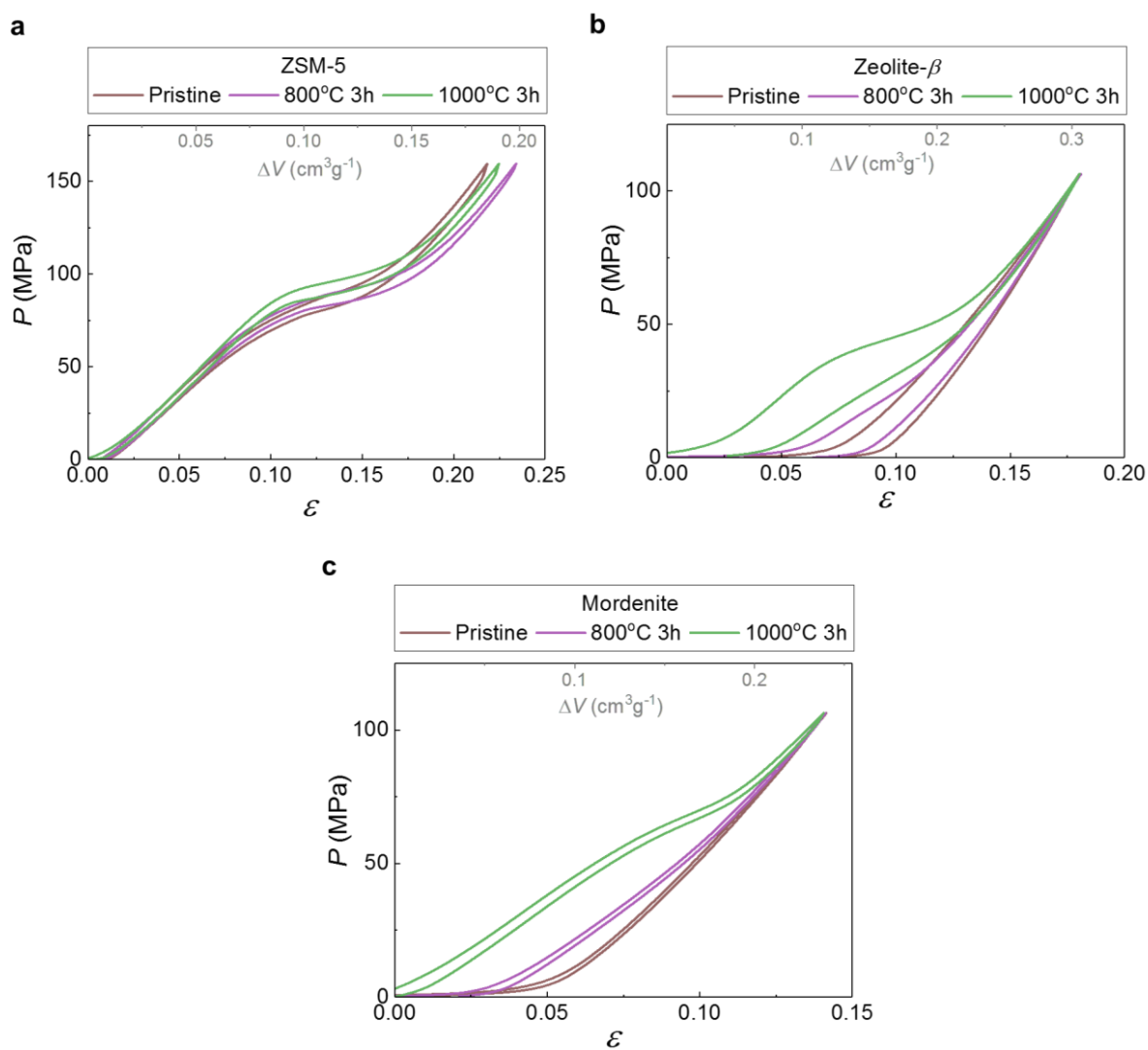

**Supplementary Figure 93 | Water intrusion and extrusion of zeolites heated at different temperatures under quasi-static loading conditions ($3\times10^{-3}$ s$^{-1}$). a**, ZSM-5, **b**, zeolite-$\beta$, and **c**, mordenite.

# Supplementary references

1. Lu, T., Yan, W. & Xu, R. Chiral zeolite beta: structure, synthesis, and application. *Inorg. Chem. Front.* **6**, 1938-1951 (2019).
2. Lively, R. P. *et al.* Ethanol and water adsorption in methanol-derived ZIF-71. *Chem. Commun.* **47**, 8667-8669 (2011).
3. Li, Y. S. *et al.* Molecular sieve membrane: supported metal–organic framework with high hydrogen selectivity. *Angew. Chem. Int. Ed.* **49**, 548-551 (2010).
4. Sun, Y., Li, Y. & Tan, J. C. Liquid intrusion into zeolitic imidazolate framework-7 nanocrystals: Exposing the roles of phase transition and gate opening to enable energy absorption applications. *ACS Appl. Mater. Interfaces* **10**, 41831-41838 (2018).
5. He, M., Yao, J., Liu, Q., Zhong, Z. & Wang, H. Toluene-assisted synthesis of RHO-type zeolitic imidazolate frameworks: synthesis and formation mechanism of ZIF-11 and ZIF-12. *Dalton Trans.* **42**, 16608-16613 (2013).
6. Khay, I. *et al.* Assessment of the energetic performances of various ZIFs with SOD or RHO topology using high pressure water intrusion-extrusion experiments. *Dalton Trans.* **45**, 4392-4400 (2016).
7. Rouquerol, J. *et al.* Liquid intrusion and alternative methods for the characterization of macroporous materials (IUPAC Technical Report). *Pure Appl. Chem.* **84**, 107-136 (2011).
8. Fadeev, A. Y. & Eroshenko, V. A. Study of Penetration of Water into Hydrophobized Porous Silicas. *J. Colloid Interface Sci.* **187**, 275-282 (1997).
9. Grosu, Y., Renaudin, G., Eroshenko, V., Nedelec, J. M. & Grolier, J. P. E. Synergetic effect of temperature and pressure on energetic and structural characteristics of {ZIF-8 + water} molecular spring. *Nanoscale* **7**, 8803-8810 (2015).
10. Park, K. S. *et al.* Exceptional chemical and thermal stability of zeolitic imidazolate frameworks. *Proc. Natl. Acad. Sci. USA* **103**, 10186 (2006).
11. Zhang, H., Liu, D., Yao, Y., Zhang, B. & Lin, Y. S. Stability of ZIF-8 membranes and crystalline powders in water at room temperature. *J. Membr. Sci.* **485**, 103-111 (2015).
12. Zhang, H., Zhao, M. & Lin, Y. S. Stability of ZIF-8 in water under ambient conditions. *Microporous Mesoporous Mat.* **279**, 201-210 (2019).
13. Zhang, H., Zhao, M., Yang, Y. & Lin, Y. S. Hydrolysis and condensation of ZIF-8 in water. *Microporous Mesoporous Mat.* **288**, 109568 (2019).
14. Cychosz, K. A. & Matzger, A. J. Water stability of microporous coordination polymers and the adsorption of pharmaceuticals from water. *Langmuir : the ACS journal of surfaces and colloids* **26**, 17198-17202 (2010).
15. Liu, X. *et al.* Improvement of hydrothermal stability of zeolitic imidazolate frameworks. *Chem. Commun.* **49**, 9140-9142 (2013).
16. Zhang, H. *et al.* Improving hydrostability of ZIF-8 membranes via surface ligand exchange. *J. Membr. Sci.* **532**, 1-8 (2017).
17. Tanaka, S. & Tanaka, Y. A Simple Step toward Enhancing Hydrothermal Stability of ZIF-8. *ACS Omega* **4**, 19905-19912 (2019).
18. Trzpit, M., Soulard, M. & Patarin, J. The pure silica Chabazite: A high volume molecular spring at low pressure for energy storage. *Chem. Lett.* **36**, 980-981 (2007).
19. Tzanis, L., Trzpit, M., Soulard, M. & Patarin, J. Energetic Performances of Channel and Cage-Type Zeosils. *J. Phys. Chem. C* **116**, 20389-20395 (2012).
20. Khay, I., Chaplais, G., Nouali, H., Marichal, C. & Patarin, J. Water intrusion–extrusion experiments in ZIF-8: impacts of the shape and particle size on the energetic performances. *RSC Adv.* **5**, 31514-31518 (2015).
21. Cai, W. *et al.* Thermal Structural Transitions and Carbon Dioxide Adsorption Properties of Zeolitic Imidazolate Framework-7 (ZIF-7). *J. Am. Chem. Soc.* **136**, 7961-7971 (2014).
22. Tiba, A. A., Tivanski, A. V. & MacGillivray, L. R. Size-Dependent Mechanical Properties of a Metal–Organic Framework: Increase in Flexibility of ZIF-8 by Crystal Downsizing. *Nano Lett.* **19**, 6140-6143 (2019).

23. Grosu, Y. *et al.* A highly stable nonhysteretic {Cu2 (tebpz) MOF+water} molecular spring. *ChemPhysChem* **17**, 3359-3364 (2016).
24. Ryzhikov, A. *et al.* High-Pressure Intrusion–Extrusion of Water and Electrolyte Solutions in Pure-Silica LTA Zeolite. *J. Phys. Chem. C* **119**, 28319-28325 (2015).
25. Ronchi, L., Nouali, H., Daou, T. J., Patarin, J. & Ryzhikov, A. Heterogeneous lyophobic systems based on pure silica ITH-type zeolites: high pressure intrusion of water and electrolyte solutions. *New J. Chem.* **41**, 15087-15093 (2017).
26. Saada, M. A. *et al.* Investigation of the Energetic Performance of Pure Silica ITQ-4 (IFR) Zeolite under High Pressure Water Intrusion. *J. Phys. Chem. C* **114**, 11650-11658 (2010).
27. Surani, F. B., Kong, X., Panchal, D. B. & Qiao, Y. Energy absorption of a nanoporous system subjected to dynamic loadings. *Appl. Phys. Lett.* **87**, 163111 (2005).
28. Beurroies, I. *et al.* Using Pressure to Provoke the Structural Transition of Metal–Organic Frameworks. *Angew. Chem. Int. Ed.* **49**, 7526-7529 (2010).
29. Zhou, X., Miao, Y.-R., Shaw, W. L., Suslick, K. S. & Dlott, D. D. Shock Wave Energy Absorption in Metal–Organic Framework. *J. Am. Chem. Soc.* **141**, 2220-2223 (2019).
30. Yot, P. G. *et al.* Mechanical energy storage performance of an aluminum fumarate metal–organic framework. *Chemical Science* **7**, 446-450 (2016).
31. Lu, G. & Yu, T. *Energy Absorption of Structures and Materials*. (Woodhead Publishing Limited, Cambridge, 2003).
32. Gibson, L. J. & Ashby, M. F. *Cellular Solids: Structure and Properties*. 2nd edn, (Cambridge University Press, Cambridge, 1997).
33. Gokulakrishnan, N. *et al.* Improved hydrophobicity of inorganic–organic hybrid mesoporous silica with cage-like pores. *Colloids Surf. Physicochem. Eng. Aspects* **421**, 34-43 (2013).
34. Martin, T. *et al.* Dissipative water intrusion in hydrophobic MCM-41 type materials. *Chem. Commun.*, 24-25 (2002).
35. Punyamurtula, V. K., Han, A. & Qiao, Y. Pressurized Flow in a Mesostructured Silica Modified by Silane Groups. *J. Fluids Eng.* **131** (2009).
36. Kong, X. & Qiao, Y. Improvement of recoverability of a nanoporous energy absorption system by using chemical admixture. *Appl. Phys. Lett.* **86**, 151919 (2005).
37. Sun, Y. *et al.* Crushing of circular steel tubes filled with nanoporous-materials-functionalized liquid. *International Journal of Damage Mechanics* **27**, 439-450 (2018).
38. Kresse, G. & Furthmüller, J. Efficient iterative schemes for ab initio total-energy calculations using a plane-wave basis set. *Phys. Rev. B* **54**, 11169-11186 (1996).
39. Vanpoucke, D. E. P., Lejaeghere, K., Van Speybroeck, V., Waroquier, M. & Ghysels, A. Mechanical properties from periodic plane wave quantum mechanical codes: The challenge of the flexible nanoporous MIL-47(V) framework. *J. Phys. Chem. C* **119**, 23752-23766 (2015).
40. Vinet, P., Ferrante, J., Rose, J. H. & Smith, J. R. Compressibility of solids. *J. Geophys. Res.* **92** (1987).
41. Vanduyfhuys, L. *et al.* QuickFF: A program for a quick and easy derivation of force fields for metal-organic frameworks from ab initio input. *J. Comput. Chem.* **36**, 1015-1027 (2015).
42. Vanduyfhuys, L. *et al.* Extension of the QuickFF force field protocol for an improved accuracy of structural, vibrational, mechanical and thermal properties of metal-organic frameworks. *J. Comput. Chem.* **39**, 999-1011 (2018).
43. Chen, J. & Martínez, T. J. QTPIE: Charge transfer with polarization current equalization. A fluctuating charge model with correct asymptotics. *Chem. Phys. Lett.* **438**, 315-320 (2007).
44. Verstraelen, T. *et al.* Minimal basis iterative stockholder: Atoms in molecules for force-field development. *J. Chem. Theory Comput.* **12**, 3894-3912 (2016).
45. Mayo, S. L., Olafson, B. D. & Goddard, W. A. DREIDING: a generic force field for molecular simulations. *J. Phys. Chem.* **94**, 8897-8909 (1990).
46. Avery, P. genXRDPattern, <https://github.com/psavery/genXrdPattern> (2018).
47. Moggach, S. A., Bennett, T. D. & Cheetham, A. K. The effect of pressure on ZIF-8: increasing pore size with pressure and the formation of a high-pressure phase at 1.47 GPa. *Angew. Chem. Int. Ed.* **48**, 7087-7089 (2009).
48. Yaff, yet another force field. Available online at http://molmod.ugent.be/software/.

49. Plimpton, S. Fast parallel algorithms for short-range molecular dynamics. *J. Comput. Phys.* **117**, 1-19 (1995).
50. Sear, R. P. The non-classical nucleation of crystals: microscopic mechanisms and applications to molecular crystals, ice and calcium carbonate. *Int. Mater. Rev.* **57**, 328-356 (2012).
51. Kalikmanov, V. I. in *Nucleation Theory* (ed V. I. Kalikmanov) 17-41 (Springer Netherlands, 2013).
52. Marbach, S., Dean, D. S. & Bocquet, L. Transport and dispersion across wiggling nanopores. *Nature Physics* **14**, 1108-1113 (2018).
53. Coudert, F.-X. Water Adsorption in Soft and Heterogeneous Nanopores. *Acc. Chem. Res.* **53**, 1342-1350 (2020).
54. Faucher, S. *et al.* Critical Knowledge Gaps in Mass Transport through Single-Digit Nanopores: A Review and Perspective. *J. Phys. Chem. C* **123**, 21309-21326 (2019).
55. Rieth, A. J., Hunter, K. M., Dincă, M. & Paesani, F. Hydrogen bonding structure of confined water templated by a metal-organic framework with open metal sites. *Nat. Commun.* **10**, 4771 (2019).
56. Hiraoka, T. & Shigeto, S. Interactions of water confined in a metal–organic framework as studied by a combined approach of Raman, FTIR, and IR electroabsorption spectroscopies and multivariate curve resolution analysis. *Phys. Chem. Chem. Phys.* **22**, 17798-17806 (2020).
57. Torrie, G. M. & Valleau, J. P. Nonphysical sampling distributions in Monte Carlo free-energy estimation: Umbrella sampling. *J. Comput. Phys.* **23**, 187-199 (1977).
58. Cnudde, P. *et al.* Light Olefin Diffusion during the MTO Process on H-SAPO-34: A Complex Interplay of Molecular Factors. *J. Am. Chem. Soc.* **142**, 6007-6017 (2020).
59. Kumar, S., Rosenberg, J. M., Bouzida, D., Swendsen, R. H. & Kollman, P. A. The weighted histogram analysis method for free‐energy calculations on biomolecules. I. The method. *J. Comput. Chem.* **13**, 1011-1021 (1992).
60. Souaille, M. & Roux, B. Extension to the weighted histogram analysis method: combining umbrella sampling with free energy calculations. *Comput. Phys. Commun.* **135**, 40-57 (2001).
61. Temelso, B., Archer, K. A. & Shields, G. C. Benchmark Structures and Binding Energies of Small Water Clusters with Anharmonicity Corrections. *The Journal of Physical Chemistry A* **115**, 12034-12046 (2011).
62. Phan, A. *et al.* Synthesis, structure, and carbon dioxide capture properties of zeolitic imidazolate frameworks. *Acc. Chem. Res.* **43**, 58-67 (2010).
63. Banerjee, R. *et al.* High-throughput synthesis of zeolitic imidazolate frameworks and application to $CO_2$ capture. *Science* **319**, 939 (2008).
64. Hayashi, H., Côté, A. P., Furukawa, H., O'Keeffe, M. & Yaghi, O. M. Zeolite a imidazolate frameworks. *Nat. Mater.* **6**, 501 (2007).
65. Fu, Y. M. *et al.* A chiral 3D polymer with right- and left-helices based on 2,2'-biimidazole: Synthesis, crystal structure and fluorescent property. *Inorg. Chem. Commun.* **10**, 720 (2007).
66. Rettig, S. J., Sanchez, V., Storr, A., Thomson, R. C. & Trotter, J. Polybis(4-azabenzimidazolato)-iron(II) and -cobalt(II). 3-D single diamond-like framework materials which exhibit spin canting and ferromagnetic ordering at low temperatures. *J. Chem. Soc., Dalton Trans.*, 3931 (2000).
67. Zhang, J. *et al.* Zeolitic boron imidazolate frameworks. *Angew. Chem. Int. Ed.* **48**, 2542 (2009).
68. Banerjee, R. *et al.* Control of pore size and functionality in isoreticular zeolitic imidazolate frameworks and their carbon dioxide selective capture properties. *J. Am. Chem. Soc.* **131**, 3875 (2009).
69. Liu, Y., Kravtsov, V. C. & Eddaoudi, M. Template-directed assembly of zeolite-like metal-organic frameworks (ZMOFs): a usf-ZMOF with an unprecedented zeolite topology. *Angew. Chem. Int. Ed.* **47**, 8446 (2008).
70. Wang, B., Côté, A. P., Furukawa, H., O'Keeffe, M. & Yaghi, O. M. Colossal cages in zeolitic imidazolate frameworks as selective carbon dioxide reservoirs. *Nature* **453**, 207 (2008).
71. Liu, Y., Kravtsov, V. C., Larsen, R. & Eddaoudi, M. Molecular building blocks approach to the assembly of zeolite-like metal-organic frameworks (ZMOFs) with extra-large cavities. *Chem. Commun.*, 1488 (2006).

72. Morris, W., Doonan, C. J., Furukawa, H., Banerjee, R. & Yaghi, O. M. Crystals as molecules: postsynthesis covalent functionalization of zeolitic imidazolate frameworks. *J. Am. Chem. Soc.* **130**, 12626 (2008).
73. Huang, X. C., Lin, Y. Y., Zhang, J. P. & Chen, X. M. Ligand-directed strategy for zeolite-type metal-organic frameworks: zinc(II) imidazolates with unusual zeolitic topologies. *Angew. Chem. Int. Ed.* **45**, 1557-1559 (2006).
74. Tian, Y. Q. *et al.* Two polymorphs of cobalt(II) imidazolate polymers synthesized solvothermally by using one organic template N,N-dimethylacetamide. *Inorg. Chem.* **43**, 4631 (2004).
75. Tian, Y. Q. *et al.* Design and generation of extended zeolitic metal-organic frameworks (ZMOFs): synthesis and crystal structures of zinc(II) imidazolate polymers with zeolitic topologies. *Chem. Eur. J.* **13**, 4146 (2007).
76. Muller-Buschbaum, K. A three-dimensional network with complete nitrogen coordination obtained from an imidazole melt. *Z. Naturforsch., B: Chem. Sci.* **61**, 792 (2006).
77. Lehnert, V. R. & Seel, F. Preparation and Crystal Structure of the Manganese(II) and Zinc(II) Derivative of Imidazole. *Z. Anorg. Allg. Chem.* **464**, 187 (1980).
78. Tian, Y. Q. *et al.* The silica-like extended polymorphism of cobalt(II) imidazolate three-dimensional frameworks: X-ray single-crystal structures and magnetic properties. *Chem. Eur. J.* **9**, 5673 (2003).
79. Sturm, M., Brandl, F., Engel, D. & Hoppe, W. Crystal Structure of Diimidazolylcobalt. *Acta Crystallogr.* **B31**, 2369 (1975).
80. Lorente, M. A. M., Dahan, F., Petrouleas, V., Bousseksou, A. & Tuchagues, J. P. New ferrous complexes based on the 2,2′-biimidazole ligand: Structural, moessbauer, and magnetic properties of [FeII(bimH2)2(CH3OH)2](OAc)2, [FeII(bimH2)3]CO3, [FeII(bimH2)2]n, and {[FeII(bim)]}n. *Inorg. Chem.* **34**, 5346 (1995).
81. Masciocchi, N. *et al.* Synthesis and ab-initio XRPD structure of group 12 imidazolato polymers. *Chem. Commun.*, 2018 (2003).
82. Tian, Y. Q. *et al.* Determination of the solvothermal synthesis mechanism of metal imidazolates by x-ray single-crystal studies of a photoluminescent cadmium(II) imidazolate and its intermediate involving piperrazine. *Eur. J. Inorg. Chem.*, 1039 (2004).
83. Wu, T., Bu, X., Zhang, J. & Feng, P. New zeolitic imidazolate frameworks: from unprecedented assembly of cubic clusters to ordered cooperative organization of complementary ligands. *Chem. Mater.* **20**, 7377 (2008).
84. Han, J. Y., Fang, J., Chang, H. Y., Dong, Y. & Liang, S. Poly[μ2-4,4′-bipyridine-di-μ2-imidazolido-cadmium(II)]. *Acta Crystallogr.* **E61**, m2667−m2669 (2005).
85. Masciocchi, N. *et al.* Extended polymorphism in copper(II) imidazolate polymers: a spectroscopic and XRPD structural study. *Inorg. Chem.* **40**, 5897 (2001).
86. Lehnert, V. R. & Steel, F. Crystal Structure of the Iron(II) Derivative of Imidazole. *Z. Anorg. Allg. Chem.* **444**, 91 (1978).
87. Rettig, S. J., Storr, A., Summers, D. A., Thompson, R. C. & Trotter, J. Transition metal azolates from metallocenes .2. Synthesis, X-ray structure, and magnetic properties of a three-dimensional polymetallic Iron(II) imidazolate complex, a low-temperature weak ferromagnet. *J. Am. Chem. Soc.* **119**, 8675 (1997).
88. Spek, A. L. & Duisenberg, A. J. M. The structure of the three-dimensional polymer poly[μ-Hexakis(2-Methylimidazolato-N, N')-Triiron(II)], [Fe3(C4H5N2)6]n. *Acta Crystallogr.* **C39**, 1212 (1983).
89. Zhang, J. P. & Chen, X. M. Crystal engineering of binary metal imidazolate and triazolate frameworks. *Chem. Commun.*, 1689 (2006).
90. Huang, X., Zhang, J. & Chen, X. [Zn(bim)2]·(H2O)1.67: A metal-organic open-framework with sodalite topology. *Chin. Sci. Bull.* **48**, 1531 (2003).
91. Wu, T. *et al.* A new zeolitic topology with sixteen-membered ring and multidimensional large pore channels. *Chem. Eur. J.* **14**, 7771 (2008).
92. Huang, X. C., Zhang, J. P., Lin, Y. Y., Yu, X. L. & Chen, X. M. Two mixed-valence copper(I,II) imidazolate coordination polymers: metal-valence tuning approach for new topological structures. *Chem. Commun.*, 1100 (2004).

93. Rettig, S. J., Storr, A., Summers, D. A., Thompson, R. C. & Trotter, J. Iron(II) 2-methylimidazolate and copper(II) 1,2,4-triazolate complexes: Systems exhibiting long-range ferromagnetic ordering at low temperatures. *Can. J. Chem.* **77**, 425 (1999).
94. Tian, Y. Q. *et al.* [Co5(im)10·2MB] ∞ : A metal-organic open-framework with zeolite-like topology. *Angew. Chem. Int. Ed.* **41**, 1384 (2002).
95. Wu, H., Zhou, W. & Yildirim, T. Hydrogen storage in a prototypical zeolitic imidazolate framework-8. *J. Am. Chem. Soc.* **129**, 5314 (2007).
96. Sun, Y. *et al.* A candidate of mechanical energy mitigation system: Dynamic and quasi-static behaviors and mechanisms of zeolite β/water system. *Mater. Des.* **66**, 545-551 (2015).
97. Sun, Y. *et al.* Experimental study on energy dissipation characteristics of ZSM-5 zeolite/water system. *Adv. Eng. Mater.* **15**, 740-746 (2013).